\documentclass[aps, 11pt, prd, a4paper, superscriptaddress, onecolumn, amsmath, preprintnumbers, nofootinbib, floatfix]{revtex4-2}

\pdfoutput=1
\usepackage[utf8]{inputenc}
\usepackage{graphicx,subfigure,bm,color,psfrag}
\usepackage{amsfonts}
\usepackage{mathtools}
\usepackage{amsmath}
\usepackage{times}
\usepackage{xspace}
\usepackage{enumerate}
\usepackage{verbatim}
\usepackage[arrowdel]{physics}
\usepackage{accents}
\usepackage{xspace}
\usepackage[dvipsnames]{xcolor}
\usepackage[pscoord]{eso-pic}
\usepackage[normalem]{ulem}
\usepackage[percent]{overpic}
\usepackage{slashed}
\usepackage{wrapfig}
\usepackage{tabu}
\usepackage{diagbox}
\usepackage{amssymb,amsthm,tikz,hyperref}
\usepackage{dsfont,epiolmec, latexsym, stmaryrd, comment}
\usepackage{slashed,ccaption}
\usepackage{mathrsfs, calligra}
\usepackage{leftidx}
\usepackage{import}
\usepackage{multirow}
\usepackage{pifont}
\usepackage{tabularx}
\usepackage{journal_macros}
\usepackage{cancel}

\hypersetup{ linktoc=all,
    colorlinks, linkcolor={palatinateblue},
    citecolor={red}, urlcolor={vividviolet}
}

\definecolor{red}{RGB}{255,0,0}
\definecolor{vividviolet}{rgb}{0.62, 0.0, 1.0}
\definecolor{palatinateblue}{rgb}{0.15, 0.23, 0.89}

\definecolor{green}{RGB}{11,98,17}
\definecolor{darkpink}{RGB}{153,0,76}
\definecolor{bluegreen}{RGB}{0,102,102}
\definecolor{greenlagan}{RGB}{0,102,0}
\definecolor{redgreen}{RGB}{102,102,0}
\definecolor{Redgreen}{RGB}{153,76,0}
\definecolor{amaranth}{rgb}{0.9, 0.17, 0.31}
\definecolor{brightpink}{rgb}{1.0, 0.0, 0.5}
\definecolor{cornflowerblue}{rgb}{0.39, 0.58, 0.93}
\definecolor{deepcarminepink}{rgb}{0.94, 0.19, 0.22}
\definecolor{radicalred}{rgb}{1.0, 0.21, 0.37}
\definecolor{grinchgreen}{RGB}{27,100,45}

\newcommand{\bcomm}[1]{#1}

\makeatletter \@addtoreset{equation}{section}

\newcommand{\Sspace}     {{{\mathbb S}}} 
\newcommand{\relBetti}[1]{{{b}_{#1}}}
\newcommand{\Rspace}     {{{\mathbb R}}} 

\newcommand{\Excursion}  {{{\mathbb E}}}

\def\be{\begin{equation}}
\def\ee{\end{equation}}
\def\beq{\begin{equation}}
\def\eeq{\end{equation}}
\def\bea{\begin{eqnarray}}
\def\eea{\end{eqnarray}}

\def\ave#1{\langle{#1}\rangle}

\begin{document}
\date{\today}
\vspace*{5mm}
\title{\vspace{-6ex}Is the Observable Universe Consistent with the Cosmological Principle?}

\let\mymaketitle\maketitle
\let\myauthor\author
\let\myaffiliation\affiliation

\author{Pavan Kumar Aluri}
\affiliation{Department of Physics, Indian Institute of Technology (BHU), Varanasi - 221005, India }

\author{Paolo Cea} 
\affiliation{INFN - Sezione di Bari, Via Amendola 173 - 70126 Bari, Italy} 

\author{Pravabati Chingangbam}
\affiliation{Indian Institute of Astrophysics, Koramangala II Block, Bangalore 560 034, India}
\affiliation{School of Physics, Korea Institute for Advanced Study, 85 Hoegiro, Dongdaemun-gu, Seoul, 02455, Korea}

\author{Ming-Chung Chu}
\affiliation{Department of Physics and Institute of Theoretical Physics, The Chinese University of Hong Kong\\Shatin, Hong Kong}

\author{Roger G. Clowes}
\affiliation{Jeremiah Horrocks Institute, University of Central Lancashire, Preston PR1 2HE, UK} 


\author{Damien Hutsem{\'e}kers}
\affiliation{F.R.S.-FNRS, Institut d'Astrophysique et de G{\'e}ophysique (B5c), University of Li{\`ege}, 4000 Li{\`ege} Belgium}

\author{Joby P. Kochappan}
\affiliation{Asia Pacific Center for Theoretical Physics, Pohang, 37673, Korea}


\author{Alexia M. Lopez} 
\affiliation{Jeremiah Horrocks Institute, University of Central Lancashire, Preston PR1 2HE, UK}

\author{Lang Liu}
\affiliation{Department of Physics, Kunsan National University, Kunsan 54150, Korea}

\author{Niels C. M. Martens}
\affiliation{Lichtenberg Group for History and Philosophy of Physics, University of Bonn, Germany}
\affiliation{Institute for Theoretical Particle Physics and Cosmology, RWTH Aachen University, Germany}

\author{C.~J.~A.~P.~Martins}
\address{Centro de Astrof\'{\i}sica da Universidade do Porto, Rua das Estrelas, 4150-762 Porto, Portugal}
\address{Instituto de Astrof\'{\i}sica e Ci\^encias do Espa\c co, CAUP, Rua das Estrelas, 4150-762 Porto, Portugal}

\author{Konstantinos Migkas}
\affiliation{Argelander-Institut f\"{u}r Astronomie, Universit\"{a}t Bonn, Auf dem H\"{u}gel 71, 53121 Bonn, Germany}

\author{Eoin \'O Colg\'ain}
\affiliation{Atlantic Technological University, Ash Lane, Sligo, Ireland}
\affiliation{CQUeST \& Department of Physics, Sogang University, Seoul 121-742, Korea}

\author{Pratyush Pranav}
\affiliation{Univ Lyon, ENS de Lyon, Univ Lyon1, CNRS, Centre de Recherche Astrophysique de Lyon UMR5574, F–69007, Lyon, France}

\author{Lior Shamir}
\affiliation{Kansas State University, Manhattan, KS 66506, United States }

\author{Ashok K. Singal}
\affiliation{Astronomy and Astrophysics Division, Physical Research Laboratory, Navrangpura, Ahmedabad - 380 009, India }

\author{M. M. Sheikh-Jabbari}
\affiliation{School of Physics, Institute for Research in Fundamental Sciences (IPM),\\ P.O.Box 19395-5531, Tehran, Iran}
\affiliation{The Abdus Salam ICTP, Strada Costiera 11, I-34014, Trieste, Italy}



\author{Jenny Wagner}
\affiliation{Bahamas Advanced Study Institute and Conferences, 4A Ocean Heights, Hill View Circle, Stella Maris, Long Island, The Bahamas}

\author{Shao-Jiang Wang}
\affiliation{CAS Key Laboratory of Theoretical Physics, Institute of Theoretical Physics, Chinese Academy of Sciences, Beijing 100190, China}

\author{David L. Wiltshire}
\affiliation{School of Physical \& Chemical Sciences, University of Canterbury, Private Bag 4800, Christchurch 8140, New Zealand}

\author{Shek Yeung}
\affiliation{Department of Physics and Institute of Theoretical Physics, The Chinese University of Hong Kong\\Shatin, Hong Kong}

\author{Lu Yin}
\affiliation{CQUeST \& Department of Physics, Sogang University, Seoul 121-742, Korea}

\author{Wen Zhao}
\affiliation{CAS Key Laboratory for Researches in Galaxies and Cosmology, Department of Astronomy, University of Science and Technology of China, Chinese Academy of Sciences, Hefei, Anhui 230026, China \\ School of Astronomy and Space Science, University of Science and Technology of China, Hefei, 230026, China}
\begin{abstract}
\centerline{\textbf{Abstract}}
\noindent
The Cosmological Principle (CP) -- the notion that the Universe is spatially isotropic and homogeneous on large scales -- underlies a century of progress in cosmology. It is \bcomm{conventionally} formulated through the Friedmann-Lema\^{\i}tre-Robertson-Walker (FLRW) cosmologies as the spacetime metric, and culminates in the successful and highly predictive $\Lambda$-Cold-Dark-Matter ($\Lambda$CDM) model. Yet, tensions have emerged within the $\Lambda$CDM model, most notably a statistically significant discrepancy in the value of the Hubble constant, $H_0$. Since the notion of cosmic expansion determined by a single parameter is intimately tied to the CP, implications of the $H_0$ tension may extend beyond $\Lambda$CDM to the CP itself. This review surveys current observational hints for deviations from the expectations of the CP, highlighting synergies and disagreements that warrant further study. Setting aside the debate about individual large structures, potential deviations from the CP include variations of cosmological parameters on the sky, discrepancies in the cosmic dipoles, and mysterious alignments in quasar polarizations and galaxy spins. While it is possible that a host of observational systematics are impacting results, it is equally plausible that precision cosmology may have outgrown the FLRW paradigm, an extremely pragmatic but non-fundamental symmetry assumption. 
\end{abstract}
\footnotetext[15]{Corresponding author: Eoin \'O Colg\'ain, email: \href{eoin.ocolgain@atu.ie}{eoin.ocolgain@atu.ie}}
\maketitle

\tableofcontents


\section{Prologue} 
\label{sec:prologue}
The Cosmological Principle (CP) is a working assumption in modern cosmology that can be simply stated as \textit{the Universe is (statistically) isotropic and homogeneous at suitably large scales}. This statement is admittedly vague, but nevertheless intuitive, and most importantly, practically very useful. Simply put, there exists a length scale beyond the reach of the rich structures that we observe in the local Universe, namely stars, galaxies and galaxy clusters, where the Universe should look the same in all directions. In other words, the Universe is \textit{isotropic}. Moreover, this statement must hold true for all observers; if observer A erects a telescope somewhere else in the cosmos, she is expected to recover a universe that looks the same as observer B's universe at an expected scale. Given enough observers seeing isotropic universes, it is once again intuitive that this guarantees there is no special place in the Universe, or alternatively that the Universe is \textit{homogeneous} \cite{bib:Hawking1973, bib:Ellis1974}. 

In using words such as ``large scales'', ``isotropy'' and ``homogeneity'', any statement of the CP {\em must} assume a notion of spacetime where on its \textit{constant time slices} one defines ``directions'' and ``positions''. In modern cosmology,  the CP is formulated as follows:  At suitably large scales the average evolution of the Universe is {\em exactly} governed by the Friedmann-Lema\^{\i}tre-Robertson-Walker (FLRW) metric, 
\begin{equation}
    \label{eq:flrw_metric}
    \textrm{d} s^2 = - c^2 \textrm{d} t^2 + a(t)^2 \textrm{d} \Sigma ^2 \;, 
\end{equation}
where $\bcomm{\textrm{d}}\Sigma\bcomm{^2}$ \bcomm{represents} \bcomm{the 3-metric of} a maximally symmetric space, independent of time and all time dependence resides in the scale factor $a(t)$. Recall that the FLRW family \eqref{eq:flrw_metric} covers all possible spacetime geometries with explicit time-dependence (hence useful as cosmologies) with at least 6 isometries, or 6 Killing vectors, see e.~g.~\cite{bib:Weinberg1972}. Of course the maximally symmetric geometries with 10 isometries, namely flat Minkowski space, de Sitter and anti-de Sitter spacetimes are also special members of the FLRW family. Explicitly, in this review we \bcomm{assume a global\footnote{\bcomm{Since we are concerned with observations, our sense of global is restricted to the observable universe. We understand the CP to assume the simplest spatial topology at the scale of the particle horizon without restrictions beyond that scale.}} average FLRW expansion history} as synonymous to the formulation of the CP. This assumption is separate from the question of whether the field equations to be solved are obtained from the Einstein-Hilbert theory minimally coupled to matter, or another diffeomorphism-invariant action. It underlies whole fields of theoretical cosmology which treat inhomogeneities perturbatively and which require a spacetime background about which to perturb. Despite its central importance, the question of how the ``suitably large scales'' of the CP are to be rigorously understood is never precisely addressed. Furthermore, the more detailed our observations of the cosmic web of large-scale structures become, the harder it is becoming to answer this question.

The goal of this review is to better come to grips with FLRW cosmologies by highlighting the key \textit{observational} implications of the underlying assumption and evaluating whether or not they are borne out in current observations. A forthcoming sister paper \cite{bib:Wiltshire2022} will address theoretical considerations, \bcomm{particularly the interdependence of the CP with additional assumptions like specific choices of FLRW metrics.
It will also show that the CP is vague in its formulation and there are several physically meaningful ways to interpret and apply it to data, one of them being a relaxation of exact homogeneity and isotropy beyond a certain distance scale to \textit{statistical} isotropy around our observational point and assuming that other observers at different locations would obtain the same results. 
Although discussions of the CP go back several decades \cite{bib:Milne1932,bib:Bondi1952}, there is no consensus about its definition.
In his 1979 book \textit{Theoretical Cosmology}, Raychaudhuri \cite{bib:Raychaudhuri1979} identifies four approaches to the CP: 
(i) philosophical; (ii) mathematical; (iii) deductive; and (iv) empirical. 
The first, philosophical, is said to have emotional appeal. 
The second, mathematical, is said to have the appeal of mathematical beauty (symmetry). 
The protagonists of the third, deductive, hope to deduce the CP as arising inevitably from arbitrary initial conditions (inflation?). 
And those affiliating themselves to the fourth, empirical, hope to maintain an open mind and be guided by the observational evidence. 
Even in 1979 the evidence from the CMB was strong, and most people today would regard it as so compelling, in spite of observations gathered in this review, as to be beyond reasonable doubt.}
See \cite{bib:Schwarz2010, bib:Maartens2011, bib:Clarkson2012, bib:Buchert2016} for \bcomm{more recent} appraisals of the FLRW assumption.  
\bcomm{Based on these aspects of the CP-definition, a more detailed study on the impact of different CP-versions and additional prior assumptions on observational tests is warranted, but beyond the scope of this work. In short, our scope here is to largely comment on tests of the CP that are currently feasible given existing data quality.}

Since the CP or FLRW \bcomm{assumption} may be a divisive topic, potentially a topic rooted in belief structures, let us begin with some comments that we can all agree upon. Modern cosmology has its origins in solutions to the Einstein field equations, which are horrendously difficult to analytically solve without symmetries.\footnote{More specifically one makes simplifying assumptions about the form of the metric. That is still the case for the Szekeres dust cosmologies which have no symmetries; i.~e., no Killing vectors.} Seen in this historic context, some assumption, FLRW or equivalent, is required even without observational data to analytically solve the field equations \cite{bib:Cotsakis2022}. Given that maximally symmetric spacetimes with 10 isometries do not constitute realistic cosmologies, the next most-symmetric and cosmologically relevant family is FLRW. Thus, the simplest assumption is to take the Universe to be described by an FLRW-type metric,  which,  based on the first cosmological models of the 20th century, is filled by (pressureless) matter, radiation and the cosmological constant as constituents. \bcomm{At certain stages, for example, observations in sections \ref{sec:Hom-scale},  \ref{sec:CMB_variations},   \ref{sec:galaxy_cluster_anisotropies} and \ref{sec:emergent_H0}, this will entail specialising to the spatially flat $\Lambda$-Cold-Dark-Matter ($\Lambda$CDM) minimal model, but staying within the FLRW assumption. In contrast to that, redshift-dependent anomalies can in principle be addressed by deviating from the minimal model and directional anomalies can only be addressed by inflating errors to \textit{absorb} anomalies. Needless to say, this runs contrary to any research program attempting to constrain cosmological parameters to \%-level accuracy, i.e. precision cosmology. Starting from these working assumptions,} one can now steadily add observations and it is plausible that observations eventually take one beyond the FLRW family, since no fundamental principle guarantees that the Universe is FLRW \textit{a priori}.\footnote{However, note that as a general result, one expects to find a close to homogeneous and isotropic Universe, \textit{if} the early Universe underwent cosmic inflation. There are various theoretical arguments, most notably Wald's cosmic no hair theorem \cite{bib:Wald1983} and its extension to include inflationary models \cite{bib:Maleknejad2012}, which argue how homogeneity  (generic Bianchi models) yield isotropy as a result of a fast, almost exponential expansion. Of course, inflation is not a fully established part of cosmology and these arguments leave the possibility of small anisotropies open. Such primordial (small) anisotropies can in principle grow to sizable anisotropies during the course of cosmic evolution after inflation.} 
It is possible that the Universe is consistent with the FLRW paradigm to a high precision.
As the FLRW cosmologies are very special relative to their non-FLRW brethren, representing a subset of measure zero, such a finding calls for a physical explanation, in particular, \bcomm{because in physics symmetries -- especially global symmetries --} are generically broken at some scale. Putting naturalness aside, objectively one can only venture beyond FLRW through i) data of sufficient quality and ii) cosmological probes that are sensitive to deviations from FLRW. Conversely, if the goal is to recover FLRW across diverse scales, it suffices to downgrade cosmological probes to the point that they are insensitive to FLRW deviations. However, if one follows the natural process, one will first observe FLRW deviations or anomalies in specific probes, before data quality improves and eventually all observables show the same signal. We are open to statistical fluctuations being the explanation, but this can only be assessed as data improves. Until we reach this stage with future sky surveys, there is value in documenting the currently observed anomalies. \bcomm{As explained earlier, some of these anomalies are strictly speaking $\Lambda$CDM anomalies, but being directional in nature, they cannot be addressed without inflating errors.}

The traditional tack in ascertaining whether the CP holds in the observable Universe involves testing the homogeneity scale. One would expect that this scale should not be smaller than typical distances to our neighboring clusters up to few 100~Mpc. A closer investigation has led to confirmation of a scale in a \bcomm{$60$--$80 \, h^{-1}\,$~Mpc window\footnote{\bcomm{These estimates are for galaxy clustering statistics at various lower redshifts, $z<0.8$ \cite{bib:Hogg2005, bib:Yadav2005, bib:Scrimgeour2012, bib:Ntelis2017, bib:Sarkar2019, bib:Andrade2022}. In $\Lambda$CDM the homogeneity scale decreases with increasing redshift. Equivalent estimates of the homogeneity scale using quasars at redshifts $0.8<z<3.2$ are consistent with the expected decreasing trend, albeit by analyses that incorporate model dependent bias corrections \cite{bib:Laurent2016, bib:Goncalves2018, bib:Goncalves2021}.}} \cite{bib:Hogg2005, bib:Yadav2005, bib:Scrimgeour2012, bib:Ntelis2017, bib:Sarkar2019, bib:Andrade2022}.} Such a low homogeneity scale has been challenged elsewhere \cite{bib:Antal2009, bib:Labini2011, bib:Park2017, bib:DeMarzo2021, bib:Kim2022}.
\bcomm{Observational claims of a low homogeneity scale typically}
involve estimating the fractal dimension, $D_2(r)$, of the two-point galaxy correlation function which should asymptotically approach $D_2\to3$ as homogeneity is approached. As one example, in the {\em WiggleZ} survey \cite{bib:Scrimgeour2012}, $D_2$ crosses homogeneity scale within 1\% of the fractal dimension threshold of $D_2=2.97$  in spheres of comoving radius $r$ that vary from $r=(70\pm4)\,h^{-1}\,$Mpc to $r=(81\pm5)\,h^{-1}\,$Mpc depending on the redshift of the sample. \bcomm{Consequently, $60$--$80 \, h^{-1}$~Mpc should be viewed as a lower bound on the homogeneity scale, which would increase if other definitions of this scale are invoked.}

Theoretically, within the flat  $\Lambda$CDM model, $N$-body simulations have led to an upper bound on the homogeneity scale of $260\,h^{-1}\,$Mpc \cite{bib:Yadav2010}. With $h \sim 0.7$, this scale equates to $\sim 370$~Mpc. This appeared to be challenged by the Sloan Great Wall at $\sim 420$~Mpc \cite{bib:Gott2005}, which understandably led to ensuing speculation and debate \cite{bib:Sheth2011, bib:Park2012}. There are claims of even larger structures \cite{bib:Clowes2013, bib:Balazs2015, bib:Horvath2014, bib:Balazs2015, bib:Horvath2015, bib:Horvath2020, bib:Lopez2022}, but whether they are consistent with the CP is open to interpretation \cite{bib:Nadathur2013, bib:Ukwatta2016, bib:Christian2020, bib:Fujii2022a, bib:Fujii2022b}. More often than not, large structures turn out to be consistent within the working cosmological model of the day \cite{bib:Marinello2016}. Seen in a historic context, the discovery of large structures may raise eyebrows, but ultimately has not succeeded in budging the CP. Note, given the nebulous nature of the opening definition, one can even periodically increase the homogeneity scale through changes in the cosmological model within FLRW, if required. For this reason, the traditional large structure claims, while interesting, are not expected to constitute robust tests of the CP. 

As stated earlier, the CP is both theoretically, as well as observationally, an extremely powerful and pragmatic assumption. At a mundane level, {simplifying assumptions} are required to solve the Einstein equations, and once the FLRW symmetries are imposed and the energy-momentum tensor associated with a homogeneous and isotropic cosmic perfect fluid  specified, the Friedmann equations reduce to a first order ordinary differential equation (ODE) with a unique integration constant, namely the Hubble constant $H_0$. As a result, within the FLRW paradigm, $H_0$ is a universal constant parameter in any FLRW cosmological model \cite{bib:Krishnan2021, bib:Krishnan2022a}. Thus, the constancy of $H_0$ over the sky provides a handle for testing isotropy and hence FLRW cosmology; concretely, one can decompose the sky in hemispheres, or smaller patches, and directly compare $H_0$ inferences in different directions.\footnote{In principle, this test can be performed with any cosmological parameter, but as explained, $H_0$ has a special or universal status. One can also test the constancy of $H_0$ in different redshift ranges, but this constitutes a test within FLRW (see \cite{bib:Krishnan2021, bib:Krishnan2022a}).} Beyond purely theoretical considerations, working in the FLRW framework has the advantage that \bcomm{it} allows one to make the most of sparse data. In principle, one can pick just one spot on the sky, record observables of interest in that direction and make inferences about the rest of the sky. 

A focus on $H_0$ is particularly timely; the discrepancy between local determinations (measurements) based on Cepheids and Type Ia supernovae (SN) \cite{bib:Riess2021} ($H_0 = (73.04 \pm 1.04)$~km/s/Mpc) and Planck's analysis of the Cosmic Microwave Background (CMB) based on the $\Lambda$CDM model \cite{bib:Planck2018b} ($H_0 = (67.36 \pm 0.54)$~km/s/Mpc) has now reached a discrepancy of $5 \, \sigma$. We caution that this anomaly, or the ``Hubble tension" \cite{bib:DiValentino2021, bib:Abdalla2022}, has not appeared overnight and the controversy has spanned a decade \cite{bib:Riess2011, bib:Hinshaw2012}. Furthermore, local $H_0$ measurements are not limited to those coming from SNe calibrated with Cepheids and there are a host of such measurements suggesting $H_0 \gtrsim 70$~km/s/Mpc \cite{bib:Huang2020, bib:deJaeger2020, bib:Pesce2020, bib:Kourkchi2020, bib:Schombert2020, bib:Khetan2021, bib:Blakeslee2021, bib:Freedman2021, bib:Riess2021, bib:Vagnozzi2020}. In other words, the cosmology community has had time to come to terms with the anomaly, in particular, whether to ignore it in the hope that systematics come to the rescue, or embrace the result and look for a resolution within FLRW. What is important is that local measurements, while not cosmological model dependent, are implicitly assuming the CP, since $H_0$ is defined as the leading term in a perturbative expansion in redshift. The cosmological inferences of $H_0$, on the other hand, are always made within a specific model, usually within the flat $\Lambda$CDM setting. 

Despite a zoo of proposed Hubble tension resolutions, one can objectively say that no simple fix works \cite{bib:DiValentino2021, bib:Abdalla2022}, stimulating solutions employing modifications to gravity~\cite{bib:Haslbauer2020}, or discontinuous evolution histories of Cepheid parameters\cite{bib:Mortsell2021, bib:Perivolaropoulos2021a, bib:Odintsov2022}, etc. 
Within this context upper bounds on $H_0$ for any FLRW cosmology may be derived from the age of the Universe and astrophysical objects in it, that leads to $H_0 \lesssim 73$ km/s/Mpc at a $2\sigma$ level \cite{bib:Krishnan2021b, bib:Vagnozzi2021}. Evidently, this is consistent with the Cepheids/SNe determination \cite{bib:Riess2021}, but the key point is that there can be larger local $H_0$ determinations, such as $H_0 \sim 76$ km/s/Mpc \cite{bib:Kourkchi2020}, which exceed this upper bound. If such higher determinations are substantiated and systematics do not come to rescue, even a combination of late and earlier (pre- and post- recombination) modifications to $\Lambda$CDM will not help and one will eventually hit the limits of the FLRW setting.  
It should be stressed here that the Hubble tension is only thought of as a discrepancy between $H_0$ values \bcomm{inferred from FLRW expansion histories with $H(z)$ determined} at different redshifts. Later we will argue that this actually extends to orientations on the sky, \textit{so the Hubble tension may be a three-dimensional problem and not a one-dimensional (redshift) problem}, as is routinely assumed.

The Hubble constant aside, the Cosmic Microwave Background (CMB) provides another excellent way to test the CP. In particular, the CMB has provided the strongest evidence for isotropy and homogeneity of the early (pre-recombination) Universe. Nevertheless, there are some important loose ends that need to be better understood. More concretely, the CMB possesses a number of anomalies with directional dependence. The most interesting from the perspective of FLRW are the hemispherical power asymmetry \cite{bib:Eriksen2003, bib:Park2003, bib:Hansen2004, bib:Eriksen2007, bib:Hansen2009}, the alignment of the quadrupole-octopole \cite{bib:deOliveira2004, bib:Schwarz2004,bib:Ralston2004}, the mirror-parity~\cite{bib:deOliveira2004, bib:Land2005, bib:Finelli2012, bib:Ben-David2012} and the point-parity anomaly \cite{bib:Land2005, bib:Kim2010, bib:Aluri2012}.
What makes some of these anomalies especially interesting is their alignment with the kinematic CMB dipole \cite{bib:Smoot1977}, which represents the largest temperature anisotropy, one that is subtracted on the assumption that it is due to our motion with respect to the CMB. One can attempt to explain these anomalies through non-trivial topology \bcomm{\cite{bib:Aurich2008, bib:Bielewicz2009, bib:Aurich2012, bib:Aurich2014, bib:Bernui2018, bib:Aurich2021, bib:Akrami2022}}, but anisotropic geometry remains a distinct possibility, which has been best explored in the context of homogeneous but anisotropic Bianchi cosmologies \cite{bib:Ellis1969}. Finally, it has recently been observed that the flat $\Lambda$CDM cosmological parameters may vary on the sky, as the CMB is masked \cite{bib:Fosalba2021, bib:Yeung2022}. It is worth noting that these analyses are consistent with, but more significant than, earlier results obtained with Wilkinson Microwave Anisotropy Probe (WMAP) data \cite{bib:Axelsson2013}. Such significant deviations have been questioned in Planck data \cite{bib:Mukherjee2016, bib:Mukherjee2018}, but they are plausible, given the CMB anomalies; as we shall see, the anisotropic patterns may even reflect the anomalies. Throughout, $H_0$ is also varying over the sky, ranging from $H_0 = (61.2 \pm 2.6)$~km/s/Mpc to $H_0 = (75.8 \pm 2.2)$~km/s/Mpc in \cite{bib:Fosalba2021}, whereas \cite{bib:Yeung2022} reports values from $H_0 = (64.4 \pm 1.3)$~km/s/Mpc to $H_0 = (70.1 \pm 1.4)$~km/s/Mpc, but in contrast to  \cite{bib:Fosalba2021}, the latter considers larger regions on the sky. Taken at face value, one may be looking at potential deviations up to $3 \, \sigma$. Since these discrepancies appear from analysis within the flat $\Lambda$CDM context, one can change the cosmological model without leaving FLRW in order to inflate errors and absorb any discrepancy. Note, this is not cost free and reduces the precision of any associated $H_0$ determination. As we shall see later in section \ref{sec:late_universe_FLRW_anomalies}, similar variations in $H_0$ are evident at \bcomm{cosmological scales in higher redshift data sets, where such anomalies are unexpected.}

The claimed variation of $H_0$ on the sky within the CMB \cite{bib:Fosalba2021, bib:Yeung2022}, if confirmed, would represent a watershed to the FLRW setting, and hence to the `standard cosmology'. 
However, it is fitting to highlight the constraints placed on a putative anisotropic universe using both Big Bang Nucleosynthesis (BBN) \cite{bib:Barrow1976} and CMB \cite{bib:Jaffe2005, bib:Jaffe2006, bib:Planck2013b, bib:Planck2016b}. As noted in the WMAP era \cite{bib:Jaffe2005}, CMB anomalies, including low-$\ell$ alignments and power asymmetry, can be explained through the introduction of heat maps from Bianchi VII$_{h}$ cosmologies \cite{bib:Collins1973, bib:Barrow1985}. This class is more predictive than Bianchi I cosmologies, which allows one to trivially overcome a low quadrupole amplitude \cite{bib:Campanelli2006, bib:Rodrigues2008, bib:Koivisto2008}. 
However, both with and without dark energy, the Bianchi VII$_{h}$ cosmology returned value of the FLRW matter density parameter $\Omega_m$ is inconsistent with Planck-$\Lambda$CDM \cite{bib:Jaffe2005, bib:Jaffe2006, bib:Bridges2007, bib:McEwen2013}. Separately, BBN is also expected to impose independent constraints \cite{bib:Barrow1976}. More recently, both physically motivated and phenomenological Bianchi VII$_{h}$ models have been explored by \bcomm{the} Planck \bcomm{Collaboration} \cite{bib:Planck2013b, bib:Planck2016b}. Concretely, when the Bianchi cosmology is fitted independently of the standard model, Planck temperature data favors its inclusion, but once the Bianchi parameters are fitted jointly with the standard cosmological parameters, there is no evidence for a physical Bianchi VII$_h$ cosmology. \bcomm{This analysis tells us that there is no physical Bianchi VII$_h$ Universe that arises as a simple deformation of the $\Lambda$CDM model, i. e. any Bianchi model favored by CMB is disconnected from $\Lambda$CDM.} These conclusions were subsequently strengthened through an analysis of tensor modes (see \cite{bib:Pontzen2007,bib:Pontzen2009}) and Planck polarization data \cite{bib:Saadeh2016}. 
In summary, there is no physical homogeneous but anisotropic Bianchi VII$_h$ cosmology that can replace the flat $\Lambda$CDM model \cite{bib:Saadeh2016}. 
Nevertheless, CMB data does allow an anisotropic component \cite{bib:Jaffe2005, bib:Jaffe2006, bib:Planck2013b, bib:Planck2016b}, which is expected given the hemispherical power asymmetry. Clearly, an anisotropic universe is not excluded, but the leading candidate (Bianchi VII$_h$) appears to be disfavored. 

\begin{figure}[htp]
\centering
\includegraphics[width=0.82\linewidth]{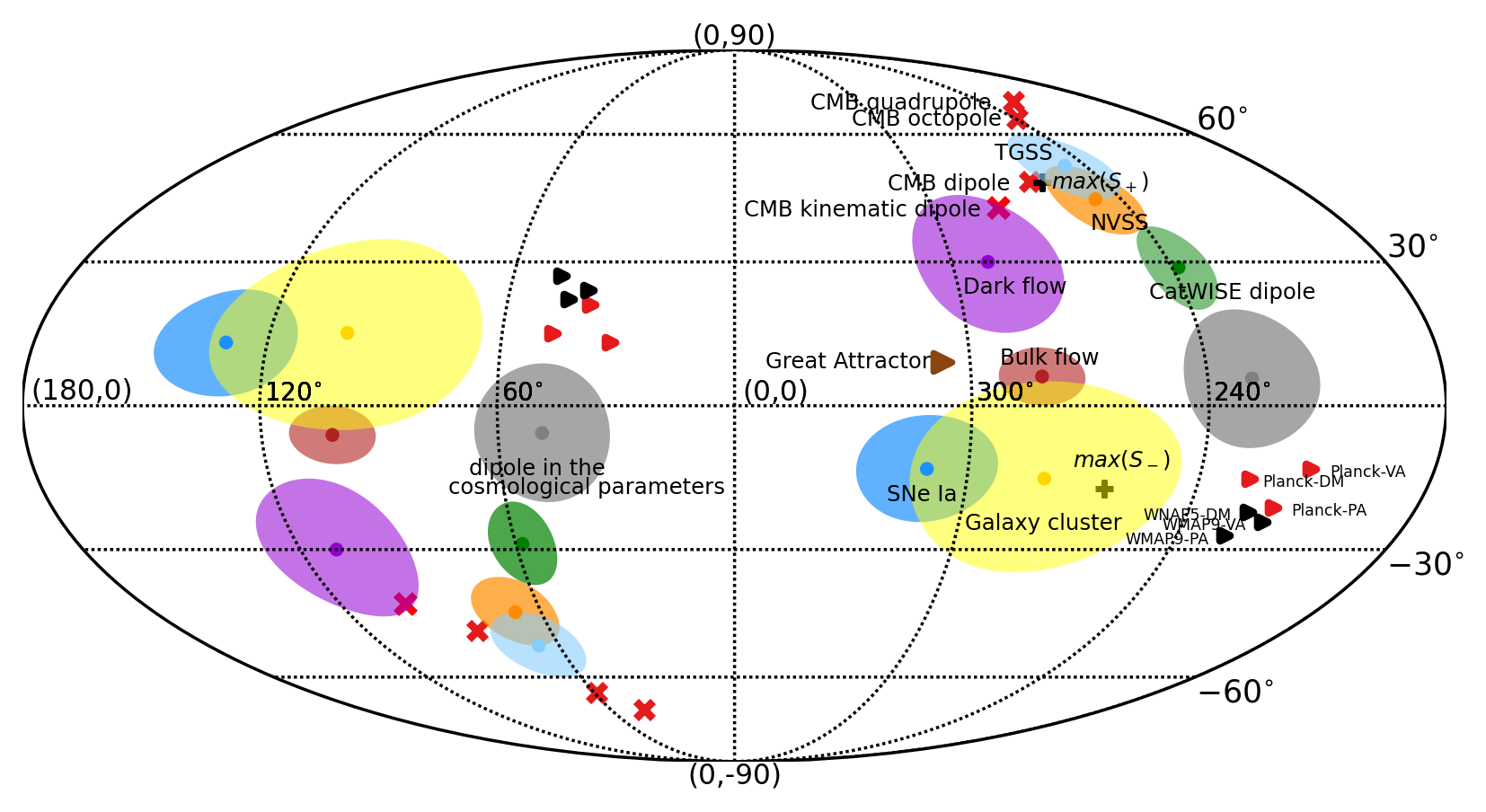}
\caption{Directions of anisotropy in the Universe in the Galactic $(l, b)$ coordinates with the galactic center in the middle, as inferred from Table~\ref{table:dipole}. Directions from the literature are shown with different markers or ellipses (data points and their $1 \, \sigma$ uncertainties) with text labels.}
\label{fig:aniso}
\end{figure}

\begin{table*}\small
		\setlength\tabcolsep{2pt}
		\renewcommand{\arraystretch}{1.5}
		\caption{Directions of anisotropy in the Universe as inferred from several data sets, the locations of the data points are shown in Fig.~\ref{fig:aniso}. 
		}
		\label{table:dipole}
		\centering
			\begin{tabular}{|c|c|c|}
				\hline
				Data Point     & $(l, b)$   & Ref.   \\
				\hline
				Galaxy cluster          & $(280^{\circ} \pm 35^{\circ},  -15^{\circ} \pm 20^{\circ})$           & \cite{bib:Migkas2021}    \\
				NVSS          & $(248^{\circ}\pm12.5^{\circ}, 44^{\circ}\pm8^{\circ})$         & \cite{bib:Singal2011}   \\
				TGSS          & $(247^{\circ}\pm14.6^{\circ}, 52^{\circ}\pm8^{\circ})$           & \cite{bib:Singal2019}  \\
				Dipole in the cosmological parameters         & $({48.8^{\circ}}^{+14.3^{\circ}}_{-14.4^{\circ}}, {-5.6^{\circ}}^{-17.0^{\circ}}_{-17.4^{\circ}})$          &  \cite{bib:Yeung2022}  \\
			    CatWISE dipole & $(238.2^{\circ}, 28.8^{\circ})$          &  \cite{bib:Secrest2021}  \\
				CMB kinematic dipole          & $(280^{\circ}, 42^{\circ})$          & \cite{bib:Naselsky2012}  \\
				\hline
				CMB dipole          & $(263.99^{\circ}, 48.26^{\circ})$          &  \cite{bib:Planck2016c}   \\
				CMB quadrupole          & $(224.2^{\circ}, 69.2^{\circ})$          &  \cite{bib:Planck2013}   \\
				CMB octopole          & $(239^{\circ}, 64.3^{\circ})$          & \cite{bib:Planck2013}   \\
				Planck-VA  (Variance Asymmetry)        & $(212^{\circ}, -13^{\circ})$          &  \cite{bib:Akrami2014}   \\
				Planck-DM (Dipole Modulation)         & $(227^{\circ}, -15^{\circ})$          & \cite{bib:Akrami2014}   \\
				Planck-PA  (Power Asymmetry)        & $(218^{\circ}, -21^{\circ})$           & \cite{bib:Akrami2014}  \\
				WMAP9-VA          & $(219^{\circ}, -24^{\circ})$          &  \cite{bib:Akrami2014}   \\
				WMAP5-DM          & $(224^{\circ}, -22^{\circ})$          &   \cite{bib:Akrami2014}   \\
				WMAP9-PA          & $(227^{\circ}, -27^{\circ})$          &   \cite{bib:Akrami2014}   \\
				Great Attractor          & $(307^{\circ}, 9^{\circ})$          & \cite{bib:Lynden1988}   \\
				SNe Ia          & $(310.6^{\circ}\pm18.2^{\circ},  -13.0^{\circ}\pm11.1^{\circ})$          & \cite{bib:Antoniou2010}   \\
				Dark flow          & $(290^{\circ}\pm20^{\circ}, 30^{\circ}\pm15^{\circ})$          & \cite{bib:Abdalla2022}   \\
				Bulk flow          & $(282^{\circ}\pm11^{\circ}, 6^{\circ}\pm6^{\circ})$          & \cite{bib:Atrio2010}  \\
				Odd mirror parity $max(S_-)$          & $(264^{\circ}, -17^{\circ})$          & \cite{bib:Gruppuso2011}  \\
				Even mirror parity $max(S_+)$          & $(260^{\circ}, 48^{\circ})$          & \cite{bib:Gruppuso2011}  \\
				\hline
			\end{tabular}
		\end{table*}

Returning to the CMB dipole, as foreseen by Baldwin \& Ellis \cite{bib:Ellis1984}, the assumed purely kinematic CMB dipole can be tested. Once the CMB dipole is subtracted, this defines the CMB as the rest frame of the Universe. More precisely, one infers that we are moving with a velocity $v = (369.82 \pm 0.11)$~km/s \cite{bib:Planck2018} relative to the CMB frame. The basic idea of Ref.~\cite{bib:Ellis1984} is to use radio galaxy counts to infer our velocity from aberration effects, namely shifts in the perceived location of the sources due to our motion. Throughout, it is assumed that the radio galaxies are isotropically distributed in line with the CP and it is imperative to ensure that the sources are at large redshifts $z \sim 1$, since one can get a spurious clustering dipole at lower redshifts, $z < 0.1$ \cite{bib:Tiwari2016}. It is worth emphasizing that, isotropy aside, these tests are essentially independent of a specific cosmological model, so there is no debate about alternative models within FLRW. To first approximation, various independent groups have reported an excess in the amplitude of the cosmic dipole \cite{bib:Singal2011, bib:Rubart2013, bib:Tiwari2016, bib:Bengaly2018, bib:Singal2019, bib:Secrest2021, bib:Siewert2021, bib:Singal2021, bib:Singal2022, bib:Secrest2022} (however, see  \cite{bib:Blake2002, bib:Horstmann2021, bib:Darling2022}), while the inferred directions coincide with the expected direction of the CMB dipole. Na\"{\i}vely, this excess in amplitude implies an anisotropic Universe at scales well beyond inferred isotropy scales (see, for example, Ref.~\cite{bib:Marinoni2012}). 
Nevertheless, systematics remain a concern. For instance, some frequency dependence in the dipole has been noted \cite{bib:Siewert2021},\footnote{This frequency dependence is driven by a larger dipole in the TGSS survey \cite{bib:Intema2017}, but the same survey exhibits also an anomalous angular power spectrum \cite{bib:Dolfi2019}.} which runs contrary to the expectation of a purely kinematic dipole, while similar tests involving SNe Ia have led to a deficit in the cosmic dipole \cite{bib:Horstmann2021}. A more recent study \cite{bib:Darling2022} succeeds in recovering the CMB dipole, but larger errors mean that the inferred dipole is consistent within $3\sigma$ with observations that claim an excess, such as, for example, \cite{bib:Secrest2021}.\footnote{As explained in the appendix of \cite{bib:Secrest2022}, the analysis combines two catalogues, at least one of which appears to be inconsistent with the kinematic interpretation of the CMB dipole.} Separately, the Baldwin-Ellis  methodology has been questioned \cite{bib:Dalang2022, bib:Murray2022, bib:Guandalin2022}. These points aside, it is worth bearing in mind that these observations, along with \cite{bib:Fosalba2021, bib:Yeung2022}, are serious claims that could potentially upend modern cosmology, but the results are preliminary. It is imperative to repeat these tests with upcoming facilities, including the LOw Frequency ARray (LOFAR) two-meter sky survey \cite{bib:Shimwell2017} and the Square Kilometer Array Observatory (SKAO) \cite{bib:Bengaly2018, bib:Bacon2020} \bcomm{(see also \cite{bib:Ghosh2023})}, in a bid to either confirm or refute existing results. 

Nevertheless, now that some groups are reporting an excess in the radio galaxy or quasar (QSO) dipole with respect to CMB expectations \cite{bib:Singal2011, bib:Siewert2021, bib:Secrest2021}, one is free to take the claim at face value and ask, whether it is confirmed or refuted by results at different redshifts? Are there synergies that make this claim more credible? To begin, any excess in the cosmic dipole appears to sit well with CMB anomalies \cite{bib:Eriksen2003, bib:Park2003, bib:Hansen2004, bib:Eriksen2007, bib:Hansen2009, bib:deOliveira2004, bib:Schwarz2004, bib:Land2005, bib:Kim2010, bib:Planck2013, bib:Planck2016, bib:Planck2018c} that are also tracking the CMB dipole direction. Moreover, at the other end of the Universe, it is well documented that there is a coherent bulk flow towards the Shapley supercluster\footnote{However, observations from the {\em Nearby Supernova Factory} find no evidence for a backside infall behind the Shapley Supercluster \cite{bib:Feindt2013}.} \cite{bib:Hoffman2017}, which is in the rough direction of the CMB dipole.
Earlier, this phenomenon was referred to as \emph{Virgo Alignment Puzzle}, when preferred directions seen in different astronomical surveys spanning the electromagnetic spectrum were found to be well aligned when studied in a consolidated manner, perhaps for the first time \cite{bib:Ralston2004}.
Such a bulk flow has implications for local $H_0$ inferences, and in line with expectations\footnote{\bcomm{See for example discussion in \cite{bib:Parnovsky2001,bib:Clarkson2012b, bib:Heinesen2020c, bib:Dhawan2022}.}}, it has been \bcomm{observationally confirmed} in the aftermath of the Hubble Space Telescope (HST) key project \cite{bib:Freedman2001} that $H_0$ varies across the sky in the local Universe \bcomm{\cite{bib:McClure2007,bib:Wiltshire2012}}. This simply underscores the difficulty in determining $H_0$ through the distance ladder \cite{bib:Huang2020, bib:deJaeger2020, bib:Pesce2020, bib:Kourkchi2020, bib:Schombert2020, bib:Khetan2021, bib:Blakeslee2021, bib:Freedman2021, bib:Riess2021}. We thus need to test whether this flow is consistent with the expectations of the flat $\Lambda$CDM model \cite{bib:Watkins2009, bib:Kashlinsky2008, bib:Lavaux2010, bib:Magoulas2016}, especially given that the most recent study suggests the flow may be larger than expected \cite{bib:Howlett2022}. Clearly, this line of inquiry should be extended to $z \sim 0.1$, where there is already an intriguing observation \cite{bib:Migkas2020, bib:Migkas2021}. \bcomm{Here, it is worth emphasising that variations in the Hubble flow or $H_0$ are allowed at low redshifts, but as the redshifts increase, such variations start to challenge FLRW.} Concretely, a 9\% spatial variation of $H_0$ in a direction consistent with the CMB dipole has been inferred across a host of galaxy cluster scaling relations \cite{bib:Migkas2020, bib:Migkas2021}. One potential interpretation of the result is a bulk flow of velocity $900$~km/s out to distances of $500$~Mpc \cite{bib:Migkas2021}. This observation is supported by a residual dipole in the ``Low z'' subsample of the Pantheon SNe data set \cite{bib:Scolnic2018}, which is enhanced by higher redshift SNe \cite{bib:Krishnan2022}, but the statistical significance is low $< 2 \, \sigma$ (see also \cite{bib:Cooke2010, bib:Antoniou2010, bib:Li2013, bib:Javanmardi2015} for earlier observations of similar features). Moreover, a recent study \cite{bib:Giles2022} replicates the original observation with the same data \cite{bib:Migkas2020}, but fails to detect anisotropies in galaxy cluster scaling relations in the SDSS DR8 sample \cite{bib:Rykoff2014}, while cautioning that the incomplete declination coverage of the SDSS DR8 sample may mask any anisotropies.

Let us stress that variations in the cosmological parameters on the sky \cite{bib:Fosalba2021, bib:Yeung2022} (supported by \cite{bib:Migkas2021, bib:Krishnan2022, bib:Luongo2022}), excesses in the cosmic dipole \cite{bib:Blake2002, bib:Singal2011, bib:Gibelyou2012, bib:Rubart2013, bib:Tiwari2016, bib:Bengaly2018, bib:Singal2019, bib:Secrest2021, bib:Siewert2021, bib:Singal2021, bib:Singal2022, bib:Secrest2022}, anomalous bulk flows \cite{bib:Watkins2009, bib:Kashlinsky2008, bib:Lavaux2010, bib:Magoulas2016, bib:Howlett2022} and anisotropies in scaling relations \cite{bib:Migkas2020, bib:Migkas2021, bib:Giles2022} constitute current results. 
Clearly, excess or residual dipoles are unexpected within FLRW, at least beyond perturbative level. The claimed cosmic dipole excesses and anomalous bulk flows have been persisting for a decade. As we will explain in section~\ref{sec:late_universe_FLRW_anomalies}, a previously reported dipole in the fine structure constant appears to have disappeared in recent data analyses, which underscores the need for dedicated surveys. This may not be surprising, since it was largely uncorroborated by other observations (however, see \cite{bib:Mariano2013, bib:Mariano2012}). 
In contrast, the CMB anomalies, which are also tracking the CMB dipole, have largely survived the transitions from WMAP to Planck, so they do not represent obvious systematics. The anomalies in galaxy cluster scaling relations are newer results, which can be addressed by mapping out distances in the local Universe to $z \sim 0.1$. Such a program is ongoing \cite{bib:Kourkchi2020} and results are expected in the near future. We gather results in Fig.~\ref{fig:aniso} to highlight the synergies between the complementary approaches and different data sets.\footnote{The 
drawing method refers to \url{https://github.com/super1010/dipole-sky}.} 
Throughout, there appears to be a significant discrepancy in $H_0$ \cite{bib:Riess2021, bib:Planck2018b}, which is difficult to resolve theoretically within FLRW \cite{bib:DiValentino2021, bib:Abdalla2022}. Furthermore, as we have seen, anisotropies are expected to translate into variations of $H_0$ across the sky and such variations are apparently already evident at different scales within the flat $\Lambda$CDM model \bcomm{\cite{bib:McClure2007, bib:Wiltshire2012, bib:Migkas2020, bib:Migkas2021, bib:Krishnan2022, bib:Luongo2022, bib:Fosalba2021, bib:Yeung2022}}. \bcomm{We remind the reader again that such directional anomalies can only be addressed within an FLRW cosmology by changing the FLRW model to inflate errors.} 
For this reason, it is timely to revisit observations that are unexpected in an FLRW universe, to evaluate their joint information content, to openly discuss their caveats, while also investigating how to improve CP tests with them in the future. As remarked, even if the Universe is only approximately FLRW and precision cosmology drives one beyond, some of the simplest models, e.~g.~Bianchi VII$_h$, are not expected to work \cite{bib:Saadeh2016}. 
In a sister paper \cite{bib:Wiltshire2022}, we detail the methodological shortcomings of current approaches to interpret observables within an FLRW model, outline options to improve on these issues, and introduce alternative cosmologies and tests to support or refute them with current and future data sets.

The structure of this document is as follows. We first introduce conventions and notation. In section~\ref{sec:Hom-scale} we review the homogeneity scale and the constraints on it set by current observations.
Then, sections~\ref{sec:early_universe_FLRW_anomalies} and \ref{sec:late_universe_FLRW_anomalies} outline observations that test the isotropy of the Universe from the perspective of the early Universe (CMB) and late Universe, respectively. Throughout, we address various caveats and conflicting observations. In \bcomm{sections~\ref{sec:alignments} and \ref{sec:test_cosmological}} we document \bcomm{challenges and} tests of the CP that are expected to become more competitive as data quality improves. In section~\ref{sec:epilogue}, we frame the discussion in terms of answers to a number of relevant questions, including the pressing question in the title.

\subsection*{Terminology and conventions} 
\label{sec:terminology}

In an FLRW setting, the line element is given by
\begin{equation}
\mathrm{d}s^2 = - c^2 \, \mathrm{d}t^2 + a^2(t) \left[\mathrm{d}r^2 + f_\mathrm{k}^2(r) \, \left(\mathrm{d}\theta^2 + \sin^2\theta\ \mathrm{d} \varphi^2 \right) \right] \;,
\label{eq:FLRW_metric}
\end{equation}
in which $c$ denotes the speed of light, $t$ is the cosmic time coordinate, $a(t)$ the (time-dependent) scale factor, $r$ the radial coordinate, and $\varphi\in [0,2\pi]$ and $\theta\in [0,\pi]$ denote the polar angles spanning the solid angle perpendicular to the radial coordinate. We work in units in which $c=1$. The constant time $t$ slices denote a maximally symmetric, constant curvature, 3 dimensional hypersurface. The curvature of this space is encoded in parameter $\text{k}$ which can take positive, negative, and zero values, respectively corresponding to a closed, open and flat universe.  
The function 
\begin{equation}
    f_\mathrm{k}(r) := \left\{ \begin{matrix} \sqrt{\mathrm{k}}^{-1} \sin(r\sqrt{\mathrm{k}}) & \ \ \mathrm{k} > 0 \\ r & \ \ \mathrm{k} = 0 \\ \sqrt{|\mathrm{k}|}^{-1} \sinh(r\sqrt{|\mathrm{k}|}) & \ \  \mathrm{k} < 0 \end{matrix} \right.
\label{eq:k_function}
\end{equation}
represents the dependence on this spatial curvature and $\mathrm{k}$ is a quantity of dimension inverse-length-squared such that $\mathrm{k} r^2$ is dimensionless. 
Eq.~\eqref{eq:FLRW_metric} is the most general metric one can write down with at least 6 Killing vectors, a spatially homogeneous and isotropic Universe; this form is fixed by the symmetries. 

Next, one should impose field equations and specify the scale factor $a(t)$ in terms of the matter/energy content of the Universe. While one can work with beyond-Einstein gravity model, here we restrict ourselves to the Einstein General Relativity (GR). 
As already mentioned, we take the CP to be equivalent to FLRW metric ansatz plus the Einstein field equations. Inserting  metric \eqref{eq:FLRW_metric} into the Einstein field equations, we obtain the Friedmann equations
\begin{align}
\left( \dfrac{\dot{a}(t)}{a(t)} \right)^2 := H(t)^2 &= \frac{8\pi G}{3} \rho - \frac{\mathrm{k}}{a^2(t)} + \frac{\Lambda}{3} \;,
\label{eq:Friedmann1} \\
\left( \frac{\Ddot{a}(t)}{a(t)}\right)= \dot H+ H^2 &= -\frac{4\pi G}{3} \left( \rho + 3p \right) + \frac{\Lambda}{3} \;,
\label{eq:Friedmann2}
\end{align}
where $dot$ denotes derivative with respect to $t$ and $H(t)$ is the Hubble \bcomm{parameter}.

We choose the initial values such that $t=0$ corresponds to today and $a(t=0)=1$. In this convention, one may define the redshift factor $z$ as
\begin{equation}\label{z--a}
    a(t)=\frac{1}{1+z} \;.
\end{equation}
This may be viewed as a change of coordinates from comoving time $t$ to $z$, in which metric \eqref{eq:FLRW_metric} takes the form
\begin{equation}
\mathrm{d}s^2 = \frac{1}{(1+z)^2} \left[- \frac{\mathrm{d}z^2}{H(z)^2}  + \mathrm{d}r^2 + f_\mathrm{k}^2(r) \, \left(\mathrm{d}\theta^2 + \sin^2\theta\ \mathrm{d} \varphi^2 \right) \right] \;,
\label{eq:FLRW_metric-z}
\end{equation}
which clearly  shows that a radial null geodesic is given by $r(z)=\int_0^z\frac{\mathrm{d}\tilde{z}}{H(\tilde{z})}$. The luminosity distance $d_\text{L}(z)$ is hence defined as 
\begin{equation}
\label{eq:dL}
d_\mathrm{L} (z) =(1+z) D(z)\;, \qquad D(z):={f_\mathrm{k}(r(z))}\;.
\end{equation}
Two other useful  quantities are the angular diameter distance $d_{\text{A}}$, 
\begin{equation}\label{dL-dA}
d_{\text{A}}:= \frac{D(z)}{1+z}=\frac{d_{\text{L}}}{(1 + z)^2} \;,   
\end{equation}
and  the proper time between two events at the same comoving spatial coordinates, at redshift $z$ and at $z=0$, $\Delta t$,
\be\Delta t=\int_0^z \frac{\mathrm{d}\tilde{z}}{(1+\tilde{z}) H(\tilde{z})} \;.
\ee
That is, $\Delta t$ is the age of Universe since the event at redshift $z$. For large $z$ this simply reduces to the age of the Universe. The important point is that $d_{\text{A}}, d_{\text{L}}, \Delta t$ are \textit{geometric} quantities, their physical relevance and significance does not come from details of the gravity theory but from the form of the metric and that photons (massless states) move on null geodesics. The fact that these quantities depend only on $z$ and not on $\theta,\phi$ is a property of the FLRW metric, a manifestation of the CP.

The Hubble diagram $H(z)$ provides the cosmic expansion rate at a celestial sphere whose radius is specified by the redshift $z$ and the Hubble constant $H_0$ is the value of expansion rate today. As a concrete example, choosing the matter sector to be composed of pressureless matter and radiation, in the presence of spatial curvature $\mathrm{k}$ and cosmological constant, one can solve the Friedmann equations to obtain $H(z)$:
\begin{equation}
H(z) := H_0 \, E(z) = H_0 \, \sqrt{\Omega_\mathrm{r} \left( 1+z\right)^4 + \Omega_\mathrm{m} \left(1+z \right)^3 + \Omega_\mathrm{k}  \left(1+z \right)^2 + \Omega_\Lambda} \;,
\label{eq:expansion_rate}
\end{equation}
where  $H_0:=H(z=0)$ and from left to right, the $\Omega_i$ denote the dimensionless density parameters for radiation (r), matter (m), curvature (k), 
$ \Omega_\mathrm{k}:=\mathrm{k}/H_0^2$ and a cosmological constant/dark energy contribution ($\Lambda$) to the total energy-momentum content at $z=0$.
As $E(z)$ is normalised such that $E(z=0)=1$, the sum of all $\Omega_i$ is supposed to yield 1. 

That $H_0$ is a constant both in $z$ and on the sky, is a generic feature of \textit{any} FLRW cosmology \cite{bib:Krishnan2021, bib:Krishnan2022a}. In a non-FLRW universe, the expansion rate at a given celestial sphere $H(z)$ is not given by a number, but, generically, by a function over the sphere. In such cases we obtain an \textit{expansion rate} which varies over the sky at $z=0$ instead of a single $H_0$. Still, in such general cases one may \emph{define} $H_0$ (and similarly $H(z)$) as the expansion rate averaged over the celestial sphere. 

These are only the most relevant equations and formulae to set a common standard notation for the subsequent sections. 
Further details on theories of gravity, cosmological models, and astrophysics at perturbation level can be found in any standard textbook on GR or cosmology, see, for instance \cite{bib:Ellis2012}.

\section{Homogeneity Scale}
\label{sec:Hom-scale}

\begin{quote}
\textit{It certainly has convinced me that we're not living in a homogeneous, isotropic [universe]. I mean these things that I really suspected in the back of my mind, I can now say publicly. I'm not sure the Robertson-Walker universe exists.} --- Vera Rubin on the discovery of the CfA Great Wall \cite{bib:Geller1989}\footnote{ \href{www.aip.org/history-programs/niels-bohr-library/oral-histories/33963}{{www.aip.org/history-programs/niels-bohr-library/oral-histories/33963}}}
\end{quote} 

 
In the study of the Large-Scale Structure (LSS) of the Universe, predictions can be made of the expected size of clustering. On small scales, one can expect lots of clustering and structures like walls, filaments and voids. However, for large-scale homogeneity, as required by the CP, one should reach a limit where no, or very little, structure exists beyond a certain size. To sharpen any debate, these notions have to be defined, which means bringing together \bcomm{Raychaudhuri's four approaches introduced in section~\ref{sec:prologue}} with a minimal set of assumptions.
 
Firstly, to talk about spatial homogeneity at a mathematical level we are dealing with spatial averages of a density field $\rho$, defined on a compact domain of a spatial hypersurface, $\Sigma_t$, according to
\begin{equation}
\ave{\rho(t)}{_{{\cal D}_R}}=\frac1{{\cal V}(t)}\left(\int_{{\cal D}_R}{\rm d}
^3x\sqrt{\det\ {}^3\!g}\,\rho(t,{\mathbf x})\right) \;,
\end{equation}
where ${\cal V}(t)\equiv \alpha R^3(t)=\int_{{\cal D}_R}{\rm d}^3x\sqrt{\det \ {}^3\!g}$ is the volume of
the domain ${\cal D}_R\subset\Sigma_t$ with radius $R(t)$, $g_{ij}$, ($1\le i,j \le3$) is the intrinsic metric on $\Sigma_t$ and $\alpha$ is a dimensionless constant determined by a choice of geometry; e.~g., $\alpha=4\pi/3$ for Euclidean spheres. Furthermore, a definition of spatial homogeneity often presupposes a notion of ergodicity and the existence of an average positive density, $\rho_0(t)$, defined by the limit
\begin{equation}
\lim_{R(t)\to\infty}\ave{\rho(t)}{_{{\cal D}_R}}=\rho_0(t)>0 \;.
\end{equation}
The homogeneity scale $\lambda_0(t)$ may then be defined\footnote{\bcomm{This definition is a simple extension of Eq.~(2.6) of \cite{bib:Gabrielli2005} from a static Euclidean geometry to a general metric on spatial hypersurfaces that evolve with time.}} by the requirement that every point in $\Sigma_t$ be contained in a domain ${\cal D}_{\lambda_0}\subset{\cal D}_R$ such that
\begin{equation}
\left|\ave{\rho(t)}_{{\cal D}_R}-\rho_0(t)\right|<\rho_0(t)\qquad
\forall\; R\bcomm{(t)}>\lambda_0\bcomm{(t)} \;.
\label{homogenscale}
\end{equation}

Observationally, the density field can only be inferred indirectly from the statistical properties of galaxy clustering. 
This inevitably entails a host of systematic issues related to flux-limited and volume-limited surveys and observational biases associated with using galaxies as tracers of the density field, see, for instance, \cite{bib:Treciokas1971} for details on the relation between particle distributions and inferred densities for the energy-momentum tensor. Consequently, any practical estimates of a homogeneity scale are not directly based on relations such as (\ref{homogenscale}), but rather on the scale dependence of galaxy-galaxy correlation functions.

\begin{figure}[htb]
\includegraphics[width=0.43\textwidth]{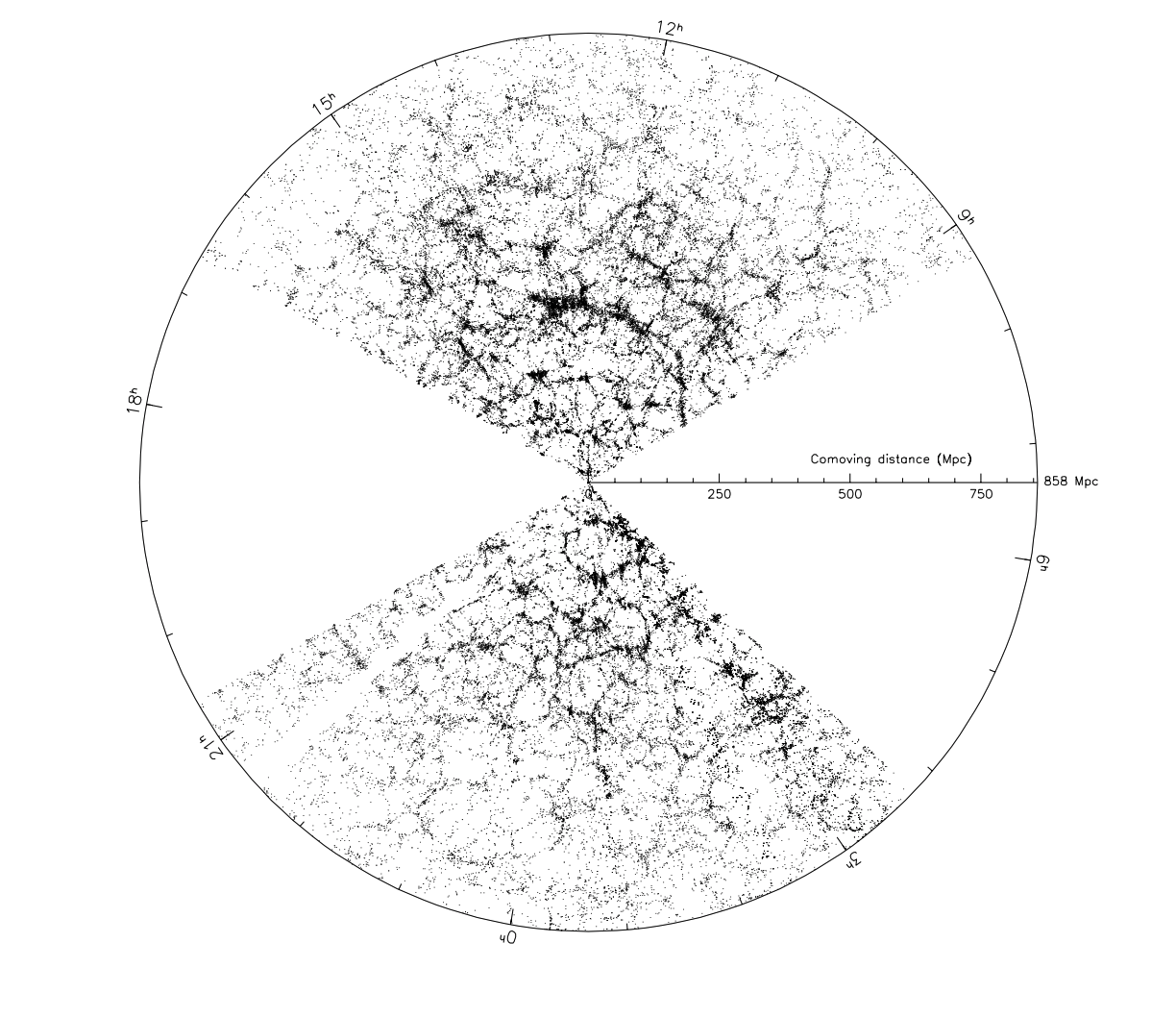}
\includegraphics[width=0.42\textwidth]{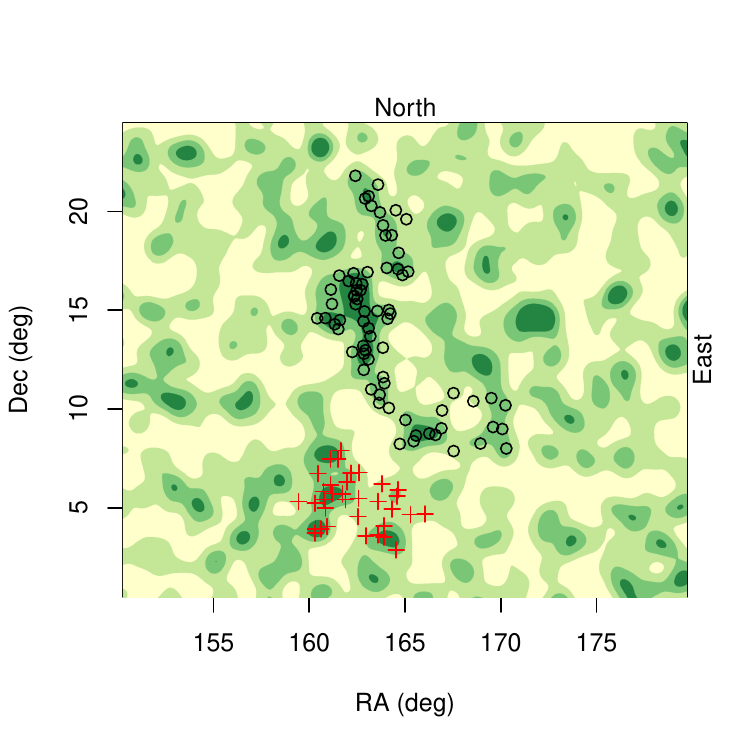}
\vspace{-2ex}
\caption{Left: The Sloan Great Wall \bcomm{(reproduced from Fig. 2 of \cite{bib:Gott2005})} is a large, wall-like filament in the relatively local Universe ($z=0.073$). Its longest dimension is $\sim450$~Mpc.
Right: The Huge-LQG (black circles) \bcomm{(reproduced from Fig. 1 of \cite{bib:Clowes2013})} and the CCLQG (red crosses) \cite{bib:Clowes1991}. The contours indicate the general distribution of DR7QSO QSOs in the same redshift interval. The CCLQG and the Huge-LQG were the `largest structures known in the Universe' in 1991 and 2013. They are adjacent on the sky and at the same redshift ($z=1.27$ for the Huge-LQG, $z=1.28$ for the CCLQG), but there is no obvious connection between them.}
\label{fig:SGW_LQG}
\end{figure}

As already noted in section~\ref{sec:prologue}, the range \bcomm{60--80}$\,h^{-1}\,$Mpc represents the lowest bound at which any notion of statistical homogeneity can be argued to emerge \cite{bib:Hogg2005, bib:Yadav2005, bib:Scrimgeour2012, bib:Ntelis2017, bib:Sarkar2019, bib:Andrade2022} \footnote{\bcomm{These studies have further limitations. The $\Lambda$CDM model is routinely assumed to transform angles or redshifts to (comoving) distances, and occasionally a study augments observational data with simulated $\Lambda$CDM data to fill in holes in the survey, e. g. \cite{bib:Scrimgeour2012}. In addition, the normalised estimators used in practice typically assume convergence to homogeneity within the scale of the survey (see \cite{bib:Heinesen2020d}).}}, based on the two-point galaxy correlation function. 
However, a definition such as (\ref{homogenscale}) actually requires that {\em all} $N$-point correlations of the galaxy distribution should also converge appropriately. Determining higher order $N$-point correlations requires ever larger survey volumes, but is becoming feasible. Using Minkowski functionals, an analysis of the Sloan Digital Sky Survey Data Release 7 (SDSS DR7) showed significant deviations ($>3\sigma$) from flat $\Lambda$CDM simulations on scales up to 500$\,h^{-1}\,$Mpc and slight deviations ($\sim2\sigma$) on scales of 700$\,h^{-1}\,$Mpc \cite{bib:Wiegand2014}. 
Future surveys will enable more refined tests of the scale of homogeneity, which is currently estimated to be $260\,h^{-1}\,$Mpc $\simeq 370$~Mpc \cite{bib:Yadav2010} in $\Lambda$CDM simulations.
Moreover, as noted in section~\ref{sec:prologue}, several very large scale structures which exceed this scale of homogeneity have already been discovered and it is predicted \cite{bib:DeMarzo2021} that even larger structures will be discovered with larger data sets out to higher $z$. 
Using bulk flows, a technique which links the local peculiar velocities to the gravitational interactions of the perturbing matter agglomerations at low redshifts, see section~\ref{sec:bulk_flows}, structures like the Laniakea Supercluster with an extension of $\sim160$~Mpc \cite{bib:Tully2014} and the South Pole Wall with an extension of $\sim430$~Mpc \cite{bib:Pomarede2020} have been found. 
Fig.~\ref{fig:SGW_LQG} showcases two further examples of this class, the Sloan Great Wall \cite{bib:Gott2005}, which has a similar extension as the South Pole Wall \bcomm{and used to be the largest known structure in the observable universe until 2013}, on the left and various Large Quasar Groups (LQG) \cite{bib:Clowes1991, bib:Clowes2013} on the right.

\subsection{The Giant Arc}
\label{sec:giant_arc}
The Giant Arc (GA) \cite{bib:Lopez2022} is the fourth largest LSS to be detected after the Hercules–Corona Borealis Great Wall \bcomm{(HCBGW)} \cite{bib:Balazs2015, bib:Horvath2020}, the Giant GRB Ring \bcomm{(GGR)} \cite{bib:Balazs2015} and the Huge-LQG \cite{bib:Clowes2013}. 
\bcomm{More details on the latter can be found in sections~\ref{sec:qso_polarization_alignments} and \ref{sec:LQG_alignments} because it was first questioned to be a contiguous structure, and observations of polarization alignments supported that the structure is more than a random clustering on the sky. The former two are discussed in section~\ref{sec:larger_structures}, as these alignments were disputed as cosmic structures in the past and still wait for further complementary corroboration.}
\bcomm{But, together}, the accumulated set of discovered LSSs raises important questions for the validity of the CP. The method used for the discovery of the GA is that of intervening MgII absorbers in the spectra of QSOs. When the light from very luminous and distant QSOs passes through intervening low-ionized gas (around galaxy haloes) a prominent MgII doublet profile can be detected in the QSO spectra. 
Thus, the QSOs can essentially be used as a probe of low luminosity matter at intermediate redshifts ($0.5 < z < 2.2$), where either galaxies are often too faint to be detected or they require large amounts of telescope time. 

SDSS contains ca.~120,000 QSO spectra in DR7 and DR12. Independently, \cite{bib:Zhu2013} have catalogued the MgII absorption systems present in the SDSS QSOs (DR7, DR12), of which there are around 64,000 MgII systems. The MgII catalogue presents both accurate spectroscopic redshifts and also the sky coordinates, allowing one to map the LSS of a large portion of the sky over a wide redshift interval in great detail.

\begin{figure}[b]
\includegraphics[width=0.67\textwidth]{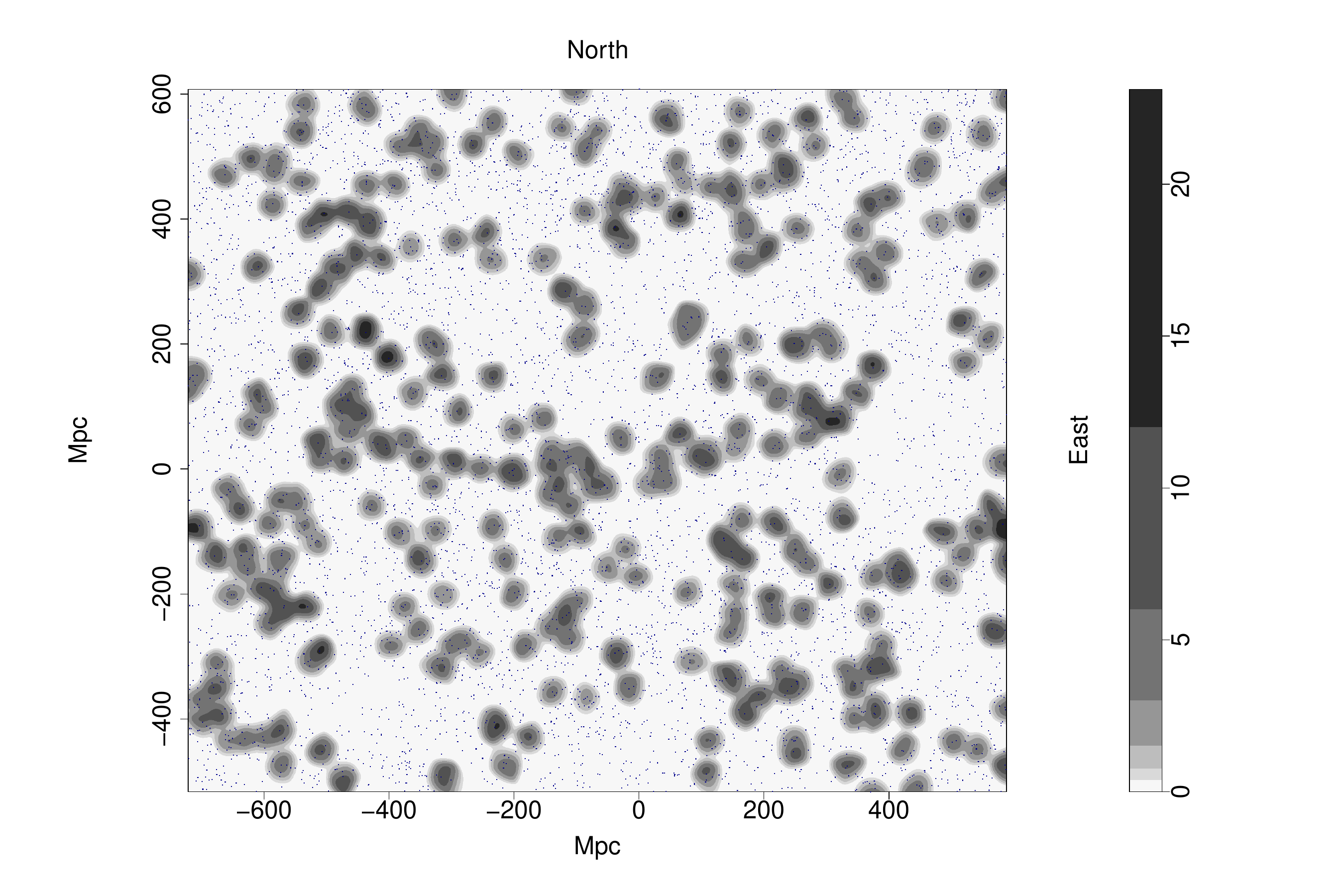}
\vspace{-2ex}
\caption{The Giant Arc: Tangent-plane distribution of MgII absorbers centered in the redshift interval $z=0.802\pm0.060$.
 The grey contours, increasing by a factor of two, represent the density distribution of the absorbers, which were smoothed using a Gaussian kernel of $\sigma=11$~Mpc, and flat-fielded with respect to the distribution of background probes. The dark blue dots represent the background probes (QSOs).
 The axes are labelled in Mpc, scaled to the present epoch.
 East is towards the right and north is towards the top.
 The GA runs west-east in the center of the figure, spanning $\sim1$~Gpc. \bcomm{Figure reproduced from Fig. 1 of \cite{bib:Lopez2022}.}}
\label{fig:GA}
\end{figure}

In the preliminary stages of testing the MgII method for LSS studies, known and documented clusters were examined with the MgII maps. One such cluster, discovered through the Sunyaev-Zeldovich (SZ) effect, appeared to have a narrow, dense band of MgII absorbers running west-east in the MgII maps. Refining the redshift a little and extending the field of view revealed what we now call the Giant Arc \cite{bib:Lopez2022}, as seen in Fig.~\ref{fig:GA}.

To validate the existence of the GA, three different statistical tests have been performed \cite{bib:Lopez2022}: a minimal-spanning-tree-type test (MST-type test); a Power Spectrum Analysis (PSA); and the Cuzick-Edwards test (CE) \cite{bib:Cuzick1990}. The MST-type test works by applying a sequence of the single linkage hierarchical clustering (SLHC) and the convex hull of member spheres (CHMS), see \cite{bib:Clowes2012} for details. This test grouped a large portion of what was visually identified as the GA into one structure with a clustering significance of $4.5 \, \sigma$, an overdensity of $\delta \rho/\rho = 1.3$, and a mass excess of $1.8 \times 10^{18} M_{\odot}$. 
The PSA test \cite{bib:Webster1976} calculates the statistic by testing for clustering on some scale, say $\lambda_c$, by considering the Fourier modes with wavelengths greater than $\lambda_c$. The test revealed significant clustering with statistical significance $4.8 \, \sigma$ on approximately $230$~Mpc scales, which likely corresponds to the width of the GA. 
Lastly, the CE test \cite{bib:Cuzick1990} was applied. This test has been used mostly in medical research as a way to understand clustering of cases in unevenly distributed populations. It is a case-control $k$-NN test that assesses the clustering of the `cases' (here MgII absorbers) in the whole field with respect to the distribution of `controls' (here QSOs). The CE test detected a $3 \, \sigma$ significance of clustering. 

Statistics aside, independent corroboration of such large-scale structures can be valuable.
\cite{bib:Lopez2022} visually inspected the DR16Q QSOs \cite{bib:Lyke2020} in the same redshift slice as the GA. The authors examined the density of the DR16Q QSOs, before superimposing them on the grey contours in Fig.~\ref{fig:GA}. This led to a striking agreement (see Fig.~13 in \cite{bib:Lopez2022}), which further supports the existence of the GA. Note that the DR16Q QSOs are completely independent, since all of the (DR7, DR12) QSOs used as probes are beyond the redshift slice. 

In summary, the Giant Arc is a discovery of an intriguing, almost symmetrical, $\sim 1$~Gpc structure found at a redshift of $z \sim 0.8$. Three different statistical tests find that the GA has significant clustering and connectivity, and independent corroboration from the DR16Q QSOs in the same redshift slice shows that there is an association between the MgII absorbers and QSOs. This discovery is the newest and one of the largest LSSs in an accumulating set and thus adds to the potential challenges to the CP and to the Standard Model of Cosmology.

\subsection{Larger structures}
\label{sec:larger_structures}

\begin{figure}[t]
\includegraphics[width=0.5\textwidth]{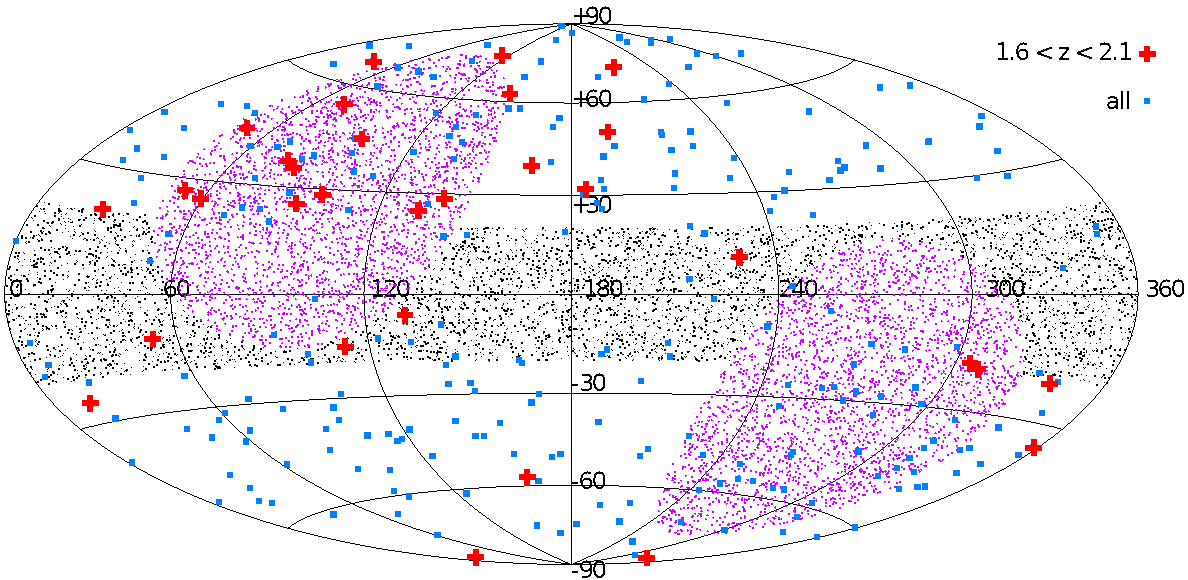}
\hfill
\includegraphics[width=0.48\textwidth]{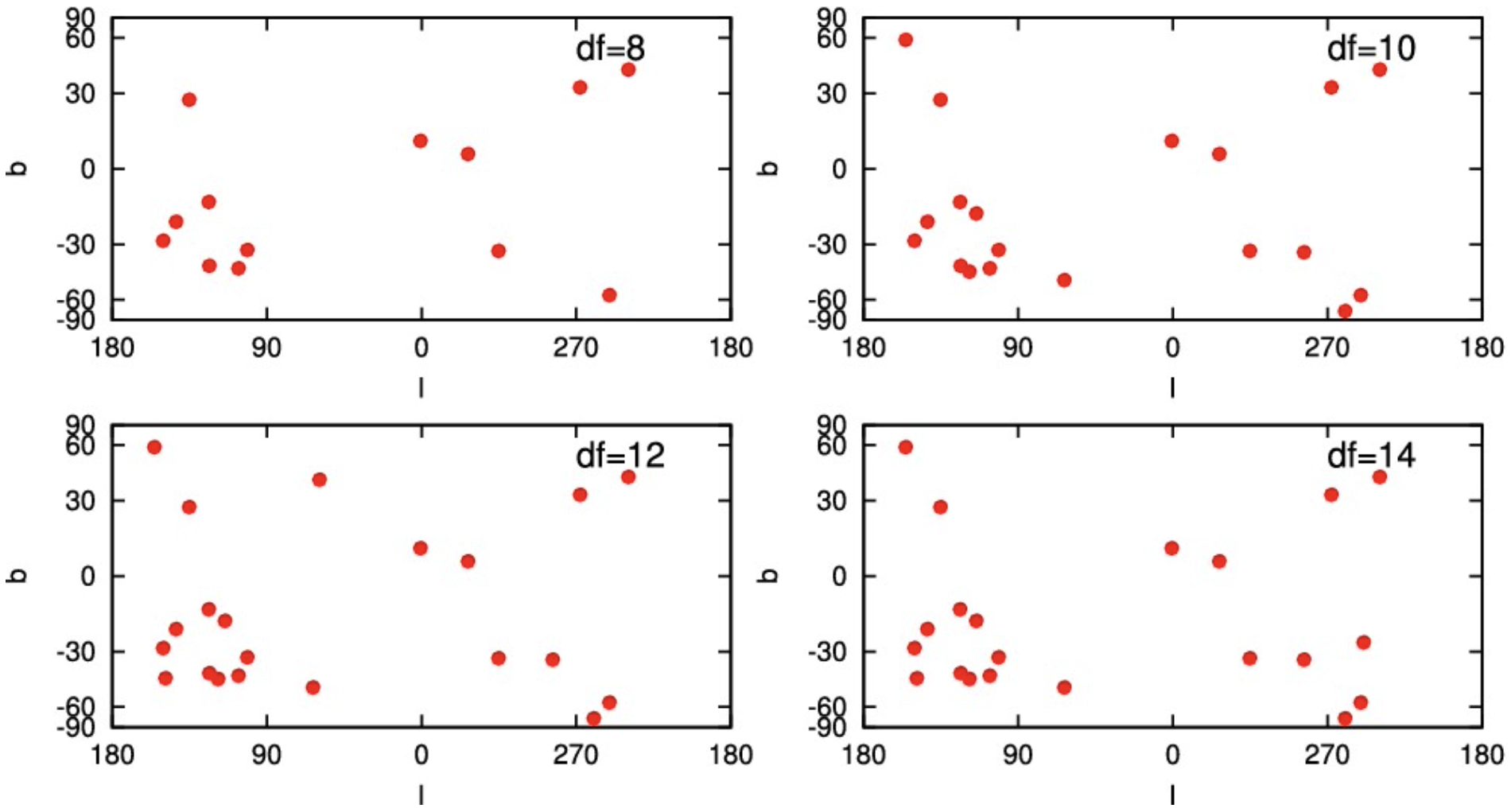}
\vspace{-2ex}
\caption{\bcomm{Left: HCBGW consisting of 31 GRBs in $1.6 \le z \le 2.1$ (red crosses) as discovered by \cite{bib:Horvath2014} (reproduced from Fig. 3 of \cite{bib:Horvath2014}) from a sample of 283 GRBs (blue dots) in Galactic coordinates, overlaid with the Galactic plane (black dotted region) and the ecliptic poles (purple dotted region). 
Right: GGR (ring-like structure in the lower left bottom of each plot) as discovered in \cite{bib:Balazs2015} (reproduced from Fig. 4 of \cite{bib:Balazs2015}). Each of the plots uses a different amount of degrees of freedom (df) to investigate the persistence of the ring-like feature when varying the order $k$ of nearest neighbours, with the dfs being equal to $k$.}}
\label{fig:GRB_structures}
\end{figure}

\bcomm{The two largest structures, known to this day, the HCBGW (see Fig.~\ref{fig:GRB_structures}, left) and the GGR (see Fig.~\ref{fig:GRB_structures}, right) were both detected by an excess of correlated GRBs. 
In 2013, the HCBGW was found due to an overdensity of about 20 GRBs in a redshift range between 1.6 and 2.1. 
These GRBs form a cluster covering an area on the sky of about 3~Gpc in length and 2.2~Gpc in width, thereby violating the expected scale of homogeneity according to \cite{bib:Yadav2010} by an entire order of magnitude.
Due to the small number of GRBs in this structure and the fact that the clustering only showed up when binning the 283 GRBs of the total sample into nine different redshift bins, the existence of the HCBGW is still debated. 
Similarly to the approaches performed for the GA, different statistical significance tests have been applied to the data to determine the significance of the HCBGW as a genuine cluster, which turned out to be around $3 \sigma$ in a two-dimensional Kolmogorov-Smirnov test, a nearest-neighbour statistic, and a bootstrap point-radius approach, as detailed in \cite{bib:Horvath2014}.
A discussion on potential biases due to the sampling strategy of the telescopes used to collect the data concludes that sampling effects are unlikely to generate the observed clustering signal. 
In a follow-up study with an extended sample of 361 GRBs, \cite{bib:Horvath2015} corroborated their findings made in \cite{bib:Horvath2014} and thus further supported the existence of the HCBGW. 
In the new GRB sample, the increase in GRBs in the HCBGW was 42\%.
An analysis of possible extinction and exposure biases by \cite{bib:Ukwatta2016} found that the HCBGW anisotropies can be explained away, even though these biases were excluded as the cause of the clustering in \cite{bib:Horvath2014} and \cite{bib:Horvath2015}. 
Challenging the statistical approaches employed in \cite{bib:Horvath2014} and \cite{bib:Horvath2015}, \cite{bib:Christian2020} used a further increased GRB sample of 520 GRBs to reproduce all analyses performed in \cite{bib:Horvath2014}. 
As the final result, \cite{bib:Christian2020} deems the statistical significance of the HCBGW as exaggerated.
Due to these contradicting results, \cite{bib:Horvath2020} not only increased the data set of GRBs further for a third existence test of the HCBGW, but they also investigated potential astrophysical origins and formation scenarios. 
On the whole, further support and corroboration studies are thus required to settle the controversies over its existence, as stated in \cite{bib:Horvath2020}.}

\bcomm{One argument in favour of the existence of the HCBGW is the discovery of the second-largest structure in the universe, the GGR, which was discovered in a similar way, scanning existing data for correlations in GRB observations to test the CP and, serendipitously finding this excess in GRBs now called the GGR.
As detailed in \cite{bib:Balazs2015}, the GGR has an extension of about 1.7~Gpc in diameter and lies in a redshift range between 0.78 and 0.86.
The ring-like structure, as shown in Fig.~\ref{fig:GRB_structures} (right), only consists of 9 GRBs, which motivated a follow-up study to find similar structures in \cite{bib:Balazs2018}, discovering three more, yet less extreme, ring-like patterns to support their original discovery.}

\bcomm{Whether or not these GRB-based largest structures exist will be subject to further sky surveys. 
One may argue that the increasing amount of data becoming available for an extended cosmic volume also increases the probability of chance alignments, such that these largest structures can be explained as a `look-elsewhere-effect'.
While this is well possible, the main criticism, however, is often related to the over-estimated statistical significance of the clustering rather than the observations themselves.
This is sometimes also called `$p$-value-fishing'. 
But regardless of whether these clusters are accepted as cosmic structures or not, several (smaller) structures have also been found that are not based on sparse, transient GRB signals and that still challenge the expected scale of homogeneity set up by \cite{bib:Yadav2010}. 
Apart from the statistical significance and the look-elsewhere-effect, there is also an ongoing debate where the scale of homogeneity should be placed. 
For instance, \cite{bib:Canay2020} claim that inhomogeneous structure growth leading to clusters like the HCBGW is possible on Gpc scales within their framework of perturbative GR, so that all observed structures would still be compatible with the $\Lambda$CDM concordance cosmology.}

\newpage
\section{Early Universe FLRW anomalies}
\label{sec:early_universe_FLRW_anomalies}

\begin{quote}
\textit{The real reason for our adherence to the Cosmological Principle is not that it is surely correct, but rather, that it allows us to make use of the extremely limited data provided to cosmology by observational astronomy.} --- Steven Weinberg
\end{quote} 

Naturally enough, Weinberg has a point and observational cosmologists need to be pragmatic. Evidently, the Universe is homogeneous and isotropic by default if one has little or no data. That being said, CMB has ushered in  a golden era of precision cosmology, allowing us to start to question even the most fundamental assumptions. Thus, our goal in this section and subsequent sections \ref{sec:late_universe_FLRW_anomalies} and \ref{sec:alignments} is to present a number of observations that appear anomalous from the perspective of FLRW. The terminology `$\Lambda$CDM tensions' has entered the cosmology lexicon \cite{bib:Perivolaropoulos2021}, so here we discuss FLRW anomalies. As detailed in sections~\ref{sec:prologue} and \ref{sec:Hom-scale}, the Universe is statistically homogeneous at scales beyond $\sim 70 \, h^{-1}\,$Mpc \cite{bib:Hogg2005, bib:Yadav2005, bib:Scrimgeour2012, bib:Laurent2016, bib:Ntelis2017, bib:Sarkar2019, bib:Goncalves2018, bib:Goncalves2021}, subject to the proviso that one only considers two-point correlations (see comments in section~\ref{sec:Hom-scale}). But this does not guarantee the constancy of the Hubble expansion rate over the sky or that the kinematic CMB dipole is recovered in the matter sector at large scales because these are prerequisites for any FLRW universe. Alternatively put, it is extremely difficult to confirm FLRW, since \textit{this is the default that one recovers either with poor quality data or insensitive cosmological probes}. For this reason, one can only hope to falsify FLRW by focusing on the anomalies. Moreover, to return to the status quo, these anomalies themselves need to be falsified or diluted in statistical significance. Our purpose here is to put them on the table, to carefully outline the assumptions and invite readers to engage with and address the anomalies.

We begin the list with a subset of anomalies seen in CMB temperature anisotropy data exhibiting directional dependence, before progressing to anomalies in the late Universe in section \ref{sec:late_universe_FLRW_anomalies}. These anomalies concern the higher order multipoles $\ell = 2, 3, \dots$ beyond the kinematic dipole $\ell =1$. Note, the latter has a much larger amplitude in the $m K$ range and has to be subtracted to process the smaller temperature anisotropies in the $\mu K$ range. It is worth noting that CMB anomalies with statistical significance $2$-$3 \, \sigma$ are well documented (see, for example, \cite{bib:Schwarz2016} for a review). But the prevailing narrative is that the anomalies are \bcomm{of marginal significance, in other words, benign, especially since one has to resort to \textit{a posteriori} statistics\footnote{\bcomm{This complaint is of less concern when one repeats an experiment and finds the same deviating results. Some trace of the CMB anomalies we document in this section can be found in both WMAP and Planck data sets. Indeed, little progress elucidating the origin of the anomalies was made by the Planck satellite and we await polarization data from LiteBIRD \cite{bib:Matsumara2014}.}}}, and do not affect cosmological parameters. 
Emerging results \cite{bib:Fosalba2021, bib:Yeung2022}, which are supported by, for example, Ref.~\cite{bib:Axelsson2013}, and contradicted elsewhere by Ref.~\cite{bib:Mukherjee2016, bib:Mukherjee2018} in the literature, suggest this may no longer be the case and the implications are profound. In short, if \bcomm{the} Planck \bcomm{Collaboration} \cite{bib:Planck2018b} has underestimated directional systematic uncertainties, or stated differently, if the flat $\Lambda$CDM model needs to be extended within the FLRW paradigm to ensure cosmological parameters have no directional dependence, this has knock on implications for both precision cosmology, in particular constraints on the neutrino masses \cite{bib:DiValentino2021b, bib:Jimenez2022}, and the Hubble and $S_8$ tensions \cite{bib:Huang2020, bib:deJaeger2020, bib:Pesce2020, bib:Kourkchi2020, bib:Schombert2020, bib:Khetan2021, bib:Blakeslee2021, bib:Freedman2021, bib:Riess2021, bib:Asgari2021, bib:Heymans2021, bib:Amon2022, bib:Abbott2022}. It is hard to imagine the findings of Ref.~\cite{bib:Fosalba2021, bib:Yeung2022} without the CMB anomalies, especially the directional anomalies \cite{bib:Eriksen2003, bib:Park2003, bib:Hansen2004, bib:Eriksen2007, bib:Hansen2009, bib:deOliveira2004,bib:Ralston2004, bib:Schwarz2004, bib:Land2005, bib:Kim2010, bib:Planck2013, bib:Planck2016, bib:Planck2018c}. So we also do due diligence by reviewing these results. A key point is that despite the kinematic CMB dipole being subtracted, some residual of this direction appears to remain imprinted through CMB anomalies, as we explain in sections \ref{sec:quadrupole_octopole} and \ref{sec:parity}. Furthermore, preferred axes noted in other astronomical surveys spanning the electromagnetic spectrum were also found to be aligned with the CMB kinematic dipole, pointing towards the Virgo cluster~\cite{bib:Ralston2004,bib:Ralston2010}.
The standard narrative says that these are coincidences. Moreover, consistency checks internal to the CMB have been performed and the results are consistent with a purely kinematic interpretation for the CMB dipole \cite{bib:Ferreira2021, bib:Saha2021}.  

In the standard inflationary scenario, both primordial scalar and tensor perturbations are random Gaussian fields. To linear order approximation, the CMB temperature fluctuations also satisfy a random Gaussian distribution and are described by a scalar field on the two-dimensional celestial sphere.
Since spherical harmonics ($Y_{lm}(\hat{n})$) form a natural basis for expanding a function defined on a sphere, CMB anisotropies are conventionally expanded as $\Delta T(\hat{n}) = T(\hat{n}) - T_0=\sum_{\ell=1}^{\infty} \sum_{m=-\ell}^{\ell} a_{\ell m} Y_{\ell m} (\hat{n})$, where $\hat{n}$ denotes the direction on the sky, $T_0\approx2.73$ Kelvin is the mean sky CMB temperature, and $a_{\ell m}$ are the corresponding expansion coefficients. For an isotropic Gaussian random field as is the CMB in the standard model, the amplitudes of the coefficients, $|a_{\ell m}|$, are distributed according to Rayleigh's probability distribution function, and the phases of $a_{\ell m}$ for $m\neq0$ are expected to be evenly distributed in the range $[0,2\pi]$. Further
its statistics are completely described by the second-order power spectrum, namely $\langle a_{\ell m}a^*_{\ell' m'}\rangle=C_{\ell}\delta_{\ell\ell'}\delta_{mm'}$, where $\langle ... \rangle$ denotes an average over the statistical ensemble of realizations, and the power spectrum $C_{\ell}$ is independent of the magnetic quantum number $m$\footnote{\bcomm{Note that CMB physics assumes Gaussian primordial fluctuations in addition to isotropy/homogeneity and that the statistical independence of modes, i.~e. $\langle a_{\ell m}a^*_{\ell' m'}\rangle \propto \delta_{\ell\ell'}\delta_{mm'}$, follows from these assumptions. See \cite{bib:Abramo2010} for discussion on the differences between these assumptions. Thus, it is in principle possible that some CMB anomalies could be due to non-Gaussianity even in the context of FLRW \cite{bib:Schmidt2012, bib:Namjoo2013, bib:Adhikari2015b}. There is only very weak evidence ($1$--$2\,\sigma$) of non-Gaussianity in the Planck data \cite{bib:Buchert2017}, however.}}. In real measurements, one typically has to construct and work with estimators. For the full-sky map, if the noise is negligible, the best unbiased estimator for $C_{\ell}$ is $\hat{C}_{\ell}=\frac{1}{2\ell+1} \sum_{m=-\ell}^{\ell} a_{\ell m} a_{\ell m}^*$.

\subsection{Quadrupole and octopole alignments}
\label{sec:quadrupole_octopole}
In the flat $\Lambda$CDM cosmological model, all CMB coefficients $a_{\ell m}$ are expected to be statistically independent. Nevertheless, $a_{2m}$ and $a_{3m}$ seem to be anomalously correlated, \bcomm{which, as explained earlier, hints at violations of fundamental assumptions, such as FLRW or Gaussianity}. Statistics such as Angular Momentum Maximization (AMM)~\cite{bib:deOliveira2004} method, Maxwell's Multipole Vectors (MMVs)~\cite{bib:Schwarz2004}, and Power Tensor (PT) method~\citep{bib:Ralston2004,bib:Samal2008} were employed to understand this spurious alignment between $\ell=2,3$ CMB modes. When using MMVs~\cite{bib:Schwarz2004}, it was found that the normals perpendicular to the planes defined by $v^{2, 1}$ and $v^{2,2}$, $v^{3,1}$ and $v^{3,2}$, $v^{3, 2}$ and $v^{3,3}$, and finally $v^{3,3}$ and $v^{3,1}$, were unusually aligned, in addition to being close to the CMB dipole direction. \bcomm{Thus,} the three octopole vectors inhabit a plane that is close to that defined by the quadrupole vectors. These alignments were further found to be robust against any galactic residuals when analyzed accounting for or excising potentially contaminated regions in the \emph{cleaned} CMB map~\cite{bib:Bielewicz2005,bib:deOliveira2006,bib:Aluri2011}. Also the quadrupole and octopole planes become more pronouncedly aligned when one takes into account the Doppler contribution to the quadrupole~\cite{bib:Copi2015,bib:Notari2015}, reducing the p-value from $0.1\%$ to $0.02\%$, thereby further reducing the probability of such a configuration arising from a statistical fluctuation. The alignment plane is perpendicular to the ecliptic, i.~e.~the plane of our orbit around the sun. This anomaly survived the transition from WMAP to Planck era, which seems to rule out systematics. 
Despite accounting for kinematic dipole contributions to higher multipoles, a dominant dipole mode measured with complex satellite scanning strategy using a detector with non-circular beam can leak power into quadrupole, and also into octopole mode to a lesser extent~\cite{bib:Santanu2012}. It was further noted that for similar detector beam shapes this leakage is more for WMAP scan strategy in comparison to Planck. This prospect can provide a plausible resolution to the anomalous alignment of low multipoles. Here we also recall that the WMAP and Planck satellites made observations from Lagrange point 2 ($L_2$) while going round the sun (in ecliptic plane) from behind the Earth. But it would be difficult to explain this in terms of a solar system foreground (see section 4.1 of \cite{bib:Burigana2013}, and also \cite{bib:Dikarev2015}).

Some of the higher order multipoles ($\ell>3$) of CMB sky were also found to be initially aligned with the quadrupole and octopole (besides CMB dipole) \cite{bib:Samal2008}. But other than $\ell=3$, the set of modes aligned closely with $\ell=2$ (and $\ell=1$) varied depending on the data release, and their significance also reduced with more data from WMAP and Planck missions \cite{bib:Samal2009,bib:Pranati2015}. Finally, as we note in section~\ref{sec:hemispherical_power_asymmetry}, the lower CMB variance in the northern ecliptic plane seems connected to this alignment.

\subsection{Hemispherical power asymmetry}
\label{sec:hemispherical_power_asymmetry}

The northern and southern ecliptic hemispheres of CMB sky seem to have distinct statistical properties. The southern hemisphere appears to be consistent with an isotropic CMB sky expected from standard $\Lambda$CDM model, suggesting that the CMB anomalies may have their origin largely in the northern hemisphere. In particular, a low variance for CMB temperature fluctuations was reported in WMAP data \cite{bib:Monteserin2008, bib:Cruz2011, bib:Gruppuso2013}, which seemed intimately connected to the quadrupole-octopole modes (see section \ref{sec:quadrupole_octopole}), as the anomaly disappeared once the quadrupole and octopole were removed \cite{bib:Cruz2011}. Follow up studies by Planck \cite{bib:Planck2013} confirmed that the low variance was localized in northern ecliptic hemisphere, with a p-value of $\sim 0.1\%$ versus a p-value of $\sim 45\%$ for the southern hemisphere. Planck also confirmed the connection between the quadrupole-octopole modes and the low variance \cite{bib:Planck2013}.

Moving along, in \cite{bib:Eriksen2003,bib:Hansen2004} tesselating the CMB sky with evenly spaced discs whose disc centers were used to estimate the power spectrum locally, revealed power asymmetry in opposite hemispheres with excess power centered at the Galactic coordinates $(l,b)=(237^\circ, -10^\circ)$ when using multipoles $\ell=2$-$40$.
At the time, it was concluded that \textit{``asymmetric distributions of power on the sky provide a serious test for the cosmological principle of isotropy"} \cite{bib:Hansen2004}. This hemispherical power asymmetry (HPA) in WMAP data was also evident when using other statistical measures, such as Minkowski functionals \cite{bib:Park2003}. This anomaly persisted in Planck data also \cite{bib:Planck2013, bib:Planck2016, bib:Planck2018c}. The significance of HPA as seen in WMAP and Planck data was evaluated using a variety of methods in both harmonic and real space \cite{bib:Lew2008, bib:Paci2010, bib:Pranati2013, bib:Flender2013, bib:Akrami2014, bib:Quartin2015, bib:Adhikari2015, bib:Aiola2015,bib:Shabbir2019}, all placing it at about a $3\, \sigma$ confidence level.

While the asymmetry was originally observed only at the largest scales, it was later shown to extend to much smaller scales in five-year WMAP data \cite{bib:Hansen2009}. Concretely, over the extended range $\ell =2$-$600$, it was demonstrated in comparison to simulations that there is significantly more power in the hemisphere centered at $(l, b) = (226^{\circ} \pm 10^{\circ}, -17^{\circ} \pm 10^{\circ})$ in Galactic coordinates. Interestingly, the authors decomposed the CMB power spectrum in bins of 100 multipoles between $\ell = 2$-$600$ and recovered the asymmetry dipole from each bin, thereby demonstrating that there is a significant alignment among dipoles recovered from distinct bins and this feature is present even at smaller scales. It was found through $\sim 10,000$ simulations that not a single realization produced a similarly strong asymmetry, while systematic effects and foregrounds were eliminated as an explanation. Later, Planck conducted a similar analysis \cite{bib:Planck2013, bib:Planck2016} by studying the power spectrum amplitude in non-overlapping directions in bins of $\Delta \ell = 8, 16, 32$ up to maximum multipoles of $\ell_\mathrm{max}=600$ and $\ell_\mathrm{max}=1500$. In \cite{bib:Planck2013} a dipole was fitted to each bin and the clustering of dipoles was studied, concluding that none of 500 simulations showed clustering larger than the real data. Nevertheless, it was noted that the significance of the anomaly reduced at higher multipoles when Doppler effects were taken into account, in line with the findings of \cite{bib:Flender2013, bib:Adhikari2015, bib:Quartin2015}. Final results quoted a mean dipole direction of $(l,b) = (218^\circ,-21^\circ)$ \cite{bib:Planck2013}, thereby recovering the earlier result reported in \cite{bib:Hansen2009}.

The hemispherical power asymmetry was also studied from different perspectives. In \cite{bib:Gordon2005}, a phenomenological model wherein the CMB sky is supposedly modulated by a dipole field was introduced as,
\begin{equation}
    \label{eq:dipole}
    \frac{\Delta T}{T}|_{\textrm{mod}}({\bf \hat{n}})=(1+\alpha \,{\bf \hat{n}}\cdot{\bf \hat{p}})\frac{\Delta T}{T}|_{\rm iso}({\bf \hat{n}})\;,
\end{equation}
where $\frac{\Delta T}{T}|_{\textrm{iso}}$ and $\frac{\Delta T}{T}|_{\textrm{mod}}$ are the isotropic and modulated CMB temperature fluctuations, respectively, along a direction ${\bf \hat{n}}$ on the sky. The parameter $\alpha$ keeps track of the amplitude of the dipole modulation and ${\bf \hat{p}}$ is the direction of the modulating dipole field. This represents a particularly simple model to explain the hemispherical power asymmetry.
Nevertheless, this simplified approach sufficiently explains the observed asymmetry in the low-$\ell$ regime, as higher order modulation effects were found to be insignificant \cite{bib:Planck2013,bib:Akrami2014}. We also note that while the dipole modulation model of (\ref{eq:dipole}) assumes $\alpha$ to be a constant, this is unlikely to be the case as variations of $\alpha$ with $\ell$ have been reported \cite{bib:Lew2008, bib:Hanson2009, bib:Bennett2011, bib:Pranati2013, bib:Quartin2015, bib:Aiola2015}.

The first studies of this model using WMAP data concluded that the model consisting of an isotropic CMB sky modulated by a dipole field gives a substantially better fit to observations than a purely isotropic model \cite{bib:Eriksen2007,bib:Hoftuft2009}. Despite the simplicity of the model given by (\ref{eq:dipole}), the preferred axis was found tobe $(l,b) = (225^\circ, -27^\circ)$ in three-year WMAP data \cite{bib:Eriksen2007} and $(l, b) = (224^\circ, -22^\circ)$ when using five-year WMAP data \cite{bib:Hoftuft2009}, in good agreement with earlier results from \cite{bib:Eriksen2003, bib:Hansen2004}. In particular, \cite{bib:Hoftuft2009} extended the multipole range from $\ell_{\textrm{max}} = 40$ to $\ell_{\textrm{max}} = 64$, finding that the significance of non-zero $\alpha$ increased from $2.8 \, \sigma$ to $3.3 \, \sigma$. In \cite{bib:Planck2013} and \cite{bib:Planck2016}, the Planck collaboration revisited these claims of hemispherical power asymmetry, noting that their findings also lead to the same conclusion. It was observed that the significance levels for the amplitude $\alpha$ compared to $\Lambda$CDM simulations varied with smoothing scale and the $5^\circ$ scale showed the highest significance ($\sim 3.5 \, \sigma$). 

Independent studies performed in harmonic-space, for example, in \cite{bib:Paci2010, bib:Pranati2013, bib:Aiola2015, bib:Shabbir2019}, confirmed the presence of power asymmetry at large angular scales as expected. The dipole modulation model was also formulated in terms of Bipolar spherical harmonics (BipoSH) \cite{bib:Hajian2003, bib:Hajian2006, bib:Planck2013, bib:Planck2016,bib:Aluri2015}. When this BipoSH estimator was applied to Planck data, once again, a dipole modulation was detected with a significance $> 3 \, \sigma$. However, higher-order multipole modulations (beyond dipole) were found to be insignificant \cite{bib:Planck2013}. Interestingly enough, a joint Bayesian analysis of the dipole modulation field when constrained along with some other low multipole anomalies using the BipoSH framework did not markedly favor the dipole modulation model over the isotropic model. A local variance estimator (LVE) was introduced in \cite{bib:Akrami2014} based on the expectation that any random patch of the CMB sky should be statistically consistent with any other randomly chosen patch of any shape. LVE was used in the analyses of Planck full-mission and legacy data~\cite{bib:Planck2016, bib:Planck2018c} on both temperature and polarization maps. Using this estimator, the Planck collaboration found that the HPA directions from E-mode polarization data were consistent with those seen in the temperature data \cite{bib:Planck2018c}. Some differences in statistical significance were observed for different component-separated polarization maps.

Overall, results across different satellites, both with and without polarization data, seem to confirm the underlying hemispherical power asymmetry in the CMB sky.

\subsection{Odd mirror parity anomaly}
\label{sec:cmbmirrorparity} 

In this section we discuss the CMB mirror parity anomaly that has an odd parity preference. In the next section, we discuss anomalous power asymmetry between even and odd multipoles arising from odd point parity preference in the observed CMB sky. As is obvious, it involves searching for an axis in the CMB sky along which maximum mirror reflection (a)symmetry may be present.
In \cite{bib:deOliveira2004}, a real space statistic
\begin{equation}
    S(\hat{n}_i) = \frac{1}{N_{pix}}\sum_{j=1}^{N_{\rm pix}} [\Delta T(\hat{n}_j) - \Delta T(\hat{n}_{ij})]^2\;,
\end{equation}
 was introduced to map the mirror parity (a)symmetry in the CMB sky, where $\hat{n}_j$ and $\hat{n}_{ij}=\bcomm{\hat{n}_j} - 2(\hat{n}_i\cdot\hat{n}_j)\hat{n}_i$ are the pixel centers of the $j^{th}$ pixel and its mirror reflection with respect to the plane defined by the normal $\hat{n}_i$, respectively. $\Delta T(\hat{n})$ is the CMB anisotropy map used (after subtracting monople and dipole), and $N_{\rm pix}$ are the number of pixels in that map. Visually, a minimum was noticed along the CMB dipole direction when applied to study a foreground cleaned CMB map derived from WMAP first-year data, but no significant mirror (a)symmetry was found. However, when the statistic was applied to $\ell=3$ mode alone, a significant mirror symmetry with a $p$-value of less than $5\%$ was found in comparison to \bcomm{$\Lambda$CDM simulations based on WMAP best-fit parameters \cite{bib:deOliveira2004}}. Those authors further reported a visual minimum along the direction of CMB dipole in the $S$-map of quadrupole and octopole alone, and their combination map ($\ell=2+3$), but not in other modes. This even mirror parity preference was independently confirmed in \cite{bib:Land2005}, but over an extended multipole range $\ell=2$-$5$ however insignificant with a $p$-value of about $10\%$.

With successive data releases from WMAP, further tests of mirror parity (a)symmetry preference in the CMB sky revealed an anomalous odd parity preference. In \cite{bib:Gruppuso2011}, the $S$-map diagnostic was modified to separately quantify even or odd mirror parity along a chosen axis as $ S_\pm(\hat{n}_i) = \sum_{j=1}^{N_{\rm pix}} [\Delta T(\hat{n}_j) \pm \Delta T(\hat{n}_{ij})]^2 $ (apart from some prefactors). Evidently, the maxima of the $S_{+}$ map correspond to the minima of the $S_{-}$ map, and vice versa. An odd mirror parity preference was reported by those authors at about 93\% confidence level for the full sky CMB map from WMAP seven-year data in the direction $(l,b)=(264^\circ,-17^\circ)$ (degrees) in Galactic coordinates, which is in proximity to the Hemispherical power asymmetry direction. Interestingly, while the even mirror parity maximum was found to be insignificant, its direction was reported to be closely aligned with the CMB kinematic dipole. Mirror parity handedness in the microwave sky was also tested in harmonic space. In a given frame, the spherical harmonic coefficients, $a_{\ell m}$, corresponding to even or odd $\ell+m$ combinations correspond to even or odd mirror parity modes. Thus a statistic, $\tilde{S}(\hat{n}) = \sum_{\ell=2}^{\ell_{\rm max}}\sum_{m=-\ell}^{+\ell} (-1)^{\ell+m} a_{\ell m}(\hat{n})/\hat{C}_\ell$, can be defined \cite{bib:Ben-David2012} where $a_{\ell m}(\hat{n})$ and $\hat{C}_\ell$ are the spherical harmonic coefficients and corresponding power spectrum as estimated in a rotated frame with $\hat{n}$ as the $z$-axis. For an isotropic CMB sky, the expectation value $\langle \tilde{S} \rangle$ is given by ``$\ell_{max}-1$'' which is subtracted to define the zero-mean mirror parity map as $S(\hat{n})=\tilde{S}(\hat{n})- \langle \tilde{S} \rangle$. In line with the findings of \cite{bib:Finelli2012}, those authors also report an unusual odd mirror parity preference in the CMB map derived from WMAP seven-year data in the same direction ($(l,b)=(264^\circ,-18^\circ)$ in Galactic coordinates) but with a $3.6 \,\sigma$ confidence level. However, the authors \cite{bib:Ben-David2012} considered only low mulitpoles in their analysis viz., $\ell=2$-$7$. By choosing higher values of $\ell_{\rm max}$, the significance was found to decrease.

With the advent of Planck era, the odd mirror parity anomaly was affirmed with a $p$-value of $0.5\%$ to $8.9\%$ in 2013 data and $1.6\%$-$2.7\%$ when using full mission data, the variations being from the specific foreground removal technique used to clean the raw satellite data for estimating the CMB map \cite{bib:Planck2013,bib:Planck2016}. In summary, an odd mirror parity anomaly was detected in the CMB sky with a $2\, \sigma$ confidence level or better with the maximum pointing in the direction $(l,b)=(264^\circ, -17.0^\circ)$ that is closer to the direction of HPA discussed in section~\ref{sec:hemispherical_power_asymmetry}, indicating a possible connection between them. Such a mirror parity in CMB data suggests a small universe with non-trivial topology. In particular, it suggests a toroidal topology for our Universe instead of an infinite universe for the realization that we exist in~\cite{bib:Starobinskij1993,bib:Stevens1993,bib:deOliveira2004}.

\subsection{Parity violation and its directionality}
\label{sec:parity}

In the standard model of cosmology, the two-point correlation function, $C(\hat{n}_1,\hat{n}_2)$, of the CMB sky is independent of the two directions, $\hat{n}_1$ and $\hat{n}_2$, individually but depends only on the angular separation between them. It is given by $C(\alpha) = \sum_{\ell=2}^\infty \frac{2\ell+1}{4\pi} C_\ell P_\ell(\cos\alpha)$, entirely in terms of the power spectrum $C_\ell$ and the separation angle $\alpha=\arccos(\hat{n}_1\cdot\hat{n}_2)$. If one were to consider antipodal correlations, we get $C(\pi) = \sum_{\ell=2}^\infty \frac{2\ell+1}{4\pi} (-1)^\ell C_\ell = \mathcal{C}^+_\ell - \mathcal{C}^-_\ell$, where $\mathcal{C}_\ell = \sum_{\ell} (2\ell+1) C_\ell/4\pi$ and $\mathcal{C}^\pm_l$ represent contributions from CMB even ($+$) or odd ($-$) multipoles alone. Significant power asymmetry between even and odd multipoles may then be interpreted as a preference for a particular parity, which is not expected in the CMB sky. Probing for such a preference or lack thereof using WMAP first-year data has indicated an odd parity preference at about $2\, \sigma$ level in the low multipole regime ($\ell=2$-$20$) \cite{bib:Land2005}.

In order to quantify this asymmetry, the following statistics are defined \cite{bib:Kim2010,bib:Kim2012}:
\begin{eqnarray}
P^{+}=\sum_{\ell=2}^{\ell_{\max}}\mathcal{D}_{\ell}\Gamma_{\ell}^{+}\,, \quad
P^{-}=\sum_{\ell=2}^{\ell_{\max}}\mathcal{D}_{\ell}\Gamma_{\ell}^{-}\,,
\end{eqnarray}
where $\mathcal{D}_\ell=\ell(\ell+1)\hat{C}_\ell/2\pi$, $\Gamma_{\ell}^{\pm}=[1\pm(-1)^l]/2$,\footnote{Alternatively, $\Gamma_{\ell}^{+}=\cos^2\left(\frac{\ell\pi}{2}\right)$ and $\Gamma_{\ell}^{-}=\sin^2\left(\frac{\ell\pi}{2}\right)$ are also used.} and $P^{\pm}$ is the total power in even ($+$) or odd ($-$) multipoles with the exclusion of the dipole ($\ell=1$) up to a maximum multipole, $\ell_{\rm max}$. Therefore, the ratio $P(\ell_{\rm max})=P^{+}/P^{-}$ can be used to assess the degree of parity asymmetry, where lower values of $P(\ell_{\rm max})$ indicate an odd-parity preference, and \textit{vice versa}. In the WMAP seven-year data, an odd parity preference in the CMB sky was found with a higher significance at about $99.5\%$ confidence level for $\ell_{\rm max}=22$ \cite{bib:Kim2010,bib:Gruppuso2011}. This is the so-called CMB parity asymmetry anomaly.
With an alternative statistic \cite{bib:Aluri2012}, defined as $Q(\ell_{\rm odd})=1/(\ell_{\rm max}-1)\sum_{\ell=3}^{\ell_{\rm odd}} \mathcal{D}_{\ell-1}/\mathcal{D}_\ell$, with the summation taken over odd multipoles, this anomaly was analyzed in more detail in the multipole range $\ell=2$-$101$ (i.e., the first 100 multipoles) to find parity asymmetry anomaly at a $3\, \sigma$ confidence level over a range of multipoles viz., $\ell_{\rm max}=18$-$33$. Considering CMB maps recovered using various cleaning procedures, incorporating additive and multiplicative unknown residual foregrounds/modulation in simulated CMB maps that could induce an odd parity preference, and also the effect of masking, the odd parity preference seen in the data was found to be robust. Further, modifying the statistic as $Q(\ell_{\rm odd})=1/(\ell_{\rm max}-\ell_{\rm cut}+1)\sum_{\ell=\ell_{\rm cut}}^{\ell_{\rm odd}} D_{\ell-l}/D_\ell$ to determine the impact of low-multipoles on the parity asymmetry, it was found that ignoring the first six multipoles i.~e. for $\ell_{\rm cut}=8$ and beyond, the anomalous nature of parity asymmetry drops below $95\%$ confidence level (see Fig.~15, 16, and 17 of \cite{bib:Aluri2012}).

The Planck collaboration also analyzed this anomaly \cite{bib:Planck2013,bib:Planck2016,bib:Planck2018c} finding a lower-tail probability of $0.1$-$0.4\%$ depending on the data release and component separated CMB map used, with the multipole range $\ell_{max}=20$-$30$ generally lying outside the $2\, \sigma$ confidence level. A similar analysis was performed on CMB polarization data using WMAP and Planck observations to find no significant odd parity asymmetry as seen in CMB temperature data \cite{bib:Gruppuso2011,bib:Planck2018c}. However, due to a low signal-to-noise ratio of WMAP and Planck in measuring CMB polarization, this may change.
This disparity in CMB power between even and odd multipoles was also found to affect the cosmological parameters thus inferred (see Fig.~6 of \cite{bib:Kim2010}).

Now, we move on to discuss the directional properties of the CMB point parity asymmetry. In order to realize this study, one needs a statistic that is explicitly directional in nature. Thus replacing the estimator $\hat{C}_{\ell}$ with a rotationally variant estimator defined as \cite{bib:Naselsky2012,bib:Zhao2014}
\begin{equation}\label{D_l}
\vspace{-1ex}
\hat{D}_{\ell}=\frac{1}{\bcomm{2\ell+1}}\sum_{m=-\ell}^{\ell} a_{\ell m} a_{\ell m}^*(1-\delta_{m0})\;,
\end{equation}
where $\delta_{mm'}$ is the Kronecker symbol, the objective of probing any directional dependence of point parity asymmetry is achieved. From the above definition, $\hat{D}_{\ell}$ is also an unbiased estimator for the power spectrum $C_{\ell}$ for any given multipole, i.e., $\langle \hat{D}_{\ell}\rangle=C_{\ell}$. Now, the estimator $\hat{D}_{\ell}$ can be studied in any coordinate system. Imagining that the Galactic coordinate system is rotated by the Euler angles $(\psi,\theta,\phi)$, the coefficients $a_{\ell m}(\psi,\theta,\phi)$ in the rotated coordinate frame are given by
\begin{equation}
\vspace{-1ex}
a_{\ell m} {(\psi,\theta,\phi)} =\sum_{\bcomm{m'}=-\ell}^{\ell} a_{\ell m'} D^{\ell}_{m m'}(\psi,\theta,\phi)\,,
\end{equation}
where $a_{\ell m}\equiv a_{\ell m}(0,0,0)$ are the coefficients defined in the Galactic coordinate system, and $D^{\ell}_{m m'}(\psi,\theta,\phi)$ is the Wigner rotation matrix. Similar to $\hat{D}_{\ell}$ in (\eqref{D_l}), now the estimator $\hat{D}_{\ell}(\psi,\theta,\phi)$ can also be defined. It is easy to find that $\hat{D}_{\ell}(\psi,\theta,\phi)$ is independent of the angle $\psi$. So here, it is sufficient to consider only two Euler angles, that are denoted by $\hat{\rm{\bf q}}\equiv(\theta,\phi)$, setting $\psi=0$. If we consider $\hat{\rm{\bf q}}$ as a vector, which labels the $z$-axis of the rotated coordinate system, then $(\theta,\phi)$ are the polar coordinates of this vector in the Galactic coordinate system. In any coordinate frame labeled by $\hat{\rm{\bf q}}$, the components $a_{\ell 0}$ are naturally symmetric around the $z$-axis ($\hat{\rm{\bf q}}$). 
\emph{So from the definition of $\hat{D}_{\ell}$, in which the $m=0$ components are excluded for all $\ell$, we can now compute the rotationally variant quantity $\hat{D}_{\ell}(\hat{\bf q})$ in the rotated frame by setting the coordinates denoted by $\hat{\bf q}$ as our new $z$-axis. Doing so repeatedly for different choices of $\hat{\bf q}$, one gets estimates of $\hat{D}_{\ell}(\hat{\bf q})$ in those different frames}. Based on these quantities, the rotationally variable parity parameter $G_1(\ell;\hat{\rm{\bf q}})$ was defined as \cite{bib:Naselsky2012,bib:Zhao2014},
\begin{equation}
\label{G_1}
G_1(\ell;\hat{\rm{\bf q}})=\frac{\sum_{\ell'=2}^{\ell}{\ell'(\ell'+1)}\hat{D}_{\ell'}(\hat{\rm{\bf q}})\Gamma_{\ell'}^{+}}{\sum_{\ell'=2}^{\ell}{\ell'(\ell'+1)}\hat{D}_{\ell'}(\hat{\rm{\bf q}})\Gamma_{\ell'}^{-}} \;.
\end{equation}
This statistic now represents the degree of parity asymmetry akin to the original parity parameter $P(\ell_{max})=P^+/P^-$, but in the rotated frame labeled by $\hat{\bf q}$.
A value of $G_1<1 (>1)$ indicates an odd (even) parity preference in the data, simultaneously making it possible to find any preferred direction associated with it, which may reveal hints on the origin of observed parity asymmetry in the CMB field. For any given $\ell$, the sky map $G_1(\ell;\hat{\rm{\bf q}})$ can be constructed by considering all directions $\hat{\rm{\bf q}}$. In practice, we pixelize $G_1$ over the full sky in HEALPix format at $N_{\rm side}=64$ and choose the directions $\hat{\rm{\bf q}}$ as the pixel centers of this $N_{\rm side}=64$ HEALPix grid.

\begin{figure}[t]
\begin{center}
\includegraphics[width=0.45\linewidth]{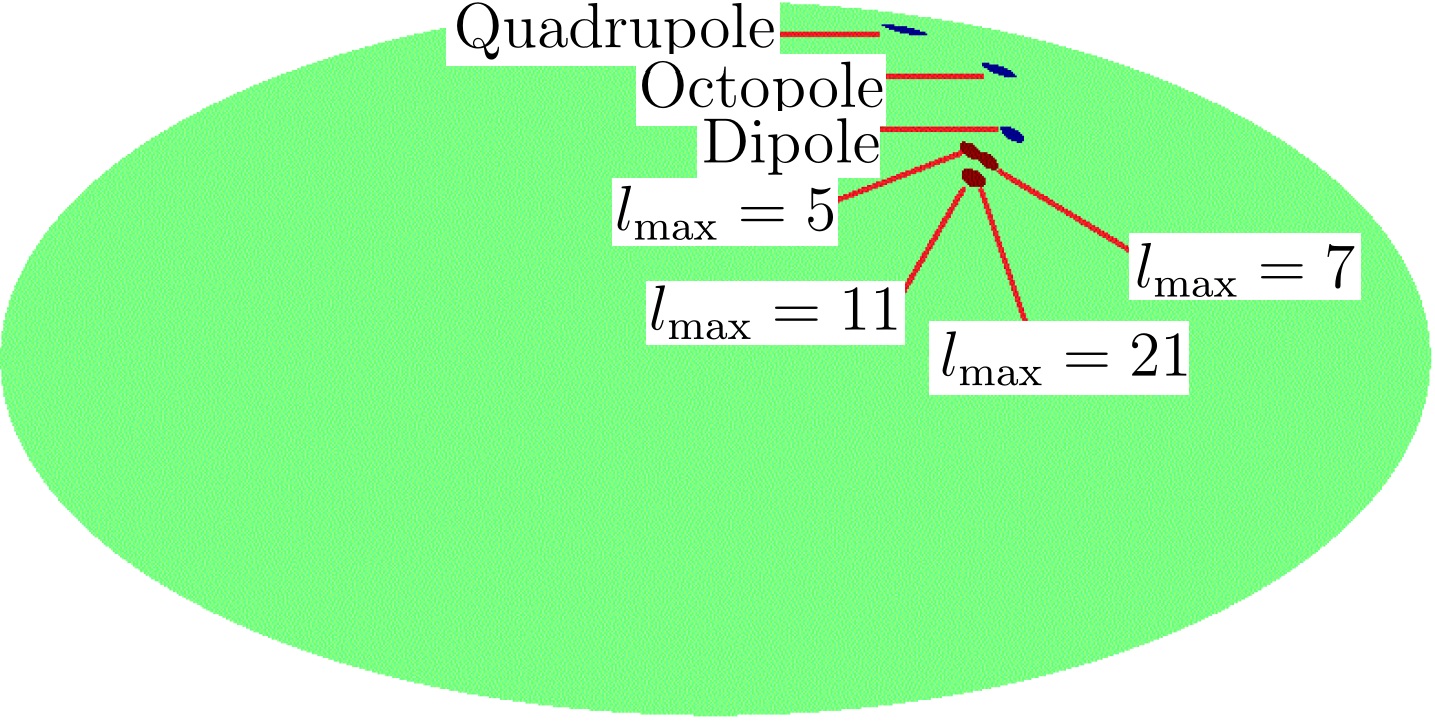}\hspace{2ex}
\includegraphics[width=0.45\linewidth]{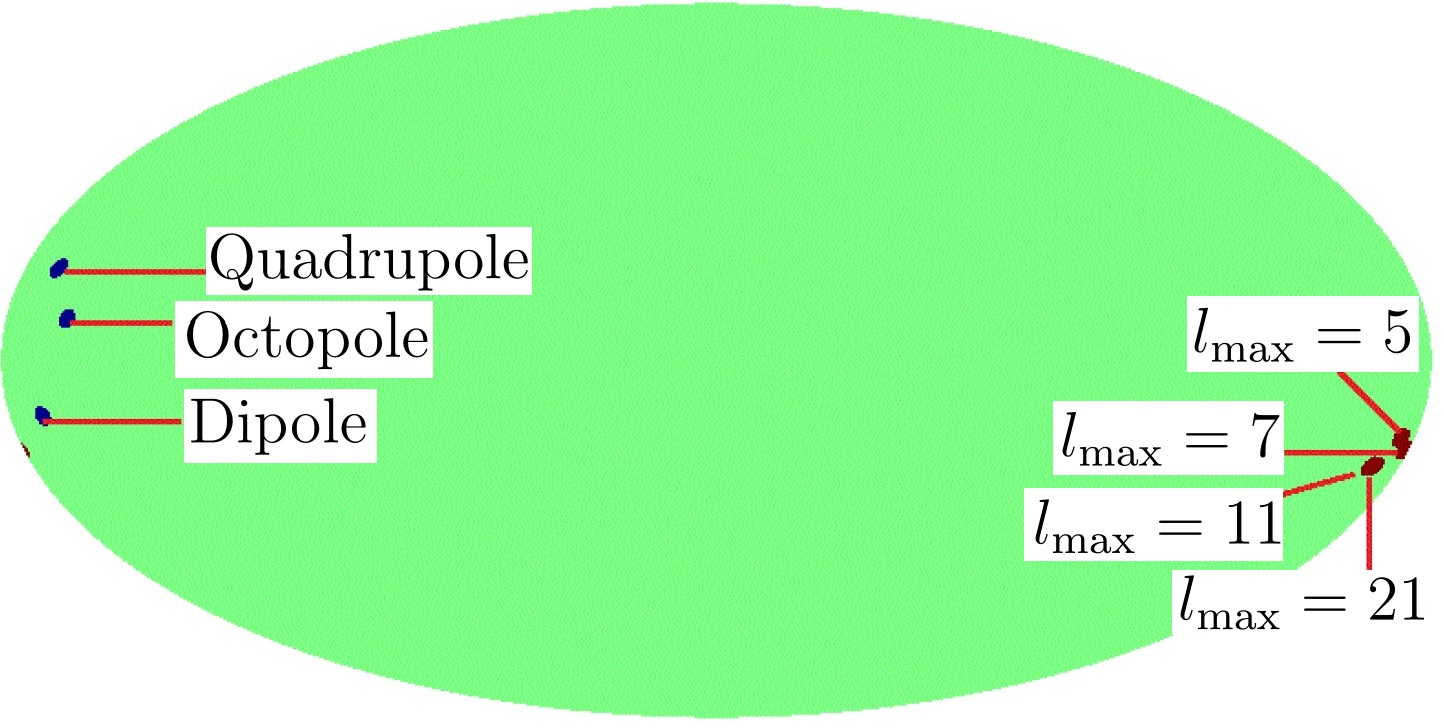}
\end{center}
\vspace{-2ex}
\caption{The preferred directions in the map of $G_1(\ell;\hat{\bf q})$ statistic in the Galactic coordinate system (left) and in the ecliptic coordinate system (right), as seen in Planck 2013 SMICA CMB map \bcomm{(reproduced from Fig. 5 of \cite{bib:Zhao2014})}. In both panels, these preferred directions for parity asymmetry are compared with the CMB kinematic dipole direction and the preferred axes of CMB quadrupole and octopole.}
\label{fig:fig10}
\end{figure}

Using the Planck 2013 SMICA CMB map, the directional parity parameter $G_1(\ell;\hat{\rm{\bf q}})$ was computed to find that $G_1(\ell;\hat{\rm{\bf q}})<1$ holds for any direction $\hat{\rm{\bf q}}$ and maximum multipole $\ell$. This is consistent with the discovery of  odd-parity preference in the Galactic coordinate frame, discussed earlier in this section. As mentioned earlier, a smaller $G_1$ value reflects a higher degree of parity violation, and indeed in maps of $G_1$ statistic its value is $<1$ everywhere, for any chosen maximum multipole $\ell$ except for the case of $\ell=3$. All the $G_1$ maps have quite similar morphologies. In all these cases (i.e., for different $\ell$), the preferred directions thus inferred, as shown in Fig.~\ref{fig:fig10}, come out to be nearly the same.  In order to test the robustness of these preferred directions, six different statistics were considered, while also applying various Galactic cuts and using CMB temperature maps derived from different component separation methods, to find that the conclusions remain unchanged \cite{bib:Naselsky2012,bib:Zhao2014,bib:Zhao2016}.

The preferred directions $\hat{\bf q}$ of parity asymmetry, shown in Fig.~\ref{fig:fig10}, for various choices of $\ell$ are compared with the CMB kinematic dipole direction. As is obvious, they are very close to each other. Further all these directions lie close to the ecliptic plane.
To quantify this trend, we define the parameter $\psi$, which is the angle between $\hat{\bf q}$ and the CMB kinematic dipole
direction, $(l, b) = (264^{\circ}, 48^{\circ})$.
For instance, in the Planck 2013 NILC map, $|\cos\psi|>0.98$ holds for all choices of maximum multipole, which means that the angular separation between the preferred directions $\hat{\rm{\bf q}}$ and the CMB dipole direction are all smaller than \bcomm{$11.5^{\circ}$ \cite{bib:Zhao2014}}. Thus it was concluded that the preferred direction in the CMB parity asymmetry strongly aligns with the CMB kinematic dipole, which is independent of the CMB maps, the directional statistics or the galactic mask used. The physical origin of this preferred direction in CMB is still unclear. However, the alignment of CMB dipole and the preferred direction of CMB parity violation strongly suggests a non-cosmological origin for this large-scale anomaly, which may be a result of unaccounted CMB dipole-related systematics or contamination.

In a different kind of analysis employing the Power tensor (PT) method~\cite{bib:Ralston2004,bib:Samal2008}, any preferred direction for CMB parity asymmetry anomaly is probed by studying collective alignments of anisotropy axes (as found using PT) of even or odd multipoles separately~\cite{bib:Aluri2017}. For any given $\ell$, PT is defined as $A_{ij}(\ell)=\sum_{mm'm''}a_{\ell m}J^i_{mm'} J^j_{m'm''}a_{\ell m''}^*$ up to some prefactors such that $\langle A_{ij} \rangle=\delta_{ij}C_\ell/3$. Here $J_{mm'}^i$ for $i=1,2,3$ are the three $(2\ell+1)$ dimensional angular momentum matrices $J_x$, $J_y$ and $J_z$. Thus, PT maps the complicated pattern of CMB anisotropies in a given $\ell$ on to an ellipsoid, whose axes lengths and frame are given by its three eigenvalues and eigenvectors. One can now associate a preferred axis with each multipole as given by the eigenvector corresponding to the largest eigenvalue of the PT matrix  (referred to as principal eigenvector, PEV for short, and denoted by $\tilde{\bf e}_\ell$). Collective alignments of PEVs can be studied using the Alignment Tensor (AT), $X(\ell_{\rm min},\ell_{\rm max}) = \sum_{\ell=\ell_{\rm min}}^{\ell_{\rm max}} \tilde{\bf e}_\ell \tilde{\bf e}_\ell^\top$ \cite{bib:Samal2008}, constructed by taking the outer product of a set of PEVs in the multipole range $\ell=\ell_{\rm min}$ to $\ell_{\rm max}$. This can also be applied to select multipole PEVs, to find their collective alignment axis as given by the eigenvector of AT corresponding to its largest eigenvalue (similar to how PEVs are defined using PT).

\begin{figure}[t]
\centering
\includegraphics[width=0.79\linewidth]{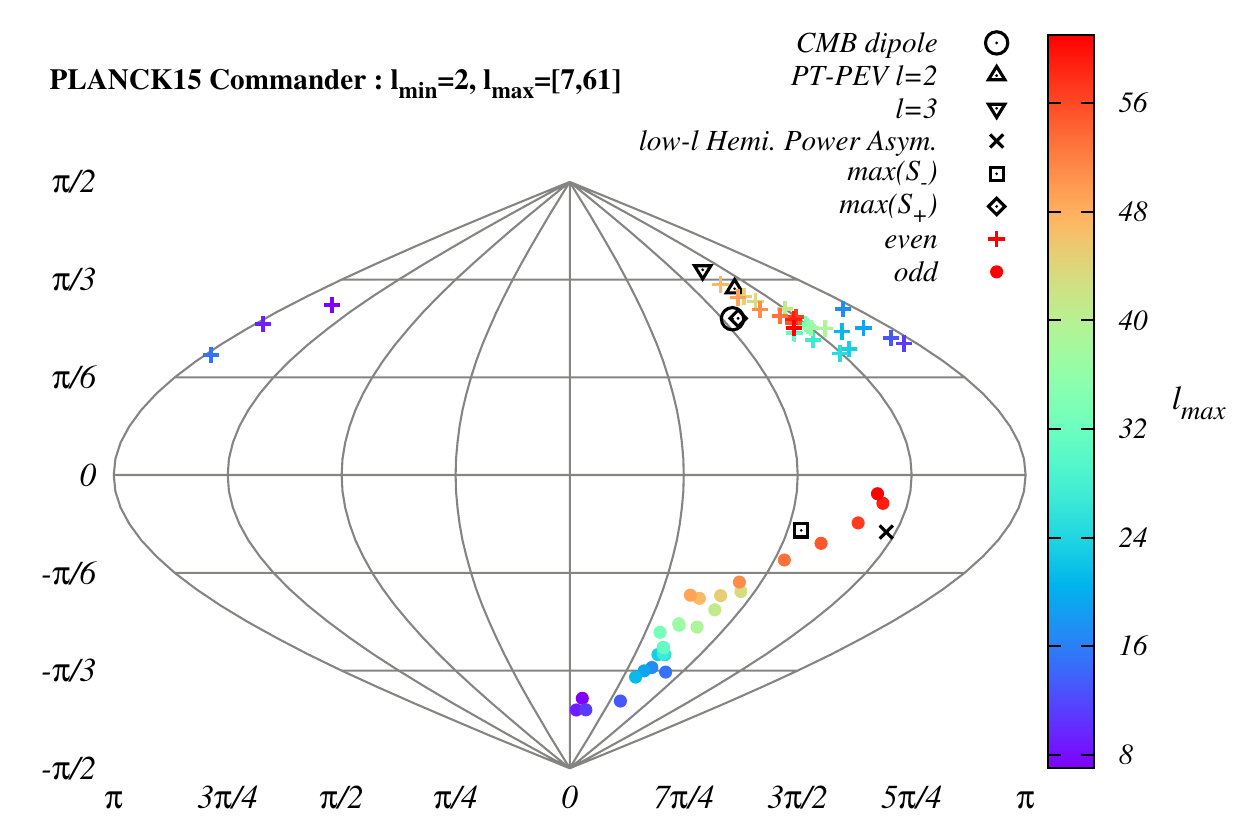}
\vspace{1ex}
\caption{Collective alignment axes of even or odd multipoles of the Planck 2015 Commander map using $\ell=2$-$61$ \bcomm{(reproduced from Fig. 1 of \cite{bib:Aluri2017})}. In a multipole bin $\ell=[2,\ell_{\rm max}]$, collective alignment axes of even or odd multipole PEVs in that bin are computed separately. The maximum multipole is varied as $\ell_{\rm max}=7$-$61$ in steps of two to keep the even/odd multipole PEV number the same for any chosen $\ell_{\rm max}$ in first constructing their respective ATs from which respective collective alignment axes are derived.}
\label{fig:pt-parity}
\end{figure}

In \cite{bib:Aluri2017}, the PT method outlined above was used to study collective alignments among even and odd multipoles separately. The results are shown in Fig.~\ref{fig:pt-parity}.
It was found that the collective alignment vectors thus computed separately for even or odd multipoles in the multipole range $[2,\ell_{\rm max}]$ span two non-overlapping regions. The odd multipole alignment axes span the lower hemisphere where preferred directions of odd mirror parity anomaly and dipolar modulation of the CMB are found as $\ell_{\rm max}$ is being varied. The even multipole PEV collective alignment axes are spread over the northern hemisphere but steadily drifting towards the CMB kinematic dipole with increasing $\ell_{\rm max}$. This is also the axis along which quadrupole and octopole are found to be anomalously aligned and the maximum of even mirror parity direction is found (though the significance of ${\rm max}(S_+)$ is low as discussed in section~\ref{sec:cmbmirrorparity}). We note that the significance of collective alignment axes of odd multipole PEVs is about $2\,\sigma$ or better up to $\ell_{\rm max}=31$ and drops afterwards, but the same for even multipoles remains insignificant for all choices of $\ell_{\rm max}$. These findings differ from \cite{bib:Naselsky2012,bib:Zhao2014,bib:Zhao2016} in that odd multipole PEV alignments (point parity \emph{asymmetry} axes) span the region that also contains preferred directions of previously detected CMB anomalies that are odd parity in nature, and the even multipole PEV collective alignments (point parity \emph{symmetry} axes) extend over the region in which preferred directions of CMB anomalies characterized by even parity fall (except for the CMB dipole direction). This observation suggests that the quadrupole and octopole may be (nearly) anti-parallel rather than parallel, as may have been thought so far. This view is quite possibly due to the fact that all multipole alignment test statistics (see section~\ref{sec:quadrupole_octopole}) return headless vectors i.e., they can only indicate an axis of anisotropy but not a direction.

\vspace{-2ex}
\subsection{Directional Variations of Cosmological Parameters}
\label{sec:CMB_variations} 

Since the hemispherical power asymmetry affects an extensive range of multipoles (see section~\ref{sec:hemispherical_power_asymmetry}), it is reasonable to ask to which extent anomalies in CMB observables have an impact on parameter inferences within the flat $\Lambda$CDM model. The prevailing narrative is that CMB anomalies represent statistical fluctuations and are benign as a result. Nevertheless, this statement requires further investigation. 
Previous efforts to quantify the variations of $\Lambda$CDM parameters over the sky yielded mixed results \cite{bib:Axelsson2013,bib:Mukherjee2016,bib:Mukherjee2018, bib:Mariano2013}. Ref. \cite{bib:Axelsson2013} demonstrated some variation in nine-year WMAP data, but  \cite{bib:Mukherjee2016,bib:Mukherjee2018} later countered that significant displacements in parameters were not expected in Planck data.

A recent analysis of both WMAP and Planck legacy temperature maps with $\ell > 30$ identified three patches, or ``horizons", on the sky where the $\Lambda$CDM parameters exhibit distinct differences \cite{bib:Fosalba2021}. The three patches are $40^\circ$-$60^\circ$ in angular diameter and centered on the directions $(l, b) = (345^{\circ}, 15^{\circ})$, $(l, b) = (240^{\circ}, -5^{\circ})$, and $(l, b) = (150^{\circ}, -40^{\circ})$. The direction of the second horizon is consistent with the hemispherical power asymmetry detailed in section~\ref{sec:hemispherical_power_asymmetry}. 
$H_0$ ranges from $(61.3\pm2.6)$~km/s/Mpc to $(76.6\pm5.4)$~km/s/Mpc, corresponding to a $2.6 \, \sigma$ discrepancy, assuming Gaussian errors. It should be noted that $H_0$ adopts lower values in the direction of the hemispherical power asymmetry, $(l, b) = (218^{\circ}, -21^{\circ})$.  

One can now define a $H_0$ tension that is intrinsic to the CMB with a significance that is comparable to the well documented early versus late Universe discrepancy as it stood in 2013 \cite{bib:Riess2011, bib:Planck2013c}. The authors confirmed that the level of variation exceeds the levels seen in 300 realizations of the $\Lambda$CDM model and estimated the discrepancy as a statistical fluctuation with probability $10^{-9}$ \cite{bib:Fosalba2021}.

In a bid to demonstrate  robustness of the result, a number of tests were performed \cite{bib:Fosalba2021}. First, the authors confirmed that similar anisotropies persisted in a high $\ell$ regime, when $\ell <450$ multipoles, including the first acoustic peak, were removed. This confirmed that the pattern is not coming from the largest scales only. Similarly, a low $\ell$ regime obtained by the removal of $\ell >1500$ multipoles did not greatly impact results. Secondly, foreground contamination was investigated by applying the analysis pipeline to the four different component separation maps produced by Planck: SMICA, Commander, SEVEM and NILC \cite{bib:Planck2018}. The patterns were found to be robust. In addition, attempts were made to model dust, but no significant effect was reported. Thirdly, in order to investigate changes in the CMB data set, the authors also applied a similar methodology to nine-year WMAP data \cite{bib:Bennett2013}. Despite the difference in resolution and the lower signal to noise ratio, similar anisotropic patterns in $\Omega_{b} h^2$, $H_0$, and $n_s$ were found.

The angular scale of the patches, $\sim 60^\circ$, coincides with the scale above which the two-point angular correlation $C^{TT}(\theta)$ is unexpectedly close to zero \cite{bib:Bennett2003, bib:Copi2015b, bib:Schwarz2016}. Moreover, the anisotropies are also comparable in size to the unexpected patterns in the quadrupole and octopole (see section~\ref{sec:quadrupole_octopole}). When coupled with the alignment of one of the patches with the dipole characterizing the hemispherical power asymmetry, this reinforces the view that CMB anomalies do indeed impact the inference of cosmological parameters within the flat $\Lambda$CDM model. Physically, \cite{bib:Fosalba2021} interpreted the patches as causally disconnected horizons across the observable Universe \cite{bib:Gaztanaga2020,bib:Gaztanaga2021,bib:Gaztanaga2021b,bib:Gaztanaga2022} and posited the existence of similar horizons in the local Universe that would account for cosmological tensions, including the Hubble tension. Variations in $H_0$ on the sky in the local Universe exist, where they are expected \bcomm{\cite{bib:McClure2007,bib:Wiltshire2012}}, and similar variations in the late Universe have been reported in \cite{bib:Migkas2020, bib:Migkas2021, bib:Krishnan2022, bib:Luongo2022} (see sections~\ref{sec:galaxy_cluster_anisotropies} and \ref{sec:emergent_H0} for further details).   

More recently, a second paper \cite{bib:Yeung2022} appeared, which also reported spatial variations of the cosmological parameters within Planck CMB temperature data. The authors applied hemispherical masks to block off half of the sky opposite to 48 different directions centering at the $N_{\mathrm{side}} = 2$ pixels of the HEALPix pixelization scheme with the ``RING" ordering on top of the original mask used in Planck. 
\texttt{PolSpice}\footnote{\href{http://www2.iap.fr/users/hivon/software/PolSpice/}{http://www2.iap.fr/users/hivon/software/PolSpice/}} was used to recalculate the CMB anisotropy cross spectra using the appropriate sky maps, masks, and beam window functions. The process then followed procedures outlined in \cite{bib:Planck2016d} to calculate the covariance matrices of different detector combinations. 

\begin{figure}[htp]
\centering
\includegraphics[width=0.85\linewidth]{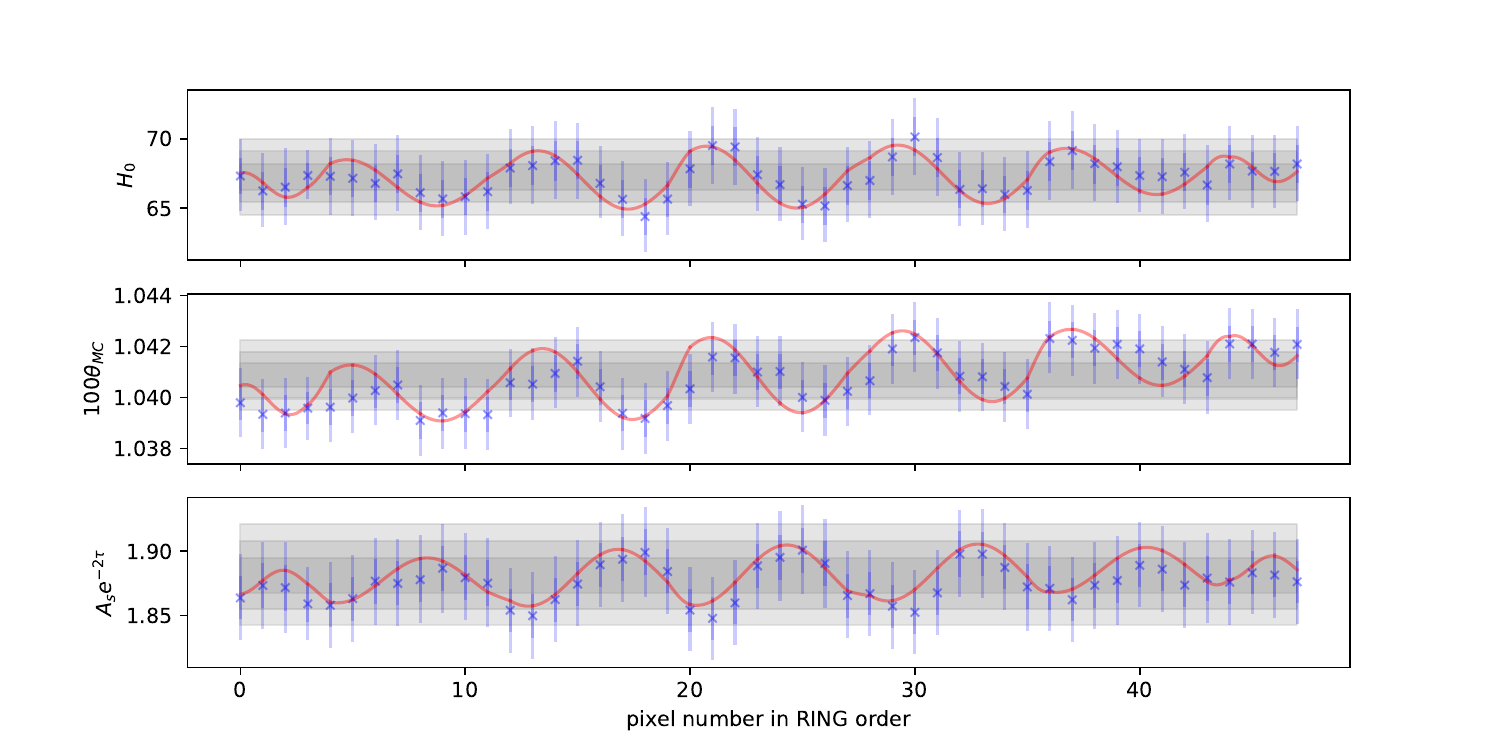}
\caption{Mean values (crosses), 68\% CI (dark blue bars), and 95\% CI (light blue bars) of cosmological parameters from the $\Lambda$CDM fitting of the Planck CMB temperature data, for 48 different half-skies centered on the $N_\mathrm{side}=2$ pixels of HEALPix \bcomm{(adapted from Fig. 2 of \cite{bib:Yeung2022})}. The grey bands are the 68\%, 95\%, and 99.7\% CI from the full-sky results. The red lines indicate the mean dipoles.}
\label{fig:cmb_scatter_all_par}
\end{figure}

\begin{figure}[htp]
    \centering
    \includegraphics[width=0.8\linewidth]{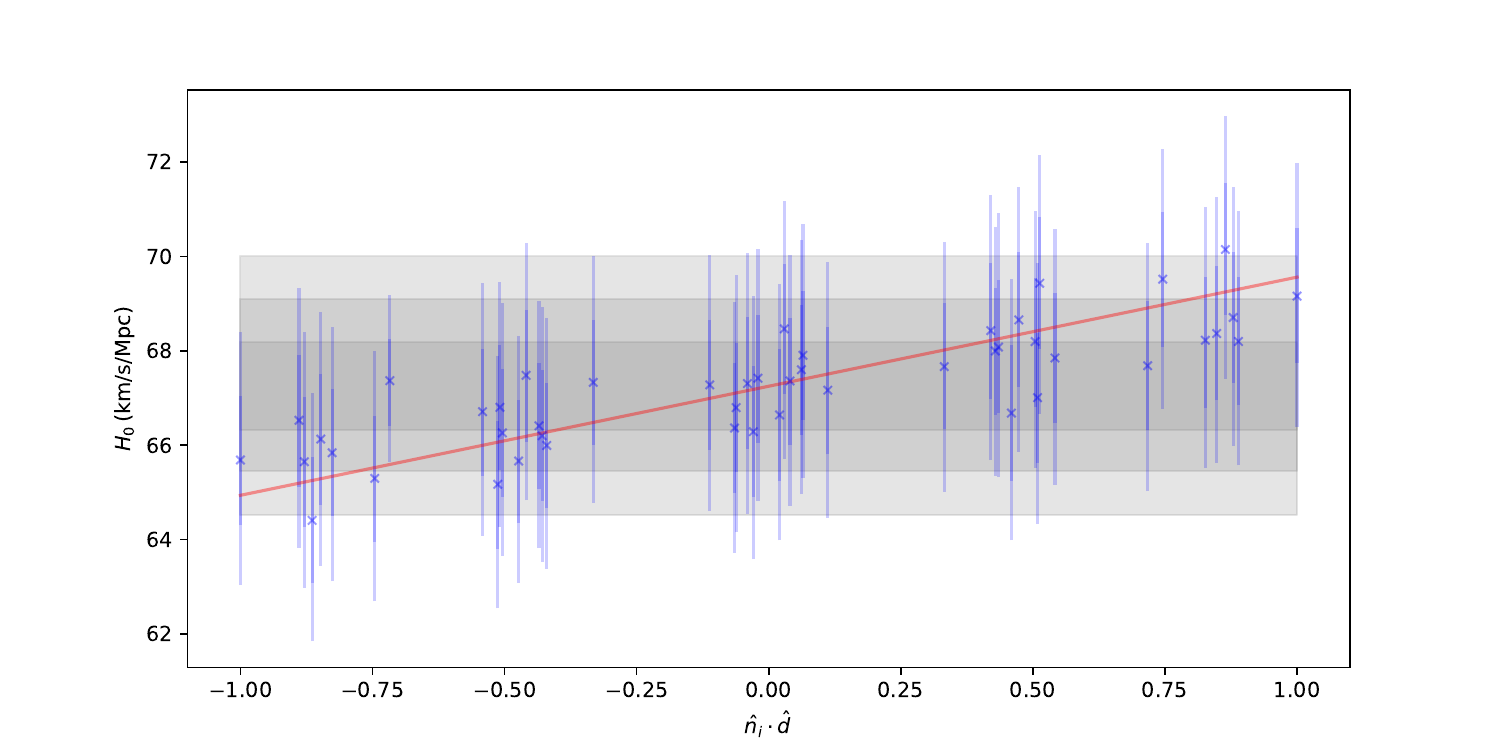}
    \vspace{-1ex}
    \caption{Best-fit values of $H_0$ for half-skies centering at directions $\hat{n}_i$ vs. $\cos\theta_i$, with $\theta_i$ the angle between $\hat{n}_i$ and the best-fit dipole direction $\hat{d}$. The grey bands are the 68\%, 95\%, and 99.7\% CI from the full-sky results. The red line represents the best-fit dipole.}
    \label{fig:cmb_dipole_H0}
    \vspace{-4ex}

\end{figure}

\begin{figure}[htp]
    \centering
    \includegraphics[width=0.78\linewidth]{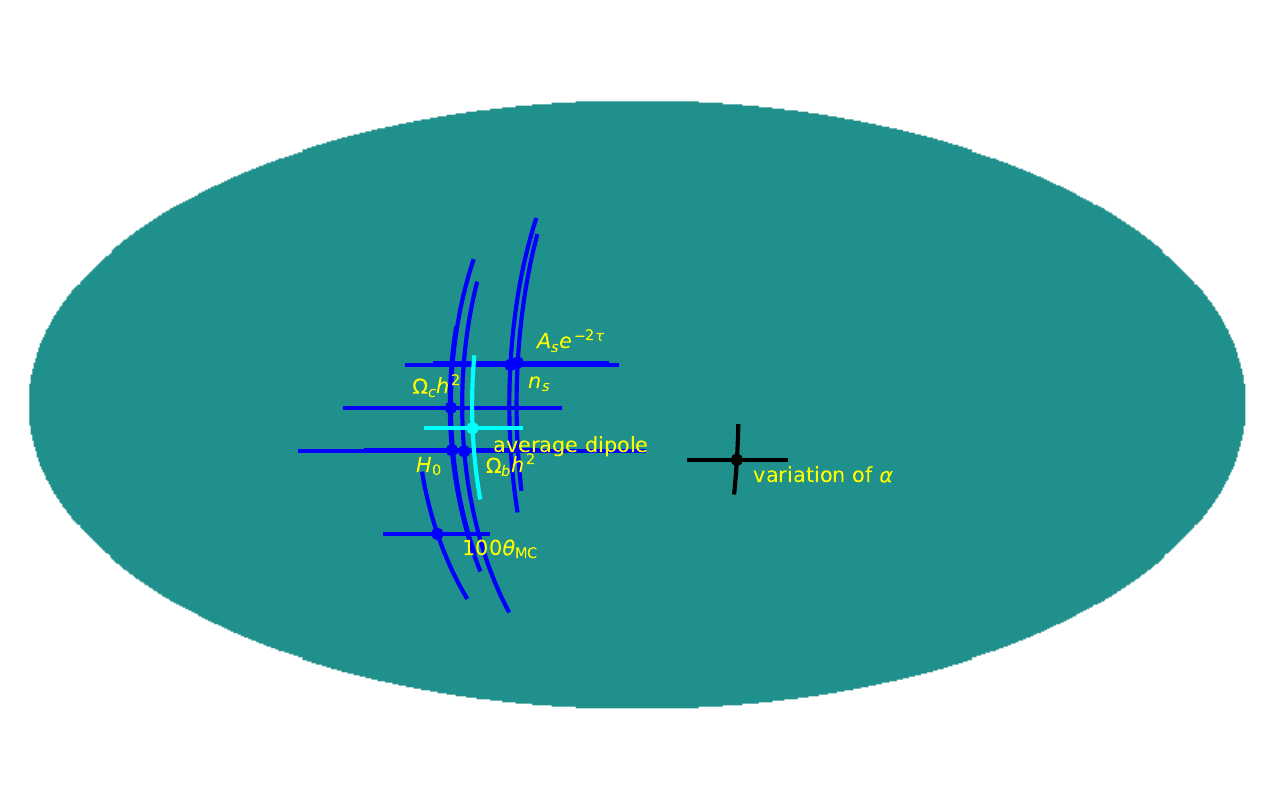}
    \vspace{-6ex}
    \caption{Directions with uncertainties of the cosmological parameter dipoles (blue), their average (cyan), and the fine structure constant variations \cite{bib:King2012} (black).}
    \label{fig:cmb_par_dipole_directions}
\end{figure}

It was found that there are 2-3~$\sigma$-level directional variations in all the parameters of the standard model of cosmology, with $H_0, 100\, \theta_{MC}$, and $A_s e^{-2\tau}$ being the most significant ones (shown in Fig.~\ref{fig:cmb_scatter_all_par}). Furthermore, in line with standard practices, these directional variations can be well fitted to a dipole form $\hat{d}\cdot \hat{n}_i$, where $\hat{d}$ is the dipole vector and $\hat{n}_i$ are the unit vectors pointing to the 48 directions shown in Fig.~\ref{fig:cmb_scatter_all_par}. For example, Fig.~\ref{fig:cmb_dipole_H0} shows the best-fit half-sky values of $H_0$ versus $\hat{d}\cdot \hat{n}_i$, which is in good agreement with a dipole variation (red line). The directional variations of $H_0$ range from $H_0 = (64.4 \pm 1.3)$~km/s/Mpc to $H_0 =(70.1 \pm 1.4)$~km/s/Mpc, thereby making them approximately $10 \%$ with a statistical significance of $3 \, \sigma$. In other words, the variations are comparable in magnitude to the Hubble tension \cite{bib:DiValentino2021, bib:Abdalla2022}. The fitted dipole directions are generally aligned, with the mean direction of $(l,b) = ({48.8^{\circ}}^{+14.3^{\circ}}_{-14.4^{\circ}}, {-5.6^{\circ}}^{+17.0^{\circ}}_{-17.4^{\circ}})$, nearly $90^{\circ}$ from the preferred direction of the fine structure constant variations in QSOs \cite{bib:King2012}, as shown in Fig.~\ref{fig:cmb_par_dipole_directions}. The cosmological parameters appear to be tracking the hemispherical power asymmetry through the antipodal direction on the sky, which means that the direction is expected to correspond to the second horizon documented in  \cite{bib:Fosalba2021}. As we discuss in section~\ref{sec:precision_spectroscopy_test} the significance of variations in the fine structure constant $\alpha$ and any dipole has decreased with new dedicated observational data. 

One may use two statistical measures to test the significance of the dipole variations of the cosmological parameters. First, one can calculate the Mahalanobis distances $D = \sqrt{d^T\Sigma^{-1}d}$ of the fitted dipoles from zero, where $\Sigma$ is the covariance. The probability $p$ of generating a dipole with $D$ larger than the fitted values from an isotropic distribution with the same covariance is then evaluated. Secondly, one can calculate the Bayes factor $K$ comparing the isotropic and dipole hypotheses. The values of $D$, $p$, and $K$ are listed in Table~\ref{table:cmd_par_dipole} for the standard cosmological parameters. Both statistical measures indicate strong evidence of a dipole component for $100\, \theta_\mathrm{MC}$. We have also performed the above analyses using the Planck FFP8 set of 100 different CMB signal and noise maps, generated assuming no parameter dipole, but for $N_{\mathrm{side}} = 1$, or 12 directions only. The results are consistent with having no dipole, showing that our analysis procedure does not bias the estimation of the parameters.
They suggest significant violations of the CP or yet unknown systematic errors in the standard CMB analysis. 


\begin{table}[htp]
    \centering
    \begin{ruledtabular}
        \begin{tabular}{llll}
            Parameter & $D$ & $p$ & $K$\\
            \hline
$\Omega_b h^2$ &  1.24 & $6.75\times 10^{-1}$ & 5.9\\
$\Omega_c h^2$ &  1.92 & $2.96\times 10^{-1}$ & 1.6\\
$n_s$ & 2.13 & $2.11\times 10^{-1}$ & 1.0\\
$H_0$ & 2.49 & $1.02\times 10^{-1}$ & 0.52\\
$100\theta_{MC}$ & 4.54 & $1.26\times 10^{-4}$ & 0.0041\\
$A_s e^{-2\tau}$ &  2.44 & $1.15\times 10^{-1}$& 1.4\\
        \end{tabular}
    \end{ruledtabular}
    \caption{Mahalanobis distance $D$ of the distribution from zero, the corresponding $p$, and the Bayes factor $K$ of the dipole variations of the cosmological parameters.}
    \label{table:cmd_par_dipole}
\end{table}

\subsection{Topological anomalies in the CMB}
\label{sec:anomalies_topo_cmb}

Observations indicate that over a few to hundreds of megaparsecs, galaxies and matter in the 
Universe aggregate in a complex pattern, commonly known as the ``cosmic web'' \cite{bib:Bond1996}. 
The cosmic web largely consists of roughly spherical clusters of matter and galaxies, inter-connected by filamentary structures, woven around three-dimensional underdense regions identified as cosmic voids. 
The complex network and arrangement of matter in the cosmic web is postulated to arise from quantum fluctuations in an otherwise homogeneous medium in the inflationary era. 
There is a treasure trove of information encoded in the structural patterns of the primordial field as well the subsequent large scale structure of the Universe that emerges from it. 
A host of past, ongoing and future ground- and space-based experiments have resulted in a massive surge of data, both for the early Universe in terms of CMB experiments, and the late time Universe in terms of large scale galaxy surveys.
This massive surge in data acquisition is only bound to keep growing significantly in the future. The amount and novelty of the cosmological data sets rightly demand 
increasingly more sophisticated methods for data reduction and analyses, in order to realize their full potential in revealing novel features and mechanisms of the Universe. 

Measures like the two-point correlation functions that have been the mainstay of cosmological analyses for many decades are insensitive to the spatial patterns in terms of information about phases, while higher order correlations become increasingly prohibitive computationally. 
Still, methods and tools arising from analysis in Fourier space, chiefly the analysis of power spectra, have been potent and ubiquitous in cosmological analyses. 
However, a description of non-Gaussian features requires information about higher order spectra, which are again computationally expensive. 

To complement these methodologies, techniques have emerged from topo-geometrical considerations in the real space, which have also been in use to describe the observed features of the cosmological fields. 
Broadly, methodologies based on topo-geometrical considerations focus on geometrical properties such as volume, area, and length of boundaries, as well as topological characteristics such as connectivity and the notion of topological holes of a manifold. 
The earliest use of topological methods to describe the matter distribution in the Universe involved the notion of genus and Euler characteristic \cite{bib:Gott1986,bib:Hamilton1986,bib:Mecke1991,bib:Park1998,bib:Park2013}, which was later extended to the full set 
of Minkowski functionals \cite{bib:Mecke1994,bib:Schmalzing1997,bib:Schmalzing1998,bib:Ducout2013,bib:Appleby2022}. 
More recently, developments in computational topology have brought the purely topological notions of homology within the ambit of applied and computational mathematics. 
Particularly powerful is its hierarchical extension ``persistent homology'', which forms the foundational method for the recently emerging discipline of ``topological data analysis" (TDA) 
\cite{bib:Carlsson2009,bib:Chazal2013,bib:Fasy2013,bib:Khasawneh2014,bib:Bubenik2015,bib:Bobrowski2012,bib:Chen2015a,bib:Pun2018,bib:Moraleda2019,bib:Kannan2019,bib:Pranav2021a}.
These methods are supported by the power of an elaborate and solid mathematical framework, and complement the existing methods for gleaning meaningful information out of the ever-growing cosmological data sets. 
It is evident that these methodologies are valuable evaluation tools from a recent proliferation of their use in the astronomical and cosmological disciplines, in a variety of contexts including structure detection and identification \cite{bib:Sousbie2008,bib:Sousbie2011,bib:Shivashankar2015,bib:Xu2019}, including detection of BAO signals \cite{bib:Kono2020}, statistical characterization of ISM 
\cite{bib:Makarenko2018}, statistical characterization of cosmological fields arising from various models, and a description of associated structures \cite{bib:Weygaert2011,bib:Park2013,bib:Chen2015a,bib:Codis2018,bib:Wilding2020,bib:Kehe2022}, and detection and quantification of non-Gaussianities \cite{bib:Cole2018,bib:Feldbrugge2019,bib:Biagetti2020}. 

\begin{figure*}
	\centering
	\includegraphics[width=\textwidth]{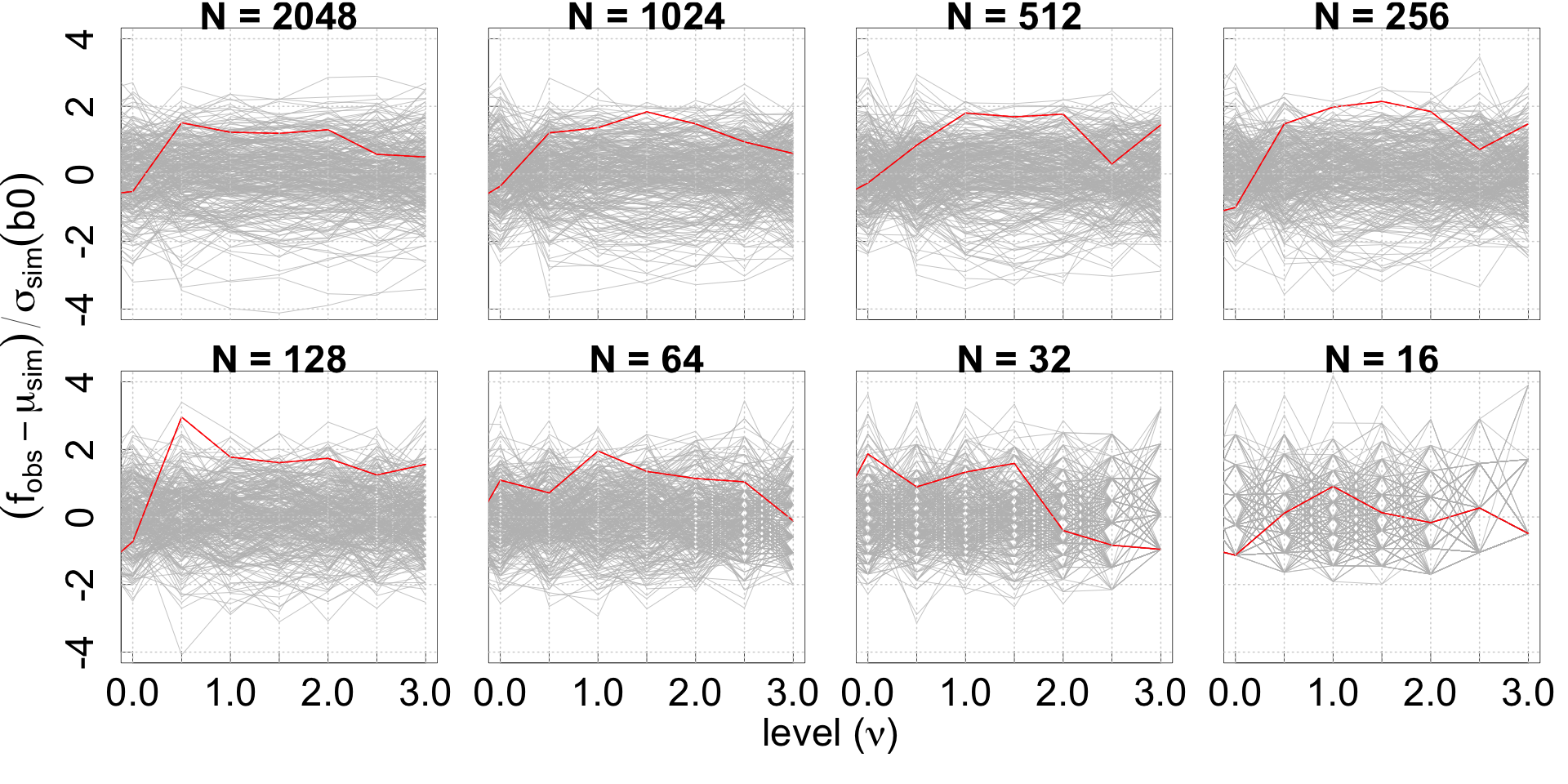} \\
	\includegraphics[width=\textwidth]{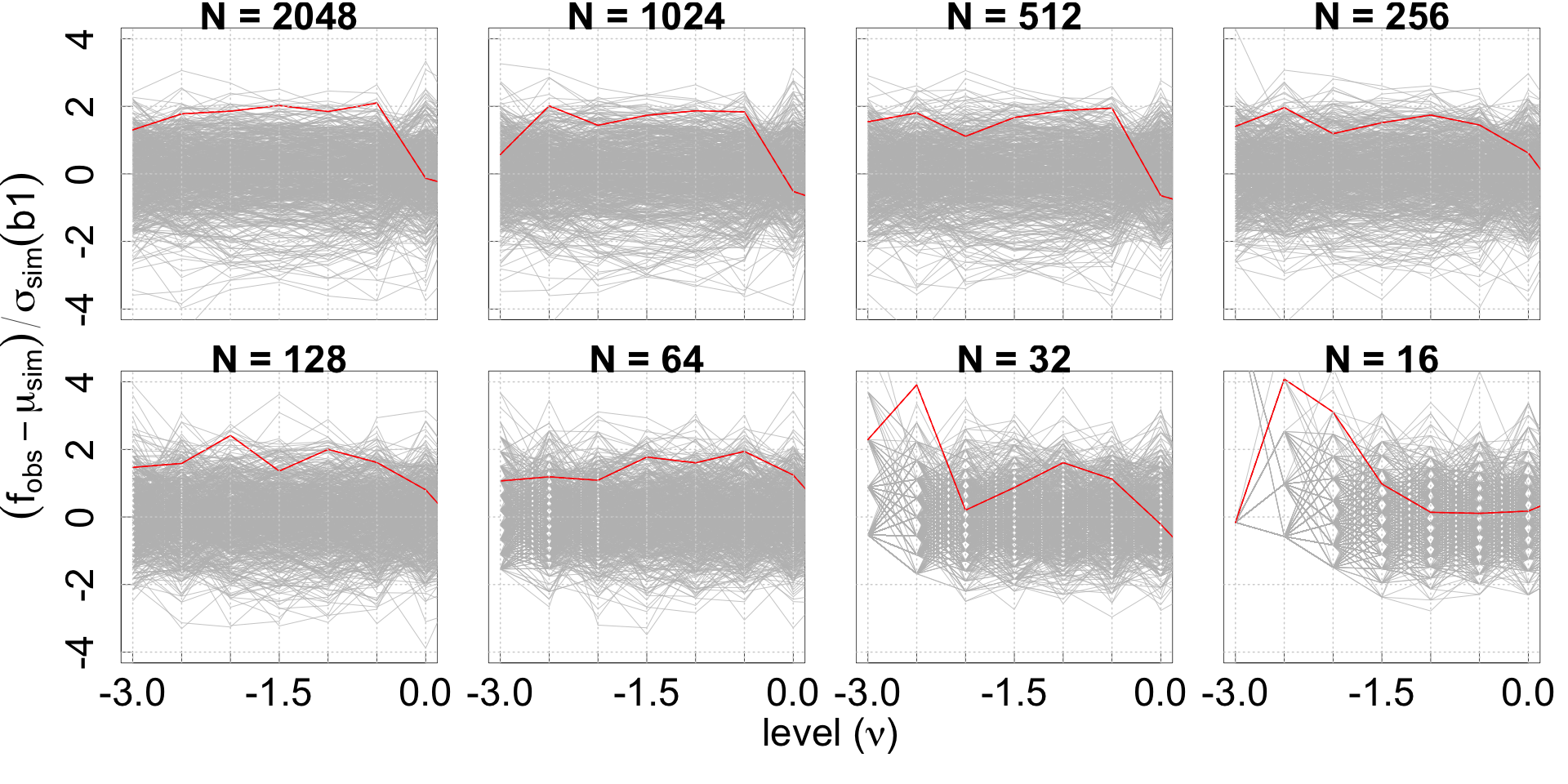}\\
	\caption{Top: Graphs for $\relBetti{0}$ for \texttt{FFP10} data set. Bottom: $\relBetti{1}$ for 
	\texttt{NPIPE} data set. The graphs present the normalized differences, with respect to the mean and the standard deviation computed from the simulations at each threshold. Each panel presents the graphs for a range of degradation and smoothing scales. The mask used is the PR3 temperature common mask. The curves from the observational map are presented in red, while the curves from individual simulations, treated as observation, are presented in gray.}
	\label{fig:b0_ffp_b1_npipe}
\end{figure*}

Topology studies the connectivity properties of a manifold, and invariance under the operations of rotation, translation and bending, but not tearing and gluing. 
We are interested in topology at the level of \emph{homology} \cite{bib:Munkres1984,bib:Edelsbrunner2002,bib:Edelsbrunner2010,bib:Pranav2017}. 
Recent developments in computational topology have brought homological topology within the ambit of applied and computational mathematics, such that TDA has become a discipline in its own right. 
Homology describes the topology of a $d$-dimensional topological space in terms of the $p$-dimensional cycles it contains, $p=0,\ldots,d$, that bound the $p$-dimensional topological holes. 
The bounding homology cycles are essentially identified by the fundamental lemma of homology, which states that the boundary of a topological cycle bounding a hole is necessarily empty. 
In three spatial dimensions, the cycles and holes have an intuitive interpretation. 
A $0$-dimensional cycle represents a connected object, and associates with the clustering properties. 
A $1$-dimensional cycle is a loop bounding a circle or a tunnel, and is associated with the percolation properties. 
A $2$-dimensional cycle is a connected surface completely bounding a cavity. 
The ``Betti numbers'' \cite{bib:Betti1871,bib:Munkres1984,bib:Edelsbrunner2010} in different dimensions count the number of independent cycle of that dimension. 

\begin{figure*}
	\centering
	\includegraphics[width=0.48\textwidth]{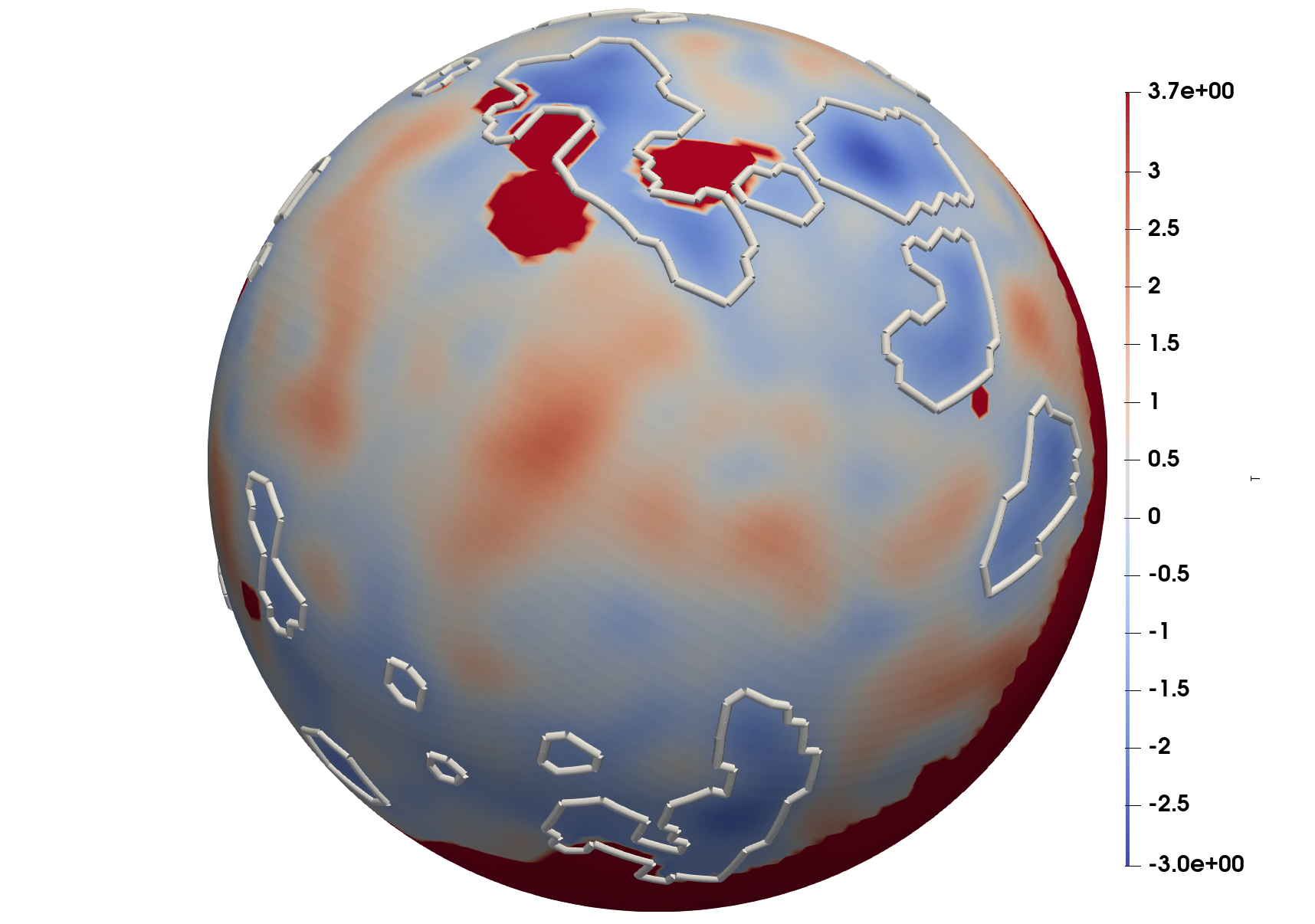} \hskip 0cm 
	\includegraphics[width=0.48\textwidth]{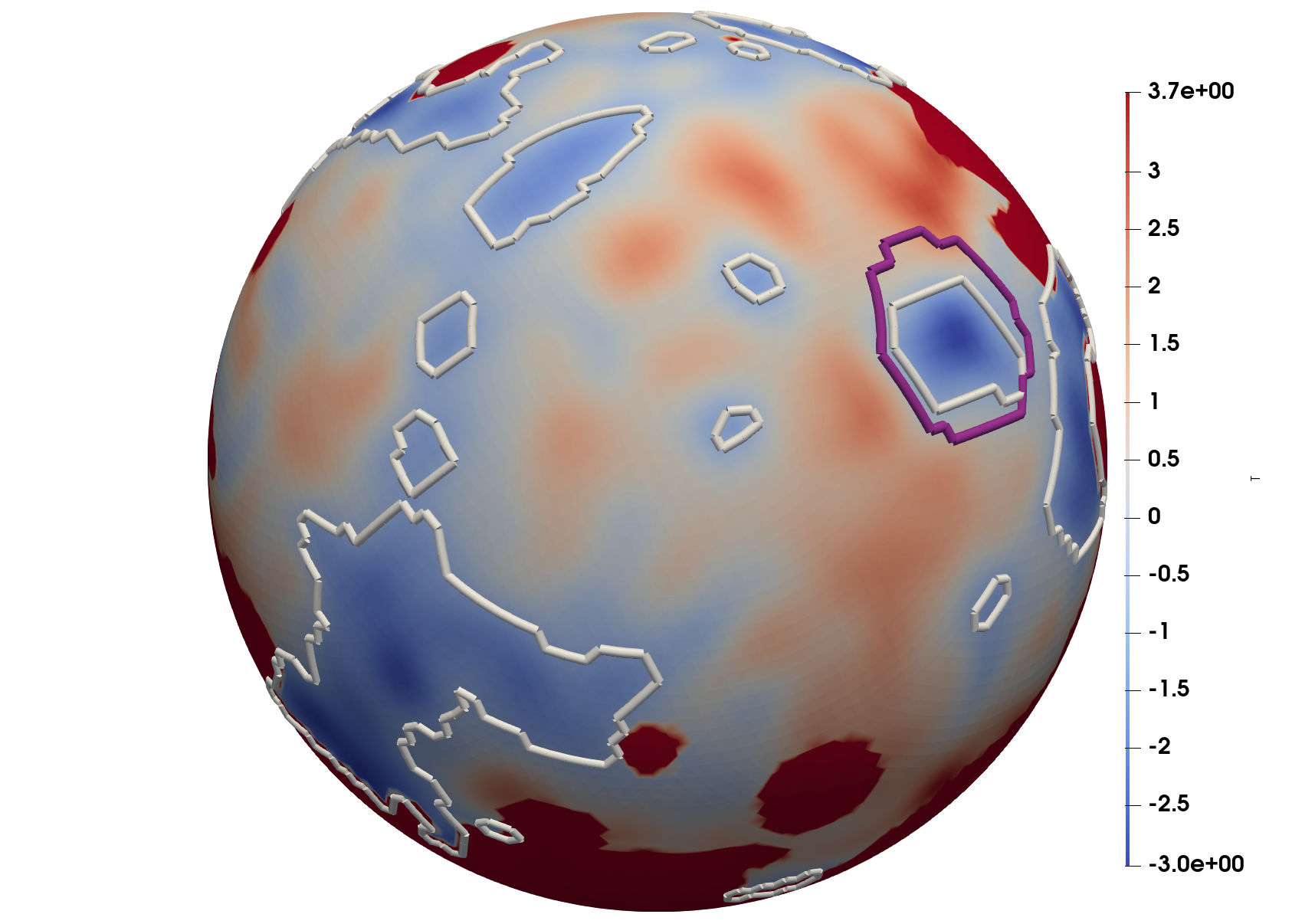}\\
	\caption{Visualization of some of the loops in the CMB sky, based on the \texttt{NPIPE} observed map smoothed with a Gaussian beam profile of $FWHM = 320'$. Left: Loops in the northern hemisphere. Right: Loops in the southern hemisphere. The cold spot loop is surrounded by an extra pink loop.}
	\label{fig:loop_visualization}
\end{figure*}

In the experimental setup of the temperature fluctuations of the CMB, the definitions are on $\Sspace^2$, in which case the relevant quantities are the number of components denoted by the $0^{th}$ Betti number $\beta_0$, and the number of independent $1$-dimensional punctures or holes on $\Sspace^2$, denoted by the $1^{st}$ Betti number $\beta_1$. 
Denoting the temperature anisotropies as a scalar function $f: \Sspace^2 \to \Rspace$, the topological descriptors are computed as a function of ``excursion set'', $\Excursion (\nu)$,
\begin{equation}
	\Excursion (\nu)  =  \{ x \in \Sspace^2  \mid  f(x) \geq \nu \}\;,
	\label{eq:excursion_set}
\end{equation}
where $\nu = (f -\mu_f)/\sigma_f$ denotes the normalized temperature threshold, and $\mu_f$ and $\sigma_f$ are the mean and the standard deviation of the field $f$. 
Keeping track of the topological changes occurring in a manifold by varying the threshold of the function, records the topological changes in a hierarchical fashion and results in the hierarchical version of homology known as persistent homology. 
The topological changes in the nested hierarchy of excursion sets are recorded in the persistence diagrams of different dimensions, from which we construct the Betti number curves. 
As the data in certain parts of the sky are not reliable, we mask these sectors, and compute the topology relative to the mask, denoting the number of components relative to the mask as $\relBetti{0}$, and the number of loops relative to the mask as $\relBetti{1}$. 

The data sets we examine are the Planck data sets from the final Data Release 4 (DR4) based on the \texttt{NPIPE} processing pipeline, as well as its predecessor Data Release 3 (DR3) based on the \texttt{FFP10} processing pipeline. 
The \texttt{NPIPE} data set is accompanied by $600$ simulations, while the \texttt{FFP10} data set contains $300$ simulations, which are based on the standard cosmological model that assumes the fluctuations to be instances of an isotropic and homogeneous Gaussian random field. 
We perform a multi-scale analysis by smoothing the original maps given at $FWHM = 5'$ and $N_{side} = 2048$, to a range of scales defined by Gaussian beam profile of $FWHM = 10', 20', 40', 80', 160', 320'$ and $640'$. 
In order to facilitate faster computations, the maps are also degraded to $N_{side} = 1024, 512, 256, 128, 64, 32$ and $16$ prior to the smoothing operation. 
The mask is subjected to an identical degrading and smoothing procedure. 
This results in a non binary mask, which is re-binarized by setting all pixels above and equal to $0.9$ to $1$, and all pixels with smaller values to $0$. 
After the pre-processing steps, performed with the aid of the \texttt{HealPix} software \cite{bib:Gorski2005}, the data is subject to the topology computation pipeline, which briefly involves tessellating the points on the sphere, computing the upper-star filtration of this tessellation, constructing the boundary matrix of the filtration, and reducing the boundary matrix to obtain the persistence 
diagrams in different dimensions. 
The Betti numbers relative to the mask are condensed from the persistence diagrams, see \cite{bib:Pranav2017,bib:Pranav2019} for details on the computational procedure.

We select some illustrative results presented in \cite{bib:Pranav2021,bib:Pranav2022,bib:Pranav2019} to examine the anomalous behavior of the topology of CMB temperature fluctuations.  
Fig.~\ref{fig:b0_ffp_b1_npipe} presents the graphs of the normalized difference between the simulations and the observations for the number of components (top) and the number of loops (bottom) as a function of the normalized temperature threshold at steps of $0.5$. 
For the number of components we choose $\nu \in[0:3]$, while for the loops we choose $\nu \in[-3:0]$, commensurate with the fact that components arise predominantly in high (positive) thresholds, 
and loops surround the low (negative) threshold regions.  
Fig.~\ref{fig:b0_ffp_b1_npipe} (top) presents the results for the 
\texttt{FFP10} data set, while Fig.~\ref{fig:b0_ffp_b1_npipe} (bottom) presents results for the \texttt{NPIPE} data set. 
Examining the graphs, we notice that the number of components exhibit the maximum significance of $2.96$ standard deviations at $N_{side} = 128, FWHM = 80'$, which is roughly a degree. The number of loops exhibit the maximum significance of $3.9$ standard deviations at $N_{side} = 32, FWHM = 320'$. 
The results are in general agreement with the results from the WMAP experiment presented in \cite{bib:Eriksen2004}, where they notice a more than $3 \, \sigma$ significance in the negative threshold of genus for $FWHM = 3.40^{\circ}$, and more than $2 \, \sigma$ deviation in the positive thresholds of genus at $FWHM = 1.2^{\circ}, 1.78^{\circ}$. 
It is important to note that the statistics are based on total numbers, and hence, the deviant behavior of the loops has only a singular contribution from the ``cold spot''. 
Fig.~\ref{fig:loop_visualization} presents a visualization of some of the loops in the CMB sky smoothed with a Gaussian beam profile of $FWHM = 320'$. 
The left and right parts present the visualization of the two different hemispheres, respectively. 
The single loop corresponding to the cold spot is marked in pink, among the many other loops that contribute to the statistics.

The results presented are from a full-sky analysis, and hence do not investigate the assumption of isotropy in the present form. 
However, there are mild to significant deviations between the observation and model, which, if cosmological, point to a possible violation of some tenet of the cosmological principle, namely, isotropy, homogeneity or Gaussianity. 
As a caveat, the statistics for the anomalous behavior of loops at $N_{side} =32, FWHM = 320'$ is based on low numbers and engendered by the fact that the deviation occurs at a low threshold. Preliminary results based on further investigations also hint at topological signatures of hemispherical asymmetry detailed in section~\ref{sec:hemispherical_power_asymmetry}; see also \cite{bib:Adler2017}.

\subsection{Testing statistical isotropy with Contour Minkowski Tensor}
\label{sec:tests_statistical_isotropy}

A new method to test the statistical isotropy (SI) of cosmological fields was proposed in~\cite{bib:Ganesan2017,bib:Chingangbam2017,bib:Chingangbam2021}. 
The method is based on measuring the morphological properties of excursion sets of smooth random fields, as introduced in section~\ref{sec:anomalies_topo_cmb} (see (\ref{eq:excursion_set}) for a definition), using the contour Minkowski tensor (CMT), $\mathcal W$, defined as  
\begin{equation}
{\mathcal W}(\nu)  := \int_{\partial Q_{\nu}} \hat n\otimes \hat n \,{\rm d}s \;,
\end{equation}
where $\nu$ denotes a chosen field threshold value, $\partial Q_{\nu}$ denotes  the  boundary of excursion sets at $\nu$, $\hat n$ denotes the unit normal vector at a point on $\partial Q_{\nu}$, and $\otimes$ is the symmetric tensor product of two vectors. 
In two dimensions, $\partial Q_{\nu}$ consists of iso-field contours and  ${\rm d}s$ denotes the infinitesimal arc-length of the contours. 
See, for instance, \cite{bib:Schroeder2010} for a general introduction. So, ${\mathcal W}(\nu)$ is a $2\times 2$ matrix.  
Let $\Lambda_1(\nu), \Lambda_2(\nu)$, such that $\Lambda_1(\nu) < \Lambda_2(\nu)$, denote the eigenvalues of ${\mathcal W}(\nu)$. For a contour that has $m$-fold symmetry with $m\ge 3$ we have $\Lambda_1= \Lambda_2$~\cite{bib:Chingangbam2021}, and any anisotropy or departure from this symmetry leads to $\Lambda_1 < \Lambda_2$. The ratio of the eigenvalues, 
\begin{equation}
\alpha(\nu) \equiv \frac{\Lambda_1(\nu)}{\Lambda_2(\nu)} \;, 
\end{equation}
provides a measure of anisotropy of the field at each $\nu$. We refer to $\alpha$ as the {\em alignment parameter} since it captures the relative alignment in a random distribution of curves. Exact isotropy gives $\alpha(\nu)=1$. However, in practical applications, since observed data is always available on finite space with finite resolution, $\alpha(\nu)$ deviates from one with the deviation becoming larger as $|\nu|\to \infty$. This sampling effect can be quantified and needs to be taken into account to infer the presence  of any true statistical anisotropy in the data.

In \cite{bib:Ganesan2017} $\alpha(\nu)$  was computed for Planck 2015 data to test for SI. For the calculation, the CMB data on the sphere was first projected onto the plane using stereographic projection and then $\mathcal W$ was calculated. The authors found that the Planck 2015 temperature data is consistent with SI. However, the E-mode polarization data at 44 and 70 GHz frequency channels deviated from SI at roughly $\sim 4 \, \sigma$. This deviation is most likely  predominantly due to two factors. The first is that the 2015 polarization data possibly contains low levels of systematic effects~\cite{bib:Planck2016c}. Secondly, stereographic projection was chosen for the analysis because it preserves angles, but it introduces size scaling of structures and hence can introduce projection errors. 
The definition of the CMT was generalized to random fields defined on curved manifolds and a method was given in~\cite{bib:Chingangbam2017} for its numerical computation directly on the sphere without requiring projection onto the plane. This method of calculation was applied to the Planck 2018 temperature data \cite{bib:Joby2019} and E-mode data~\cite{bib:Kochappan2021}. Both temperature and E-mode data were found to be consistent with SI. Hence we conclude that the $4\, \sigma$ deviation from SI obtained with the Planck 2015 E-mode data is caused by the combined effect of residual systematics in the data and stereographic projection. For the analysis of the E-mode data, the authors also used the so-called $\mathcal{D}$ statistic method to the CMB E-mode data and found the results to be in good agreement with the CMT analysis.

\begin{figure}
    \centering
    \includegraphics[width=3in,height=2in]{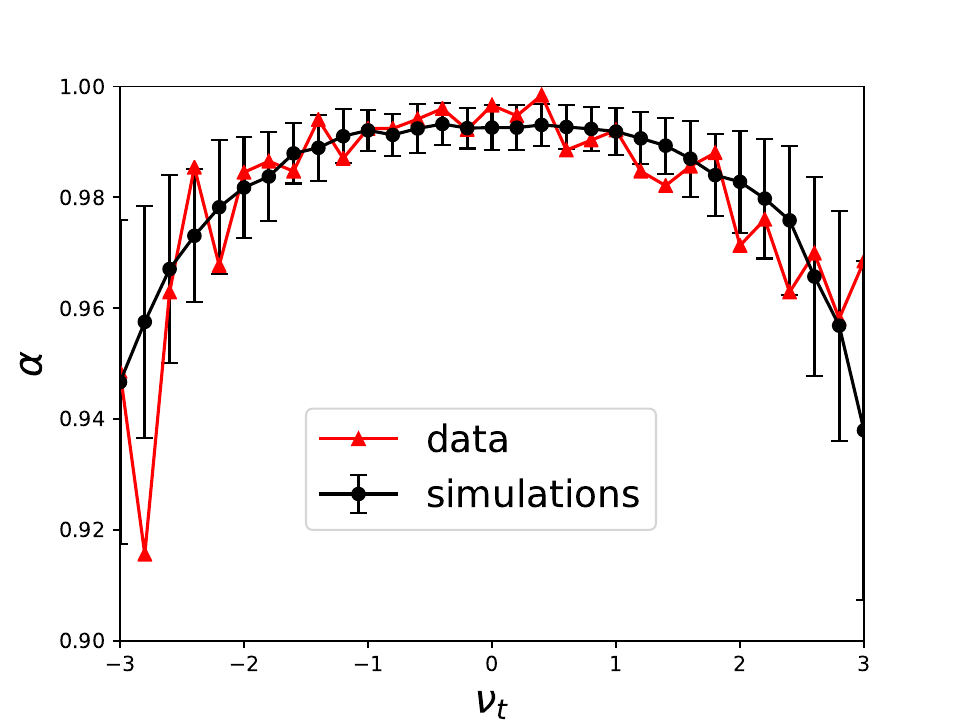} \vskip 0.2cm
    \includegraphics[width=3in,height=2in]{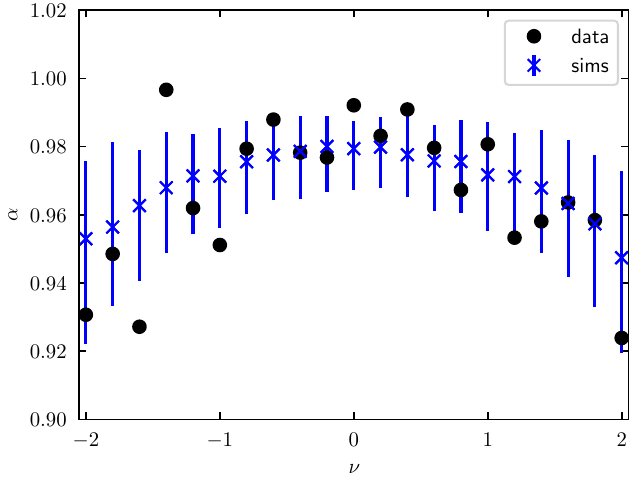} \hskip 0cm \includegraphics[width=3in,height=2in]{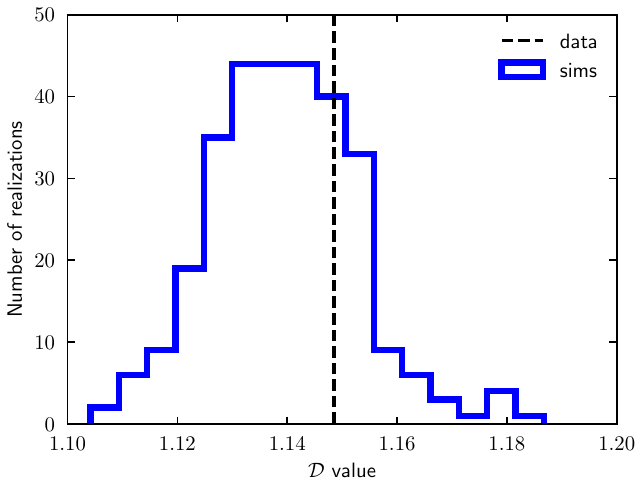}
    \caption{Top : Comparison of $\alpha$-values from the Planck 2018 CMB temperature data and simulations \bcomm{(reproduced from Fig. 4 of \cite{bib:Joby2019})}. Bottom left: Comparison of $\alpha$-values from the Planck 2018 CMB E-mode data and simulations \bcomm{(reproduced from Fig. 8 of \cite{bib:Kochappan2021})}. Bottom right: The $\mathcal{D}$-value from the Planck 2018 CMB E-mode data and the histogram from the corresponding simulations \bcomm{(reproduced from Fig. 9 of \cite{bib:Kochappan2021})}. The above results indicate statistical isotropy of the CMB temperature fluctuations.}
    \label{fig:tmf_cmb}
\end{figure}

The top panel of Fig.~\ref{fig:tmf_cmb} shows the results of the CMT analysis of the Planck 2018 CMB temperature data. The red triangles represent the SMICA temperature map, while the black dots represent the FFP8 SMICA simulations, and error bars are estimated from the simulations. It is clear that the SMICA temperature map is consistent with the simulations, from which the authors conclude that the CMB temperature data is statistically isotropic. The bottom left panel compares the $\alpha$-values from the data and simulations for the CMB E-mode. While there exist a few points where the data deviates from the simulations by more than $1 \, \sigma$, the combined p-value across all the thresholds is 0.54. The CMT analysis finds the E-mode data to be consistent with SI. The $\mathcal{D}$ statistic value from the CMB E-mode data and the corresponding FFP10 simulations are shown in the bottom right panel of Fig.~\ref{fig:tmf_cmb}. Again, the data is consistent with SI, similar to what is found using the CMT. 

\subsection{Some resolved anomalies}
\label{sec:some_resolved_anomalies}

Now, we review some features seen in CMB maps that were deemed ``anomalous'' in earlier studies, but later found to be not so.

\emph{CMB cold spot:} First-year data release from NASA's WMAP satellite revealed the CMB anisotropies in high resolution over the entire sky for the first time. The cleaned CMB sky thus provided or derived using various foreground removal schemes and enabled testing the Gaussian nature of the CMB field expected in the standard cosmological model. While some studies confirmed Gaussianity of the CMB sky, many authors reported non-Gaussian features in cosmic microwave maps. In particular, an anomalously low temperature region, the so-called \emph{CMB cold spot}, was found in the southern galactic hemisphere, located at $(l,b)=(210^\circ,-57^\circ)$ and $10^\circ$ (degrees) in angular size, in a multi-scale analysis of CMB maps using spherical Mexican hat wavelets (SMHW) with a low probability of occurrence of $0.1$-$0.4\%$~\cite{bib:Vielva2004,bib:Cruz2005,bib:Cruz2006}. It was independently confirmed in \cite{bib:Mukherjee2004,bib:McEwen2005,bib:Cayon2005} using different statistics and methods. But, when using different kernels such as a Gaussian or top hat profile to analyze the CMB sky, no such anomalously cold region was found, concluding that the cold spot detection may be due to a \emph{fortuitous choice} of SMHW that happened to match the cold spot profile~\cite{bib:Zhang2010}.

The cold spot anomaly was also studied by WMAP and Planck collaborations themselves confirming its presence~\cite{bib:Bennett2011,bib:Planck2013,bib:Planck2016,bib:Planck2018c}. However, the anomalous nature of the cold spot region was resolved when mean temperatures in concentric rings of finite width were computed at different radii centered at $(l,b)=(210^\circ,-57^\circ)$ where the cold spot region was originally found to be situated. By profiling cold regions in CMB simulations, the cold spot in the CMB data was found to be not an anomalously cold region~\cite{bib:Nadathur2014}. Instead, it is an apparently cold region in the CMB anisotropy field embedded in a hot ring like structure at about $15^\circ$ (degrees) away from the cold spot center characterized by having excess kurtosis (see also \cite{bib:Cayon2005} in which a ring like structure surrounding the cold spot region was found, but the CMB cold spot region itself was not found to be anomalous). It was further argued that such a low temperature region arising from a super void is also not supported as voids of the size required to explain the CMB cold spot region are not uncommon in the standard model~\cite{bib:Nadathur2014}.

\emph{Quadrupolar anisotropy:} A significant quadrupolar anisotropy at $9\,\sigma$ level was found in the CMB maps derived from WMAP five-year data in \cite{bib:Groeneboom2010} (see also \cite{bib:Groeneboom2009}). The authors of \cite{bib:Hanson2009} suggested beam systematics as the likely origin owing to its preferred direction being aligned with the ecliptic coordinate system. The WMAP team reanalyzed this feature using seven-year data from WMAP observations confirming the presence of a significant quadrupolar anomaly~\cite{bib:Bennett2011}, using a model independent framework based on Bipolar Spherical Harmonics (BipoSH)~\cite{bib:Hajian2003,bib:Hajian2006}.

\begin{figure}
    \centering
    \includegraphics[width=0.41\linewidth]{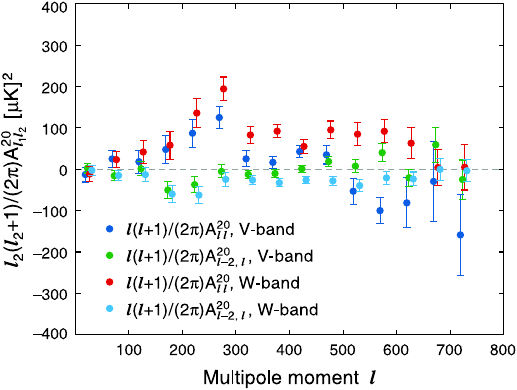}\hspace{2ex}
    \includegraphics[width=0.54\linewidth]{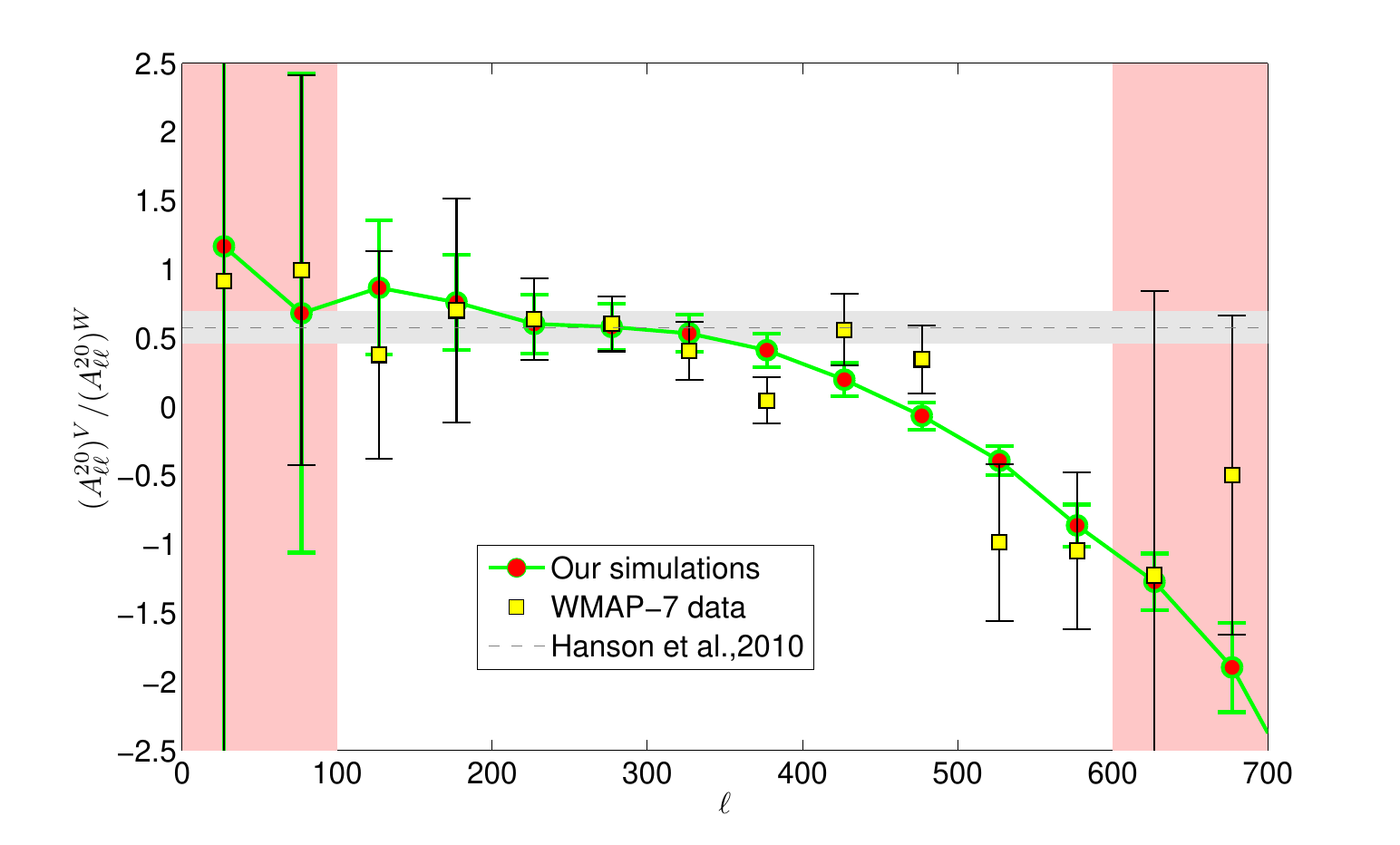}
\caption{\emph{Left:} Significant quadrupolar feature seen in WMAP seven-year CMB sky from V and W frequency bands (\bcomm{reproduced from Fig. 16 of  \cite{bib:Bennett2011}}). \emph{Right:} Through a detailed investigation of the effect of asymmetric beams and WMAP scan strategy, the observed quadrupolar like feature could be explained completely \bcomm{(reproduced from Fig. 3 of \cite{bib:Das2016})}.}
\label{fig:WMAP7yrQuadAniso}
\end{figure}

In general, the two-point correlation function is related to BipoSH coefficients, which in turn are related to the spherical harmonic coefficients $a_{\ell m}$, as
\begin{equation}
C(\hat{n},\hat{n}') = \sum_{L M \ell \ell'} A^{LM}_{\ell \ell'} \{Y_{\ell}(\hat{n}) \otimes Y_{\ell'}(\hat{n}')\}\,,\quad
A^{LM}_{\ell \ell'} = \sum_{mm'} \langle a_{\ell m} a_{\ell' m'} \rangle C^{LM}_{\ell m \ell' m'}\,,
\end{equation}
where $\{...\}_{LM}$ are the BipoSH basis~\cite{bib:Varshalovich1988}, \bcomm{$C^{LM}_{\ell m \ell' m'}$ denote Clebsch-Gordan coefficients} and $A^{LM}_{l_1 l_2}$ are the BipoSH coefficients with $L$ (and $|M|\leq L$) indicating the type of anisotropy such as, for example, dipolar ($L=1$) or quadrupolar ($L=2$) anisotropy, underlying the CMB sky.
For a statistically isotropic CMB sky $\langle A^{LM}_{\ell \ell'} \rangle = (-1)^\ell \Pi_\ell C_\ell \delta_{\ell \ell'}\delta_{L0}\delta_{M0}$ where $\Pi_\ell=\sqrt{2\ell+1}$ (and $\Pi_{\ell \ell'...} = \Pi_\ell \Pi_{\ell'}...$), and $C_\ell$ is the CMB angular power spectrum. The multipole indices $L$, $\ell$ and $\ell'$ satisfy the triangular inequality condition among them. A suitable normalization can be chosen to define BipoSH spectra as $\alpha_\ell^{LMD}=\Pi_L/(\Pi_{\ell \ell'}C^{L0}_{\ell 0 \ell' 0})A^{LM}_{\ell\ell-D}$ such that $\langle \alpha_\ell^{000} \rangle = C_\ell$~\cite{bib:Bennett2011}. The index $D$ follows from the symmetry properties of BipoSH coefficients such that it is sufficient to compute them for $M\geq 0$, and $\ell'=\ell+D$ or $\ell'=\ell-D$ with $0\leq D \leq L$. The BipoSH coefficients can also be computed for a specific model or type of anisotropy violation in the CMB sky (for example, see~\cite{bib:Aluri2015}). With this BipoSH machinery, the WMAP team found a highly significant quadrupolar component with non-zero BipoSH coefficients $A^{2M}_{l_1 l_2}$ as shown in Fig.~\ref{fig:WMAP7yrQuadAniso}.

The WMAP team noted the symmetry of estimated BipoSH spectra and, again, attributed the observed quadrupolar feature to beam systematics as the likely source of its origin, in agreement with the conclusions of \cite{bib:Hanson2009}. Following this, making a detailed investigation of the effect of detector beam asymmetries coupled with WMAP scan strategy, based on the BipoSH framework, the authors of \cite{bib:Das2016} showed an exact match with the observed amplitudes of $A^{2M}_{l_1 l_2}$ seen in WMAP seven-year data (see also~\cite{bib:Hanson2010}). Thus it is crucial to account for all the systematic (and foreground) artifacts before any claims of statistical anisotropy of the CMB sky or deviations from standard model expectations are made.

\subsection{Summary} 
\label{sec:summary_early_universe}

Evidently, beyond their discovery in WMAP data, a number of CMB anomalies have persisted in Planck data. Moreover, as is clear from section~\ref{sec:hemispherical_power_asymmetry}, the hemispherical power asymmetry has been recovered numerous times through different statistics since its discovery in WMAP data \cite{bib:Eriksen2003}. Throughout, the statistical significance of these results has remained in the $2$-$3 \, \sigma$ window.
Hence, one could argue that the progress in understanding the CMB anomalies from WMAP to Planck, a time interval spanning almost two decades, has been marginal at best. At this level of statistical significance, the discrepancies could be discounted as statistical fluctuations. As we touched upon in sections \ref{sec:tests_statistical_isotropy} and \ref{sec:anomalies_topo_cmb}, there is value in introducing new statistics that probe CMB isotropy, since there are underlying asymmetries, which allows one to gauge the sensitivity of the statistic to these well-documented anomalies.  

Some of the low multipole CMB anomalies taken together were studied to tease out any correlation among them~\cite{bib:Muir2018}. In particular, those authors considered the missing power on large angular scales $S_{1/2}$, the power in quadrupole and octopole modes ($C_2$ and $C_3$), the low variance anomaly ($\sigma_{16}^2$, as computed from CMB maps at HEALPix $N_{\rm side}=16$), the point parity anomaly $P(\ell_{\rm max})$ (but referred to as $R_{27}$ for the choice $\ell_{\rm max}=27$), the anti-podal correlations from the two-point correlation itself $C(\alpha=\pi)$, the quadrupole-octopole alignment $S_{\rm QO}$, and the dipolar amplitude of the hemispherical power asymmetry estimated using a local variance estimator ($A_{LV}$) i.e., a total of eight anomalous features. The covariances inferred from extensive simulations as well as those provided by Planck, revealed that the the first six and the last two are found to be largely uncorrelated. As expected, the $S_{1/2}$, $C_2$, $C(\pi)$ and $\sigma^2_{16}$ are all positively correlated. When a principal component analysis of their covariance is performed (which will return a map of combinations of eight features analyzed), almost 90\% of their covariance weight is accounted for by the first four components, where the first two correspond to a lack of large angle correlations and the next two being the $\ell=2,3$ alignment and the HPA anomaly.

While earlier studies with WMAP data \cite{bib:Axelsson2013} suggested small variations in parameters ($\sim 1 \, \sigma$), recent studies with Planck data \cite{bib:Fosalba2021, bib:Yeung2022} highlight variations up to $\sim 3 \, \sigma$ in cosmological parameters including $H_0$. However, such conclusions are contradicted by Ref. \cite{bib:Mukherjee2016, bib:Mukherjee2018}. In  \cite{bib:Fosalba2021}, this disagreement was discussed and attributed to the use of an approximation, namely a Taylor expansion, to relate the power spectra to cosmological parameters. Nevertheless, as we have seen, the variations of the cosmological parameters on the sky seem to have the right patterns and be of the appropriate scale to be picking up the hemispherical power asymmetry (section~\ref{sec:hemispherical_power_asymmetry}) and potentially the anomalous quadrupole-octopole (section~\ref{sec:quadrupole_octopole}).     

Clearly, if one disregards half the CMB data, one can expect the statistical errors in the resulting cosmological parameters to increase by a factor of $\sim \sqrt{2}$, but it is not anticipated that the original central value falls outside of the new errors. Contrary to this expectation, \cite{bib:Yeung2022} considers hemispherical skies and yet $H_0$ can vary between $H_0 =(64.4 \pm 1.3)$~km/s/Mpc and $H_0 =(70.1 \pm 1.4)$~km/s/Mpc. The original central value thus falls outside of both of these intervals, while the errors seem consistent with the removal of half the data. In effect, this implies that the Planck $H_0$ value may simply be an average of (mildly) discrepant $H_0$ determinations within the flat $\Lambda$CDM model. Based on analyses in  \cite{bib:Fosalba2021, bib:Yeung2022}, within the flat $\Lambda$CDM model, the `CMB Hubble tension' may be in the $2.6 \, \sigma$ to $3\, \sigma$ window and we comment more on the implications in section~\ref{sec:epilogue}.

\newpage
\section{Late Universe FLRW anomalies}
\label{sec:late_universe_FLRW_anomalies}

Having introduced early Universe anomalies in the CMB in section~\ref{sec:early_universe_FLRW_anomalies}, here we turn our attention to the late Universe. Throughout, it is worth bearing in mind that a number of discrepancies or tensions have arisen between Planck inferences of cosmological parameters based on the CMB \cite{bib:Planck2018b} and determinations of the same quantities in the late Universe. These $\Lambda$CDM tensions, which primarily involve $H_0$ \cite{bib:Huang2020, bib:deJaeger2020, bib:Pesce2020, bib:Kourkchi2020, bib:Schombert2020, bib:Khetan2021, bib:Blakeslee2021, bib:Freedman2021, bib:Riess2021}, $S_8$ \cite{bib:Asgari2021, bib:Heymans2021, bib:Amon2022, bib:Abbott2022}, and to a lesser extent curvature \cite{bib:Planck2018, bib:DiValentino2020}, and more recently $A_{ISW}$ \cite{bib:Kovacs2022, bib:Kovacs2022b}, are often framed as early versus late Universe tensions \cite{bib:Verde2019}, see \cite{bib:Abdalla2022} for a recent review.\footnote{This distinction between ``early" and ``late" may no longer be so clear given that one can infer a Planck value for $H_0$ that is largely agnostic of early Universe physics, e. g.  \cite{bib:Lin2021, bib:Philcox2022}.} 
Notably, this $\Lambda$CDM tension debate happens essentially within FLRW. If there is a mismatch, since Planck makes inferences within the flat $\Lambda$CDM model, one can change the cosmological model within the FLRW class to alleviate the anomalies, meaning that the tensions are \textit{cosmological model dependent}. That being said, once one has enough anomalies, resolving the discrepancies within FLRW is expected to be tricky, see e.~g.~\cite{bib:Krishnan2021b, bib:Krishnan2022}. This partially motivates the current review. As highlighted in section~\ref{sec:CMB_variations}, it is also possible that errors are underestimated and the tensions are less pronounced than currently claimed.  

The above comments serve as a prelude to FLRW anomalies in the late Universe. As mentioned in the introduction, the homogeneity scale has been inferred from large scale structure and the value $\sim 70 \, h^{-1}\,$Mpc, $\approx 100$~Mpc, is representative \cite{bib:Hogg2005, bib:Yadav2005, bib:Scrimgeour2012, bib:Laurent2016, bib:Ntelis2017, bib:Sarkar2019, bib:Goncalves2018, bib:Goncalves2021} if one just focuses on two-point galaxy correlations (see comments in section \ref{sec:Hom-scale}). In truth, the local Universe has been well mapped out to distances of $\sim 200$ Mpc or redshift $z \sim 0.05$ by the COSMICFLOWS program \cite{bib:Tully2016, bib:Kourkchi2020}, revealing a host of intriguing structures.
The latter are at odds with an \textit{exact} homogeneous and isotropic universe. Astronomers and cosmologists may differ on this point and we leave the door open to cosmologists treating the documented local Universe as a perturbation within FLRW, so that it is \textit{statistically} homogeneous and isotropic. This point aside, it is worth noting that in the aftermath of the HST key project \cite{bib:Freedman2001}, \cite{bib:McClure2007} plotted local $H_0$ values on the sky, reinforcing the fact that $H_0$ varies on the sky in the local Universe. This sits uneasily with the notion that $H_0$ is an integration constant \cite{bib:Krishnan2021} and underscores the difficulty in determining $H_0$ through the distance ladder. One can determine the (average) \textit{rate of expansion}, but this is only required to equate with $H_0$ within the FLRW setup. A secondary point is that evidence for a coherent bulk flow towards the Shapley supercluster at a distance of $\sim 200$ Mpc has emerged \cite{bib:Hoffman2017} (see also \cite{bib:Qin2021, bib:Howlett2022}). Thus, the local Universe is anisotropic in a well defined sense, but this anisotropy may be viewed as a local perturbation within an FLRW universe.   

The relevant question now is whether or not this well-documented bulk flow converges to the expectations of the flat $\Lambda$CDM model or more generally an FLRW cosmology? Recently, \cite{bib:Howlett2022} has noted the presence of a larger than expected bulk flow relative to $\Lambda$CDM expectations at depths of $140 \, h^{-1}\,$Mpc. 
Yet, unexpectedly large bulk flows in the CMB dipole direction, or towards the Shapley supercluster have a long record already \cite{bib:Watkins2009, bib:Kashlinsky2008, bib:Lavaux2010, bib:Magoulas2016}, which we review in section~\ref{sec:bulk_flows}. It is imperative to push the program of mapping out peculiar velocities to larger distances $\sim 400$-$500$~Mpc, to not only confirm or refute convergence to $\Lambda$CDM expectations, but also in order to test the recovery of the anticipated FLRW universe. As we discuss in section~\ref{sec:galaxy_cluster_anisotropies}, there are some unexpected anisotropies in galaxy cluster scaling relations \cite{bib:Migkas2018, bib:Migkas2020, bib:Migkas2021} cautioning that this may not happen. Translated into the Hubble constant, one encounters $\sim 10 \%$ variations of $H_0$ on the sky \cite{bib:Migkas2021}, which echoes well with
\bcomm{observations at lower redshifts \cite{bib:McClure2007,bib:Wiltshire2012},} where it may be more readily digested. The reported Hubble tension is also at $10\%$ level, e.~g.~see \cite{bib:Abdalla2022}. 

In addition to the debate about bulk flows and whether they converge to $\Lambda$CDM or FLRW expectations in the local Universe, the cosmic dipole finds itself at the heart of another fascinating discussion. Recall that the CMB is analyzed under the assumption that the Universe is isotropic. Doing so, one finds that the largest temperature anisotropy is the dipole and this is subtracted on the basis that it is simply due to relative motion. This subtraction not only fixes our velocity with respect to the CMB, but also fixes the CMB as the Universe's rest frame. As highlighted in \cite{bib:Ellis1984}, this motion can be cross-checked with observations at lower redshifts which are supposedly located in the same rest frame as the CMB. 

To perform the Baldwin-Ellis test \cite{bib:Ellis1984}, one must leverage large surveys of radio galaxies or QSOs or maybe SNe, which, provided they are at suitably large scales, are required to be isotropically distributed in line with the FLRW paradigm. Therefore, simply by assuming isotropy, one sets up a direct comparison between our velocity with respect to CMB and our velocity with respect to these distant sources as inferred from aberration effects. If FLRW is correct, the velocities are the same and the distant sources share the same frame as the CMB. Interestingly, as we review in sections~\ref{sec:radio_dipole} and \ref{sec:QSO_dipole}, preliminary results indicate an excess in the cosmic dipole amplitude as inferred by observations \cite{bib:Blake2002, bib:Singal2011, bib:Gibelyou2012, bib:Rubart2013, bib:Tiwari2016, bib:Bengaly2018, bib:Singal2019, bib:Siewert2021}, even up to high statistical significance $\sim 5 \, \sigma$ \cite{bib:Secrest2021, bib:Secrest2022}. In contrast to $\Lambda$CDM or cosmological tensions, this test assumes no specific cosmology\footnote{Needless to say, some astrophysics is assumed, for example a power law for the spectral emission. This can of course evolve with redshift \cite{bib:Dalang2022}, but such evolution warrants a physics explanation.}, only an isotropic universe, so this discrepancy is \textit{cosmological model independent}. As a result, any discrepancy could be potentially fatal to the FLRW paradigm, unless systematics come to the rescue. If substantiated, the most na\"{\i}ve or natural interpretation is an anisotropic universe, a feature that would be difficult to hide in complementary cosmological probes. In section~\ref{sec:emergent_H0} we illustrate how this could already be impacting the $\Lambda$CDM Hubble diagram, in particular $H_0$.

\subsection{Bulk flows} 
\label{sec:bulk_flows}

Assuming a flat Minkowskian spacetime in our immediate cosmic neighborhood, the blue and redshifted spectra of neighboring galaxies were interpreted as approaching and receding peculiar velocities with respect to our position and already led to the first Hubble diagrams almost 100 years ago. 
On top of a general FLRW cosmology, these radial peculiar velocities in the vicinity of an observer at $z=0$ are then given as
\begin{equation}
    v_\mathrm{pec} = c z - H_0 D \;,
    \label{eq:v_pec}
\end{equation}
in which $c$ is the speed of light, $H_0$ is the Hubble constant, $z$ is the redshift of the moving object with respect to the observer, and $D$ its distance.\footnote{Due to the proximity to $z=0$ all distance measures introduced in section~\ref{sec:terminology} yield the same distance. {Eq.~(\ref{eq:v_pec}) also assumes the Newtonian velocity addition approximation, which can lead to systematic errors at the redshifts of typical surveys if percent level precision is required \cite{bib:Calcino2017}.}} 
Consequently, any observed peculiar velocity in our local environment in which the expansion of the background $\Lambda$CDM cosmology is linear, is often interpreted as a simple Doppler shift on a Euclidean space. 
As noted in \cite{bib:Wiltshire2012} it is possible to define peculiar velocities by (\ref{eq:v_pec}) independently of a cosmological model by direct spherical averages of the data, provided the average is predominantly linear, that is on scales $z\lesssim0.05$. 

Coherent peculiar velocities of several objects in joint motion observed in our cosmic neighborhood up to $z \approx 0.1$ are usually called ``bulk flows''. 
Within $\Lambda$CDM, \bcomm{the} frame of reference in which they are defined to be moving can be given by the cosmic rest frame of the CMB, see, for instance, \cite{bib:Gunn1988} for an early overview of measurements. 
This is also the frame, in which the observed isotropy of
the Hubble expansion should only show a minimum of statistical fluctuations.
Extending the quasi-Newtonian picture in our cosmic neighborhood to higher redshifts, the latter is also often called ``Hubble flow'' because it accounts for the motion of all cosmic structures due to the expansion of the Universe within $\Lambda$CDM.
As also detailed in section~\ref{sec:Hom-scale}, complex structural patterns at low redshifts deviate from the statistical homogeneity of the $\Lambda$CDM Hubble flow.
Their redshifts can thus be interpreted as peculiar velocities on top of the background, caused by their mutual gravitational interactions. 

From (\ref{eq:v_pec}), it is obvious that any measurement of a bulk flow velocity requires the maximum distance to its outer-most part to be known and the frame of reference with respect to which it is considered.
Ref.~\cite{bib:Tully2008}, for instance, distinguishes between three different reference frames, the CMB frame, the so-called Local Sheet frame which extends out to $D\approx 6.25~h^{-1}~\mbox{Mpc}$, and our Local Group (LG) frame \bcomm{within} the Local Sheet. 
\bcomm{The Local Group is the largest group of galaxies to which we are gravitationally bound, extending to about 3~Mpc. Its two largest members are the Milky Way and Andromeda. Its barycenter defines the Local Group reference frame. The Local Sheet is a denser filamentary expanding region assumed to be one boundary of the Local Void and almost all bright galaxies in our local environment are part of it. With a
kinematic interpretation the local expansion sheet frame is interpreted as a peculiar velocity of 66~km~s$^{-1}$ between the Local Goup the Local Sheet reference frames.}
Unless stated otherwise, we refer to the CMB frame, but note that redshifts as observed from our position then have to be transformed into this frame, see \cite{bib:Davis2019} and section~\ref{sec:SN_dipole} for details. 

If the CP holds, one would expect the bulk flows to converge to the Hubble expansion at some observed redshift or distance.
Intuitively, as phrased in \cite{bib:Osborne2011}, we look for the minimum distance from us at which peculiar velocities of individual objects come to rest with respect to the CMB and are only subject to cosmic expansion (see also section~\ref{sec:Hom-scale} for further details). 
As discussed below, several anomalies questioning this convergence have surfaced,  including bulk flow amplitude excesses discovered in our cosmic neighbourhood, which shift the transition from bulk to Hubble flow further and further into the distant Universe. 

The most prominent and polarizing anomaly, called the `dark flow', was claimed in \cite{bib:Kashlinsky2008}, after the same team set up their measurement approach detailed in \cite{bib:Kashlinsky2000}. 
Using the kinematic Sunyaev-Zel'dovic effect (kSZe), it is concluded in \cite{bib:Kashlinsky2008, bib:Kashlinsky2009} that an all-sky bulk flow of about 782 X-ray-selected and X-ray-flux-limited galaxy clusters with respect to the CMB observations of the 3-yr WMAP data did not converge to the Hubble flow on scales $\leq 300~h^{-1}~\mbox{Mpc}$, which was also the limiting distance of the survey.  
The velocity of the flow was estimated to be in the range of $600$--$1000~\mbox{km/s}$ and found to be lying in the direction of the motion of our LG with respect to the CMB. 
The authors later extended their sample to more clusters, added spectroscopic redshifts to check distance-dependency effects and used the 5-yr and 7-yr WMAP data, coming to the same conclusion of a dark flow out to even 800~Mpc, as shown in \cite{bib:Kashlinsky2010} and \cite{bib:Kashlinsky2011}.

These results were not reproducible by several other independent analyses as detailed below, summarised, and further discussed in \cite{bib:Kashlinsky2012}.
Possible causes for the \bcomm{unreproducibility} were already investigated in \cite{bib:Haehnelt1994}.
It was possible that the matched-filter approach used by \cite{bib:Kashlinsky2008} under-estimated the biases in the kSZ-signal coming from fluctuations in the CMB, as argued in \cite{bib:Keisler2009}.
Yet, as analysed in detail in \cite{bib:Atrio2010}, \cite{bib:Keisler2009} did not account for all biases when setting up their simulation to determine the significance of the bulk flow, resulting in over-estimated error bars and thus a reduced significance.

There is also a dependence of the resulting signal on the assumed cosmology, as mentioned in \cite{bib:Haehnelt1994}. 
To break the degeneracy between the kSZe and the thermal Sunyaev-Zel'dovic effect (tSZe), \cite{bib:Haehnelt1994} furthermore suggest to perform the observations around 220~GHz, in order to obtain a clear signal of the kSZe only. 
The original study \cite{bib:Kashlinsky2008}, however, used observations at lower wavelengths and had to deblend both signals from each other, increasing the error bars of the resulting kSZ signal and potentially mistaking a tSZ signal for a bulk flow, as suggested in \cite{bib:Osborne2011}. 
As a reaction, \cite{bib:Atrio2015} repeated the analysis using the first year Planck data including the 217~GHz filter band for which the tSZe becomes negligibly small and checked the frequency-independence of their results as a corroboration for measuring the kSZe and not the tSZe. 
Still, the dark flow persisted.
A measurement of the tSZe in the original data, containing all SZ signals before filtering out the kSZe, showed a correlation to the bulk flow of the kSZe further hinting at the dark flow to be a true signal instead of a statistical fluke.
Besides, as \cite{bib:Osborne2011} also shows, different matched filters can have a strong impact on the resulting kSZ signal and, basing their analysis only on 100 simulations compared to the 4000 of \cite{bib:Atrio2010}, the significance is consequently affected as well.

Then, Planck \cite{bib:Planck2013d} could not reproduce the significance of bulk flow of \cite{bib:Kashlinsky2011}, observing the kSZe on the CMB with 1321 Meta Catalogue of X-ray detected Clusters of galaxies.
According to \cite{bib:Planck2013d}, a comparison to the WMAP data used in \cite{bib:Kashlinsky2008} showed that the dipole signal should not be attributed to the clusters' peculiar motion, but rather to residuals in the filtered map. 
Yet, as outlined in \cite{bib:Atrio2013}, the method to calculate the uncertainties used in \cite{bib:Planck2013d} may be biased high, thus reducing the significance. 

As mentioned in section~\ref{sec:prologue}, \cite{bib:Migkas2021} uses ten different scaling relations of multi-band observations from up to 570 clusters to constrain anisotropies in cosmological parameters.
From an analysis of the tSZe, they find that the variations in $H_0$ can be interpreted as a $\sim~800$-$900$~{km/s} bulk flow in a similar direction as the motion of our LG out to 500~Mpc at $4 \, \sigma$ significance level (see also Fig.~\ref{fig:BF-plots}). 
Thus, the controversy over an excess bulk flow amplitude measured for X-ray selected clusters at intermediate redshifts has been revived anew independently of the dark-flow collaboration.

In a similar manner as the scaling relations for X-ray clusters, flux observations of standardized SNe Ia across the sky and in different redshift bins can be used to measure bulk flows. 
Yet, the data sample has not been sufficient so far to obtain significant statements, see for instance \cite{bib:Colin2011, bib:Dai2011, bib:Turnbull2012, bib:Wiltshire2012, bib:Appleby2015}. 
The most recent analysis, performed with the Pantheon+ data set detailed in \cite{bib:Brout2022}, concludes that the residual signals in the Hubble diagram in $0.01 < z < 0.03$ after correcting for known local bulk flows and the CMB dipole could be caused by the coupled motion of the Milky Way and other nearby galaxies that has not been modeled correctly yet.  

As noted in \cite{bib:Planck2013d}, their data set of galaxy clusters in the neighborhood of the LG is too sparse to employ a kSZe-based measurement of bulk flows in our cosmic neighborhood. 
At these distances, we thus have to resort to other methods to measure bulk flows. 
One such general technique is based on measuring the redshifts and distances of galaxies independent of each other and employing (\ref{eq:v_pec}) to infer the amplitude and direction of bulk flows in different distance shells around us. 

Using photometry of the brightest cluster galaxies as distance indicators in addition to redshift observations, the bulk flow of the LG with respect to the reference frame out to $z \approx 0.03$ has been constrained by their cluster data set which is limited to $z \le 0.05$ \cite{bib:Lauer1994}.  
They find a larger than expected flow and no convergence to the CMB rest frame even for the bulk flow of their cluster data set. 
Yet, as  noted in \cite{bib:Hudson2004}, the distance measure used in \cite{bib:Lauer1994} was contaminated by filamentary or patchy dust for at least 16\% of the clusters in the sample, so that the excess in bulk-flow amplitude could have been over-estimated.
While \cite{bib:Hudson2004} agrees to exclude convergence to the CMB frame by a depth of $60h^{-1}~\mbox{Mpc}$ as do multiple other data sets, they also exclude excess flows of order 600~{km/s} at distances between 60 to $120h^{-1}~\mbox{Mpc}$ due to sparse sampling effects in this range.

\begin{figure}[htp]
    \centering
    \includegraphics[width=0.77\linewidth]{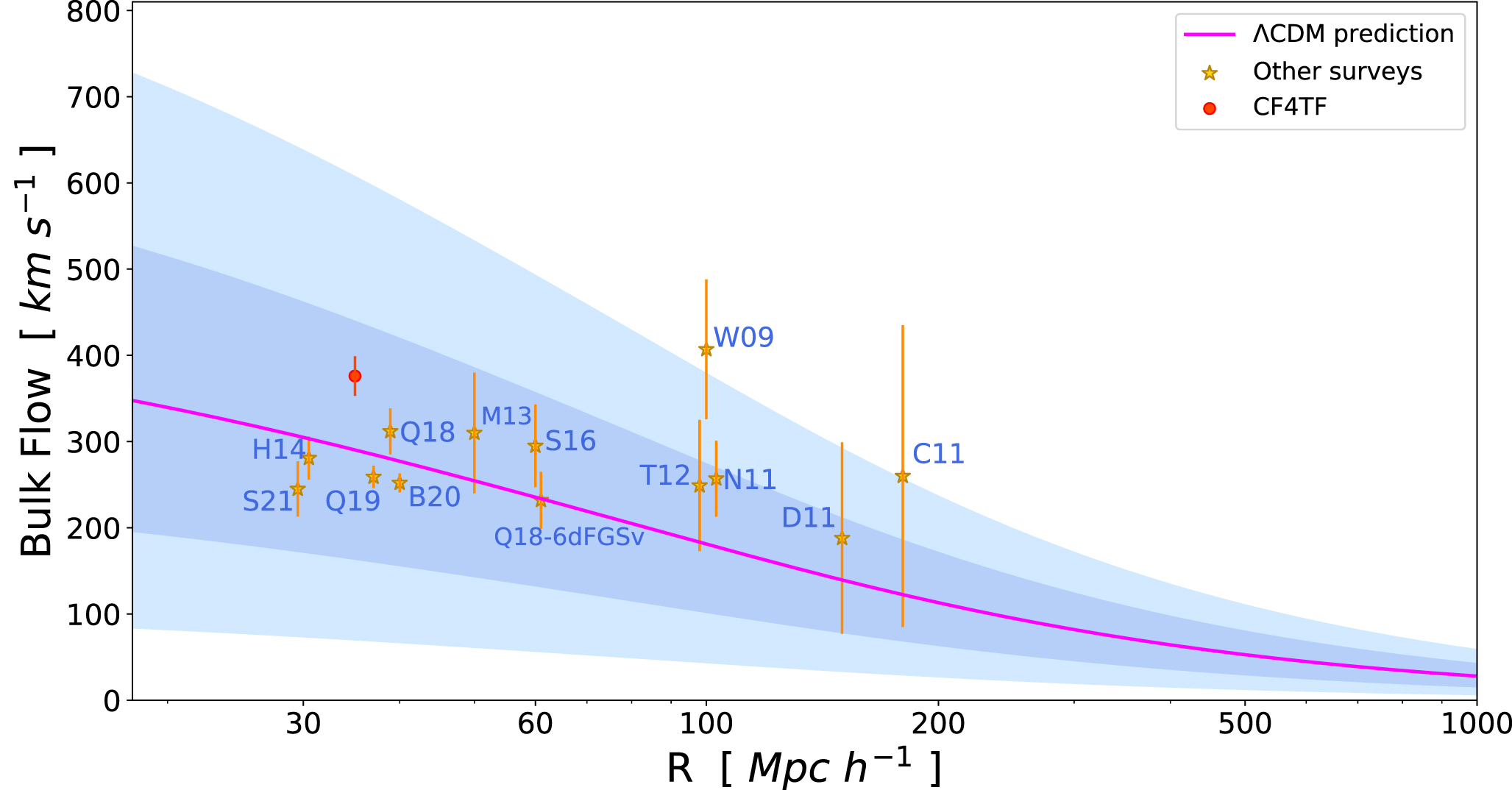}
    \caption{Fig.~9 from \cite{bib:Qin2021}: Comparison of reconstructed bulk flows in our local neighborhood. The pink curve is the $\Lambda$CDM prediction calculated from a spherical top-hat window function. The shaded areas indicate the 1$~\sigma$ and 2$~\sigma$ cosmic variance. The yellow stars are the measurements from \cite{bib:Watkins2009} (W09); \cite{bib:Colin2011} (C11); \cite{bib:Dai2011} (D11); \cite{bib:Nusser2011} (N11); \cite{bib:Turnbull2012} (T12); \cite{bib:Ma2013} (M13); \cite{bib:Hong2014} (H14); \cite{bib:Scrimgeour2016} (S16); \cite{bib:Qin2018} (Q18); \cite{bib:Qin2019} (Q19); \cite{bib:Boruah2020} (B20); \cite{bib:Stahl2021} (S21). The red dot is measurement of \cite{bib:Qin2021}.}
    \label{fig:bulk_flows_literature}
\end{figure}

The brightest-cluster-galaxy photometry can be replaced by the Tully-Fisher relation \citep{bib:Tully1975} as a distance indicator to spiral galaxies or by the Fundamental-Plane of elliptical galaxies \citep{bib:Djorgovski1987, bib:Dressler1987} to determine their distance to us. 
Both empirically derived distance indicators can increase their precision by jointly evaluating several galaxies in the same gravitationally bound structure, such that the derived bulk flow also becomes more precise. 
In addition, complementary data like SNe Ia can be added to a joint data set, see for instance \cite{bib:Tully2016}.
Mapping out the three-dimensional flows in the local Universe including attractors like the Shapley Supercluster or the Dipole Repellor out to scales of $200h^{-1}~\mbox{Mpc}$ and to a resolution of a few megaparsecs has been achieved with these local distance measures and joint data sets in numerous works, as detailed in \cite{bib:Courtois2013, bib:Hoffman2017,bib:Qin2021} and references therein.\footnote{See \url{https://tinyurl.com/yeypxsn9} for videos of bulk-flow and structure reconstructions based on COSMICFLOWS.} 
Most of the works conclude that the reconstructed bulk flows are in agreement with $\Lambda$CDM predictions. 
See Fig.~\ref{fig:bulk_flows_literature} for different approaches and data sets, all recalibrated to be comparable with each other, as detailed in \cite{bib:Qin2021}. 

Thus, we can conclude, on the one hand, that bulk flows in our observable volume mostly agree within $3 \, \sigma$ with the expectations from $\Lambda$CDM, either as obtained in simulations or as predicted by reconstructed density maps based on linear perturbation theory and redshift surveys, see for instance \cite{bib:Boruah2020}. 
On the other hand, observations from multiple, complementary data sets obtained in the last three decades also reveal that the dynamical structures out to $200h^{-1}~\mbox{Mpc}$ are more complex than originally assumed for an FLRW universe and that there is no scale of homogeneity around 50-80$h^{-1}~\mbox{Mpc}$. 
The anomalous bulk flow of \cite{bib:Migkas2021} remains to be explained. 
It would not be surprising to find a biasing effect that accounts for the excess amplitude, as has been the case for previously discovered anomalies and one possible suggestion is given in section~\ref{sec:galaxy_cluster_anisotropies}.

However, calibrating these anomalies away a posteriori, for instance, by accounting for clustering of structures, Malmquist biases, sparse sampling effects, or intrinsic fluctuations in the observed signals, could be considered a human bias towards an expected result. 
That being stated, blind analyses should be performed in the future to check for such biases as well. 

Finally, an alternative perspective is considered by \cite{bib:Wiltshire2012,bib:McKay2016} who studied the {\sc composite} \cite{bib:Watkins2009,bib:Feldman2010} and Cosmicflows-II \cite{bib:Tully2013} samples by averages in radial spherical shells on one hand, and angular averages of the type introduced by \cite{bib:McClure2007} on the other. A best fit spherically averaged linear Hubble relation was determined, with a comparison of local rest frames to determine that for which the expansion is most uniform. For the {\sc composite} sample the spherically averaged linear Hubble law was found to be significantly more uniform (with $\ln B>5$ Bayesian evidence) for an observer at rest in the LG frame\footnote{The Local Group is the largest bound structure within which conventional kinematics of non-expanding motion is obtained at our location, and represents a natural scale in the ``fitting problem'' \cite{bib:Ellis1987,bib:Wiltshire2007} if non-kinematic differential expansion is admitted as a possibility on local cosmological scales \cite{bib:Bolejko2016}. A lack of data in the Zone of Avoidance in the galactic plane means that one can perform local boosts in the galactic plane without affecting the likelihood, so the LG fame is actually one of a degenerate set of local Lorentz frames that could qualify as the frame one in which local expansion is most uniform \cite{bib:McKay2016}, but the CMB frame is excluded from this set.} as compared to the conventional CMB frame \cite{bib:Wiltshire2012}. The residual dipole in this frame is generated by structures in the range $40\lesssim r\lesssim60\,h^{-1}\,$Mpc, in a direction consistent with the CMB dipole in the LG frame. While \cite{bib:Wiltshire2012} suggested an explanation in terms of a $0.5$\% non-kinematic anisotropy on scales up to $65\,h^{-1}\,$Mpc -- which would circumvent the issue of convergence of bulk flows to the CMB frame -- it was subsequently demonstrated that the same results can be reproduced in $N$-body simulations \cite{bib:Kraljic2016} and perturbed $\Lambda$CDM models \cite{bib:Bengaly2019a}. From this viewpoint, the uniformity of expansion in the LG frame is due to our participation in a large scale coherent bulk flow on $\gtrsim100\,h^{-1}\,$Mpc scales, on top of which a dipole due to the more nearby structures is superposed. Uncertainties are too large to distinguish the conventional kinematic and non-kinematic hypotheses; though potentially future peculiar velocity surveys might do so.

\subsection{Galaxy cluster anisotropies}
\label{sec:galaxy_cluster_anisotropies}
\vspace{-1ex}
Galaxy clusters can be an excellent probe for scrutinizing the isotropy of the \bcomm{\textit{local}} Universe. They are the largest gravitationally bound systems in the cosmos and  contain up to thousands of galaxies and vast amounts of dark matter. Their intracluster medium (ICM) is filled with hot ionized plasma which strongly emits X-ray radiation. Owing to this emission, one can measure numerous X-ray properties of clusters which trace the density and temperature of the ICM. The ICM can be also indirectly  observed in the microwave regime through the SZ effect which appears as a spectral distortion of the CMB toward a cluster's direction. The amplitude of this distortion also traces the thermal state of the ICM. Finally, clusters can be also observed through their galaxy members in optical and infrared wavelengths. Galaxy clusters exhibit tight correlations among many of their properties which scale with mass, giving rise to the so-called ``cluster scaling relations'' (e.~g., \cite{bib:Kaiser1986,bib:Giodini2013}). These relations have been well-established observationally for more than two decades now and are being used for both cluster physics and cosmology studies.

Recently, \cite{bib:Migkas2018, bib:Migkas2020, bib:Migkas2021} introduced and employed a novel and powerful method to probe the isotropy of the local Universe ($z\lesssim 0.3$) using galaxy cluster scaling relations. The key  property for this test is the cluster temperature $T$ whose observational determination does not depend on any cosmological assumptions, i.~e., it is \emph{cosmology-independent}. On the other hand, the inference of many other cluster properties from direct observables (e.g, flux and apparent angular cluster size) strongly depend on the fiducial cosmology through the cluster distance, and most notably on $H_0$. Such \emph{cosmology-dependent} cluster properties are the X-ray luminosity $L_{\text{X}}$, the total integrated Compton parameter $Y_{\text{SZ}}$ measured in microwaves, the infrared luminosity of the brightest cluster galaxy $L_{\text{BCG}}$, and the X-ray effective radius $R_{50\%}$. By studying the \bcomm{angular} variation of \bcomm{observational} scaling relations between $T$ and cosmology-dependent cluster properties, one can draw conclusions about the directionality of $H_0$ and the existence of bulk flows. Both an $H_0$ anisotropy and a large bulk flow would result in a directional-dependent change of, e.~g., $L_{\text{X}}$ and $Y_{\text{SZ}}$, while $T$ would remain unaffected. This method is equivalent to considering clusters as standardizable candles \emph{and} rulers. 
The deviation of the cluster luminosity, total thermal energy or physical size from their average values for a given cluster sample is a strong function of the measured cluster temperature. Thus the values \bcomm{of} cosmology-dependent cluster properties (e.~g., X-ray luminosity) can be standardized based on the cosmology-independent temperature. The major advantage of this methodology is that a single galaxy cluster sample can provide constraints on cosmic isotropy in multiple, nearly independent ways, and across the electromagnetic spectrum, by utilizing numerous different cluster properties.

\begin{figure}
\centering
 \includegraphics[width=0.75\textwidth]{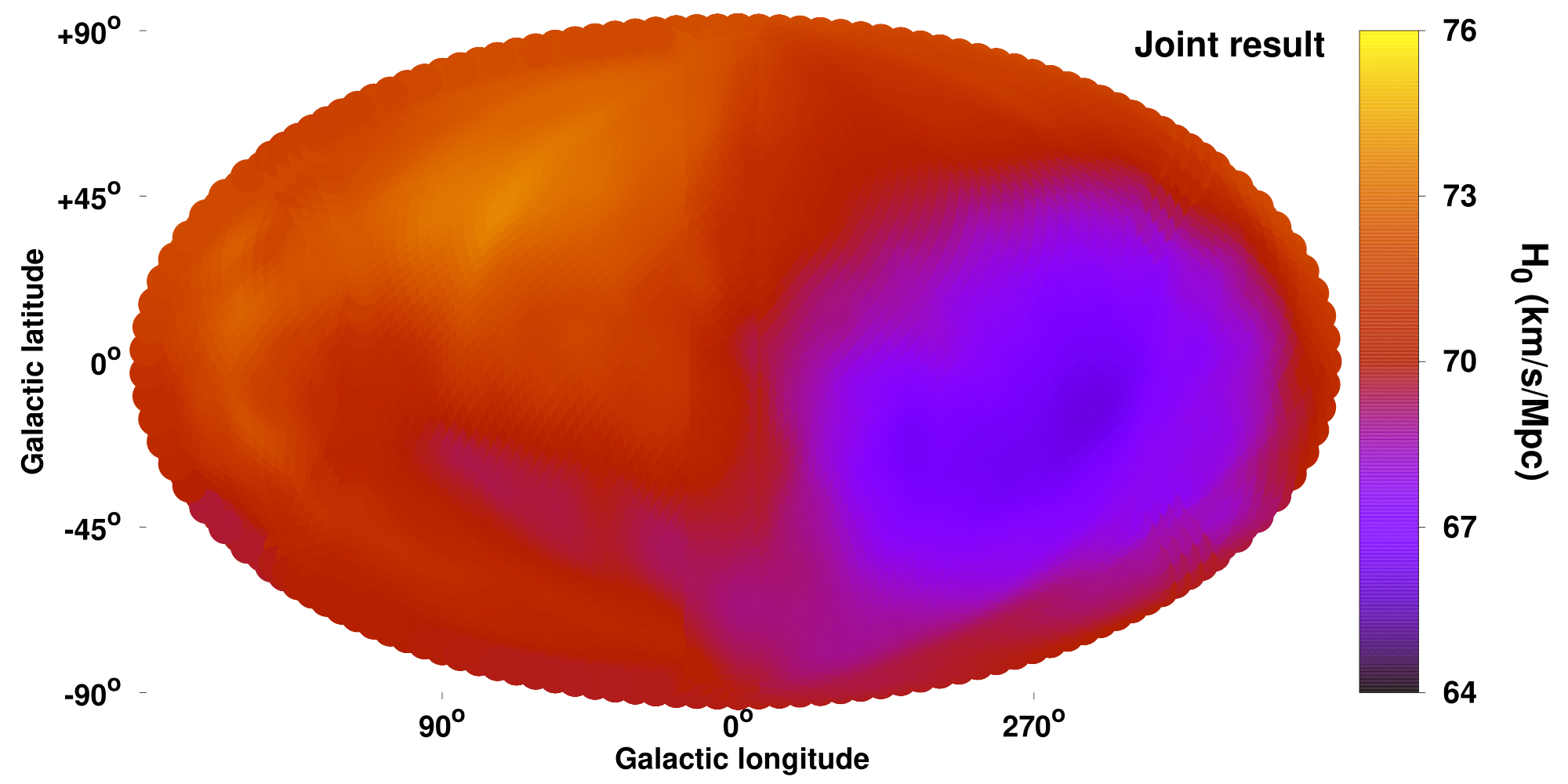}
 \caption{\bcomm{Observed $H_0$ angular variation as obtained in  (reproduced from Fig. 7 of \cite{bib:Migkas2021}) using 481 galaxy clusters with X-ray, SZ, and infrared data. The high (orange) and low (purple) $H_0$ regions share a $5.4$-$\sigma$ tension.}}
 \label{fig:migkas_H0_anisotropy}
\end{figure}

In \cite{bib:Migkas2021} all the above-mentioned cluster properties were used for up to 570 galaxy clusters and two different, independent cluster samples. The \bcomm{angular} distribution of these clusters is nearly uniform across the extragalactic sky, while they cover the redshift range $z\approx [0.01,0.60]$ with a median $z\approx 0.1$. This allowed for the construction of ten multi-wavelength cluster scaling relations. Owing to the different sensitivities of these relations on the underlying cosmology and systematic biases, one can robustly isolate the cosmological signal from any observed anisotropies. By jointly analyzing X-ray, microwave, and infrared scaling relations, \cite{bib:Migkas2021} detected a 9\% dipole $H_0$ anisotropy, pointing toward $(l,b)\sim (280^{+35^{\circ}}_{-35^{\circ}},-15^{+20^{\circ}}_{-20^{\circ}})$, see Fig.\ref{fig:aniso} for comparison to other reported anisotropy directions. By utilizing isotropic Monte Carlo simulations, the statistical significance of the observed anisotropy was assessed at $\gtrsim 5.4 \sigma$, constituting the strongest-ever evidence for an anisotropy in the local Universe. The all-sky $H_0$ \bcomm{angular} variation map as obtained in \cite{bib:Migkas2021} is displayed in Fig.~\ref{fig:migkas_H0_anisotropy}. Furthermore, the anisotropy was consistently detected in all scaling relations separately, independently of the used wavelength. 

An exhaustive list of potential systematic biases that could result in artificially observed anisotropies was tested thoroughly. For instance, the effects on the anisotropies that selection biases, Galactic absorption, the Zone of Avoidance gap, different cluster populations, and numerous other systematics could have, were carefully assessed. Overall, no test was able to alleviate the observed tension; the case for a cosmological origin of the anisotropies was eventually strengthened. 

The direction of the cluster anisotropies is consistent with many past results, as summarized in \cite{bib:Migkas2020} and \cite{bib:Migkas2021}. The agreement is stronger when the results are compared with other low redshift probes. More distant probes, such as QSO, tend to point to a slightly different direction, more aligned with the CMB dipole. The partial disagreement between these probes and galaxy clusters can be attributed to two different factors; firstly, the influence of local motions on the cluster analysis and the lack of such effects for more distant sources, and secondly to the large direction uncertainties of cluster scaling relation anisotropies.

\begin{figure}[ht]
\centering
               \includegraphics[width=0.495\textwidth]{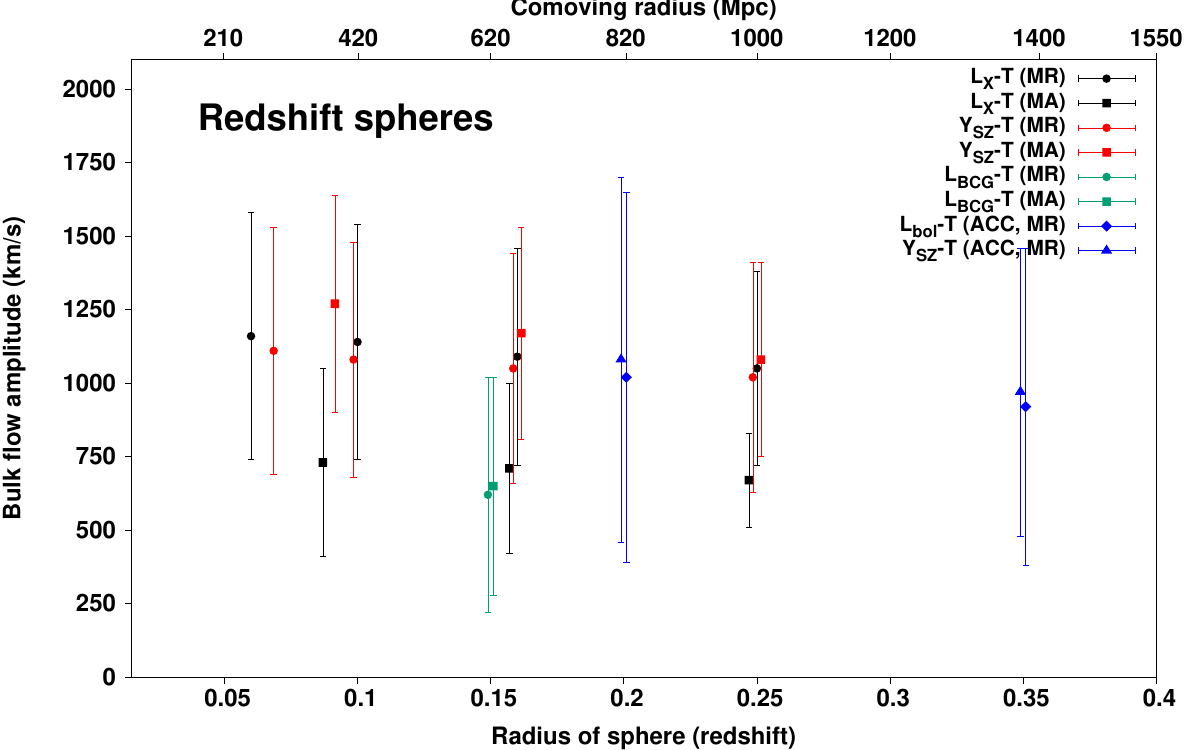}
                \includegraphics[width=0.495\textwidth]{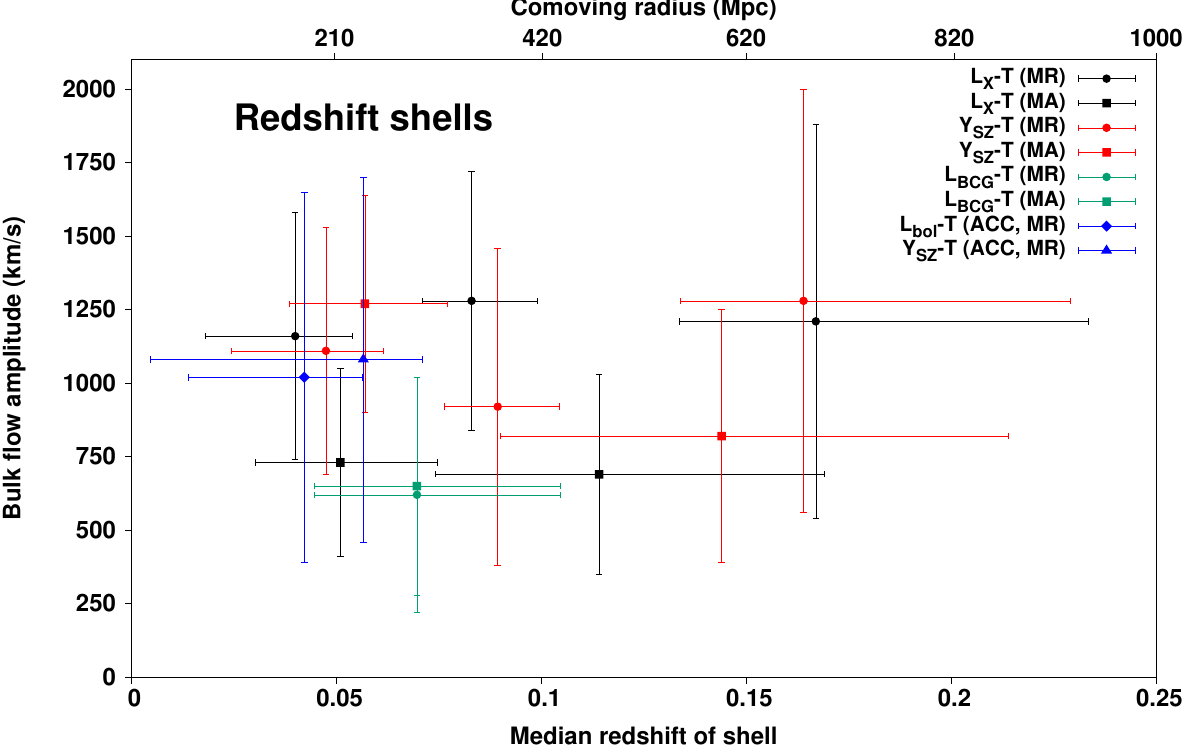}
               \caption{Left: Bulk flow amplitude as a function of the redshift/distance radius of the increasing spherical volumes. Different data point shapes and colors correspond to different scaling relation, samples, and statistical approaches. Right: Bulk flow amplitude as a function of the median redshift/distance of each shell, together with the standard deviation of the redshift distribution. \bcomm{Figures reproduced from Fig. 13 of \cite{bib:Migkas2021}.}}
        \label{fig:BF-plots}
\end{figure}

Alternatively to an $H_0$ anisotropy, a large bulk flow could explain the findings. The vast majority of used clusters lie at $z\lesssim 0.2$. As a result, one cannot distinguish between an $H_0$ anisotropy and a large-scale, strong bulk flow. Thus, \cite{bib:Migkas2021} quantified the amplitude and scale of such a flow motion that would explain the apparent cluster anisotropies. They found that a $\sim 900~$km/s bulk flow seems to exist, extending out to $\gtrsim 500$~Mpc, toward $(l,b)\sim (275^{\circ}, -10^{\circ})$. Different cluster scaling relations, methodologies, and cluster samples, agree on the detected bulk flow motion. The detected bulk flow does not seem to fade away up to the scales that the used cluster samples extend, although it is slightly more statistically significant for scales below $z<0.12$. The detected bulk flow amplitudes as a function of redshift spheres and shells are visualized in Fig.~\ref{fig:BF-plots}. Such large bulk flows strongly contradict the $\Lambda$CDM predictions. 

The amplitude of the detected bulk flow is higher than past studies which were based on galaxy surveys. However, galaxy-based bulk flow studies cover a lower redshift range than clusters. As such, the different results are not directly comparable. Nevertheless, the (statistically mild) disagreement leaves enough space for future investigations. Another interpretation of the cluster bulk flow analysis is that the obtained results trace our motion with respect to the local cluster rest frame. As discussed in \cite{bib:Migkas2021}, the amplitude of this motion is consistent with results from sources number counts at much higher scales, e.~g.~\cite{bib:Secrest2021}, with a slightly shifted motion direction. Overall, it suggests that the rest frame of galaxy clusters at $z\lesssim 0.2$ is not the same as the CMB rest frame. 

Until more high-$z$ clusters are obtained, it is not clear if an $H_0$ anisotropy or a large bulk flow is the origin of the observed cluster anisotropies. A possible bias that could single-handedly alleviate the observed tension is a systematic overestimation of the temperature toward the preferred axis of the anisotropy. However, the performed tests did not reveal any such systematic \cite{bib:Migkas2021}. Moreover, there is no obvious physical reason why such a temperature overestimation should be present in independent cluster samples. Nevertheless, future cluster surveys and larger cluster samples with more high-$z$ clusters (e.~g., the eROSITA All-Sky Survey cluster sample) will allow us to better understand the origin of the observed cluster anisotropies.

\subsection{Radio galaxy dipole}
\label{sec:radio_dipole}

The conventional wisdom is that the dipole anisotropy seen in the CMB arises from a solar peculiar velocity, $v=370$~{km/s}, in the direction RA $=168^{\circ}$, DEC $=-7^{\circ}$ \cite{bib:Lineweaver1996,bib:Hinshaw2009,bib:Planck2018}. There may be other contributions to the dipole from the integrated Sachs-Wolfe effect and primordial fluctuations, but they are expected to be small. Accordingly, one expects this solar peculiar velocity to also get reflected in an equivalent anisotropy in the sky distribution of distant radio galaxies \cite{bib:Ellis1984b}, assuming they are at large enough scales to share the same rest frame as the CMB (see sections~\ref{sec:Hom-scale} and \ref{sec:bulk_flows}).

Due to the assumed isotropy of the Universe -- \`a la CP -- an observer stationary with respect to the comoving coordinates of the cosmic fluid should find the average number counts of distant radio galaxies as well as the sky brightness to be uniform over the sky. However, an observer moving with a peculiar velocity $v$ $\ll c$ will find a dipole as a combined effect of aberration and Doppler boosting. The induced dipole can be quantified through changes in the number of sources $N$ per unit solid angle $\Omega$:   
\begin{equation}
\label{eq:radio_dipole}
\left( \frac{\textrm{d} N}{\textrm{d} \Omega} \right)_{\textrm{obs}} = \left( \frac{\textrm{d} N}{\textrm{d} \Omega} \right)_{\textrm{rest}}  \left( 1 + \left[2+{x(1+\alpha)}\right]\, \frac{v}{c} \right) + O \left(\frac{v^2}{c^2} \right) \;, 
\end{equation}
where ${\cal D}=\left[2+{x(1+\alpha)}\right]\frac{v}{c}$ is the amplitude of the dipole, $\alpha\approx 0.8$ is the spectral index \footnote{\bcomm{The measured flux $S$ of an observed radio source is modelled as a power law,
$S \propto \nu^{-\alpha}$, where $\nu$ denotes frequency and $\alpha$ is the spectral index.}} and $x\approx 1$ is the index in the integrated number counts above flux density $S$, usually following a power law, $N(>S)\propto S^{-x}$ for radio galaxies (see \cite{bib:Rubart2013} for a derivation). 
While (\ref{eq:radio_dipole}) yields a good estimate, the assumptions may be an oversimplification for the observational precision current data allows. 
Therefore, they have been investigated further since first usage and, consequently, the observed dipole is now a sum of several terms, see, for instance, \cite{bib:Dalang2022} and detailed below.
The contribution of ${\cal D}$ needs to be determined in this sum. 

The choice of $x=1$ in \cite{bib:Ellis1984b} amounts to assuming that source counts in a clumpy  and in a smooth universe coincide \cite{bib:Schneider1992} \bcomm{independent} of lensing effects.
To further investigate this bias, \cite{bib:Tiwari2015, bib:Siewert2021, bib:Murray2022} vary the power-law behavior.
In addition, (\ref{eq:radio_dipole}) does not account for any redshift evolution effects in the population of sources, as pointed out in \cite{bib:Dalang2022}, particularly targeting a potentially varying spectral index $\alpha(z)$ (see \cite{bib:Maartens2018} for a general derivation of the redshift-dependent kinematic dipole).
Biasing effects from inhomogeneous distributions of luminous galaxies within dark matter haloes and structural clustering also need to be taken into account, as detailed in \cite{bib:Rubart2014, bib:Nusser2015,bib:Tiwari2016}. In \cite{bib:Nusser2015, bib:Tiwari2016} it was argued that the intrinsic dipole is dominant in the redshift range $z < 0.1$. In order to avoid this spurious contribution, sources at higher redshift, $z \sim 1$, are required. These conditions are met by the QSO sample in section~\ref{sec:QSO_dipole}. Based on Planck results \cite{bib:Planck2018} for the velocity of the sun with respect to the CMB, the expected kinematic radio dipole amplitude is $ {\cal D} = 0.46 \times 10^{-2}$  (assuming $\alpha = 0.75$ and $x =1$) \cite{bib:Siewert2021} with negligible error, thereby matching the CMB prediction.  

Given (\ref{eq:radio_dipole}), one can extract the dipole through the analysis of source number counts. To do so, one employs an estimator, which traditionally has been linear \cite{bib:Crawford2009, bib:Singal2011}, but as demonstrated in \cite{bib:Siewert2021} (see also \cite{bib:Rubart2013}), a quadratic estimator \cite{bib:Bengaly2019b} performs better at overcoming directional biases arising from incomplete skies. Concretely, \cite{bib:Siewert2021} concludes that using the appropriate estimator one can reliably distinguish a kinematic dipole of order $10^{-3}$ from a purely random sky with at least $10^6$ point sources. Moreover, positional offsets were estimated to be in the $3$-$5$ degree range for masked sky regions. 

On the data side, studies have made use of the TIFR GMRT Sky Surveys first alternative data release (TGSS-ADR1) \cite{bib:Intema2017}, the Westerbork Northern Sky Survey (WENSS) \cite{bib:Rengelink1997}, the NRAO VLA Sky Survey (NVSS) \cite{bib:Condon1998} and the Sydney University Molonglo Sky Survey (SUMSS) \cite{bib:Mauch2003, bib:Murphy2007}. The frequency range of these surveys stretches from $147$ MHz up to $1.4$ GHz. The fraction of the sky $f_{\textrm{sky}}$ covered and the total number of sources $N_{\textrm{total}}$ in each survey are $f_{\textrm{sky}} = 0.89$, $N_{\textrm{total}} = 640017$ (TGSS-ADR1), $f_{\textrm{sky}} = 0.25$, $N_{\textrm{total}} = 229420$ (WENSS), $f_{\textrm{sky}} = 0.82$, $N_{\textrm{total}} = 1773484$ (NVSS) and $f_{\textrm{sky}} = 0.20$, $N_{\textrm{total}} = 210412$ (SUMSS). We refer the reader to \cite{bib:Siewert2021} for further details on the data sets and masking to account for unobserved regions. 
\begin{figure}
\includegraphics[width=100mm]{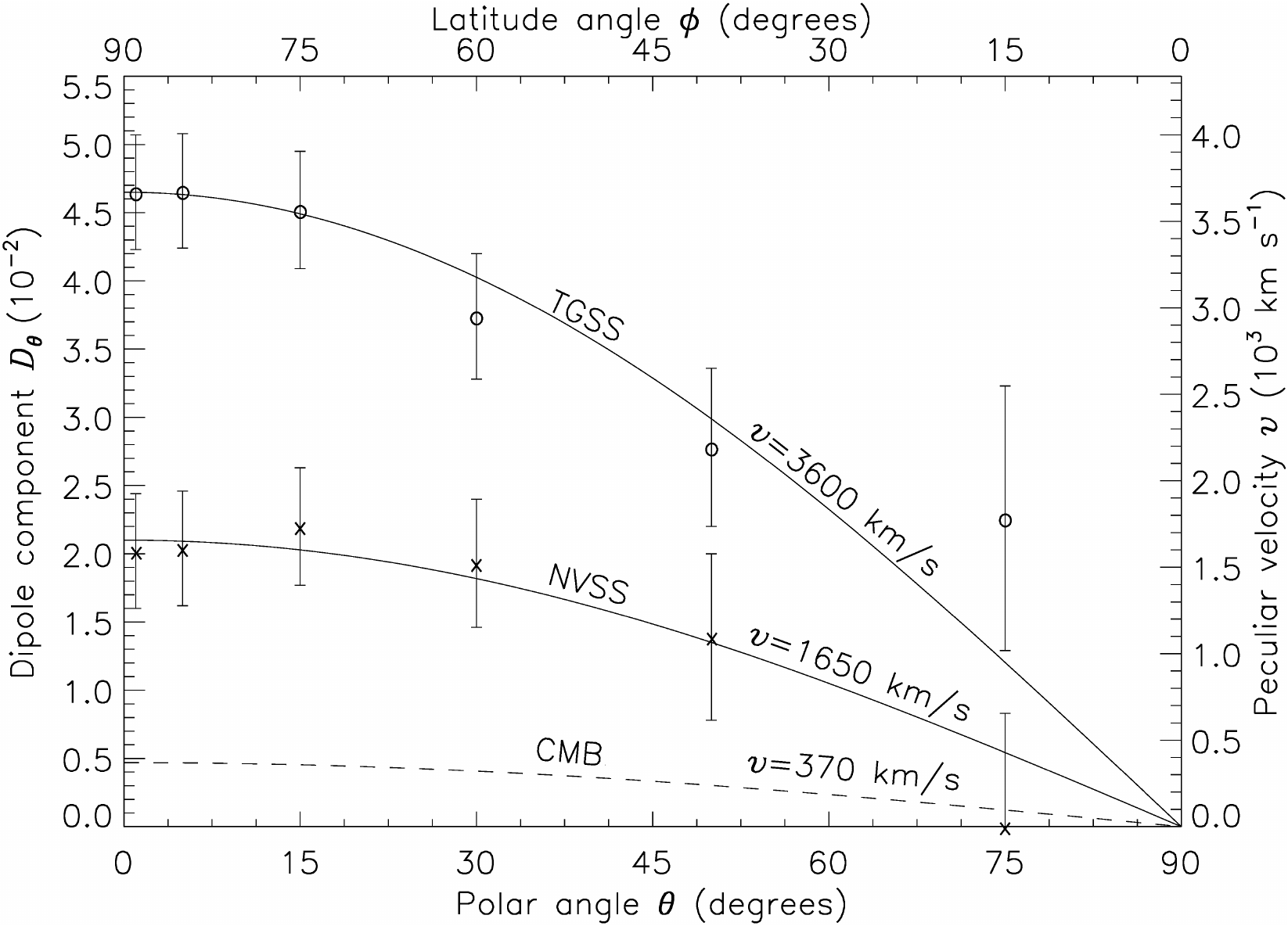}
\caption{Plots of the dipole components ${\cal D}_{\theta}$, observed for the TGSS (o) and  NVSS (x) data \bcomm{(adapted from Fig. 2 of~\cite{bib:Singal2019})} for various zones of the sky between the great circle and a parallel circle at $\theta$, the angle with respect to the CMB dipole direction. The uncertainties expected in random (binomial) distributions are shown as error bars. 
The continuous curves, corresponding to the indicated peculiar velocities, show the expected ($ \cos \theta$) behavior for ${\cal D}_{\theta}$, which seems to fit well the observed values. The dashed curve shows the plot, expected for the CMB value, $v=370$~km/s.}
\label{fig:FAKS1}
\end{figure}

Let us come to the results across different groups \cite{bib:Blake2002, bib:Singal2011, bib:Rubart2013, bib:Tiwari2016, bib:Bengaly2018, bib:Singal2019, bib:Siewert2021} and surveys \cite{bib:Intema2017, bib:Rengelink1997, bib:Condon1998, bib:Mauch2003, bib:Murphy2007}, as reviewed and reanalyzed by \cite{bib:Siewert2021}. As can be seen from Table 9. of \cite{bib:Siewert2021}, irrespective of the masking and flux threshold, the returned dipole for TGSS is ${\cal D} \sim (6  \pm 0.5 ) \times 10^{-2}$, for WENSS ${\cal D} \sim (3 \pm 1 ) \times 10^{-2}$, for SUMSS ${\cal D} \sim (3.5 \pm 1 ) \times 10^{-2}$ and from NVSS one finds ${\cal D} \sim (2 \pm 0.5) \times 10^{-2}$. When compared to the expected CMB dipole, the cosmic radio dipole is larger by a factor of approximately four to ten and a statistical significance that depends on the errors. It should be noted that despite differences in estimator, masking and other systematics, good agreement is found on the amplitudes with earlier results, including earlier analysis of TGSS \cite{bib:Bengaly2018, bib:Singal2019}, WENSS \cite{bib:Rubart2013} and NVSS \cite{bib:Singal2011, bib:Rubart2013, bib:Tiwari2016, bib:Bengaly2018}. Throughout, the CMB dipole direction is also recovered in line with expectations. 

Overall, WENSS, SUMSS and NVSS surveys return consistent amplitudes, which allows one to combine them. Colin et al. \cite{bib:Colin2017} found a dipole with $v = 1729 \pm 187~\mbox{km/s}$ towards RA =$149^{\circ}\pm2^{\circ}$, DEC $=-17^{\circ} \pm 12^{\circ}$ in the combined SUMSS plus NVSS catalogue. This velocity exceeds the CMB expectation by a factor of approximately 5. Since the TGSS survey leads to the largest amplitude, it appears to be an outlier. The vast majority of TGSS sources have a counterpart in the NVSS catalogue, but the angular spectrum of TGSS sources has significantly more power than that of the NVSS in the multipole range $2 \leq l \leq 30$ \cite{bib:Dolfi2019} (see also \cite{bib:Bengaly2018}). In  \cite{bib:Tiwari2019} through comparison to GLEAM/TGSS \cite{bib:Hurley2017}, the low $\ell$ power excess is traced to large scale flux offsets, a systematical effect. Nevertheless, in \cite{bib:Siewert2021} these samples are combined and cross-matched and one recovers an intermediate value ${\cal D} \sim 4 \times 10^{-2}$, which suggests that the observed dipoles in both samples are real. One could speculate that the difference in amplitude between TGSS and NVSS may be explained in terms of frequency dependence arising from local voids and structures, e.~g.~\cite{bib:Rubart2013}, but it can be argued that such an effect would be too small \cite{bib:Colin2017, bib:Siewert2021}. Of course, throughout one should be mindful that systematics may be at play \cite{bib:Dolfi2019, bib:Tiwari2019}. Recently, \cite{bib:Secrest2022} has shown that the TGSS sample can be corrected for two prominent systematics, namely a position-dependent flux calibration problem \cite{bib:Hurley2017} and a position-dependent RMS noise. This leads to a lower dipole amplitude that is completely consistent with NVSS and WISE dipoles \cite{bib:Secrest2022}. 

In summary, there appears to be some consensus regarding the excess in the radio galaxy dipole with respect to CMB expectations, but a tangible difference in amplitude between TGSS and NVSS surveys needs to be better understood, especially given the overlap in sources \cite{bib:Dolfi2019}. The likely explanation is systematics. 
Fig.~\ref{fig:FAKS1} shows the plots of the observed dipole amplitude, ${\cal D}_{\theta}$, for both TGSS and NVSS data as a function of $\theta$, measured with respect to the CMB dipole direction. As can be clearly seen, the peculiar velocity in either case turns out to be above the CMB value. As discussed, the peculiar velocity estimated from the (uncorrected) TGSS data is much higher than the NVSS data. Could this difference in the dipoles be due to some direction-dependent completeness problems, in either or both catalogues? An argument against this possibility is that not only the directions of the dipoles agree with that of the CMB, the number counts also show an overall $\cos\theta$ dependence in different independent sky slices (Fig.~\ref{fig:FAKS1}), as would occur only either for a genuine dipole or in a rather contrived case. In short, the $\cos \theta$ dependence is non-trivial, and as explained above, once the TGSS survey is corrected for known systematics, the magnitude of these dipoles can be brought into line \cite{bib:Secrest2022}.

Overall, the prospects of a discovery are bright. In the not too distant future, the LOw Frequency ARray (LOFAR) two-meter sky survey \cite{bib:Shimwell2017}, the Square Kilometer Array Observatory (SKAO) \cite{bib:Bacon2020, bib:Bengaly2018}, and the Evolutionary Map of the Universe (EMU) \cite{bib:Norris2011} will either confirm or refute the existing radio galaxy dipole excesses.

\subsection{QSO dipole}
\label{sec:QSO_dipole}

Apart from radio galaxies, the Baldwin-Ellis test \cite{bib:Ellis1984b} has also been conducted with QSOs. Since QSOs are typically associated with cosmological scales, $z \gtrsim 1$, contamination from the local clustering dipole is expected to be low \cite{bib:Tiwari2016}. Recently, \cite{bib:Secrest2021} constructed a flux-limited, all-sky sample of $1.36$ million QSOs observed by the Wide-field Infrared Survey Explorer (WISE) \cite{bib:Wright2010}. WISE surveyed the sky at $3.4 \mu$m, $4.6 \mu$m, $12\mu$m, and $22 \mu$m ($W1, W2, W3$, and $W4$), and being a space mission, the observational systematics are expected to differ from the radio galaxies in section~\ref{sec:radio_dipole}. From WISE observations a reliable QSO catalogue based on mid-infrared color alone can be constructed, as demonstrated in \cite{bib:Secrest2015}. 

\begin{figure}[htp]
\centering
\includegraphics[width=0.51\linewidth]{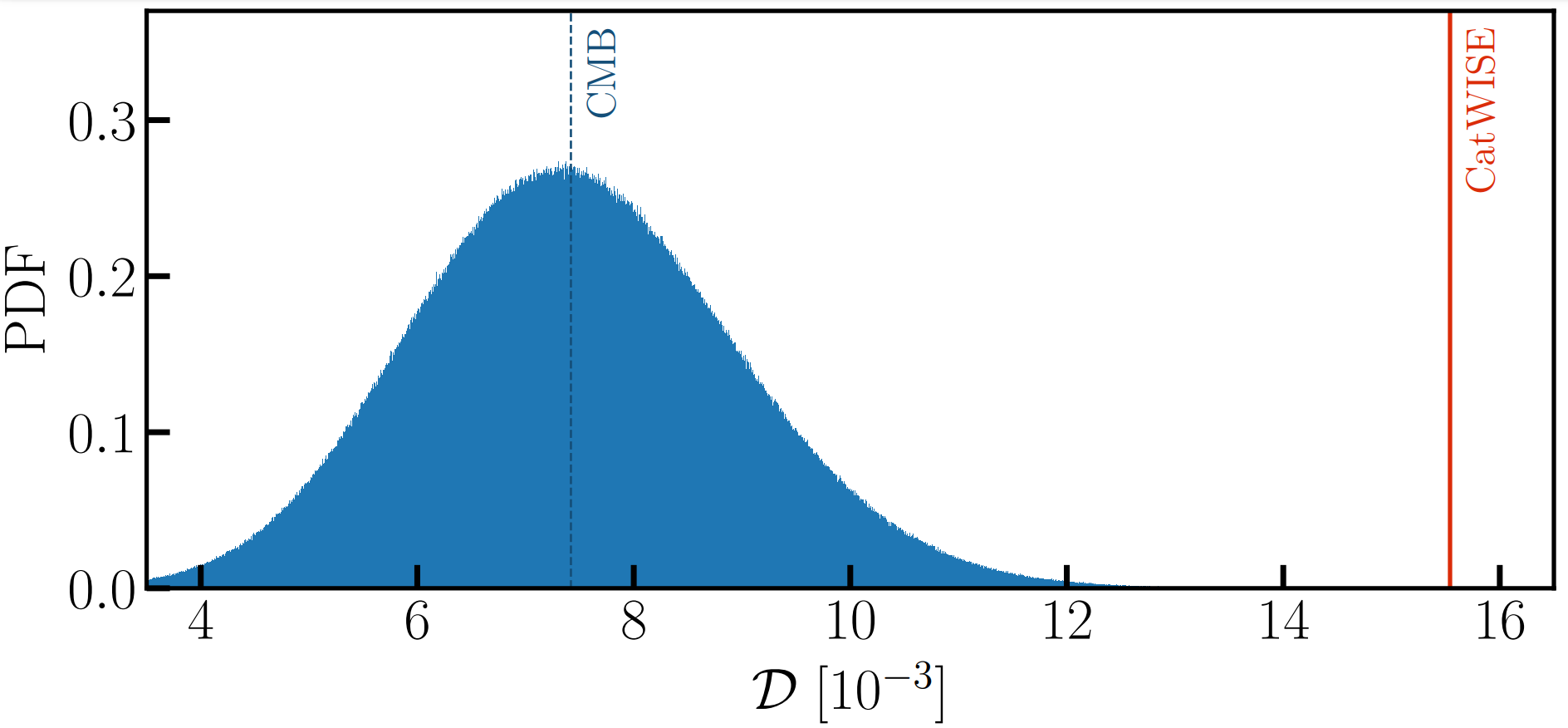}
\includegraphics[width=0.4\linewidth]{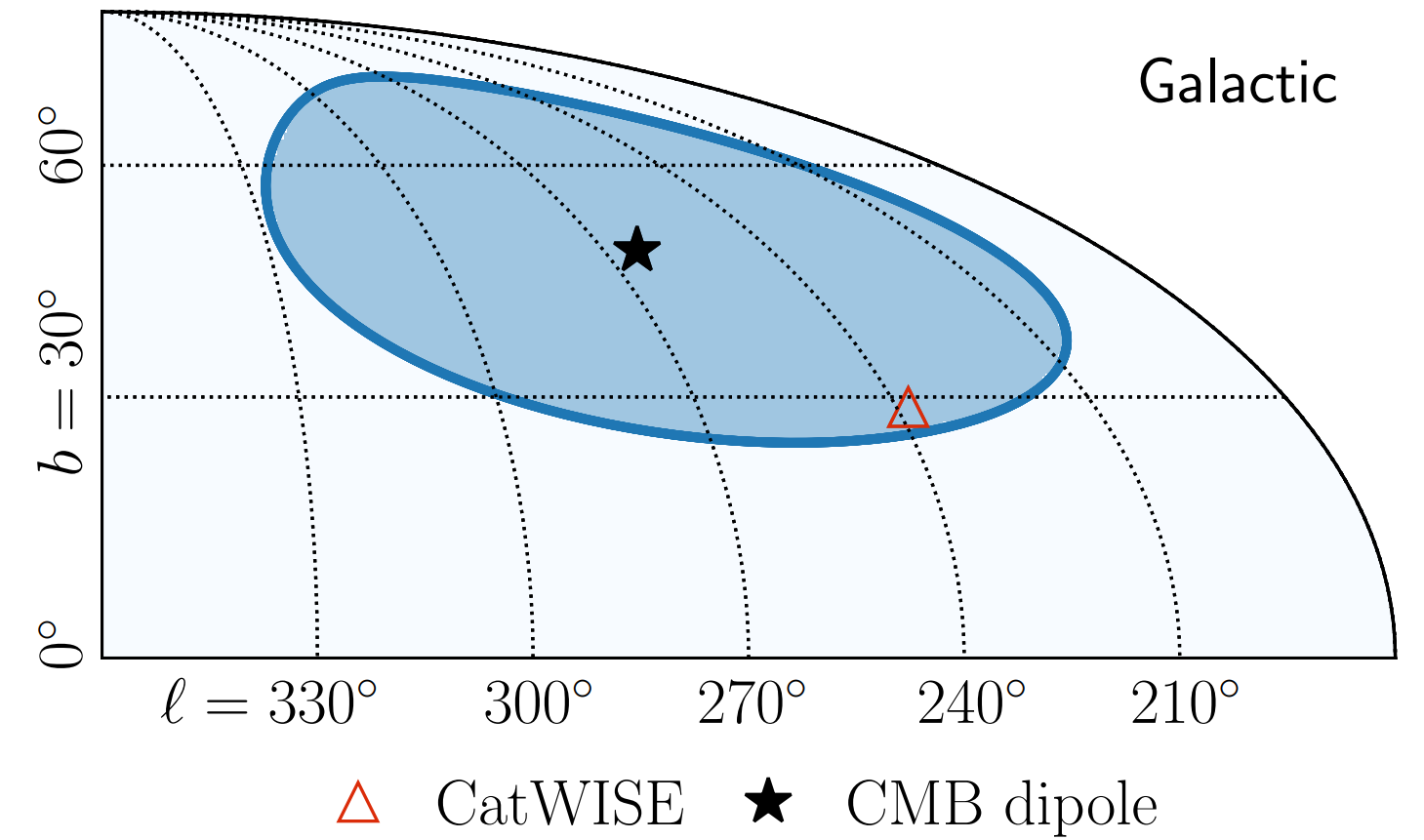}
\caption{Reproduced from Fig.~4 of \cite{bib:Secrest2021}. Left: The cosmic dipole amplitude obtained from the CatWISE QSO sample ${\cal D} = 0.01554$ in heliocentric frame (red solid line) compared to the one obtained from $10^7$ simulations based on the kinematic interpretation of the CMB dipole (blue distribution). The median of the simulated distribution is ${\cal D}_\mathrm{sim} \approx 0.007$ (blue dashed line). Right: Direction of the CatWISE dipole (red triangle) compared to the direction of the CMB dipole with the $2 \, \sigma$ null hypothesis uncertainty region around it, assuming the dipole is a purely kinematic effect with respect to the cosmic rest frame shared by the CMB and the QSOs (blue area).}
\label{fig:Secrest_dipole}
\end{figure}

Concretely, starting from the recent CatWISE2020 data release \cite{bib:Eisenhardt2020},  \cite{bib:Secrest2021} corrected $W1$ and $W2$ for dust maps, while imposing cuts on $W1$, $W2$ to ensure AGN-dominated emission following a power-law distribution, i. e. $S_{\nu} \propto \nu^{-\alpha}$. Masks were introduced to remove regions of the sky with low quality photometry and all sources with $|b| < 30^{\circ}$ were removed to exclude the Galactic plane, leaving a final sample of 1,355,352 QSOs. In addition, the redshift distribution of the sample was estimated by cross matching the CatWISE2020 QSOs with SDSS \cite{bib:Annis2014} in the range $ -42^{\circ} < \textrm{RA} < 45^{\circ}$, but outside of $|b| < 30^{\circ}$, which succeeded in producing matches for 14,193 QSOs with a mean redshift of $z = 1.2$. Of this subsample, $99 \%$ was found to lie beyond $ z= 0.1$ \cite{bib:Tiwari2016}. A novel estimator was employed to this data set yielding a dipole of magnitude $\mathcal{D} = 0.01554$ in the heliocentric frame- more than twice the CMB expectation - in the direction $(l, b) = (238.2^{\circ}, 28.8^{\circ})$ \cite{bib:Secrest2021}. The authors also checked the consistency of the results across different estimators to address concerns about estimator bias raised in \cite{bib:Rubart2013}. The statistical significance of the result was estimated at $4.9 \, \sigma$ based on mock sky simulations whereby only 5 out of $10^{7}$ simulations led to a larger dipole amplitude. The result is encapsulated in Fig.~\ref{fig:Secrest_dipole}, reproduced from \cite{bib:Secrest2021}. Despite some of the QSOs being below $z=0.1$, it was found that any clustering dipole amplitude within flat $\Lambda$CDM is exceeded by a factor of $\sim 65$ \cite{bib:Secrest2021}. In follow up analysis, \cite{bib:Secrest2022} presented a joint analysis of NVSS and WISE samples and showed that the null hypothesis, namely agreement with the CMB dipole, was rejected at $5.1 \, \sigma$. It has recently been confirmed that the angular power spectrum of the QSO sample \cite{bib:Secrest2021} shows excess power for lower multipoles $\ell \lesssim 10$ \cite{bib:Tiwari2022}, thereby providing another perspective on the anomaly.

\begin{figure}
\includegraphics[width=120mm]{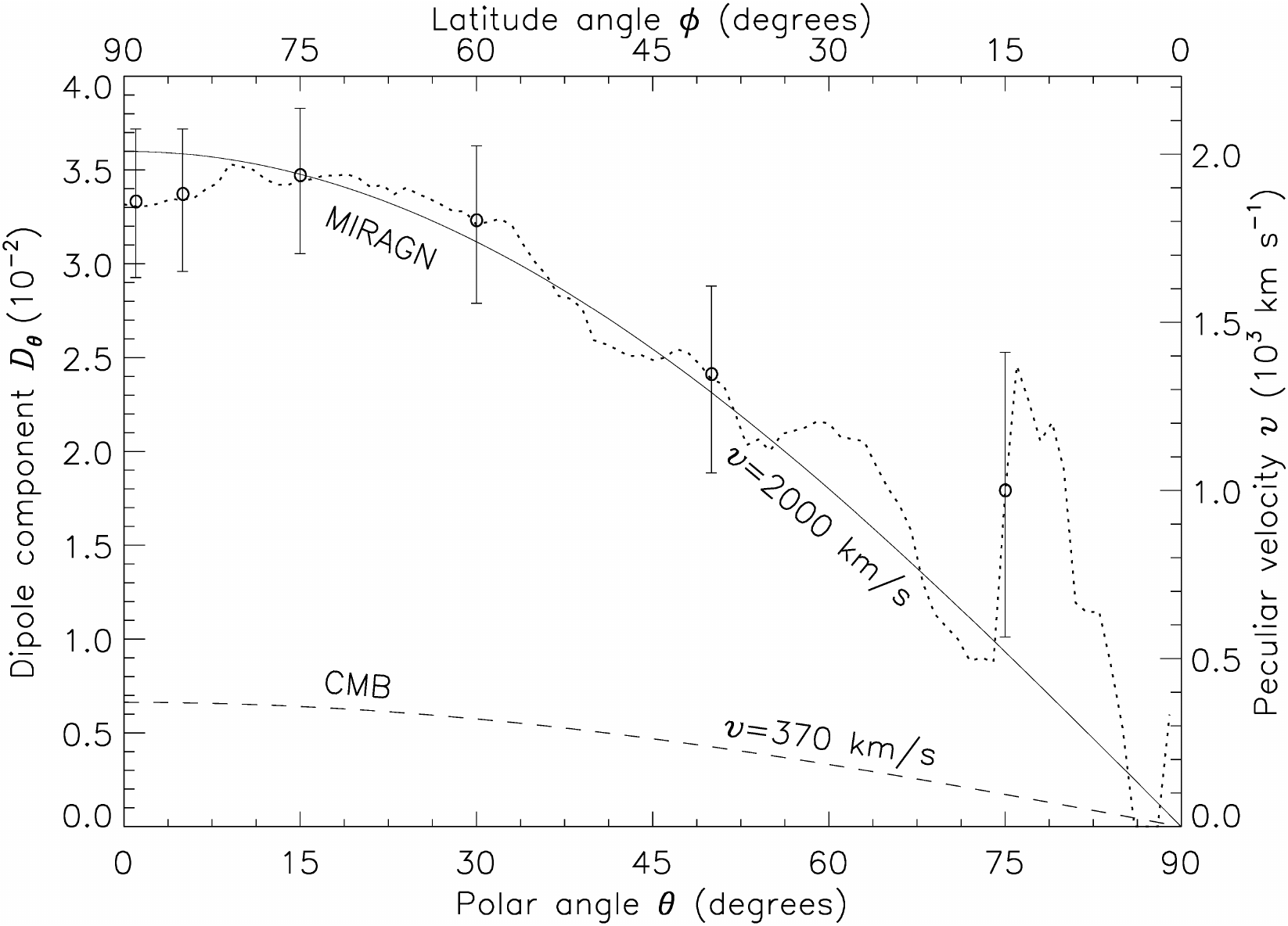}
\caption{Dipole components in heliocentric frame ${\cal D}_{\theta}$ and the equivalent peculiar velocity component $v \cos\theta$, computed for various zones of the sky between the great circle and a parallel circle at $\theta$, the~angle with respect to the determined dipole direction, RA~$=148^{\circ}$, DEC~$=22^{\circ}$ \bcomm{(adapted from Fig. 8 of \cite{bib:Singal2021b})}. The~latitude angle $\phi$ is measured from the great circle at $\theta=90^\circ$. The~observed values in  various sky strips are plotted as circles (o), with~the error bars calculated for a random (binomial) distribution. 
The continuous curve, corresponding to a peculiar velocity of $(2.0\pm 0.2) \times 10^{3}~\mbox{km/s}$, shows the expected ($\propto \cos \theta$) behavior for ${\cal D}_{\theta}$, which is a best fit to the data observed in different sky strips, shown by the dotted curve. For a comparison, the expectation for the CMB value, $v=370~\mbox{km/s}$, is shown by dashed curve, which lies way below the values observed for the MIRAGNs.}
\label{fig:FAKS2}
\end{figure}

At the same time, a determination of Solar peculiar motion from a dipole anisotropy, seen in the number density in another sample of 0.28 million AGNs, selected from the earlier Mid Infra Red Active Galactic Nuclei (MIRAGN) sample comprising more than a million sources~\cite{bib:Secrest2015}, turned out to be a factor of four larger than that inferred from the CMB dipole ({Fig.~\ref{fig:FAKS2}}), but along the anticipated direction \cite{bib:Singal2021b}. It should be noted that the conversion factor from $\cal D$ to peculiar velocity $v$ here is different in comparison with that in Fig.~\ref{fig:FAKS1}, because of different indices $x$ and $\alpha$ for the radio and the mid-infrared populations. In contrast to \cite{bib:Secrest2021}, a larger dipole was reported, but this may be due to the smaller sample size and its visibly less uniform nature. The source of this discrepancy warrants further investigation. 

To safeguard against any possible contamination due to such a differential increase, the MIRAGN sample \cite{bib:Secrest2015} was restricted to an upper limit of magnitude, $W1<15.0$. Once the basic survey has completeness at higher magnitudes, the number density distributions in the sky at lower magnitudes remain unaffected since any deeper coverage adds sources only at fainter levels, namely at higher $W1$ magnitudes. This was verified from a detailed examination of the original MIRAGN sample \cite{bib:Secrest2015} data at different $W1$ levels \cite{bib:Singal2021b}.  
Further confirmation came from the integrated number counts, which showed a constant slope of 0.68 for $W1 \lesssim 15.0$ (corresponding to an index $x=2.5\times0.68=1.7$, in $N(>S)\propto S^{-x}$). The slope flattens abruptly for fainter sources ($W1 \gtrsim 15.5$), indicating the incompleteness above those magnitude levels. Accordingly, the results should be reliable for the upper magnitude limit, chosen at $W1=15.0$ for the sample. A lower bound $12.0 <  W1$ was also imposed to prevent contamination from local bulk flows at low redshifts, but given the small number of low redshift sources, this did not greatly affect the result \cite{bib:Singal2021b}. 

These checks aside, there remains a difference of a factor of $\sim 2$ between the QSO dipoles \cite{bib:Secrest2021, bib:Singal2021b}. One possibility to examine would be whether the color difference $W1-W2$ criteria \cite{bib:Stern2012} are robust enough to pick AGNs to sufficient completeness, as one wants to study number density variations to approximately one part in 1000, and whether that could be the reason for the difference of a factor of $\sim 2$ for the dipole strengths in the two samples. This uncertainty could be avoided if one had spectroscopic redshift measurements to ensure that the sources in question are indeed  QSOs.  
In fact, for $\sim 1.2 \times 10^5$ objects in the million strong sample~\cite{bib:Secrest2015}, the spectroscopic redshifts are available. But these form a rather incomplete sample with a highly non-uniform sky coverage and the number count method could not be applied to that for estimating the dipole. More recently, \cite{bib:Secrest2022} examines the difference between the QSO dipoles  \cite{bib:Secrest2021, bib:Singal2021b} and explores systematics that could account for the discrepancy. In particular, it was found that masking stripes of reduced sensitivity along certain ecliptic lines of longitude; regions of steep drop off in source density near the ecliptic poles; and regions of lower source density at lower Galactic latitudes, that any discrepancy in magnitude disappears \cite{bib:Secrest2022}. 

Separately, a study of the dipole anisotropy in a homogeneously selected SDSS-DR12 sample of 103,245 distant QSOs has shown a redshift dipole along the CMB dipole direction, implying a velocity $\sim 6.5$ times though in a direction  opposite to, but nonetheless parallel to, the CMB dipole~\cite{bib:Singal2019b}. However, one can employ another alternate method to extract the QSO dipole. 
Recently, Solar peculiar motion has been determined from the Hubble diagram of SNe Ia \cite{bib:Singal2021}. The peculiar motion of the Solar system can be determined from the observed magnitude vs. redshift, $m-z$, Hubble diagram of QSOs as well. The observer's peculiar motion causes a systematic shift in the $m-z$ plane between sources lying in the forward hemisphere (along the velocity vector) and those in the backward hemisphere (in the opposite direction), providing a measure of the peculiar velocity. The methodology is similar to the Type Ia SNe dipole in section \ref{sec:SN_dipole}. Accordingly, from a sample of $\sim 1.2 \times 10^5$ mid-infrared QSOs with measured spectroscopic redshifts, a peculiar velocity $\sim 22$ times larger than the CMB dipole was inferred, but with direction matching within $\sim 2 \sigma$ of the CMB dipole~\cite{bib:Singal2022}. 
Previous findings from number count, sky brightness or redshift dipoles observed in samples of distant AGNs or SNe Ia had yielded values 2 to 10 times larger than the CMB value. 
Yet, this is by far the largest value obtained for a peculiar motion, although the direction in all cases agreed with the CMB dipole direction (see also section~\ref{sec:emergent_H0}).

As for the peculiar motion determination from the Hubble plot, while it is true that the sky distribution of QSOs in this sample is very uneven, this does not give rise to any systematic effects in the Hubble diagram or in the ensuing peculiar motion determination (see also section~\ref{sec:emergent_H0}). The sources in the forward and backward hemispheres show opposite displacements in the observed magnitudes $m$ at a given redshift $z$, otherwise these sources have not been subject to any other discrimination. As the peculiar motion here is determined from their relative deviations in the $m-z$ plane, their different number densities in various parts of the sky at different magnitudes due to a sample incompleteness or non-uniformity in the sky coverage, would not affect the results systematically.

\subsection{SNe Ia dipole}
\label{sec:SN_dipole}

One can also explore the cosmic dipole with SNe Ia, independent of the CMB, by assuming the SNe Ia form the cosmic rest frame. In order to compare with the CMB-based cosmic rest frame, the SNe Ia rest frame is set up with respect to the latter. Then, the observed distance modulus $\mu(z)$ in the heliocentric reference frame with redshift $z=z_\textrm{helio}$
\begin{equation}
\mu(z) = 5 \log_{10} \left[ \frac{\bar{d}_\textrm{L} (\bar{z})}{\textrm{Mpc}}\right]+25 + 5 \log_{10} \left[ 1 + 2 \left(\frac{v_\textrm{e}}{c} \right) \cos \theta - \left( \frac{v_\textrm{o}}{c} \right) \cos \varphi \right]
\label{eq:m-z_relation}
\end{equation}
consists of the luminosity distance as measured in a homogeneous and isotropic spacetime, $\bar{d}_\textrm{L}(\bar{z})$, in which neither the observer nor the observed light emitters (SNe) are moving and $\bar{z}$ is the corresponding redshift in that reference frame, here the CMB frame.
The term with the velocity of the emitters, $v_\textrm{e}$, arises as the peculiar motion of the SNe with respect to the CMB and the term including the observer's velocity $v_\textrm{o}$ accounts for the aberration and Doppler boost due to the observer's motion with respect to the CMB frame. 
In addition $\theta$ and $\varphi$ denote the angles between the line of sight and peculiar motions of emitters and observer, respectively (see \cite{bib:Horstmann2021} and \cite{bib:Sasaki1987} for details).
As already noted in \cite{bib:Sasaki1987}, there is a factor of $2$ between $v_\textrm{e}$ and $v_\textrm{o}$ because $z$ depends on $v_\textrm{e} - v_\textrm{o}$, while $\mu(z)$ depends on $2 v_\textrm{e} - v_\textrm{o}$. 
In the CMB frame, $v_\textrm{o} = 0$ and $v_\textrm{e}$ becomes a negligible correction at higher redshifts. 
For lower redshifts, accounting for $v_\textrm{e}$ in (\ref{eq:m-z_relation}) is tricky. 
On the one hand, one can use complementary observations as, for instance, provided by \cite{bib:Carrick2015} for $z < 0.1$, to correct the distance modulus that only accounts for the boosted observer frame. Yet, this comes at the cost that one has to fix a cosmic reference frame first before inferring the peculiar velocities relative to that frame. 
Consequently, inconsistencies in this inference loop may easily arise. 
On the other hand, one may choose to use the approach in \cite{bib:Huterer2017} to include the peculiar velocities as part of the covariance matrix accounting for all correlations and uncertainties in the data set.\footnote{Here, one should take care not to account for the same uncertainties and errors twice. Off-the-shelf data sets like the Pantheon \cite{bib:Scolnic2018} already contain certain corrections. See, for instance \cite{bib:Singal2021}, in which the employed SNe data set is restored to its original form first before the necessary corrections are applied.}  
This method avoids the inconsistency loop of the first ansatz, yet, as pointed out in \cite{bib:Mohayaee2020}, may lead to a contribution to the covariance matrix that does not reflect our local Universe by not accounting for peculiar motions of the Milky Way or the LG, for example. 
 
Obviously, great care is also required to insert the observed redshifts into the modeling equations (see further discussion in \cite{bib:Davis2019}). 
As explained in \cite{bib:Steinhardt2020}, it may be more accurate to use the redshifts from the spectra of the host galaxies instead of the SNe directly. 
This can lead to differences in Pantheon redshifts \cite{bib:Scolnic2018} (see also \cite{bib:Rameez2019, bib:Rameez2021}), so \cite{bib:Steinhardt2020} have an independent catalogue of heliocentric redshifts for the Pantheon data set. 
Working in heliocentric frame allows one to determine $v_\textrm{o}$ with SNe independent of the radio galaxies in section~\ref{sec:radio_dipole}. 

Using (\ref{eq:m-z_relation}) and the Pantheon sample with heliocentric redshifts in \cite{bib:Steinhardt2020}, \cite{bib:Horstmann2021} recovers $v_\textrm{o}/c = 0.82^{+0.17}_{-0.17} \times 10^{-3}$ which are lower than the value estimated by Planck, $v_\textrm{o}/c = 1.2336^{+0.004}_{-0.004} \times 10^{-3}$ when including $v_\textrm{e}$ of the SNe as proposed in \cite{bib:Huterer2017}. 
The value marginally increases to $v_\textrm{o}/c = 0.96^{+0.17}_{-0.15} \times 10^{-3}$ with peculiar velocity corrections as set up in \cite{bib:Carrick2015}. 
Thus, one is looking at a $2.4 \, \sigma$ and $1.7 \, \sigma$ deficit, respectively, with respect to the value inferred from the CMB. Obviously, a $v_\textrm{o}$ even lower than the one from Planck is in conflict with the excesses reported in radio galaxy and QSO dipoles in sections~\ref{sec:radio_dipole} and \ref{sec:QSO_dipole}. Nevertheless, the CMB dipole direction is recovered in line with expectations.  

The result by \cite{bib:Horstmann2021} is in disagreement with other SNe dipole results. 
Previously, \cite{bib:Singal2021} had determined the cosmic dipole for a JLA-based set of SNe using a different method and without corrections for $v_\textrm{e}$ because all SNe with $z\le0.06$ were excluded from the evaluation. 
With this redshift restriction, $v_\textrm{e}$ corrections should not greatly affect the results, but a follow up comparison between \cite{bib:Singal2021} and \cite{bib:Horstmann2021} is in preparation.

Going back to heliocentric redshifts by reverting the corresponding corrections previously applied to the sample, \cite{bib:Singal2021} recovers the CMB dipole direction, RA $=173^{\circ}\pm12^{\circ}$, DEC $=10^{\circ}\pm9^{\circ}$, the direction being within $\stackrel{<}{_{\sim}}2\sigma$ of the CMB dipole direction. 
The inferred velocity $v_\textrm{o}/c = (5.34 \pm 1.67) \times 10^{-3}$ \cite{bib:Singal2021} is larger than the CMB value by a factor of four. The result agrees well with earlier results from the NVSS radio galaxy survey \cite{bib:Singal2011, bib:Rubart2013, bib:Tiwari2016, bib:Bengaly2018} and the QSO dipole \cite{bib:Secrest2021}, but is at odds with \cite{bib:Horstmann2021}. One potential explanation for this disagreement is the different handling of redshifts, which has been subject of some criticism \cite{bib:Rameez2019, bib:Rameez2021, bib:Steinhardt2020}, and also Fig.~5 of \cite{bib:Singal2021}. 
Since the treatment of redshifts has been greatly improved in the new Pantheon+ sample \cite{bib:Peterson2021, bib:Scolnic2021}, it is a timely exercise to revisit all these results so that the redshifts can be eliminated as a source of disagreement.

\subsection{Variations of the Hubble constant in the Hubble diagram}
\label{sec:emergent_H0}

Allan Sandage famously framed cosmology as the search for two numbers, the Hubble constant, $H_0$, and the deceleration parameter, $q_0$, \cite{bib:Sandage1970}. Remarkably, $q_0$ turned out to have the ``wrong" sign \cite{bib:Riess1998, bib:Perlmutter1999} and various groups, following in the footsteps of Vesto Slipher, Georges Lema\^{\i}tre, and Edwin Hubble, are closing in on a purely local, cosmology independent determination of $H_0$ based on the distance ladder \cite{bib:Huang2020, bib:deJaeger2020, bib:Pesce2020, bib:Kourkchi2020, bib:Schombert2020, bib:Khetan2021, bib:Blakeslee2021, bib:Freedman2021, bib:Riess2021}. Throughout, the working assumption is that there exists a unique $H_0$. Within FLRW, a unique $H_0$ is a given, since $H_0$ arises as an \textit{integration constant} when one solves the Friedmann equations. In other words, for FLRW cosmologies there is always a universal constant that is \textit{model independent}. In contrast, matter density $\Omega_{\text{m}}$, curvature $\Omega_{\text{k}}$, and the other parameters of the cosmological model, represent further modeling assumptions (see discussion in \cite{bib:Krishnan2021}). 

\begin{figure}
    \centering
    \includegraphics[width=0.455\textwidth]{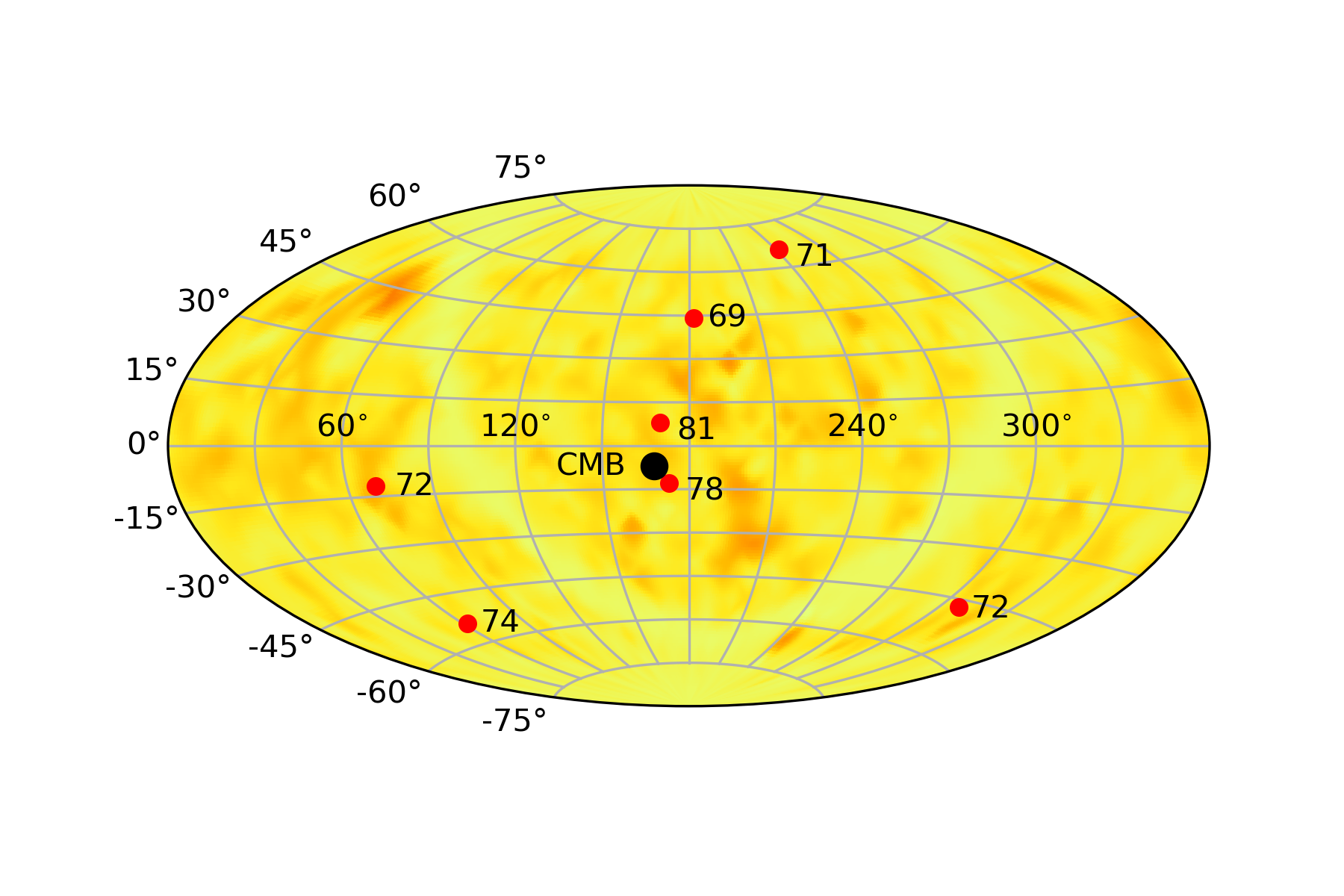}
    \includegraphics[width=0.505\textwidth]{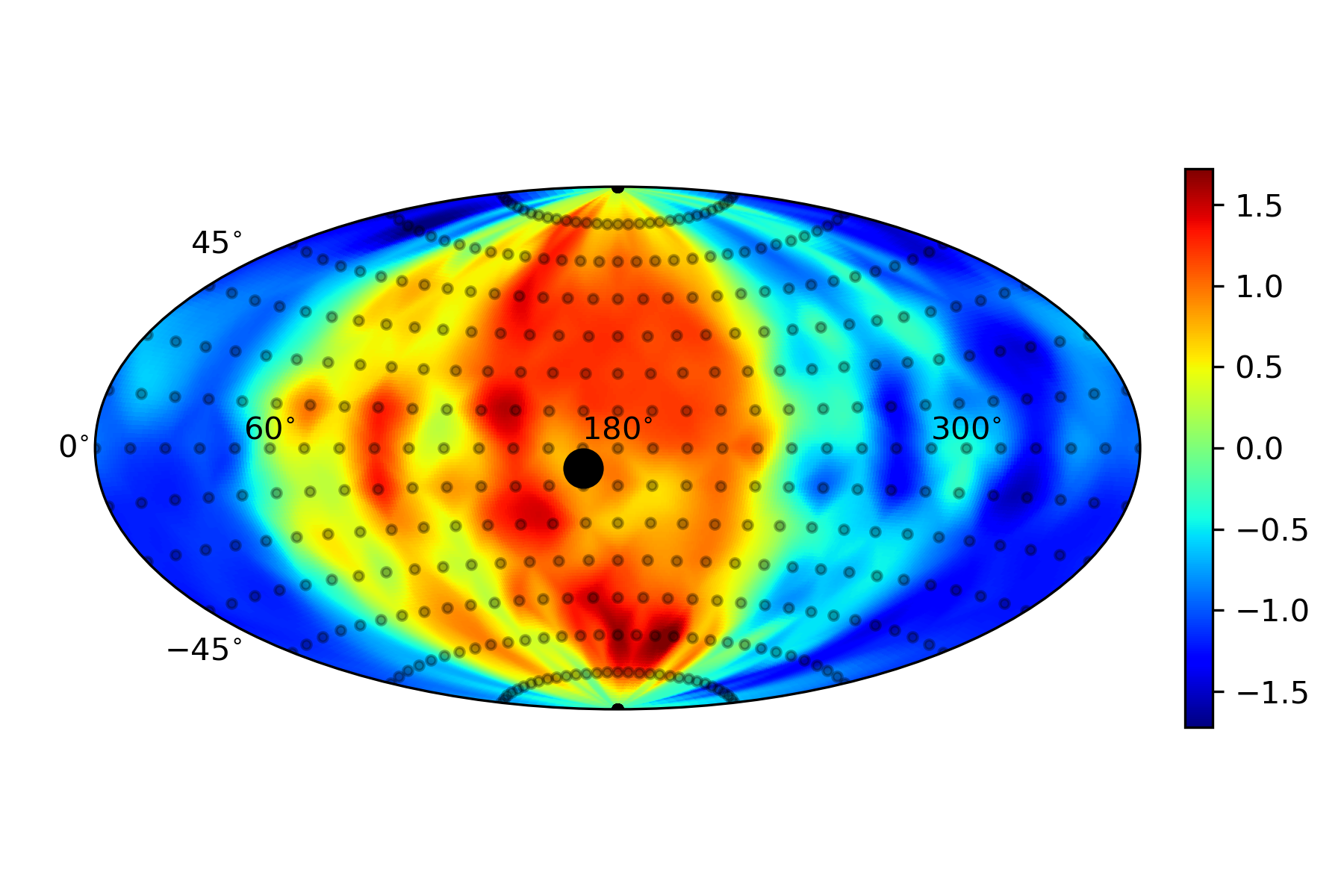}\\
    \caption{Left: $H_0$ as determined by strong gravitational lensing on the sky; larger values are evidently correlated with the CMB dipole (black dot). Right: Similar trend in the Pantheon SNe data set, the sky is decomposed into regions where $H_0$ is larger (red) and smaller (blue). Figures reproduced from \bcomm{Fig. 1 and Fig. 2 of \cite{bib:Krishnan2022}}.}
    \label{fig:H0variation}
\end{figure}

What happens if there is a twist not only in the $q_0$ narrative, but one in the $H_0$ narrative too? First, in the local Universe it is well documented that $H_0$ varies on the sky \bcomm{\cite{bib:McClure2007,bib:Wiltshire2012}}. At greater depths, anomalies in galaxy cluster scaling relations appear to confirm this feature \cite{bib:Migkas2020, bib:Migkas2021} (see section~\ref{sec:galaxy_cluster_anisotropies}), while even in the CMB $H_0$ may vary on the sky \cite{bib:Fosalba2021, bib:Yeung2022} (see section~\ref{sec:CMB_variations}). 
A further corroborating observation may be traced to strong gravitational lensing time delay cosmography \cite{bib:Wong2020, bib:Millon2020} and the realization that $H_0$ may evolve with lens redshift.\footnote{
It is easy to argue analytically that both $H_0$ and $\Omega_{\text{m}}$ should evolve with effective redshift in the flat $\Lambda$CDM model in the late Universe \cite{bib:Colgain2022, bib:Colgain:2022a}. This explains earlier observations made in \cite{bib:Krishnan2020, bib:Dainotti2021, bib:Dainotti2022, bib:Horstmann2021}.}  
This is one interpretation within FLRW, more accurately flat $\Lambda$CDM. However, there is another natural perspective, especially in light of the fact that there appears to be discrepancy in the cosmic dipole, as documented in sections~\ref{sec:radio_dipole}, \ref{sec:QSO_dipole}, and \ref{sec:SN_dipole}. When plotted against the sky, the H0LiCOW/TDCOSMO lenses \cite{bib:Wong2020, bib:Shajib2020} show a trend, as illustrated in Fig.~\ref{fig:H0variation}, whereby $H_0$ within flat $\Lambda$CDM is larger in the direction of the CMB dipole \cite{bib:Krishnan2021b}. Given the small size of the sample, such a configuration can arise simply by chance with probability $p=0.12$ \cite{bib:Krishnan2022}. Taken at face value, this is a suggestive fluke. 

Nevertheless, the same trend, i. e. $H_0$ larger in the CMB dipole direction/hemisphere, has been found within Pantheon SNe Ia \cite{bib:Scolnic2018}. One simply has to split the SNe sample in hemispheres and scan over the sky, as shown in Fig.~\ref{fig:H0variation}, where at each point one fits the flat $\Lambda$CDM model independently in the hemispheres, to find that $H_0$ is larger in the CMB dipole direction \cite{bib:Krishnan2022}. Interestingly, the observation is consistent with earlier SNe results spanning a decade \cite{bib:Cooke2010, bib:Antoniou2010, bib:Li2013, bib:Javanmardi2015, bib:Krishnan2022}.\footnote{In a more recent paper \cite{bib:Zhai2022} a larger $H_0$ in the CMB dipole direction has been reported using Cepheids and Type Ia SNe in the range $0.023 < z < 0.15$. The paper reports a variance of $\Delta H_0 \sim 4$ km/s/Mpc, but with a lower $p$-value, $p \sim 0.3$, which may be expected given the fewer local SNe.} The same anisotropic expansion can be recovered at greater statistical significance ($\sim 4 \, \sigma$) by fitting a dipole ansatz - motivated by earlier theoretical predictions \cite{bib:Tsagas2011, bib:Tsagas2015} - to the CMB dipole direction \cite{bib:Colin2019}. Since the flat $\Lambda$CDM model has no directional dependence, it is difficult to extract a precise dipole, nevertheless some signature of one emerges. Here, it should be stressed that as discussed in the previous section, the Pantheon data set is in ``CMB frame" by construction, so if the sample is tracking the kinematic CMB dipole, this is unexpected. As explained in \cite{bib:Krishnan2022}, through a weighted sum one can assign a statistical significance to this trend.
One finds that it can arise as a fluke with probability $p = 0.065$ \cite{bib:Krishnan2022}. When combined with strong lensing time delay cosmography, the significance of the observation is $1.7 \, \sigma$ \cite{bib:Krishnan2022}. The significance is low, but in the backdrop there are claims of a discrepancy in the magnitude of the cosmic dipole, so this is a potential realization of the same feature.  

\begin{figure}
    \centering
    \includegraphics[width=0.7\textwidth]{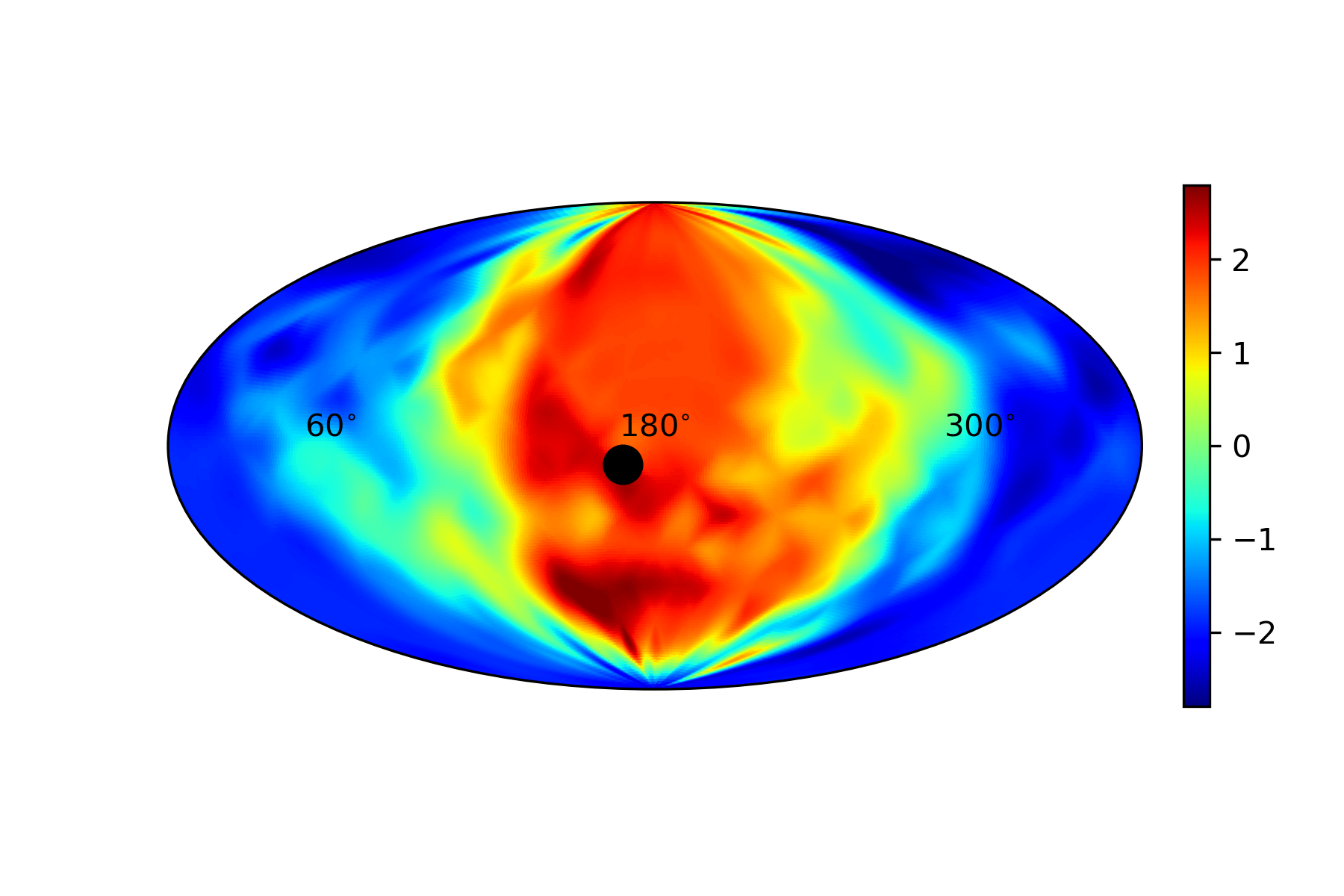}
    \caption{Variation in the absolute value of the fitting parameter $\beta$ (\bcomm{reproduced from Fig. 4 of \cite{bib:Luongo2022}}) across the sky in equatorial coordinates. The Hubble constant $H_0$ is degenerate with $\beta$, such that increases in $\beta$ correspond to increases in $H_0$, $\Delta \beta \propto \Delta H_0/H_0$. The black dot denotes the CMB dipole direction.}
    \label{fig:QSO_H0}
\end{figure}

One now has strong motivation to extend the analysis to \textit{any} cosmological data set. This motivation comes not just from Fig.~\ref{fig:H0variation}, but as explained, from CMB \cite{bib:Fosalba2021, bib:Yeung2022}, anomalous bulk flows \cite{bib:Lauer1994, bib:Hudson2004, bib:Kashlinsky2008, bib:Kashlinsky2010, bib:Atrio2015}, galaxy cluster anisotropies \cite{bib:Migkas2018, bib:Migkas2020, bib:Migkas2021} and cosmic dipole discrepancies \cite{bib:Blake2002, bib:Singal2011, bib:Gibelyou2012, bib:Rubart2013, bib:Tiwari2016, bib:Colin2017, bib:Bengaly2018, bib:Siewert2021}, which all point to curiosities concerning the CMB dipole. It should be noted that here the analysis simply involves the Hubble diagram, so individual data systematics aside, the observations are expected to be robust. 
Extending the analysis to high redshift QSOs \cite{bib:Risaliti2015, bib:Risaliti2019, bib:Lusso2020} and GRBs \cite{bib:Demianski2017}, one finds a strong dipole \cite{bib:Luongo2022}, as illustrated in Fig.~\ref{fig:QSO_H0}, where we have limited the redshift range to $0.7 < z < 1.7$ in order to avoid any potential evolution in fitting parameters \cite{bib:Khadka2021}. Nevertheless, even with evolution, one can argue that evolution in fitting parameters with redshift is not expected to mimic directional dependence (see appendix~A of \cite{bib:Luongo2022}). Despite differences in sample and methodology, there is a clear preferential direction in the direction of the CMB dipole, in line with the findings of \cite{bib:Singal2022}. Finally, observe that the QSOs are so deep in redshift that any corrections for a kinematic CMB dipole are negligible, so these QSOs must reside in CMB frame, assuming such a frame exists.  

Overall, when data sets are combined one has a statement at $2$-$3 \, \sigma$, while QSOs on their own point to an anisotropic Hubble expansion in the direction of the CMB dipole at a significance of $\sim 2 \, \sigma$.\footnote{The Risaliti-Lusso QSOs are discrepant with Planck-$\Lambda$CDM, but this discrepancy becomes apparent beyond $z \sim 1$ \cite{bib:Risaliti2019, bib:Lusso2020}. More concretely, the luminosity distance $d_\textrm{L}(z)$ is smaller, but this may yet be consistent with Lyman-$\alpha$ BAO \cite{bib:Delubac2015, bib:duMas2020}, HST SNe Ia \cite{bib:Dainotti2021} and QSO Hubble diagrams based on alternative techniques \cite{bib:Solomon2022}.} Admittedly, given the relatively smaller size and increased scatter of GRB samples, results remain inconclusive in the sense that they are sample dependent.

\subsection{Precision spectroscopy tests of the stability of fundamental couplings}
\label{sec:precision_spectroscopy_test}

Tests of the universality of physical laws, specifically testing the stability of nature's dimensionless fundamental constants are some of the cornerstones of the ongoing search for the new physics which is required to explain the recent acceleration of the Universe. The canonical way to carry out such tests is through high resolution astrophysical spectroscopy measurements of the fine-structure constant $\alpha$, the proton to electron mass ratio $\mu$, the proton gyromagnetic ratio $g_p$, or combinations thereof. These are complemented by local laboratory tests using atomic clocks. For a recent review of the field see \cite{bib:Martins2017}.

Particle physics experiments have established beyond doubt that fundamental couplings \textit{run} with energy. Similarly, in many extensions of the standard model they will also unavoidably \textit{roll} in time (or equivalently redshift) and \textit{ramble} in space (i.~e., they will depend on the local environment). In particular, this will be the case in theories with additional spacetime dimensions, such as string theory. 
Part of the interest in the field has been due to the 2011 indication of a spatial variation of the fine structure constant $\alpha$, at the $4.2 \, \sigma$ level of statistical significance \cite{bib:Webb2011}. Such a variation would violate the CP in the sense that any universe with a spatial variation of $\alpha$ is obviously not homogeneous and isotropic. In what follows we outline developments in the past decade and the current status of the field.

Constraints on $\alpha$ at a given redshift are usually expressed relative to the present-day laboratory value $\alpha_0$, specifically via
$(\Delta\alpha/\alpha)(z)\equiv(\alpha(z)-\alpha_0)/\alpha_0$, which competitive measurements determine at the parts per million (ppm) level. Direct high-resolution spectroscopy measurements of $\alpha$ are done (mainly at optical wavelengths) in low-density absorption clouds along the line of sight of bright quasars, typically with wavelength resolution $R=\lambda/\Delta\lambda\sim10,000$-$150,000$ (with the precise value being different for different measurements and spectrographs). Currently available measurements fall into two subclasses:

\begin{itemize}
\item The \textit{Archival} data set of \cite{bib:Webb2011}. This is a data set of almost 300 measurements from the VLT UVES spectrograph and the Keck HIRES spectrograph, up to redshift $z\sim4.18$. The data was originally taken for other purposes, typically with resolution $R\sim50\,000$, and subsequently reanalyzed by the authors for the purpose of measuring $\alpha$. The archival nature of the data is relevant, because $\alpha$ measurements require particularly careful wavelength calibration procedures, which rely on additional data coeval with the QSO observations. Such additional data are not ordinarily taken for standard observations (which do not have the stringent requirements for $\alpha$ tests) and cannot be obtained \textit{a posteriori}. Moreover, both spectrographs are now known to suffer from significant intra-order and long-range distortions \cite{bib:Whitmore2015}. Such limitations may be partially mitigated \cite{bib:Dumont2017}, but cannot be fully eliminated. The result is an irreducible systematic uncertainty at the level of about 3 ppm.

\item The \textit{Dedicated} data set, comprising about 30 measurements obtained for the specific purpose of constraining $\alpha$, where ancillary data enabled a more robust wavelength calibration procedure, or using more modern spectrographs that do not suffer from the limitations of VLT-UVES or Keck-HIRES. This includes measurements listed in Table~1 of \cite{bib:Martins2017} and more recent ones from the Subaru telescope \cite{bib:Murphy2017} and the X-SHOOTER \cite{bib:Wilczynska2020}, HARPS \cite{bib:Milakovic2021} and ESPRESSO \cite{bib:Welsh2020,bib:Murphy2022} spectrographs. The original spectra have resolution $R\sim50\, 000$-$150\, 000$ (with the upper end of the range due to ESPRESSO), with the exception of the X-SHOOTER data which only has $R\sim10\, 000$; these latter, which are the first direct measurements of $\alpha$ in the IR part of the electromagnetic spectrum, extend the redshift range up to $z\sim7.06$ but their sensitivity is only at the level of tens of ppm, so they do not carry significant weight in the analysis that follows.
\end{itemize}

\begin{table}
\begin{center}
\caption{Basic properties of the two sets of high-resolution spectroscopy measurements of the fine structure constant $\alpha$. Constraints are given at the one sigma ($68.3\%$) confidence level, while upper limits are given at the two sigma ($95.4\%$) confidence level.}
\label{tab:tablealpha1}
\begin{tabular}{| c | c | c |}
\hline
Parameter & Archival & Dedicated \\
\hline
Weighted mean $\Delta\alpha/\alpha$ (ppm) & $-2.16\pm0.85$ & $-0.23\pm0.56$ \\
Weighted mean redshift $z_{\rm eff}$ & $1.50$ & $1.29$ \\
\hline
Pure dipole A (ppm) & $9.4\pm2.2$ & $<2.9$ \\
Logarithmic dipole A (ppm) & $9.9\pm2.3$ & $<3.3$ \\
\hline
\end{tabular}
\end{center}
\end{table}

It is well known that the Archival and Dedicated data sets are discrepant; an earlier analysis can be found in \cite{bib:Martins2017}. A simple way to see this is to assume that there is a unique astrophysical value of $\Delta\alpha/\alpha$, which we estimate by taking the weighted mean of all the values in each data set. Similarly, we can identify an effective redshift of the sample by taking the weighted mean of the redshifts of the measurements in the set. The first two rows of Table~\ref{tab:tablealpha1} report these results. We see that while the effective redshifts are comparable (reflecting the prevalence of stringent measurements around $z\sim1$), the archival data has a preference for a negative variation at more than two standard deviations, while the Dedicated value is consistent with the null result.

\begin{figure}
\begin{center}
\includegraphics[width=0.45\textwidth]{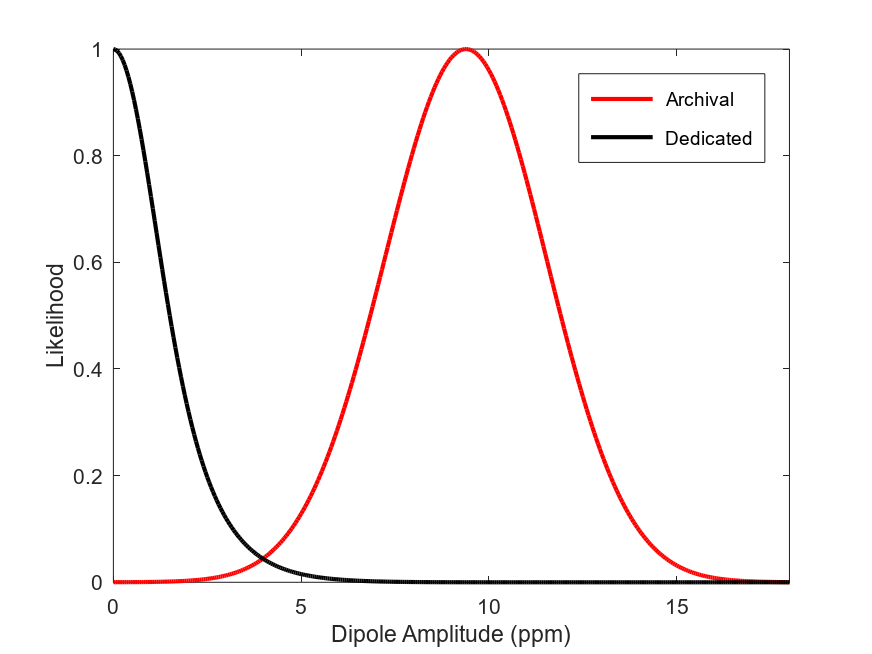}
\includegraphics[width=0.45\textwidth]{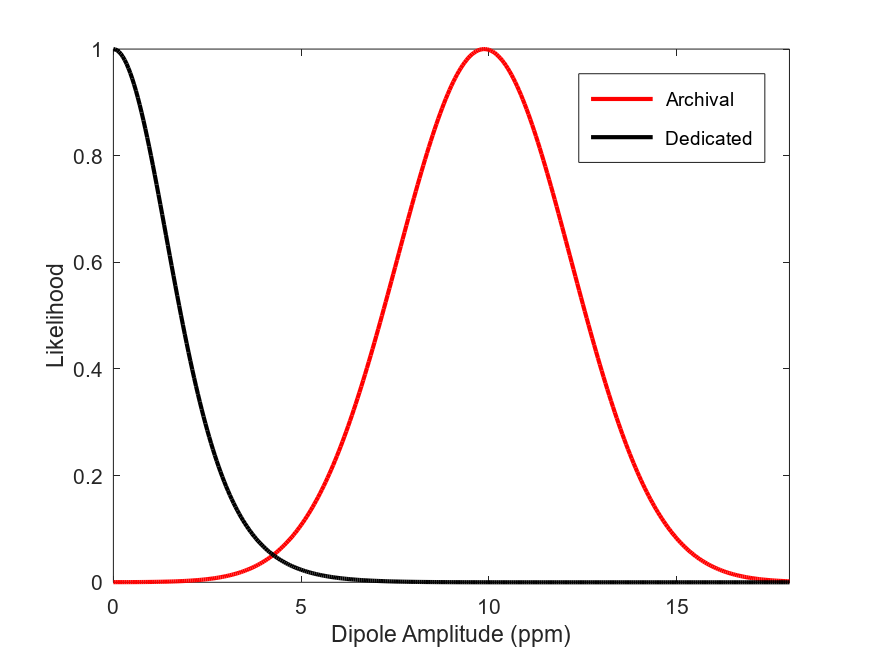}
\end{center}
\caption{Posterior likelihood constraints on the amplitude (in parts per million), with the angular coordinates marginalized, of a putative dipole in the astrophysical measurements of $\alpha$ \bcomm{(reproduced from Fig. 1 of \cite{bib:Martins2022})}. The left panel shows the result for a pure spatial dipole, while the right panel shows the result for a dipole with a further redshift dependence, cf.~(\ref{puredipole}) and (\ref{redshiftdipole}) respectively. The red and black lines correspond to the Archival and Dedicated spectroscopic data sets, described in the text.}
\label{fig:figalpha1}
\end{figure}

As for spatial variations, which are more relevant to the CP and its possible violations, we can constrain the amplitude $A$ of a pure spatial dipole for the relative variation of $\alpha$
\be\label{puredipole}
\frac{\Delta\alpha}{\alpha}(A,\Psi)=A\cos{\Psi}\;,
\ee
which depends on the orthodromic distance $\Psi$ to the North Pole of the dipole (the locus of maximal positive variation) given by
\be\label{ortho}
\cos{\Psi}=\sin{\theta_i}\sin{\theta_0}+\cos{\theta_i}\cos{\theta_0}\cos{(\phi_i-\phi_0)}\;,
\ee
where $(\theta_i,\phi_i)$ are the Declination and Right Ascension of each measurement and $(\theta_0,\phi_0)$ those of the North Pole. For comparison we further consider the case with a logarithmic redshift dependence in addition to the spatial variation
\be\label{redshiftdipole}
\frac{\Delta\alpha}{\alpha}(A,z,\Psi)=A\, \ln{(1+z)}\, \cos{\Psi}\;;
\ee
such a logarithmic dependence is physically well motivated, since it corresponds to the matter era behavior for dilaton-type scalar fields \cite{bib:Martins2017}, and this parametrization has the further advantage of not requiring any additional free parameters. Constraints on the amplitude $A$, marginalizing over the angular coordinates, are shown in the bottom part of Table~\ref{tab:tablealpha1} and also Fig.~\ref{fig:figalpha1}. One confirms that for the Archival data there is a statistical preference, at just over four standard deviations, for a dipole with an amplitude of about 9 ppm, while in the Dedicated data there is no preference for a dipole, and the amplitude is constrained to be less than about 3 ppm at the $95.4\%$ confidence level. We note that the latter constraint improves the analogous ones reported in \cite{bib:Martins2017} by about a factor of two, which highlights the steady gains in sensitivity and the impact of the growing number of dedicated measurements.

\begin{figure}
\begin{center}
\includegraphics[width=0.45\textwidth]{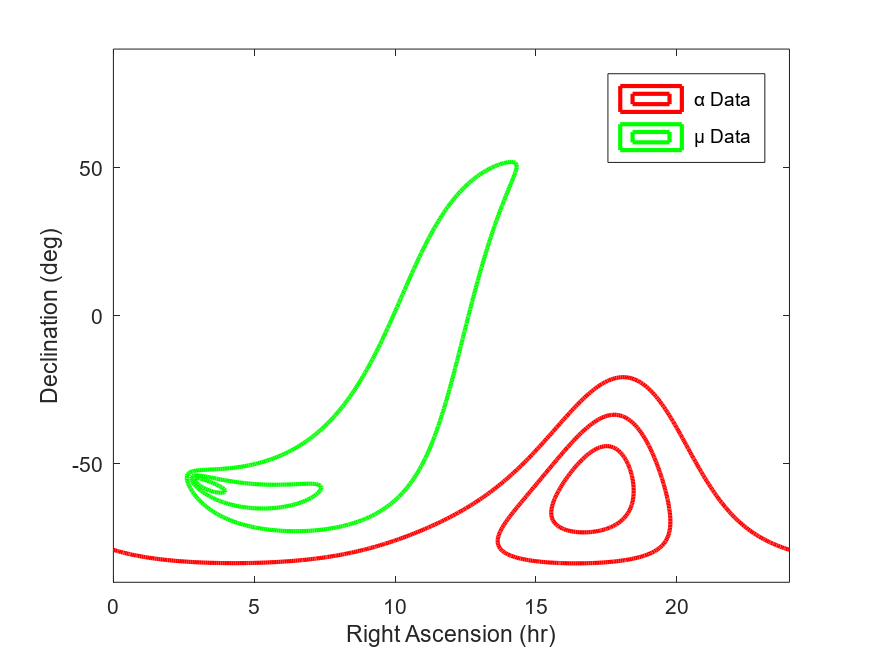}
\includegraphics[width=0.45\textwidth]{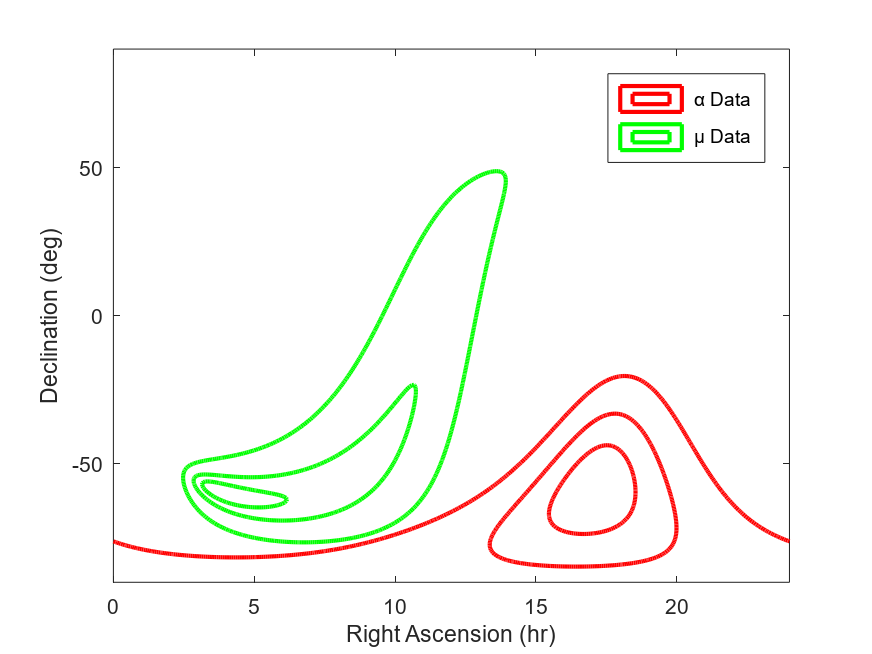}
\end{center}
\caption{Constraints on the North Pole direction, with the amplitude marginalized for a pure spatial dipole (left) and for one with redshift dependence (right). Red lines correspond to the Archival $\alpha$ data set, and green lines correspond to the $\mu$ data. 1-, 2-, and 3-$\sigma$ contours are shown in all cases. \bcomm{Figures update earlier Fig. 2 of \cite{bib:Martins2017} to include more recent data.}}
\label{fig:figalpha2}
\end{figure}

A complementary test can also be done considering analogous high resolution spectroscopy astrophysical tests of the stability of the proton to electron mass ratio, $\mu$. A compilation of such measurements can be found in Table~2 of \cite{bib:Martins2017}; this data set is somewhat heterogeneous, including optical and radio measurements, but with comparable sensitivity to the $\alpha$ measurements. In this case an analysis along the same lines as that for $\alpha$ shows a preference for a dipole at just under $2 \, \sigma$, with 2 $\sigma$ upper limits on the dipole amplitude being $A_\mu<2$~ppm and $A_\mu<4.5$~ppm, respectively for a pure and a redshift-dependent dipole. (The two limits differ a lot more here than in the $\alpha$ case because the most stringent constraints on $\mu$ are all at low redshifts, $z<1$.) Importantly, in physically motivated extensions of the standard model, where $\alpha$ and $\mu$ both vary, the two variations are related: generically, $(\Delta\mu/\mu)=\kappa_{\mu\alpha}(\Delta\alpha/\alpha)$, where the coefficient $\kappa_{\mu\alpha}$ is model dependent but a true constant for a given model. It follows that if the $\alpha$ Archival dipole is real, one would expect to see a $\mu$ dipole aligned with it. Fig.~\ref{fig:figalpha2} plots the best-fit directions for the Archival and $\mu$ dipoles, showing that there is no such alignment. The overall conclusion therefore is that \textit{there is currently no robust evidence for spacetime variations of $\alpha$, up to the few ppm level of sensitivity.}

\subsection{Summary}
\label{sec:summary_late_universe}

Challenges to the class of FLRW cosmologies, and $\Lambda$CDM in particular, from observations at low redshifts $z \sim {\cal{O}}(1)$ yield a similar picture to CMB anomalies detailed in section~\ref{sec:summary_early_universe}.
Most of them coincide, even though the inference of cosmological parameters is complicated at lower redshifts by local inhomogeneities and their motions in the late Universe. 

At redshifts $z \lesssim 0.1$, observations are mostly sensitive to $H_0$, such that bulk flows and the mass density contrast are determined on top of a linear Hubble expansion (see section~\ref{sec:bulk_flows} and, for instance, \cite{bib:Nusser2016} for details on the implementation).
Being only able to measure a combination of bulk flow and Hubble expansion effects, there is a degeneracy between the velocity of the flow $v$ and the motion caused by the expansion of the Universe, which is locally represented by $H_0$.\footnote{
As noted in \cite{bib:Horstmann2021}, attention has to be paid not to use peculiar velocity field reconstructions in cosmological parameter estimations because the reconstruction of a peculiar velocity field requires a cosmology to be fixed beforehand. Such approaches can thus yield consistency tests at best.}
As shown in Fig.~21 in \cite{bib:Tully2016}, we can choose from a small range of $v$ and $H_0$ at low redshifts.
Yet, our choice is bounded by physical consistency. 
For instance, we will not choose a $v$-$H_0$ combination which only allows for inflows of local structures. 
One can try to establish this consistency, however, $H_0=68$~km/s/Mpc, as required by \cite{bib:Planck2018}, implies an out-flowing bulk with $v\approx 700~\mbox{km/s}$ at $z=0.05$, which seems unrealistically large \cite{bib:Tully2016}. 
Other bulk flow measurements, as summarized in Fig.~\ref{fig:bulk_flows_literature}, are still in accordance with those expected in a $\Lambda$CDM cosmology, yet, there is a general trend to larger than expected bulk flows, in particular with increasing distance from us, see, for instance, \cite{bib:Howlett2022}. 
As bulk flows are only reconstructed for low redshifts, $H_0$ is the most relevant parameter of any model in the FLRW class degenerate with the flow velocity. 
The impact of the next to leading order dominant parameters remains to be investigated. 

Peculiar velocities on top of a linear Hubble expansion or a fully parametrized background cosmology can also be used as tracers of the density contrast. 
Therefore any observed bulk flow can be converted into gravitationally interacting matter agglomerations and be used as a probe for the homogeneity scale (see section~\ref{sec:Hom-scale}). 
All observations performed so far clearly show that homogeneity is not achieved on scales of 50-$80\,h^{-1}\,$~Mpc, as previously expected for a $\Lambda$CDM universe. 
This is clearly a lower bound on the homogeneity scale. Within $\Lambda$CDM,  $N$-body simulations allow for larger upper bounds \cite{bib:Yadav2010} and observations \cite{bib:DeMarzo2021} also hint at larger values. 

Determining bulk flows at higher redshifts from the kSZe has been found difficult \cite{bib:Haehnelt1994}, such that a larger than expected bulk flow derived from galaxy clusters by \cite{bib:Kashlinsky2008,bib:Kashlinsky2009} could be explained by systematics. 
Yet, as shown in \cite{bib:Migkas2021}, scaling relations including tSZe and X-ray observations, still hint at a $\sim900~\mbox{km/s}$ bulk flow out to 500~Mpc in a similar direction as the LG motion. 
This $4 \, \sigma$ significant result continues to challenge the CP because many systematic biases have already been excluded as causes. 
Alternatively, the results can be interpreted as a variation of $H_0$ across the sky, leading to a $\gtrsim 5.4\sigma$ significant anisotropy in the local Universe (see Fig.~\ref{fig:migkas_H0_anisotropy}).

Increasing the redshifts of sources up to $z\sim 1$, studies of the cosmic dipole as inferred from flux-limited radio galaxy surveys, performed by different groups with different data sets and evaluation techniques, also show deviations from expectations based on the CP. 
These deviations, however, are mostly in agreement with each other, concluding that the dipole amplitude is larger than expected (see Appendix~A in \cite{bib:Siewert2021} for an encompassing literature overview). 
Yet, the dipole direction coincides with the CMB dipole direction. Similar results have been obtained for QSOs and sample sizes over $1.3\times 10^6$ objects \cite{bib:Secrest2021}. 
Possible biases due to clustering of structures causing part of the excess amplitude was shown to depend on the observed frequency and that several unrealistically large structures may be required or several local voids, see \cite{bib:Rubart2014, bib:Rameez2018, bib:Siewert2021}, which themselves may be in tension with the CP. Moreover, this frequency dependence is largely driven by the TGSS sample, which is anomalous in a number of ways and presumably has been impacted by systematics \cite{bib:Dolfi2019, bib:Hurley2017, bib:Tiwari2019, bib:Secrest2022}. 

A very intriguing effect can be found for the cosmic dipole as determined by Type Ia SN: While \cite{bib:Singal2021} recovers an amplitude excess compatible with those from QSO and radio galaxy number counts, \cite{bib:Horstmann2021} find an amplitude lower than CMB expectations, which is in tension with the dipole amplitude excesses. Different methodological approaches requiring a different treatment of the redshifts could cause such a discrepancy and should be investigated further. 
As suggested in \cite{bib:Horstmann2021} and already implemented in some approaches (e.~g.~\cite{bib:Singal2021}), heliocentric redshifts are optimal to be used and should not be converted into other frames of reference for further evaluation, see e.~g.~\cite{bib:Rameez2021} for further details.\footnote{However, note that cosmology is only consistent in ``the CMB frame''. Assuming such a frame exists, since subtracting the CMB dipole fixes CMB as the rest frame of the Universe. As a result, conversion from heliocentric to CMB redshifts is unavoidable if one is studying cosmological parameters.} These issues aside, SNe have some advantages over galaxies as cosmic dipole probes because they do not require completeness in number count coverage and observations have higher precision such that fewer SNe are required to reach a precision in the inferred amplitude and direction that is comparable to those inferred by about two orders of magnitude more galaxy counts. 
Moreover, any (improved) variant of the Baldwin-Ellis test \cite{bib:Ellis1984} is subject to different biases than the tests based on the magnitude-redshift relation in the Hubble diagram (\ref{eq:m-z_relation}), such that an agreement of resulting deviations from CP acerbates the tensions. 
The same can be said for the arising anisotropies in the supposedly constant $H_0$ across the sky as determined from strong lensing and SNe as shown in section~\ref{sec:emergent_H0} and the correlations to the early Universe anomalies detailed in section~\ref{sec:early_universe_FLRW_anomalies}.

It will be interesting to see which conclusions about the validity of FLRW or the CP can be found from variations of fundamental constants, as detailed in section~\ref{sec:precision_spectroscopy_test}, as soon as the necessary precision in observations has been reached. The disappearance of the fine structure dipole reminds us of the need for dedicated studies of these anomalies. Anomalies can in principle go away with better quality data.

\section{Mysterious alignments} 
\label{sec:alignments} 

In this section, we document a series of intriguing alignments that are surprising in a statistically homogeneous and isotropic universe. One curious aspect of some of these alignments is their observation over large -- potentially gigaparsecs -- scales and axes that overlap with the CMB dipole direction. If not due to experimental systematics or interstellar physics, one exciting possibility is that these features are cosmological in origin. This certainly motivates future dedicated studies of the features and their statistics, since if cosmological, they are expected to provide important clues on how the Universe has evolved. Recently, evidence for parity violating physics in Large Scale Structure \cite{bib:Hou2022, bib:Philcox2022a} and the CMB \cite{bib:Minami2020, bib:Diego-Palazuelos2022} have been found. It would be interesting to see if there are any connections between these observations.

\subsection{QSO polarization alignments}
\label{sec:qso_polarization_alignments}

\begin{figure}
\includegraphics[height=5.0cm]{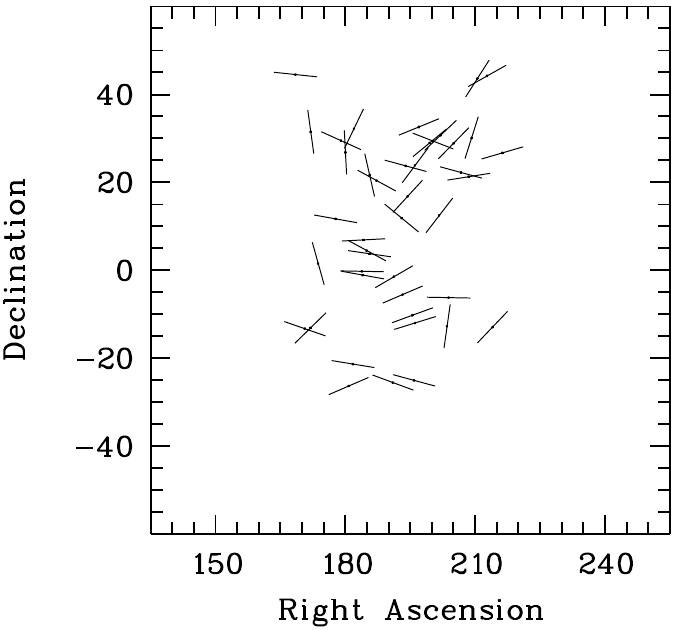}
\includegraphics[height=5.0cm]{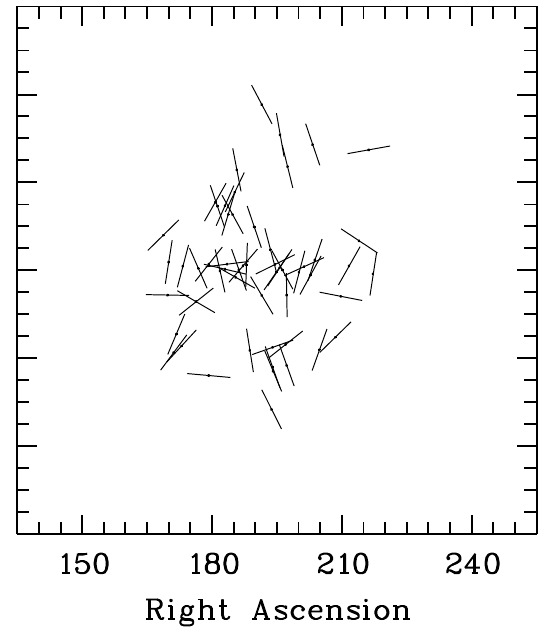}
\caption{Maps of QSO polarization vectors \bcomm{(reproduced from Fig. 7 of \cite{bib:Hutsemekers2005})} in the North Galactic hemisphere: QSOs with redshifts $0.0 \leq z < 1.0$ (left), QSOs with redshifts $1.0 \leq z \leq 2.3$ (right). At low (resp. high) redshifts the mean polarization angle is $79^{\circ}$ (resp. $8^{\circ}$).}
\label{fig:polali}
\end{figure}

\begin{figure}  
\includegraphics[height=4.8cm]{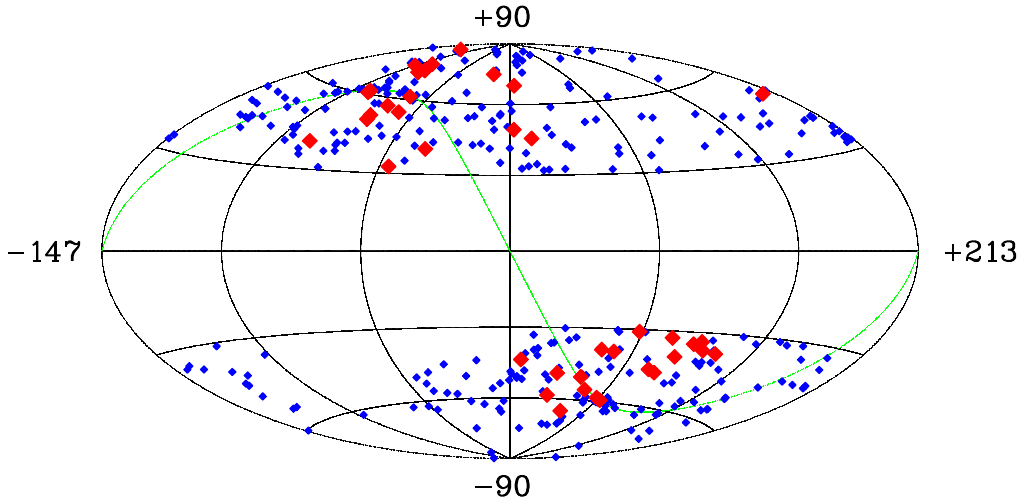}
\caption{Hammer-Aitoff projection of the 355 QSO positions on the sky (blue dots) in Galactic coordinates \bcomm{(adapted from Fig. 16 of \cite{bib:Hutsemekers2005})}. Red, large symbols indicate the location of the most significant polarization alignments. The green line gives the location of the celestial equator.}
\label{fig:polaxis}
\end{figure}

QSO\footnote{We indifferently use the term QSO for quasars, QSOs, and their low-luminosity counterparts, Seyfert and radio galaxies.} light is known to be linearly polarized at optical wavelengths with levels typically around 1\% \cite{bib:Moore1984,bib:Stockman1984,bib:Berriman1990}. Since polarization is usually related to the object's morphology, in particular the \bcomm{Very Long Baseline Interferometry (VLBI)} jet direction \cite{bib:Rusk1985}, it came as a surprise to find coherent orientations (alignments) of QSO polarization vectors in large patches of the sky \cite{bib:Hutsemekers1998}. Statistical tests based on three-dimensional nearest-neighbor analysis showed that the polarization vectors were not randomly oriented on spatial scales reaching one Gpc at redshift $z \simeq 1.5$, with a significance level (p-value) around 1\%. This result was based on a sample of 170 polarized QSOs, observed at high galactic latitudes ($|b| \geq 30^{\circ}$) to minimize the contamination by interstellar polarization. Subsequent observations increased the sample size to 217 \cite{bib:Hutsemekers2001} and finally 355 \cite{bib:Hutsemekers2005} polarized QSOs with redshifts up to $z \simeq 2.5$, distributed all over the sky. Statistical analyses of theses samples confirmed the existence of Gpc-scale alignments of QSO polarization vectors with even lower significance levels, between 10$^{-3}$ and 10$^{-5}$ depending on the test \cite{bib:Jain2004,bib:Hutsemekers2005,bib:Pelgrims2014}. Fig.~\ref{fig:polali} illustrates two regions of QSO polarization alignments. Fig.~\ref{fig:polaxis} shows that the regions where the alignments are the most significant are located in opposite directions on the sky, possibly defining a ``polarization alignment axis'', along which the mean polarization angle is found to rotate as a function of the cosmological distance \cite{bib:Hutsemekers2005}. While a more complete sky coverage is needed to definitely prove the existence of such a preferred direction, it appears to be close to the CMB dipole direction, among other coincidences \cite{bib:Ralston2004,bib:Shurtleff2013}.
Such unexpected and intriguing results require careful inspection of possible contamination by instrumental and interstellar polarization. Instrumental polarization was measured to be very small \cite{bib:Sluse2005}. Moreover, data obtained with different instruments do agree within uncertainties. Although the adopted cutoff on the polarization degree, $p \geq 0.6\%$, ensures that the measured polarizations are essentially intrinsic to the QSOs \cite{bib:Berriman1990,bib:Cabanac2005}, some data may still be affected by interstellar polarization, even at high galactic latitudes. Several tests \cite{bib:Sluse2005,bib:Hutsemekers2005,bib:Payez2010,bib:Pelgrims2019} have shown it is unlikely that interstellar polarization is at the origin of the observed alignments. This can also be inferred from Fig.~\ref{fig:polali}: should interstellar polarization (or any local contamination) be responsible for the observed alignments, one would expect the same systematic effect at all redshifts and no difference between the mean polarization angles at low and high redshifts.

To explain the observed polarization alignments, several effects that could affect light as it propagates through the Universe have been invoked \cite{bib:Brans1975,bib:Jain2002,bib:Das2005,bib:Morales2007,bib:Payez2011,bib:Ciarcelutti2012}, with the constraint that a single, local slab of matter or field cannot reproduce the redshift dependence of the mean polarization angle (Fig.~\ref{fig:polali}); whatever the mechanism, a huge coherent structure (e.~g.~a magnetic field) is still needed at a redshift around one.
Alternatively, the QSO axes themselves, or the \bcomm{supermassive black hole (SMBH)} spin axes, could be aligned on large scales, possibly related to structure formation, large-scale magnetic fields, cosmic strings, or rotation of the Universe, among other scenarios \cite{bib:Li1998,bib:Poltis2010,bib:Godlowski2011,bib:Codis2015,bib:Slagter2018,bib:Korotky2020,bib:Tiwari2021}. To provide some clues, polarization alignments have been searched for at radio wavelengths where polarization is also known to be related to the jet orientation \cite{bib:Rusk1985}, in particular, polarizations from the 8.4~GHz JVAS/CLASS survey that contains more than 4000 flat-spectrum radio sources with polarized flux $>$ 1 mJy out of which about one third have reliable redshifts. Different cuts of the sample led to the following results: no evidence for polarization alignments on Gpc scale \cite{bib:Joshi2007}, significant alignments on 150~Mpc scale \cite{bib:Tiwari2013}, and finally significant alignments on Gpc scale when only considering the QSO subsample \cite{bib:Pelgrims2015}. Although the latter alignments remain difficult to interpret either as a bias in the data or a real physical effect, it is striking to note that regions of radio polarization alignments fall within the previously identified regions of optical polarization alignments \cite{bib:Pelgrims2015}. Since QSO axes could be related to the large-scale structure in which they are embedded, as do galaxies on smaller scales \cite{bib:Tempel2013,bib:Zhang2013,bib:West2017}, the polarization of QSOs belonging to the Huge-LQG \cite{bib:Clowes2013}, first introduced in section~\ref{sec:Hom-scale}, has been measured. Interestingly, the Huge-LQG is not far from the ``polarization alignment axis''. The QSO spin axes, as derived from the polarization orientation, were found to align with their host large-scale structures on scales reaching 500~Mpc \cite{bib:Hutsemekers2014}. A similar trend was found when comparing QSO radio polarization vectors to the host structure orientations in a sample of LQGs \cite{bib:Pelgrims2016}. Alignments of LQG axes themselves are found on even larger scales \cite{bib:Friday2020}, possibly making the link between QSO polarization alignments at various scales. These results suggest that polarization alignments are related to QSO morphological alignments. Searches for coherent orientations of QSO jets were carried out in a series of recent studies \cite{bib:Taylor2016,bib:Contigiani2017,bib:Panwar2020,bib:Osinga2020,bib:Blinov2020,bib:Mandarakas2021,bib:Marcha2021}, with various results that depend on the sample, its size, the availability of redshift information, and the method of analysis. The main result is that significant jet alignments were found in regions of the sky with sizes ranging from 30~Mpc to more than one Gpc, some of these regions overlapping the regions of polarization alignments \cite{bib:Mandarakas2021}.

All these studies, sensitive to different systematic effects, independently point to the existence of regions in the sky where the QSO axes are coherently oriented. This coherent orientation occurs on various scales, the largest ones challenging the hypothesis of large-scale isotropy. Homogeneous, complete all-sky surveys are definitely needed to fully explore the largest scales at which alignments are detected, and if they really occur along a preferred axis close to the CMB dipole axis. Finally, assuming one can standardize QSOs \cite{bib:Risaliti2015, bib:Risaliti2019}, and they represent a viable rung of the distance ladder, it is timely to recall the anisotropies in the QSO Hubble diagram from section~\ref{sec:emergent_H0}.

\subsection{Large Quasar Group alignments} 
\label{sec:LQG_alignments} 

LQGs (see also section~\ref{sec:Hom-scale}) are large-scale agglomerations of QSOs that appear to be statistically significant departures from uniformity. The first LQG was found in \cite{bib:Webster1982}, the second in \cite{bib:Crampton1987}, and the third in \cite{bib:Clowes1991} (CCLQG). Many more examples have been found since, mostly in data
from SDSS, with the most extreme example currently being that of \cite{bib:Clowes2013} (Huge-LQG).

In 1991, the CCLQG was the largest structure known in the Universe; its long dimension
is $\sim 630$~Mpc. In 2013, the Huge-LQG was the largest structure known in the Universe;
its long dimension is $\sim 1240$~Mpc. The sizes given are proper sizes for the present epoch.
Both greatly exceed the scale of homogeneity as established in \cite{bib:Yadav2010}, which is $\sim 370$~Mpc.
A curious feature is that the CCLQG and the Huge-LQG are at the same mean redshift, $z = 1.28$ and $z= 1.27$ respectively, and are adjacent on the sky. They are shown in Fig.~\ref{fig:SGW_LQG} (right). 
Another curiosity, presumably a coincidence, is their proximity to the CMB dipole. They are, however, not that close to the QSO dipole of \cite{bib:Secrest2021} discussed in section~\ref{sec:QSO_dipole}. 

\begin{figure}[htb]
\includegraphics[width=70mm]{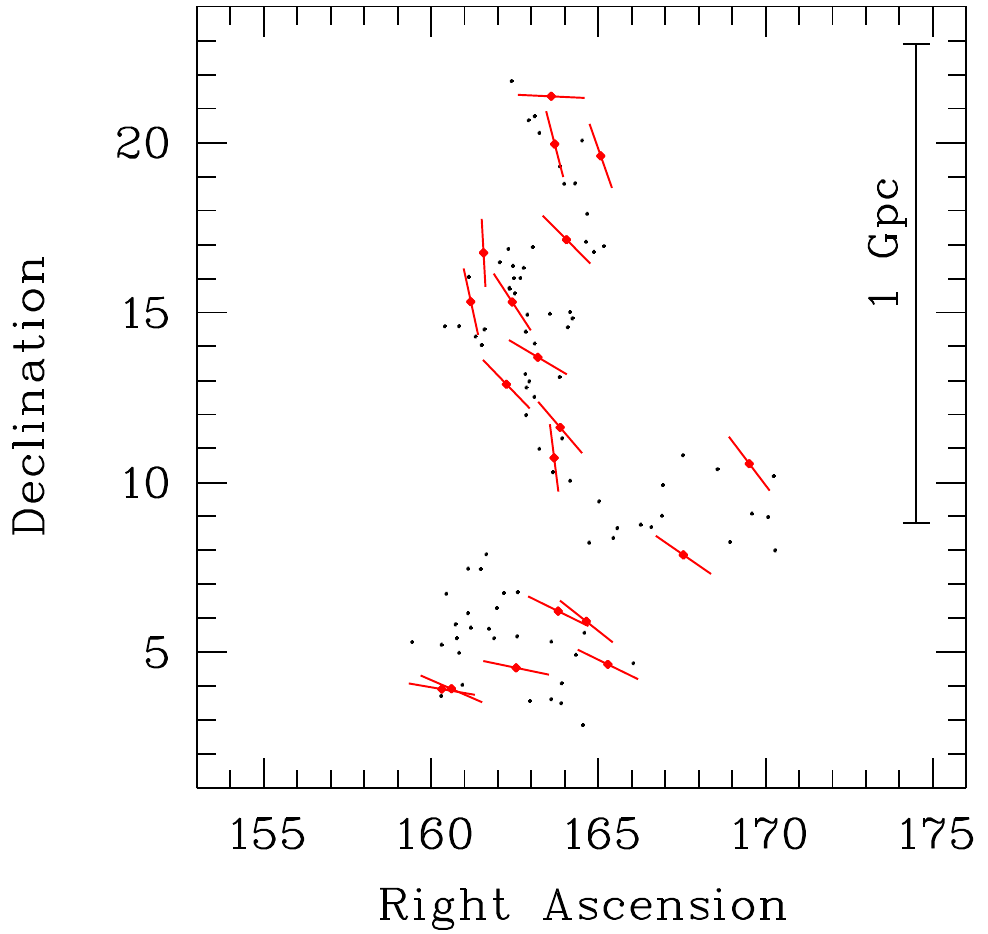}
\caption{Polarization vectors for 19 QSOs
from the Huge-LQG (upper set) and the CCLQG (lower set) \bcomm{(reproduced from Fig.~6 of \cite{bib:Hutsemekers2014})}. It shows a clear association of the polarization
vectors with the overall directions of the two LQGs.}
\label{fig:LQG_alignment}
\end{figure}

Fig.~\ref{fig:LQG_alignment} is reproduced from Fig.~6 of \cite{bib:Hutsemekers2014}. It shows the polarization
vectors for QSOs within the Huge-LQG and the CCLQG. They tend to be either parallel or orthogonal to the overall directions of the two LQGs. Note that the polarization angles
in this figure have been rotated according to $\theta^{\prime} = \textrm{mod} (\theta, 90^{\circ})+90^{\circ}$, and thus constrained
to $[90^{\circ}, 180^{\circ})$, to transform polarization orientation to spin orientation. An interpretation
of these polarization results for LQGs is that the axes of the QSOs (i.~e.~axes of their accretion disks or the SMBH spin axes)
are aligned across Gpc-scales.
Conceivably, the result that the polarization vectors of QSOs are correlated in general areas could arise from the presence of LQGs, not known at the time, especially if the orientations of the LQGs are themselves correlated across very large scales of $\gtrsim 1$~Gpc.
This possibility led \cite{bib:Friday2022} to investigate the orientation of the LQGs themselves. Some unexpected, intriguing, and ``tantalising" results emerged.

\begin{figure}[htb]
\includegraphics[width=84mm]{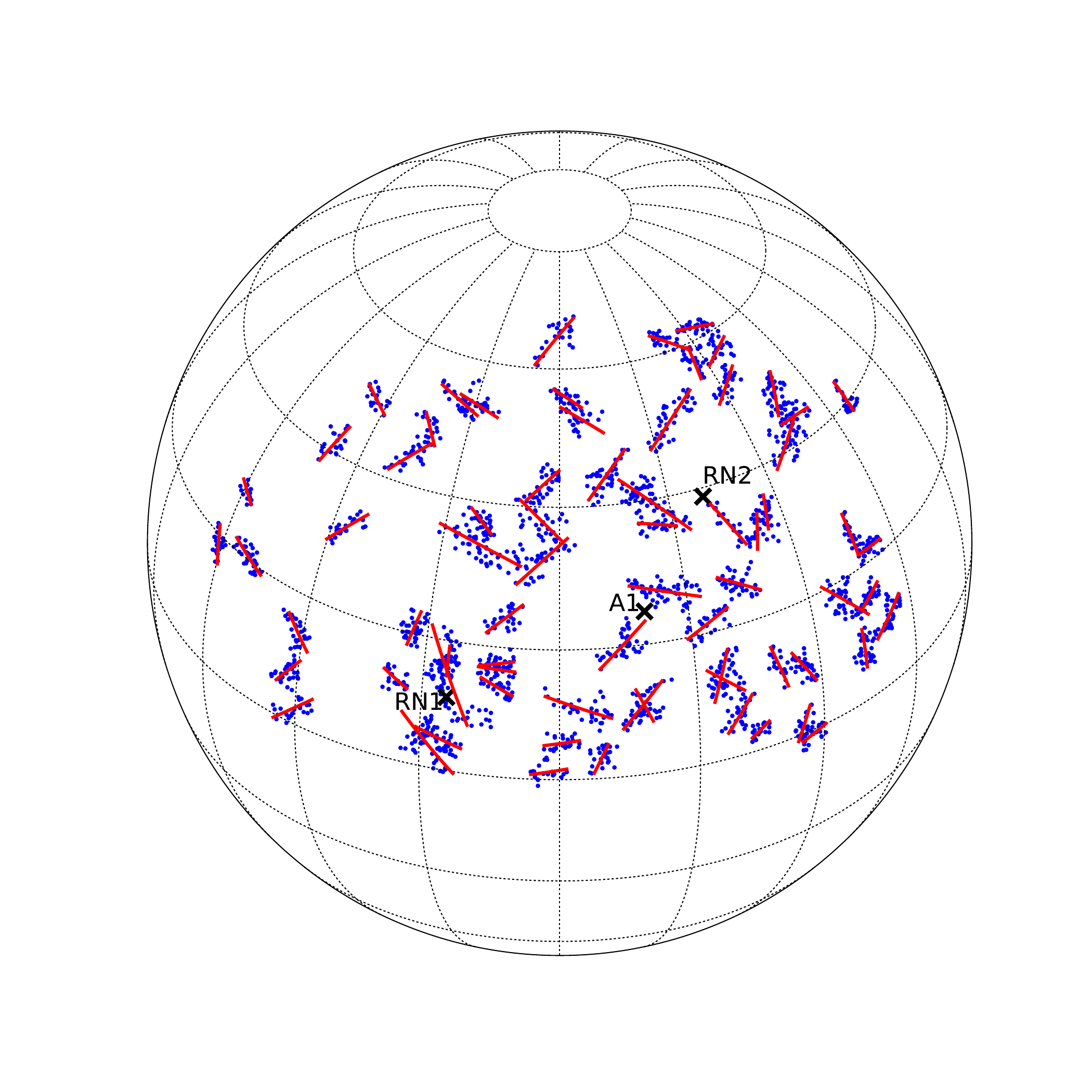}
\vspace{-5ex}
\caption{71 LQGs with redshifts $1.0 \leq z \leq 1.8$ and their position angles (PAs) on the celestial sphere \bcomm{(reproduced from Fig. 1 of \cite{bib:Friday2022})}. The blue dots indicate the member QSOs of the LQGs, and the red lines correspond to the LQG axes, fitted by orthogonal distance regression (ODR). The black crosses correspond to the centers of the A1, RN1, RN2 regions from \cite{bib:Hutsemekers1998} and \cite{bib:Pelgrims2015}. A1 is used below as a parallel-transport destination. The projection is centered on $(\alpha, \delta) = (180^{\circ}, 35^{\circ})$; parallels and meridians are separated by $20^{\circ}$; $\alpha$ increases to the right.}
\label{fig:LQG_ortho}
\end{figure}

Fig.~\ref{fig:LQG_ortho} shows a sky map for 71 of the LQGs that were discussed in \cite{bib:Clowes2012, bib:Clowes2013}. The 71 are a subset of a total set of 398, formed by restricting to membership $m \geq 20$ and significance parameter $ \geq 2.8 \, \sigma$. 
The members of the LQGs are shown as blue dots in the figure. The red lines indicate the orientations of the LQGs, determined by orthogonal distance regression (ODR) as the position angles (PAs) with respect to celestial north.

\begin{figure}
\includegraphics[height=5.0cm]{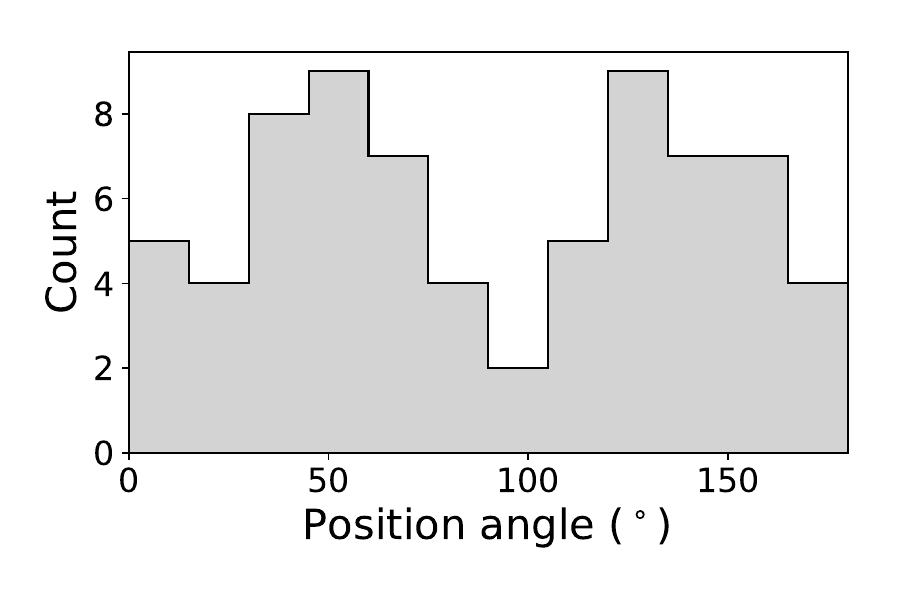}
\includegraphics[height=5.0cm]{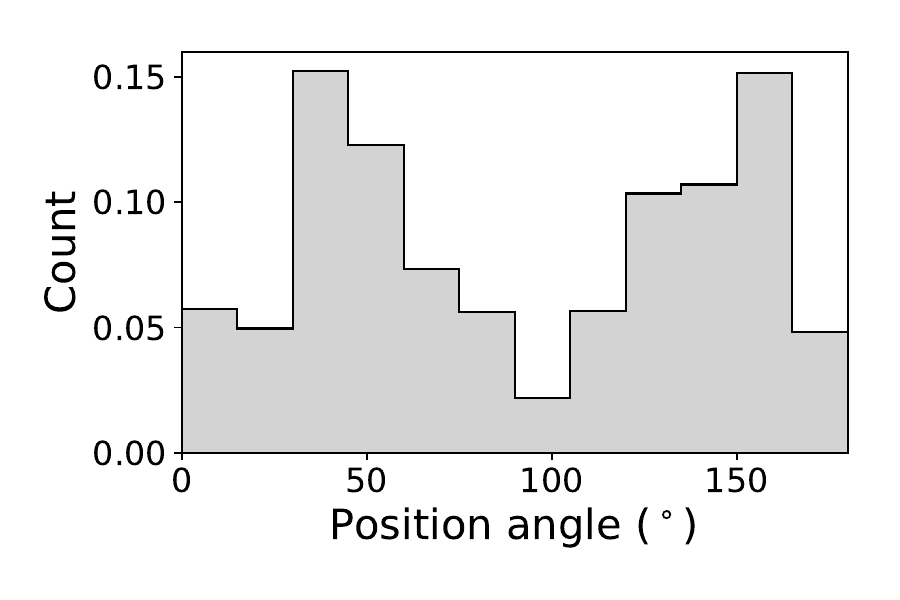}
\caption{LQG PAs parallel-transported to and measured at the center of the A1 region with $15^{\circ}$ bins \bcomm{(reproduced from Fig. 2 of \cite{bib:Friday2022})}: unweighted, $\textrm{PA} = 54^{\circ} \pm 2^{\circ}, 136^{\circ} \pm 3^{\circ}$ (left panel), ODR goodness-of-fit weighted, $\textrm{PA} =52^{\circ} \pm 2^{\circ}, 137^{\circ} \pm 3^{\circ}$ (right panel). Hartigans' dip statistic  returns 97\% confidence.}
\label{fig:PA_hist}
\end{figure}

Fig.~\ref{fig:PA_hist} shows the LQG PAs, after parallel transport, both unweighted (left panel) and weighted by orthogonal distance regression goodness-of-fit (right panel). 
The parallel transport is to the center of the A1 region of \cite{bib:Hutsemekers1998}, with measurements there. 
In both cases, the data are bimodal, with peaks separated by $\sim 90^{\circ}$. Hartigans' dip statistic \cite{bib:Hartigan1985} for continuous unweighted PA data recovers two peaks between $\sim 36^{\circ}-83^{\circ}$ and $\sim 114^{\circ}-156^{\circ}$, with 97 per cent confidence \cite{bib:Friday2022}. 
The means are then estimated as $\textrm{PA} = 54^{\circ} \pm 2^{\circ}, 136^{\circ} \pm 3^{\circ}$ (unweighted) and $\textrm{PA} = 52^{\circ} \pm 2^{\circ}, 137^{\circ}\pm3^{\circ}$ (ODR goodness-of-fit weighted). 
One can repeat the analysis with a different destination for parallel transport, but the bimodal distribution is robust. Full details of the statistical analyses are given in \cite{bib:Friday2020} and \cite{bib:Friday2022}.
In particular, the correlated orientations of the LQGs - aligned / orthogonal - are shown to be most significant for angular scales of $\sim 30^{\circ}$, corresponding to physical scales at the present epoch of $\sim 1.6$~Gpc.

In summary, as also shown in \cite{bib:Hutsemekers2021}, the distribution of PAs is bimodal, with orientations tending to be either aligned or orthogonal. 
Interestingly, $\theta$ is remarkably similar to the $\sim 42^{\circ}, 131^{\circ}$ for radio QSO polarizations \cite{bib:Pelgrims2015} in regions coinciding with the LQG sample. The LQGs are most significantly aligned or orthogonal across $\sim 30^{\circ}$, which is equivalent to $\sim 1.6$~Gpc at the present epoch. Finally, no such aligned/orthogonal effects were found in mock LQG samples from simulations \cite{bib:Marinello2016, bib:Kim2011}. The results could imply a cellular structure to the Universe or potential challenges to the CP.  

\vspace{-1ex}
\subsection{Dipole in the distribution of galaxy spin directions}
\label{sec:galaxy_spin_dipole}

The spin direction of a spiral galaxy is a visual cue that depends on the perspective of the observer. A galaxy that spins clockwise when observed from Earth may appear as a galaxy that spins counterclockwise to an observer located in a different place in the Universe. Therefore, in a homogeneous and isotropic universe, the number of galaxies that spin in a certain direction is expected to be the same, within statistical error, as the number of galaxies that spin in the opposite direction, regardless of the location of the observer. For instance, if a certain part of the sky seems to have a higher number of galaxies spinning clockwise, an observer in the other side of the Universe would see in that part of the sky a higher number of galaxies spinning counterclockwise. That would violate cosmological homogeneity, according to which all observers should see the same Universe regardless of their location in it.  

That, however, has been a matter of discussion in the past several decades, with numerous studies showing different results and opposite conclusions regarding the nature of the distribution of spin directions of spiral galaxies. The first observations suggesting that the number of galaxies spinning clockwise (`Z') can be different from the number of galaxies spinning counterclockwise (`S') were based on small sets of galaxies. One of the first studies that proposed such asymmetry was based on a small set of 264 Sab-Sc galaxies, separated into 145 galaxies that spin counterclockwise, and 119 galaxies that spin clockwise \cite{bib:MacGillivray1985}. A similar experiment with a larger catalog of more than $6 \times 10^3$ galaxies suggested that the distribution was random, with no preference for a certain spin direction \cite{bib:Iye1991}. These experiments were based on visual inspection of the galaxies, such that the spin directions of the galaxies were determined by the curve of the galaxy arms. Spiral galaxies with leading arms can exist, and a notable example for such galaxy is NGC~4622. However, galaxies with leading arms are rare, and in any case are expected to be distributed evenly between clockwise and counterclockwise galaxies. Therefore, the curves of the arms provide a reliable indication of the spin direction of the galaxy.  

The deployment of autonomous digital sky surveys enabled the collection of far larger catalogs of galaxy images that were not available in the pre-information era. The first digital sky survey that provided data for such analysis was SDSS. While SDSS provided images of a high number of galaxies, the bottleneck for using that image data was the need to annotate the large number of galaxies by their spin direction. An experiment with $\sim1.5 \times 10^4$ SDSS galaxies annotated by five undergraduate students showed that galaxies spinning in opposite directions form a statistically significant dipole axis \citep{bib:Longo2011}. Another related study found links between spin directions of galaxies that are too far from each other to have gravitational interactions, and defined these links as ``mysterious" \citep{bib:Lee2019}. An analysis of $\sim1.3 \times 10^4$ SDSS galaxies also suggested a non-random distribution \citep{bib:Shamir2016}. On the other hand, an attempt to annotate the galaxies by using a large number of volunteers who annotated the data through a web-based user interface showed that the distribution of the galaxies was random \citep{bib:Land2008}. One downside of that experiment was that the annotation was heavily biased by human perception. Since the galaxy images were not mirrored randomly to correct for that bias, the resulting annotations showed a very strong asymmetry as high as $\sim$15\%. When that bias was noticed, a small subset of galaxies was annotated again such that the images were mirrored. That process showed an excessive number of galaxies that spin counterclockwise, but due to the relatively small number of galaxies that were mirrored, the difference was determined to be statistically insignificant. 

Manual annotations of galaxies by their spin directions introduce two major obstacles: The first is that the throughput of manual annotation is limited, and therefore cannot provide an effective solution to the analysis of very large databases of galaxy images. The other downside is that manual annotation can be systematically biased, and the complex nature of human perception makes it very difficult to correct. These limitations were met with the use of automatic annotation of the galaxies. While pattern recognition algorithms have been shown to be subjected to bias for the purpose of annotation of galaxy images \citep{bib:Dhar2022}, model-driven algorithms showed nearly symmetric \citep{bib:Davis2014,bib:Hayes2017} and fully symmetric \citep{bib:Shamir2011} ability to annotate the spin direction of a galaxy automatically. 

The ability to annotate galaxies automatically allowed analyses based on far larger catalogs that are not subjected to the bias of the human perception. Such analysis showed clear statistical signal of non-random distribution and dipole axis alignment in data taken from SDSS \citep{bib:Shamir2012}, Pan-STARRS \citep{bib:Shamir2020}, Hubble Space Telescope \citep{bib:Shamir2020b}, DES \citep{bib:Shamir2022}, and DESI Legacy Survey \citep{bib:Shamir2021}; all data sets show very similar profiles of asymmetry, and dipole axes well-within $1\, \sigma$ statistical error from each other \citep{bib:Shamir2022b}. The axes were observed in data collected from telescopes located in the Northern hemisphere \citep{bib:Shamir2020}, and nearly identical patterns were observed when using data collected by telescopes in the Southern \citep{bib:Shamir2021} hemispheres. 

The profile of asymmetry was consistent also when the data sets had no overlap in the galaxies. For instance, a data set of SDSS galaxies and a data set of Pan-STARRS galaxies provided nearly identical locations of the dipole axis even when the two data sets are completely orthogonal \citep{bib:Shamir2020}. A related experiment separated the galaxies by redshift ranges, showing that galaxies in none-overlapping redshift ranges exhibit a consistent profile of asymmetry \citep{bib:Shamir2022}.

On the other hand, other studies that used automatic analysis suggested that the distribution of the spin direction of galaxies is random \citep{bib:Hayes2017,bib:Iye2021}. Applying a nearly symmetric galaxy image annotation algorithm \citep{bib:Davis2014} to SDSS galaxies led to the conclusion that the distribution of the spin directions of galaxies was random \citep{bib:Hayes2017}. It should be mentioned, however, that applying the algorithm to the data annotated by Galaxy Zoo without prior assumptions led to statistical signal of $2.52 \, \sigma$ or stronger, which is not considered random. Only after selecting the galaxies by applying machine learning and explicitly removing the attributes that were reported in \citep{bib:Shamir2016} as attributes that correlate with the asymmetry, the signal became insignificant. Another study suggested that the non-random distribution is the result of duplicate objects in the data set \citep{bib:Iye2021}. As explained in \citep{bib:shamir2022e,bib:Shamir2021,bib:Shamir2022c}, the catalog used in \citep{bib:Iye2021} was a catalog of photometric objects \citep{bib:Shamir2017}, and was therefore used for photometric analysis rather than analysis of the distribution of galaxies by their spin direction. The method used in \citep{bib:Iye2021} was three-dimensional, where the distance of the galaxy was determined by its redshift. Since the galaxies in \citep{bib:Shamir2017} did not have spectra, the photometric redshift was used. The inaccuracy of the photometric redshift is far greater than the asymmetry signal, and therefore the lower statistical signal can be expected \citep{bib:shamir2022e,bib:Shamir2021,bib:Shamir2022c}. When analyzing the same catalog without using the photometric redshift, and without limiting the redshift to nearby galaxies, the distribution is non-random and statistically significant \citep{bib:shamir2022e,bib:Shamir2022c}. Access to the exact same data used in the experiments and instructions and code \citep{bib:shamir2022d} for reproducing the results are provided in \url{https://people.cs.ksu.edu/~lshamir/data/iye_et_al}.

\begin{figure}  
\includegraphics[width=10.0cm]{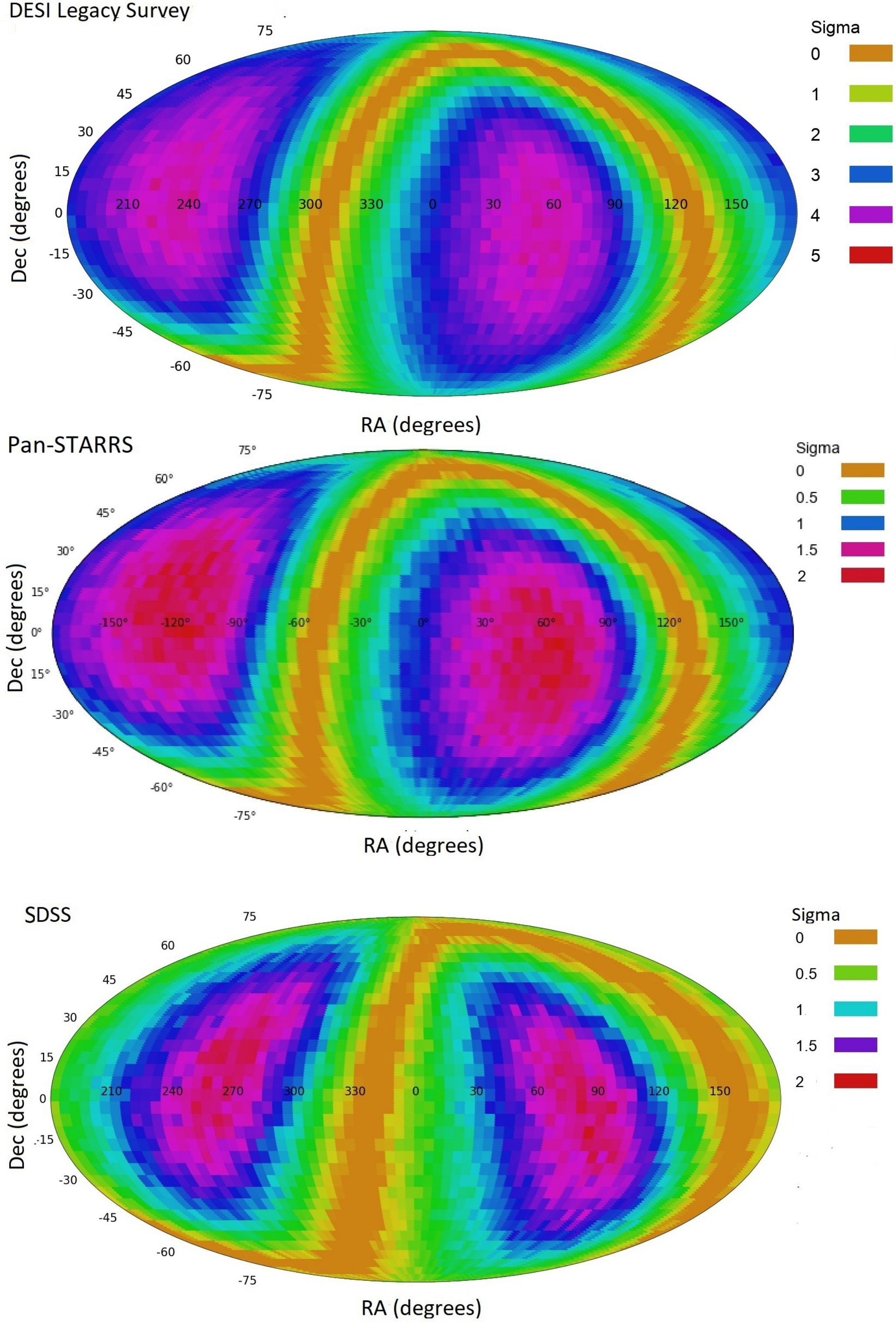}
\caption{Statistical strength of a dipole axis in galaxies imaged by Pan-STARRS, SDSS, and DECam \bcomm{(reproduced from Fig. 8 of \cite{bib:Shamir2022b})}. The location of the most likely axis agrees across different sky surveys, and the most likely axis peaks at close proximity to the CMB Cold Spot. The largest DESI Legacy Survey data set of $\sim8 \times 10^5$ galaxies provides the strongest signal of 4.7$\sigma$ to occur by chance.}
\label{fig:decam_sdss_panstarrs_normalized}
\end{figure}

Currently, the most recent analyses of the distribution of galaxy spin directions suggest that the asymmetry exhibits a possible dipole and quadrupole alignment \citep{bib:Shamir2020,bib:Shamir2021,bib:Shamir2021}. Analysis of $\sim 8 \times 10^5$ galaxies acquired by the DESI Legacy Survey showed a most likely dipole axis that peaks at very close proximity to the CMB Cold Spot \citep{bib:Shamir2021}, which agrees with an analysis of $\sim10^6$ galaxies from four different telescopes \citep{bib:Shamir2022c}. Fig.~\ref{fig:decam_sdss_panstarrs_normalized}, taken from \citep{bib:Shamir2022b}, shows the statistical strength of a dipole axis in galaxy spin directions. The SDSS galaxies were normalized such that their redshift distribution is similar to the redshift distribution of the DESI Legacy Survey galaxies. The figure shows agreement in the most likely dipole axis.

Analyses of galaxies separated by their redshift ranges showed that the asymmetry becomes stronger as the redshift gets higher \citep{bib:Shamir2020,bib:Shamir2020b,bib:Shamir2022b,bib:Shamir2022}. Another observation is that the position of the most likely axis changes consistently with the redshift \citep{bib:Shamir2020,bib:Shamir2020b,bib:Shamir2022b,bib:Shamir2022}, and the change gets smaller at higher redshifts. This provided an indication of an axis that is not Earth-centered. An analysis of $\sim9 \times 10^4$ galaxies with spectra showed an axis that peaks at roughly 300~Mpc from Earth \citep{bib:Shamir2022}.


\section{Tests of the \protect\bcomm{FLRW metric}} 
\label{sec:test_cosmological}

Ever since the discovery of the accelerated expansion of the Universe \cite{bib:Riess1999, bib:Perlmutter1999}, the introduction of the cosmological constant $\Lambda$ has been deeply troubling from a theoretical perspective. It should come as no surprise that the Cosmological Constant Problem \cite{bib:Weinberg1989} remains unsolved. Moreover, in recent years we have seen that local $H_0$ determinations are universally biased higher than Planck-$\Lambda$CDM \cite{bib:DiValentino2021, bib:Abdalla2022}, which ostensibly is at odds with common wisdom that $\Lambda$ is a placeholder for a dark energy EFT \cite{bib:Heisenberg2022, bib:Heisenberg2022b, bib:Lee2022, bib:Vagnozzi2020}. 
When these observations are taken together, one starts to see a physical case for replacing $\Lambda$ with a radial inhomogeneity, which has no problem fitting any Hubble diagram \cite{bib:Celerier2007, bib:Mustapha1997, bib:Ishak2007}. That being said, any inhomogeneous cosmology would need to compete with flat $\Lambda$CDM from a Bayesian perspective. 

The standard model of cosmology incorporates inhomogeneities as perturbative corrections to the FLRW model, and their nonlinear evolution via Newtonian $N$-body simulations. There are two broad approaches to testing the consistency of this paradigm. One approach is to look at consistency of predictions relating to the inhomogeneities: the details of the anisotropy of galaxy number counts, the convergence of bulk flows, the amplitude of the ISW effect, local curvature perturbations \cite{bib:Tian2021}, as discussed in section~\ref{sec:late_universe_FLRW_anomalies}. 

An alternative approach is to directly test the average FLRW evolution itself. In particular, regardless of the \bcomm{metric} theory of gravity involved, the FLRW metric incorporates a spatially constant curvature, which can be rewritten in terms of observable quantities that depend on redshift, resulting in a variety of possible null tests, see section~\ref{sec:terminology}. We begin with such tests in section~\ref{sec:omk_tests}.

\subsection{Curvature}
\label{sec:omk_tests}

In order to distinguish inhomogeneous cosmologies from homogeneous dark energy models, Clarkson, Bassett, and Lu (CBL) introduced a model independent test \cite{bib:Clarkson2008}. The advantage of this approach is that it only assumes FLRW, but does not assume General Relativity, so it can also be applied to modified gravity models. The basic idea is to start from the definition of the luminosity distance \eqref{eq:dL}, which is a geometric quantity and independent of the gravity model.
One can then recast (\ref{eq:dL}) as \cite{bib:Clarkson2007}, 
\begin{equation}
\Omega_{\text{k}} = \frac{[H(z) D^{\prime}(z)]^2-1}{[H_0 D(z)]^2} \;, 
\label{eq:omk}
\end{equation}
where \textit{prime} now denotes derivative with respect to redshift $z$. Since the FLRW curvature parameter is independent of redshift, one can now differentiate this expression to get 
\begin{equation}
\mathcal{C} (z) = 1 + H^2 (D D^{\prime \prime} - (D^{\prime})^2) + H H^{\prime} D D^{\prime} \;,
\label{eq:C}
\end{equation}
where for any FLRW model, $\mathcal{C} (z) = 0$ at all redshifts. Moreover, any determination of $\mathcal{C}(z) \neq 0$ for a given $z$ cannot be explained by changes in the dark energy model or deviations from General Relativity, so one is immediately led to conclude a change in the cosmological geometry. Arguably, this is the great strength of the CBL test.\footnote{It should be noted that \eqref{eq:C} relies both on isotropy and homogeneity. However, one can extend this test to anisotropic cases by replacing $H(z), D(z)$ in \eqref{eq:C} with expansion rate and luminosity distance in  a specific direction in the sky.}

Furthermore, spatial curvature does not in general obey a conservation law in general relativity \cite{bib:Buchert2008}. It can be shown that the Lema\^itre-Tolman-Bondi (LTB) models \cite{bib:Lemaitre1933,bib:Tolman1934,bib:Bondi1947} and models with backreaction of inhomogeneities \cite{bib:Larena2009,bib:Wiltshire2009,bib:Lavinto2013} generically lead to $\mathcal{C}(z) \neq 0$. Hence, for this test to be effective, it is imperative that one determines the Hubble parameter $H(z)$ independently of $D(z)$ \cite{bib:Shafieloo2009}. One can in principle exploit differential ages from passively evolving galaxies, i.~e.~cosmic chronometers \cite{bib:Jimenez2001, bib:Moresco2016}, to extract the Hubble parameter directly, $H(z) = - t^{\prime}(z)/(1+z) $. 
It is important to properly account for systematic uncertainties in these techniques, e. g. \cite{bib:Moresco2020, bib:Kjerrgren2021}, before this test becomes competitive.

Another route to separate $H(z)$ and $D(z)$ is to {\em independently} extract the radial and transverse BAO scales in galaxy clustering statistics, effectively a variant of the Alcock-Paczy\'nski test \cite{bib:Alcock1979}. This has the advantage of two separate measurements on the same survey; although the radial determination of $H(z)$ is subject to the systematics of redshift-space distortions, which must be carefully accounted for. Current methodology for such analysis often applies Fourier space techniques assuming a flat FLRW model. Thus to test the Friedmann equation it is imperative to revisit the analysis. Given any particular non-FLRW model, this is possible on a model-by-model basis, as has been demonstrated by the extraction of the BAO scale for the Timescape model in BOSS data \cite{bib:Heinesen2019}. Furthermore, a general model-independent approach has been developed and tested on a variety of toy models \cite{bib:Heinesen2020a}. 

\begin{figure}[htb]
\includegraphics[width=100mm]{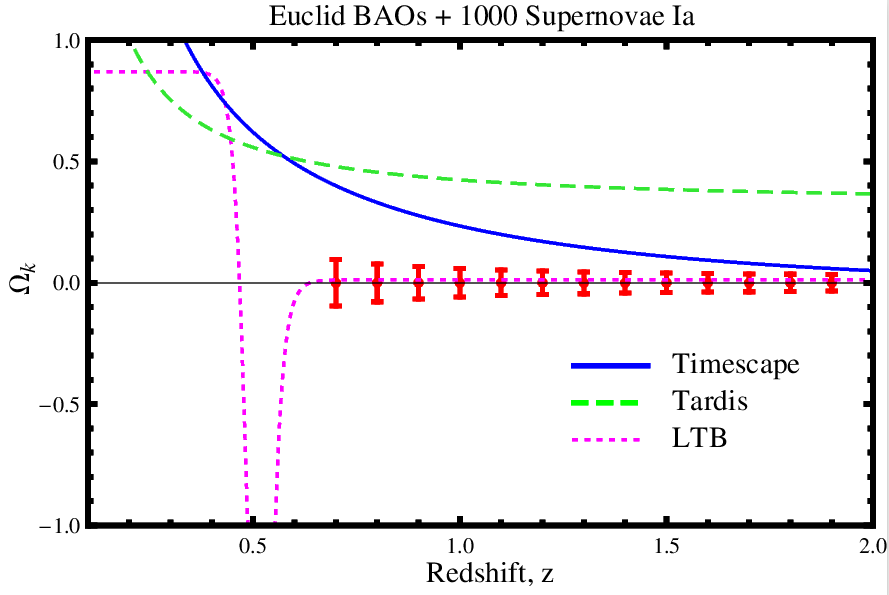}
\caption{Expected error bars for flat $\Lambda$CDM for the curvature {test statistic} $\Omega_{\text{k}}(z)$ (\ref{eq:omk}) \cite{bib:Clarkson2007} with Euclid {BAOs} combined with an additional 1000 SNe  \bcomm{(adapted 
from Fig.~10 of \cite{bib:Sapone2014})}. The 
purple dotted
line corresponds to the curvature parameter in a representative LTB model. The 
green dashed line indicates the Tardis model \cite{bib:Lavinto2013}. The 
blue solid line corresponds to the Timescape cosmology \cite{bib:Wiltshire2009, bib:Duley2013}.}
\label{fig:omk_test}
\end{figure}

One drawback of (\ref{eq:C}) is that determining and measuring second derivative $D^{\prime \prime}(z)$ is expected to be difficult. However, for practical purposes since $|\Omega_{\text{k}}|<0.01$ for concordance values, it is sufficient to compute $\Omega_{\text{k}}(z)$ directly, see, for instance, \cite{bib:Larena2009, bib:Wiltshire2009, bib:February2010, bib:Mortsell2011, bib:Lavinto2013, bib:Rana2017, bib:Wang2017, bib:Marra2018, bib:Marra2018, bib:Wei2018,bib:Rustagi2019, bib:Heinesen2020b, bib:Wang2021, bib:Zhang2022, bib:Nesserris2022}; or just the numerator of (\ref{eq:omk}), ${\cal B}:=[H(z) D^{\prime}(z)]^2-1$. While determination of ${\cal B}(z)$ might be viewed simply as a test for flatness \cite{bib:Yang2021}, FLRW models with $\Omega_{\text{k}}\ne0$ lead to ${\cal B}(z)$ functions which can be easily  distinguished from those of specific non-FLRW models \cite{bib:Wiltshire2009,bib:Lavinto2013}, while avoiding the problem that as $z\to0$ generically $D(z)\to 0$ in the denominator of (\ref{eq:omk}) at a rate that makes $\Omega_{\text{k}}(0)$ singular (c.f.\ Fig.~\ref{fig:omk_test}). 

An alternative test of FLRW was presented in \cite{bib:Rasanen2015} employing SN-based distances and strong gravitational lensing observations simultaneously. 
The approach exploits the distance sum rule \cite{bib:Peebles1993}, 
\begin{equation}
\label{eq:dls}
    \frac{d_\mathrm{ls}}{d_\mathrm{s}} = \sqrt{1 + \Omega_{\text{k}} d_{\text{l}}^2}- \frac{d_\mathrm{l}}{d_\mathrm{s}} \sqrt{1 +\Omega_{\text{k}} d_\mathrm{s}^2} \;,
\end{equation}
where once again $\Omega_{\text{k}}$ is the curvature parameter of an exact FLRW metric, $d_\mathrm{s}$ denotes the distance to the lensed source object, $d_\mathrm{l}$ denotes the distance between observer and gravitational lens, and $d_\mathrm{ls}$ is the distance between the lens and the source, as can be retrieved from modeling a lens after measuring the distances between multiple images coming from the background source.\footnote{These distances are dimensionless $d(z_1,z_2):=(1+z_2)H_0 d_\mathrm{A}(z_1, z_2)$ with angular diameter distance $d_\mathrm{A}$, $z_1 < z_2$.}
This expression, (\ref{eq:dls}), can be rewritten to extract $\Omega_{\text{k}}$,
\begin{equation}
\label{eq:sum_omk}
    \Omega_{\text{k}} = \frac{d_\mathrm{l}^4+d_\mathrm{s}^4+d_\mathrm{ls}^4-2 d_\mathrm{l}^2 d_\mathrm{s}^2 - 2 d_\mathrm{l}^2 d_\mathrm{ls}^2 - 2 d_\mathrm{s}^2 d_\mathrm{ls}^2}{4 d_\mathrm{l}^2 d_\mathrm{s}^2 d_\mathrm{ls}^2} \;. 
\end{equation}
The test then reduces to measuring the three distances, ($d_\mathrm{l}, d_\mathrm{s}, d_\mathrm{ls})$ for two different redshifts $(z_\mathrm{l}, z_\mathrm{s})$. If the Universe is described by an FLRW metric, then $\Omega_{\text{k}}$ is a constant. In contrast to (\ref{eq:omk}), the alternative consistency condition (\ref{eq:sum_omk}) does not involve any derivatives of distance. The condition (\ref{eq:sum_omk}) has been extensively studied \cite{bib:Liao2017, bib:Denissenya2018, bib:Qi2019, bib:Qi2019b, bib:Li2019, bib:Liao2019, bib:Collett2019, bib:Cao2019, bib:Wang2020, bib:Zhou2020, bib:Kumar2021, bib:Liu2020}. In the literature one finds close variants of these curvature tests, including the identification of invariants through Noether's theorem \cite{bib:Arjona2021} and a  comparison of proper and transverse comoving distances \cite{bib:Yu2016}.

As an advantage, the test is fully independent of $H_0$ and SN-based distances only rely on a minimum of assumptions (see \cite{bib:Wagner2019} for an overview of approaches and analyses of biases and uncertainties). Yet, as already noted in \cite{bib:Rasanen2015}, the largest source of biases and uncertainties comes from the strong gravitational lens modeling that is required to obtain $d_\mathrm{ls}$. 
This issue has become a source of criticism for time-domain cosmography to obtain $H_0$ \cite{bib:Kochanek2020, bib:Kochanek2021} and its impact on (\ref{eq:sum_omk}) also needs to be investigated. 
Formalism-intrinsic degeneracies as recently detailed in \cite{bib:Wagner2018} and \cite{bib:Wagner2019b} need to be taken into account to arrive at accurate values of $\Omega_{\text{k}}$.
But given that our understanding of the mass density profiles of strong lenses increases \cite{bib:Wagner2020}, prospects are looking promising to assemble a golden sample of suitable lenses to decrease the error bounds on $\Omega_{\text{k}}$ in (\ref{eq:sum_omk}). 

Throughout these curvature tests, the goal is to either rule out or confirm evolution in $\Omega_{\text{k}}$. This gives one a handle on cosmological models beyond FLRW. Euclid \cite{bib:Laureijs2009, bib:Laureijs2011} is expected to give tight constraints on $\Omega_{\text{k}}$ (see Fig.~\ref{fig:omk_test}, reproduced from \cite{bib:Sapone2014}).
While the results in the literature are more or less consistent with $\Omega_{\text{k}} = 0$, one recent study using the distance-sum-rule test finds a preference for a closed universe \cite{bib:Li2018}.
While it may be tempting to join the dots with recent CMB results favoring $\Omega_{\text{k}} < 0$ \cite{bib:Planck2018b, bib:DiValentino2020}, if $\Omega_{\text{k}}(z)$ varies then one must carefully consider the redshifts at which particular estimates are made. In this light,
an intriguing observation \cite{bib:LHuillier2017} is that while being $2\sigma$ consistent with flat $\Lambda$CDM, $\Omega_{\text{k}}(z)$ appears to differ in BOSS DR12 BAO data \cite{bib:Cuesta2016} between the LOWZ and CMASS samples at mean redshifts of $z=0.32$ and $z=0.57$ respectively. While systematic issues may remain, which must also be carefully revisited for non-FLRW models \cite{bib:Heinesen2020a}, we can expect considerable advances in precision for tests of emerging spatial curvature in the coming decade. 

\subsection{Distance Duality Relation} 
\label{sec:distance_duality_relation}

In the aftermath of the discovery of the BAO feature \cite{bib:Eisenstein2005}, it was noted that one can test the relation between luminosity distance $d_{L}$ and angular diameter distance $d_\mathrm{A}$, $d_\mathrm{L} = (1+z)^2 d_\mathrm{A}$, \eqref{dL-dA}  Distance Duality Relation (DDR) or Reciprocity Relation \cite{bib:Etherington1933,bib:Avgoustidis2009, bib:Maartens2011}. This relation is expected to hold provided i) photons travel on null geodesics of the spacetime which is described by a pseudo-Riemannian geometry, as required by the equivalence principle, and ii) photon number is conserved.\footnote{Thereby precluding the decay of photons to other particles including axions. Such couplings have been proposed as a means to explain the late-time accelerated expansion \cite{bib:Csaki2002, bib:Csaki2002b, bib:Bassett2004} and breakdown in the DDR \cite{bib:Bassett2004b, bib:Burrage2008}.}$^,$\footnote{{
For non-FLRW cosmologies, in general, the $d_\text{A}/d_\text{L}$ ratio is a function of $z$ as well as angle on the sky. This ratio is a geometric one which is fixed by the geometry and is independent of the gravity theory as long as photons follow a null geodesic path.} } 

To test the validity of the DDR, recall that BAO constrain the following length scales, 
\begin{equation}
L_{\parallel} = \frac{r_\mathrm{s}}{D(z)}, \quad L_{\perp} = \frac{r_\mathrm{s}}{(1+z) d_\mathrm{A}(z)} \;, 
\end{equation} 
where $r_\mathrm{s}$ is the radius of the sound horizon and $D(z)$ is the comoving distance defined in \eqref{eq:dL}, for FLRW $D(z) = (1+z)^{-1} d_\mathrm{L} = (1+z) d_\mathrm{A}$. One can then define the quantity 
\begin{equation}
    \zeta(z) = 1 - \frac{L_{\parallel}}{L_{\perp}} \;,
\end{equation}
which, if $\zeta(z) \neq 0$ for some redshift, does not only rule out FLRW, but a much larger class of cosmological spacetimes on the assumption that photons follow null geodesics and  do not decay. See \cite{bib:Bengaly2022, bib:Renzi2022} for studies of the DDR. Note, the introduction of torsion is expected to lead to violations of both the curvature and DDR tests \cite{bib:Bolejko2020}. Within the context of cosmology, one way to do this is teleparallel gravity (see \cite{bib:Bahamonde2021} for an up-to-date review). A void or LTB model is also expected to lead to appreciable differences to the BAO feature \cite{bib:February2013},\footnote{See \cite{bib:Zumalacarregui2012} for an earlier study reporting stringent BAO constraints on LTB models in a geometric approximation that does not take into account the anisotropic growth of structure in void models.} which can be tested through future large scale surveys including SKAO and Euclid. When confronted with current data, LTB models have been largely unsuccessful in explaining Hubble tension \cite{bib:Kenworthy2019, bib:Camarena2022} and any contrast in mass density is well constrained \cite{bib:Camarena2021}.

\subsection{CMB} 
As highlighted by Goodman \cite{bib:Goodman1995}, one can probe inhomogeneities through observations of the Sunyaev-Zel’dovich effect in clusters of galaxies, excitation of low-energy atomic transitions, and the accurate thermal spectrum of the CMB. More concretely, galaxy clusters with their hot ionized intra-cluster gas, act via electron scattering of CMB photons like giant mirrors, allowing us to compare the CMB spectrum seen in the ``mirror" from the spectrum obtained along unobstructed lines of sight. The kinematic Sunyaev-Zeldovich (SZ) effect \cite{bib:Sunyaev1980} on the CMB temperature anisotropies probes the monopole (thermal SZ) and dipole (kinetic SZ) seen by the cluster in a perturbed Friedmann model. This method has been used to place constraints on void models providing a putative replacement for dark energy \cite{bib:Garcia2008, bib:Yoo2010, bib:Zhang2011, bib:Zibin2011} (see also section~\ref{sec:bulk_flows}). However, see \cite{bib:Clarkson2011} for a different perspective on these CMB constraints. 

A second approach involves observations of the CMB spectral distortions due to Compton scattering from inside the light cone \cite{bib:Goodman1995}. While the kSZ approach relies on the knowledge of the free electron perturbation power, the calculation of the Compton $y$ distortion requires only background information and is therefore more certain. Constraints on LTB void models using this approach have been placed in a number of papers \cite{bib:Caldwell2008, bib:Moss2011b, bib:Zibin2011, bib:Zibin2011b}.

\subsection{Other tests}
Generalizing the earlier ideas of \cite{bib:Sandage1962, bib:McVittie1962} and later \cite{bib:Loeb1998}, it was proposed that the cosmological drift of redshifts\footnote{To perform this test requires 10 to 20 years of observations of spectral lines in the Lyman-$\alpha$ forest at redshifts $2<z<5$, to observe changes at the order of $10^{-10}$ of the spectral frequencies over such periods. This is a challenge for the future, but is feasible with the CODEX spectrograph proposed for the European Extremely Large Telescope.} in non-FLRW spacetimes could be combined with distances to test FLRW \cite{bib:Lake2007,bib:Uzan2007, bib:Amendola2008, bib:Uzan2008,bib:Yoo2011,bib:Koksbang2016,bib:Koksbang:2019,bib:Heinesen2021}. Separately, it has been suggested that large scale deviations from isotropy could be explored through ``cosmic parallax" \cite{bib:Quercellini2009}, namely observations of the changes in angular separation for sources at cosmic scales.  
It has also been suggested that homogeneity can be probed by comparing the look-back time using radial BAO, and the time along galaxy world lines using stellar physics \cite{bib:Heavens2011}. Any significant deviation in these two numbers would signal a breakdown in homogeneity. Indeed, cosmic time has been exploited as means to assess the viability of void models \cite{bib:Lan2010}. Finally, the cosmic neutrino background has been proposed as a test of FLRW, but such a test may only be feasible in the (far) future \cite{bib:Jia2008}.

\newpage
\section{Epilogue} 
\label{sec:epilogue}

\begin{quote} 
\textit{…in a period of normal science, scientists tend to agree about what phenomena are relevant and what constitutes an explanation of these phenomena, about what problems are worth solving and what is a solution of a problem. Near the end of a period of normal science a crisis occurs - experiments give results that don't fit existing theories, or internal contradictions are discovered in these theories. There is alarm and confusion. Strange ideas fill the scientific literature. Eventually there is a revolution. Scientists become converted to a new way of looking at nature, resulting eventually in a new period of normal science. The ``paradigm" has shifted.} -- Steven Weinberg on Kuhnian shifts
\end{quote} 

The working assumption that the geometry of the Universe is described by the FLRW metric (\ref{eq:FLRW_metric}) represents the bedrock of modern cosmology. The origin of this assumption can be traced to the Einstein equations, where metric simplifications are a prerequisite for analytic solutions. Since one can extend the Einstein equations beyond the strict remit of Einstein GR through modified gravity theories, modern cosmology has outgrown GR in recent years, but not the FLRW assumption. Within the FLRW class, the flat $\Lambda$CDM model may be regarded as a minimal model.
While luminous matter and radiation clearly exist, and constraints for the properties of dark matter on large scales tighten \cite{bib:Pardo2020}, the cosmological constant $\Lambda$ is still a placeholder for missing physics, for instance, as some sort of dark energy. Although $\Lambda$ arguably has physics pedigree stretching back a century \cite{bib:ORaifeartaigh2017}, its theoretical limitations are clear; the cosmological constant problem \cite{bib:Weinberg1989} remains a serious outstanding issue. Moreover, it is traditionally assumed that $\Lambda$ can be replaced with a (scalar) dark energy effective field theory with equation of state $w_{\phi} > -1$.
This replacement drives $H_0$ to lower values, no matter which dark energy model is chosen with this constraint on the equation of state. 
However, in recent years the \textit{universal} biasing of locally measured values of $H_0$ \cite{bib:Huang2020, bib:deJaeger2020, bib:Pesce2020, bib:Kourkchi2020, bib:Schombert2020, bib:Khetan2021, bib:Blakeslee2021, bib:Freedman2021, bib:Riess2021, bib:Vagnozzi2020} to higher values than Planck-$\Lambda$CDM value \cite{bib:Planck2018b} has cast doubt on the viability of these dark energy models \cite{bib:Banerjee2021, bib:Heisenberg2022, bib:Heisenberg2022b, bib:Lee2022}. 
It is still possible that systematics are at play, but $\Lambda$ or a more general dark energy model, remain deeply challenging from a theoretical perspective.

The phenomenologically successful flat $\Lambda$CDM model is observationally supported by the CMB, BAO and SNe Ia. As we have discussed, the CMB is analyzed under the assumption that the Universe is isotropic. This entails the definition of our motion with respect to the CMB, and redshift corrections are required to place observables in the `CMB frame'. Some of the observations in section~\ref{sec:late_universe_FLRW_anomalies} already raise questions about the CMB frame being the global cosmic rest frame, but this idea can be tested with future surveys. BAO cosmological constraints assume not only isotropy, but also homogeneity and potentially a fiducial cosmology, i. e. flat $\Lambda$CDM. Nevertheless, a $\gtrsim 2 \, \sigma$ anomaly exists in the curvature parameter $\Omega_{\text{k}}$ between these two data sets \cite{bib:Planck2018b, bib:DiValentino2020}, so the CMB and BAO are no longer fully in agreement \cite{bib:Alam2021}. Finally, over the last decade, even when transferred into the CMB frame, SNe Ia have consistently returned hints of an excess anisotropy in the CMB dipole direction/hemisphere \cite{bib:Cooke2010, bib:Antoniou2010, bib:Li2013, bib:Javanmardi2015}, but at low statistical significance. 
Separately, in addition to discrepancies in $H_0$ \cite{bib:Huang2020, bib:deJaeger2020, bib:Pesce2020, bib:Kourkchi2020, bib:Schombert2020, bib:Khetan2021, bib:Blakeslee2021, bib:Freedman2021, bib:Riess2021} and $S_8$ \cite{bib:Asgari2021, bib:Heymans2021, bib:Amon2022, bib:Abbott2022}, there are a host of other well documented $\Lambda$CDM tensions \cite{bib:Perivolaropoulos2021}. While systematics may resolve a number of these anomalies, if some continue to persist, patching them within the FLRW framework by changing the flat $\Lambda$CDM model via new sectors beyond (dark) matter, dark energy and radiation and/or interactions among these sectors, may become an impossible task, see e.~g.~\cite{bib:Krishnan2021b, bib:Krishnan2022}. Provided anomalies persist, the FLRW assumption itself becomes a safer bet as the origin of the tensions. Physics reminds us time, and time again, that symmetries are typically broken unless some fundamental physical principle protects them.\footnote{Needless to say, symmetries that were historically assumed to solve the Einstein equations, and later assumed to make the best use of sparse cosmological data, are not fundamental from the perspective of physics.} Historically,\footnote{See \cite{bib:Ellis1999} for a systematic overview of models sorted by symmetry assumptions.} cosmology has moved beyond the Einstein static Universe \cite{bib:Einstein1917} and the Einstein-de Sitter Universe \cite{bib:Einstein1932}. 

Against this backdrop, this review is a timely exercise. We have documented observations that are at odds with the working assumption that the Universe is homogeneous and isotropic at large scales and have discussed systematics that could be at play. 
But there remains a rich set of findings based on different, uncorrelated probes which jointly suggest that the CP may already be violated in current observations. We finish by touching upon some open questions that arise from the evaluation of all these data sets and the results obtained. 
Detailed summaries of individual findings for the early and the late universe are contained in sections~\ref{sec:summary_early_universe} and \ref{sec:summary_late_universe}, respectively.\\

\noindent \textit{Does the CMB frame provide a cosmic rest frame?} 

The presumed statistical isotropy of the CMB supports the idea that the Universe, or more precisely the early Universe, is homogeneous and isotropic. 
One can always assume that the dipole in the CMB temperature anisotropies is due purely to the motion of the central observer and hence move to a `dipole-free-frame' (DFF) by a local Lorentz boost. The question of statistical isotropy then becomes a question of whether the higher multipoles of the CMB in this DFF are statistically isotropic. If yes, such a frame is called the CMB frame. The next question is whether all distributions of matter in the Universe are isotropic in the CMB frame, showing (statistical) spherical symmetry \cite{bib:Planck2013e, bib:Ferreira2021, bib:Saha2021, bib:Ferreira2021b}. In particular, as a necessary condition, one should ask: does the CMB DFF match the matter DFF? This latter question may be tested by studying the magnitude of the cosmic dipole in the heliocentric frame \cite{bib:Ellis1984}. Observational systematics aside, consistency demands that the CMB dipole direction is recovered -- otherwise FLRW would be badly broken. Alternatively, one can adopt a cosmological model, e.~g.~flat $\Lambda$CDM, and check that there are no variations in cosmological parameters across the sky within the assumed CMB frame. There are now enough independent existing anomalies in the late Universe, as documented in section~\ref{sec:late_universe_FLRW_anomalies}, that serious questions about the `CMB frame' being the global cosmic rest frame are raised. Systematics may be at play across a host of observables, but either way, further investigation is required to resolve the tensions.  \\

\noindent \textit{Does the CMB support FLRW at higher orders beyond the dipole?} 

Putting aside the question of whether matter and radiation cosmic dipoles agree, even if a common cosmic rest frame exists, there is a glaring need to better understand anomalies that now stretch back almost two decades. This raises the pressing, yet daunting question: \textit{do these residual asymmetries impact the inference of cosmological parameters within the flat $\Lambda$CDM model?} 
Independent studies, outlined in section~\ref{sec:CMB_variations}, suggest that this is the case \cite{bib:Fosalba2021, bib:Yeung2022} and intriguingly the emergent dipoles in cosmological parameters tend to align with hemispherical power asymmetry. This alignment is physically compelling. On the other hand, it is possible that such large variations are precluded \cite{bib:Mukherjee2016, bib:Mukherjee2018}, so further studies are warranted. Furthermore, given documented CMB anomalies from section~\ref{sec:early_universe_FLRW_anomalies}, it is no surprise Planck has confirmed that the CMB prefers a phenomenological Bianchi component, but the cosmological parameters are discordant with the current model \cite{bib:Jaffe2005, bib:Jaffe2006, bib:Planck2013b, bib:Planck2016, bib:Saadeh2016}. As a result, a simple \textit{homogeneous but anisotropic} model extending the $\Lambda$CDM model is strongly disfavored by Planck. 

The global anisotropic expansion of Bianchi universes is just one simple example of generic models with a non-kinematic differential expansion \cite{bib:Bolejko2016}, however. The intriguing planar alignment of quadrupole and octopole (section~\ref{sec:quadrupole_octopole}), and especially the plane normals with the CMB dipole, not only \bcomm{suggest a contradiction of} the isotropy assumption, but they may also point to some misinterpretation of the CMB dipole. Setting the anomalies aside, the CMB anisotropy spectrum appears to be consistent with the kinematic interpretation of the dipole \cite{bib:Planck2013e, bib:Ferreira2021, bib:Saha2021, bib:Ferreira2021b} \footnote{See \cite{bib:Domenech2022} for an analysis of superhorizon isocurvature modes and the possibility of a non-kinematic contribution to the CMB dipole.}.
However, recovery of the boost to the CMB frame via special relativistic modulation and aberration of the anisotropy spectrum is sensitive to the multipole range used. Imposing a high multipole cutoff, $\ell_{\rm max}$, one finds that the dipole direction moves across the sky from the hemispherical power asymmetry direction to its final direction as $\ell_{\rm max}$ is increased \cite{bib:Planck2013e}. This angular scale dependence may be a smoking gun of a non-kinematic contribution to the dipole. Further studies of models for non-kinematic anisotropies \cite{bib:Bolejko2016} in relation to both the large angle CMB multipoles and to the Ellis--Baldwin test \cite{bib:Ellis1984} could shed light on this important question.

In summary, the CMB is to first approximation consistent with FLRW, but loose ends remain.\footnote{See \cite{bib:Kashlinsky2022} for a recent analysis on a possible non-kinematic, primordial nature of the CMB dipole.} While a third satellite experiment, LiteBIRD \cite{bib:Matsumara2014}, is expected to throw further light on CMB anomalies, dedicated studies of the cosmic dipole in the late Universe through SKAO \cite{bib:Bacon2020} will also be illuminating. \\

\noindent \textit{Does precision cosmology based on FLRW give an accurate picture of the Universe?} 

Precision cosmology has the aim of determining the parameters of an FLRW cosmological model as tightly as possible by using an increasing amount of high-quality data. 
Yet, the FLRW assumption averages over the celestial sphere, suppressing any information about directional dependency potentially contained in these observables. 
Knowing that the Universe is not homogeneous and isotropic on all scales and that our observations cover multiple scales within it \cite{bib:Ellis1987}, one may ask whether the aim of percent-precision cosmology inferred from this data can yield an accurate representation of the Universe as well.

Let us start with an analogy.  
Recall that the CMB monopole was initially discovered serendipitously by Penzias and Wilson \cite{bib:Penzias1965}, before subsequent discoveries led to the dipole \cite{bib:Partridge1967,bib:Smoot1992, bib:Lineweaver1996} and the higher order multipoles through a series of satellite experiments with increasing measurement precision, including COBE \cite{bib:Bennett1996}, WMAP \cite{bib:Hinshaw2012} and Planck \cite{bib:Planck2018}, and complementary terrestrial experiments, such as the Atacama Cosmology Telescope (ACT) \cite{bib:Choi2020} and South Pole Telescope (SPT) \cite{bib:Dutcher2021}. 
In the same vein, there is no \textit{a priori} reason that forbids a Universe with a dipole, or even higher multipoles, in its expansion and our knowledge thereof is only a question of observational capabilities. 
Succinctly put, if we assume systematics like intrinsic source properties and its evolution over redshift can be controlled, any confirmed excess in the radio galaxy or QSO dipole relative to the CMB in the directions of \cite{bib:Singal2011, bib:Siewert2021, bib:Secrest2021, bib:Secrest2022}, would immediately imply a non-kinematic anisotropy, which, to leading order, would be seen as a dipole. 

It should be noted that we may already be at the limits of FLRW in a well defined sense. Despite Planck having determined $H_0 = (67.36 \pm 0.54)$~km/s/Mpc \cite{bib:Planck2018}, recalling the CMB anomalies, it is compelling that $H_0$ may vary between $H_0 = (64.4 \pm 1.3)$~km/s/Mpc and $H_0 =(70.1 \pm 1.4)$~km/s/Mpc when one analyzes the CMB subject to half-sky masks \cite{bib:Yeung2022}, as reported in section~\ref{sec:CMB_variations}. 
If true, this implies that Planck is averaging over $H_0$ values that are ostensibly discrepant along an axis aligned with the hemispherical power asymmetry. The relevant question now is, has Planck failed to allow for directional systematics? If so, would $\sim 10$\% rather than $\sim 1$\% precision be a better reflection of the current status of cosmology within the FLRW framework? Then, one could ask what is the status of the Hubble tension \cite{bib:Verde2019, bib:DiValentino2021, bib:Abdalla2022}? The key point here is that if there are underlying asymmetries, yet we impose FLRW, care must be taken to make sure that errors are sufficient to absorb variations attributable to differences in orientation. 
\\

\noindent \textit{Are there synergies in the FLRW anomalies?} 

It is easy to identify points of agreement. The anisotropies in cluster scaling relations \cite{bib:Migkas2020, bib:Migkas2021} can be explained in terms of a potential bulk flow with velocity $900$~km/s out to distances of $500$~Mpc, or $z \sim 0.1$, in the direction of the CMB dipole, as explained in \cite{bib:Migkas2021}. This observation is only credible if anomalous bulk flows exist that do not converge to the expectations of flat $\Lambda$CDM at lower redshifts. 
As is clear from Fig.~\ref{fig:bulk_flows_literature}, local bulk flows are typically consistent with $\Lambda$CDM predictions, but as the depths increase, this is no longer guaranteed. Indeed, recent results point to bulk flows at depths of 140$\,h^{-1}$~Mpc that are larger than expected \cite{bib:Howlett2022} (see also \cite{bib:Watkins2009, bib:Kashlinsky2008, bib:Lavaux2010, bib:Magoulas2016}) in a direction consistent with the Shapley supercluster. 
Independent observations supporting related anomalous bulk flows can also be found in the SNe literature \cite{bib:Colin2011, bib:Dai2011, bib:Turnbull2012, bib:Wiltshire2012} and these may be the basis of further claims of an anisotropy in the Hubble diagram \cite{bib:Cooke2010, bib:Antoniou2010, bib:Li2013, bib:Javanmardi2015}. Without these complementary and seemingly overlapping observations of larger than expected bulk flows, the anisotropies in scaling relations in galaxy clusters \cite{bib:Migkas2020, bib:Migkas2021} would be less compelling.
The cosmic dipoles inferred from WENSS, NVSS and SUMSS surveys return consistent excesses relative to the CMB expectation, as reviewed in \cite{bib:Siewert2021}. In addition, the values are more or less consistent with the CatWISE QSO dipole \cite{bib:Secrest2021}. WENSS and SUMSS are smaller surveys, which give rise to larger errors, but already the agreement between NVSS and CatWISE QSOs is suggestive. Moreover, as is clear from \cite{bib:Siewert2021}, the TGSS survey leads to a larger dipole, but there are indications that systematics may be at play \cite{bib:Dolfi2019, bib:Hurley2017, bib:Tiwari2019}. Recently, some of these discrepancies in amplitudes have been tackled \cite{bib:Secrest2022}, where it was shown that the amplitude of the NVSS/CatWISE dipole can be recovered from TGSS and MIRAGN through a careful treatment of systematics. Furthermore, one common feature of the late Universe is that cosmological probes, whether analyzed in the heliocentric frame, where a dipole is expected, or in the CMB frame, where one is not, are hinting at a putative anisotropy in the same direction, namely in the direction of the CMB dipole. Neglecting a host of systematics, the simplest physical interpretation of these results is a non-kinematic anisotropy in a direction consistent with the CMB dipole direction. Finally, angular variations in the flat $\Lambda$CDM parameters noted in the CMB \cite{bib:Fosalba2021, bib:Yeung2022} consistently return dipoles in the obvious direction, namely along the dipole picked out by the hemispherical power asymmetry \cite{bib:Eriksen2003, bib:Park2003, bib:Hansen2004, bib:Eriksen2007, bib:Hansen2009, bib:Planck2013, bib:Planck2016, bib:Planck2018c}. 
\\

\noindent \textit{Are there points of disagreement in FLRW anomalies?} 

There are a number of noticeable points of disagreement. While the dipoles from NVSS \cite{bib:Singal2011, bib:Rubart2013, bib:Tiwari2016, bib:Bengaly2018} and CatWISE QSOs \cite{bib:Secrest2021} show good agreement on an excess amplitude, the TGSS dipole is larger \cite{bib:Bengaly2018, bib:Singal2019}, but the survey appears anomalous \cite{bib:Dolfi2019, bib:Hurley2017, bib:Tiwari2019}. Moreover, even if there is an amplitude excess, one needs to understand why related studies report none and are consistent with CMB expectations for the cosmic dipole \cite{bib:Blake2002, bib:Darling2022}.\footnote{See appendix of \cite{bib:Secrest2022} where it is argued \cite{bib:Darling2022} combines two catalogues that are either in tension or inconsistent with the CMB dipole. The claim is that consistency in the joint sample may be a coincidence.} A further concern, as mentioned above, is that our motion inferred from different observables may not coincide with each other~\cite{bib:Rubart2013,bib:Tiwari2015,bib:Colin2017,bib:Bengaly2018,bib:Singal2019,bib:Siewert2021,bib:Singal2019b,bib:Secrest2021,bib:Singal2021b,bib:Singal2021,bib:Singal2022}, even though they all seem to move along the same direction.
Large genuine differences in dipole amplitudes are rather disconcerting since a genuine solar peculiar velocity cannot vary from one data set to another. If statistically significant, discordant dipoles exist, this might imply that the Baldwin-Ellis test \cite{bib:Ellis1984} is flawed as currently implemented, e.~g.~\cite{bib:Dalang2022}, or one may instead have to look for some other cause for the genesis of these dipoles, including that of the CMB. 
As regards SNe Ia in section~\ref{sec:SN_dipole}, the dipole inferred from the Pantheon data set exhibits a \textit{deficit} with respect to CMB expectations \cite{bib:Horstmann2021}, when the redshifts are corrected to host galaxy redshifts \cite{bib:Steinhardt2020}, but without these corrections, an \textit{excess} persists \cite{bib:Singal2021}. 
There is clear disagreement here, but this discrepancy can be resolved in the near future by revisiting the analysis with the newer Pantheon+ sample \cite{bib:Brout2022}. 
One interesting point of difference concerns the direction of the prevailing asymmetries between the early and late Universe. From the perspective of the CMB, where the CMB dipole is removed, the most prominent asymmetry is the hemispherical power asymmetry (section~\ref{sec:hemispherical_power_asymmetry}). As can be seen from Fig.~\ref{fig:aniso}, it does not point in the same direction as the CMB dipole, which appears to be the direction tracked by late Universe probes in section~\ref{sec:late_universe_FLRW_anomalies}. 
Here, one should draw a distinction between late Universe observables in the CMB and heliocentric frame. In the latter, the recovery of the CMB dipole is expected. Interestingly, the radio galaxy and QSO dipoles all recover the CMB dipole direction, $(l, b) = (264^{\circ}, 48^{\circ})$ within the errors, but it is possible that there is some inclination towards the hemispherical power asymmetry at $(l, b) = (218^{\circ}, -21^{\circ})$ (see Table~9 of \cite{bib:Siewert2021}). 
Regardless, despite the difference in the prevailing asymmetry between the early and late Universe, one has already removed the dipole from the CMB, so some difference is expected. 

\newpage
\noindent \textit{Are dedicated surveys warranted?} 

Yes, dedicated surveys are expected to provide immense value. As explained in section~\ref{sec:precision_spectroscopy_test}, erstwhile observations supporting both variations in the fine structure constant $\alpha$ and a related dipole have largely disappeared with dedicated observational results \cite{bib:Martins2017, bib:Murphy2017, bib:Milakovic2021, bib:Welsh2020, bib:Murphy2022}. 
In addition, a lot of observational data is collected for other purposes than cosmology, so further processing or cuts are necessary and it is possible that the resulting data sets are never of the quality required. To bring this point home, there is a factor of two discrepancy in the magnitude of QSO dipoles between the MIRAGN \cite{bib:Secrest2015} and the CatWISE samples \cite{bib:Secrest2021}. This difference may have already been explained by \cite{bib:Secrest2022}, but it is important to understand how the size and homogeneity of the sample impact results.
Therefore, dedicated surveys are warranted, especially for the CP, since it underpins all of modern cosmology. To that end, a sky survey at multiple frequencies, carried out with LOFAR \cite{bib:Shimwell2017} or the upcoming SKAO \cite{bib:Bacon2020} could be used for an independent investigation of the controversial dipole anisotropy with superior sensitivity. 
As shown in \cite{bib:Maartens2018}, the HI intensity mapping survey of SKA phase 1, which still suffers from foreground contamination (see \cite{bib:Bacon2020}), and later the billion-galaxy HI galaxy survey of SKA phase 2, which is ideal for this task, can further probe the dipole anisotropy. 
These surveys allow one to analyze the cosmic matter dipole dependent on redshift out to $z \sim 3$ to separate the properties of the sources and their evolution from those of the Universe. 
SKAO is expected to provide a large number of radio sources at sub-$\mu$Jy levels \cite{bib:Wilman2008,bib:Schwarz2015,bib:Bengaly2019b}.
However, going deeper than sub-mJy flux levels, a substantially increasing fraction of very different populations of radio sources, e.~g., nearby normal galaxies, star burst galaxies and even galactic sources \cite{bib:Windhorst2003,bib:Padovani2011,bib:Luchsinger2015} may enter the sample as well.
Therefore, SKAO surveys at a few mJy levels or above would be more fruitful to keep the sample as homogeneous as possible and comprise mostly powerful radio galaxies and QSOs.\footnote{This is particularly important for the method of \cite{bib:Ellis1984}, which does not account for source properties or their evolution over redshift.} An investigation of radio dipoles at different frequencies from these surveys might reveal whether the difference in dipoles from surveys at widely separated frequency bands like that seen in the NVSS and TGSS dipoles is genuinely wavelength dependent or the frequency dependence found in \cite{bib:Siewert2021} is merely an artifact of the shortages in data acquisition mentioned above. 
Evidently, the validity of the CP could be corroborated if multiple-frequency radio source surveys yield dipoles consistent with the CMB dipole, though it would still remain to be explained why the presently determined dipoles are inconsistent. 
Separately, in section~\ref{sec:alignments}, we have recorded a number of intriguing alignments that appear at odds with the CP. Once again it is worth stressing that these results rest upon archival observations and there is a need for follow up dedicated studies. 
\\

\noindent \textit{Is the Universe consistent with the Cosmological Principle?} 

As already outlined in section~\ref{sec:prologue}, in order to have computable and predictive cosmological modeling, it is natural to start with the CP as the working assumption. Yet, since the CP is \textit{not} a fundamental physical symmetry, but rather an assumption imposed when we interpret observations, there is nothing that prevents it being violated at a higher precision and/or when one does not assume the same behavior for a physical observable over the celestial sphere, i.~e.~not averaging over the sky. The results highlighted in this review suggest that we have already reached the level of precision where violations of the CP may show in results. 
Systematics may be at play across a host of observables, but as the number of observations grow, this possibility seems less likely. If the Universe is not FLRW, but we view it through the prism of FLRW, cosmological tensions are inevitable. Interestingly, a host of fascinating observational tensions exist, including the $H_0$ tension \cite{bib:Huang2020, bib:deJaeger2020, bib:Pesce2020, bib:Kourkchi2020, bib:Schombert2020, bib:Khetan2021, bib:Blakeslee2021, bib:Freedman2021, bib:Riess2021}, the $S_8$ tension \cite{bib:Asgari2021, bib:Heymans2021, bib:Amon2022, bib:Abbott2022}, potentially a curvature tension \cite{bib:Planck2018, bib:DiValentino2020}, and an $A_{ISW}$ tension \cite{bib:Kovacs2022, bib:Kovacs2022b}. 
Finally, even if the best quality data set underpinning modern cosmology, namely the CMB, is considered, and the results of \cite{bib:Fosalba2021} and \cite{bib:Yeung2022} are correct -- the well documented hemispherical power asymmetry suggests they may well be -- then one is looking at $\sim10$\% variations in cosmological parameters, most notably $H_0$, despite Planck errors at the $\sim 1$\% level. There is no immediate contradiction here. 
If one is content to do cosmology at the $\sim 10$\% error level, FLRW may be a valid approximation. Depending on how the anomalies in this review are resolved, the limitations of FLRW will either be evident or it may be possible to reconcile the ``precision cosmology'' program with FLRW to the percent level. 


\section*{Acknowledgements} 
We thank the participants of the Asia Pacific Center for Theoretical Physics (APCTP) workshop ``A Discussion on the Cosmological Principle", 25-28 October 2021, for extensive discussion on the topic of FLRW. In particular, we thank Stephen Appleby for hosting the workshop and laying the foundations for this review. We also thank Yashar Akrami, Fernando Atrio-Barandela, Thomas Buchert, David Camarena, Guillem Dom\`{e}nech, Pablo Fosalba, Enqrique Gazta\~{n}aga, Asta Heinesen, Istv\'an Horv\'ath, Alexander Kashlinsky, Michael J. Longo, Roy Maartens, Roya Mohayaee, Subir Sarkar, Dominik Schwarz, Nathan Secrest, Glenn Starkman, Christos Tsagas and Evgeny Zavarygin for correspondence, discussion and comments. 

\begin{itemize} 
\vspace{-1.5ex}
\item 
The work of CJAPM was funded by FEDER through COMPETE 2020 (POCI), and by Portuguese funds through FCT through project POCI-01-0145-FEDER-028987 and PTDC/FIS-AST/28987/2017; the author also acknowledges FCT and POCH/FSE (EC) support through Investigador FCT Contract 2021.01214.CEECIND/CP1658/CT0001.
\vspace{-0.75ex}
\item 
E\'OC was supported by the National Research Foundation of Korea grant funded by the Korea government (MSIT) (NRF-2020R1A2C1102899).
\vspace{-0.75ex}
\item 
MMShJ would like to acknowledge SarAmadan grant No. ISEF/M/400121.
\vspace{-0.75ex}
\item 
LY was supported by the National Research Foundation of Korea funded through the CQUeST of Sogang University (NRF-2020R1A6A1A03047877).
\vspace{-0.75ex}
\item PP wishes to acknowledge the support of ERC advanced grant ARThUs (grant no: 740021).
\vspace{-0.75ex}
\item DH acknowledges support from the F.R.S.–FNRS (Belgium) under grants IISN 4.4503.19 and PDR T.0116.21.
\vspace{-0.75ex}
\item WZ acknowledges support from the National Key R\&D Program of China Grant No. 2021YFC2203100, NSFC No. 11903030 and 11903033, the Fundamental Research Funds for the Central Universities under Grant No. WK2030000036 and WK3440000004.
\vspace{-0.75ex}
\item SJW is supported by the National Natural Science Foundation of China Grants No.~12105344 and the Science Research Grants from the China Manned Space Project with No. CMS-CSST-2021-B01.
\end{itemize}

\section*{Author contributions}

Writing of the manuscript was lead by E.~\'O Colg\'ain, M.~M.~Sheikh-Jabbari, and J.~Wagner with discussion and input from all the authors. 
Specific contributions include R.~G.~Clowes, A.~M.~Lopez, D.~L.~Wiltshire for section \ref{sec:Hom-scale}, P.~K.~Aluri, P.~Chingangbam, M.~Chu, J.~P.~Kochappan, P.~Pranav, S.~Yeung, P.~W.~Zhao for section \ref{sec:early_universe_FLRW_anomalies}, C.~J.~A.~P.~Martins, K.~Migkas, A.~K.~Singal, L.~Yin for \ref{sec:late_universe_FLRW_anomalies}, R.~G.~Clowes, D.~Hutsem\'ekers, A.~M.~Lopez, and L.~Shamir for section \ref{sec:alignments} and finally D.~L.~Wiltshire for section \ref{sec:test_cosmological}. 
All authors reviewed the final manuscript.

\section*{Data availability statement}
No new data were created or analysed in this study.

\bibliographystyle{cpstyle3}
\bibliography{refs}

\end{document}